# The Hubble Arp Galaxy Survey

Julianne J. Dalcanton [1,2] Meredith J. Durbin [3,2] and Benjamin F. Williams [2]

[1] Center for Computational Astrophysics, Flatiron Institute, 162 Fifth Ave, New York, NY 10010, USA
[2] University of Washington, Department of Astronomy, Box 351580, Seattle WA 98103, USA
[3] Department of Astronomy, University of California Berkeley, Berkeley, CA 94720, USA

## ABSTRACT

The typical galaxy in the local universe is expected to be in a self-regulated quasi-equilibrium, displaying a settled morphology that falls within the Hubble Sequence. The Arp and Arp-Madore catalogs are filled with striking examples of galaxies that defy these expectations, making them useful targets for studying the astrophysics that controls dramatic, but short-lived, episodes of disequilibrium that mark galaxies' evolution. In this paper, we greatly expand the available Hubble Space Telescope (HST) imaging of galaxies drawn from the Arp and Arp-Madore catalogs. We present new optical $F606W$ images and point-source photometry for 216 systems, whose sizes are well-matched to the Advanced Camera for Surveys' (ACS) wide field of view. Essentially none of the sample had been previously observed with Hubble. The resulting images display rich morphologies, revealing a variety of massive stars, HII regions, stellar clusters, dust lanes, tidal tails, backlit galaxies, and occasional chance superpositions. We provide a pedagogical guide for interpreting highly-resolved optical galaxy images, which has general application beyond this atlas. The atlas images also provide a superb starting point for more detailed studies with high-resolution imaging in other wavelengths, and spectroscopy to track kinematics and the interstellar medium (ISM). Areas of obvious scientific relevance include feedback and star formation in merging and interacting galaxies, resolved stellar populations at the extremes of stellar density, the properties of young massive stars and stellar clusters, the physics of the cold ISM and dust, and stellar and gas dynamics.

## 1. INTRODUCTION

The universe is a very odd place filled with strange and wonderful things. Few compendiums capture this truth like Arp's Atlas of Peculiar Galaxies (H. Arp 1966) – the end result of a systematic compilation of some of the weirdest well-resolved galaxies in the sky. Starting with Vorontsov-Velyaminov's (1959) list of 355 galaxies identified visually from the new Palomar Sky Survey plates (B. A. Vorontsov-Velyaminov 1959), Arp began photographing peculiar galaxies with the 48 and 200 inch telescopes on Palomar. The initial list was culled, and then supplemented with his colleagues contributions of odd galaxies they'd made note of through the years. The end result was a collection of 338 striking galactic systems in the northern hemisphere. It was subsequently extended to the southern hemisphere using even deeper data from the UK Schmidt Survey, resulting in the larger H. C. Arp & B. Madore (1987) "Catalog of Southern Peculiar Galaxies and Associations", containing close to 6,500 systems[4], supplanting some earlier, but smaller, collections by E. L. Agüero (1971) and (J. L. Sérsic 1974).

While the overall catalogs span systems as diverse as one-armed spirals, dramatic mergers with large tidal tails, and dusty ellipticals, they are all well-resolved, dramatic manifestations of important galactic phenomena, the vast majority of which are driven by interactions or other breakdowns of more familiar equilibria seen in typical galaxies. Their photometric properties are consistent with this interpretation, showing evidence for being far more affected by recent starbursts than typical galaxies (R. B. Larson & B. M. Tinsley 1978). In an era when an increasing number of models and observations argue for star formation in galaxies being naturally "self-regulating", galaxies that depart from the norm are of particular interest.

[4] For conciseness, we will use the term "Arp galaxy" to refer to systems drawn from either of the two catalogs.

Unsurprisingly, the dominance of interacting systems in the Arp atlases has long made Arp galaxies targets of intensive observing campaigns. The HST archive contains ∼4600 individual imaging exposures of targets from the original Arp atlas (roughly 2/3 taken with WFPC2, and the remainder split roughly evenly between ACS and WFC3), many of which have resulted in dramatic press release images and been used in 100's of scientific papers. These same galaxies will undoubtedly continue to be important targets in the JWST+ALMA era, both for their being superb local analogs of the interacting galaxies that will dominate high-redshift observations, and for JWST and ALMA's ability to peer through highly obscured star formation in the Arp galaxies themselves.

However, in spite of their popularity as HST targets, the observations of Arp galaxies are far from systematic or complete. Systems that were targeted early with HST became popular for follow-up observations, which in turn begat even more intensive study. In contrast, other systems have never been looked at with HST, in spite of appearing equally compelling in ground-based imaging. Before the work presented in this paper began, out of the 338 systems in the original Arp atlas, more than 60% had never been observed with HST. [5] The fraction of southern Arp & Madore systems that had been observed was even smaller.

The work presented here was designed to populate the archive with HST imaging of previously unobserved galaxies from the Arp atlases, and thus to establish a much larger database from which to draw promising candidates for targeted follow-up observations (e.g., JWST, ALMA, JVLA, integral field spectroscopy, etc), or in-depth analysis with simulations and/or with surveys at other wavelengths (e.g., LSST, Euclid, Roman, etc). The sample galaxies are also are natural complements to observations from SNAP-14840's similar program targeting more "typical" NGC galaxies.

Finally, these images have significant scientific value on their own, carried solely by the morphology revealed in high-resolution optical imaging. Interpreting such imaging is a fading art in the current era of survey astronomy. We try to remedy this situation by providing a highly pedagogical guide to interpreting these images, demonstrating how physical intuition for a given system can be developed solely from a careful inspection of galaxy morphology.

The structure of the paper is as follows. In Section 2 we present the observations and the description of atlas images and accompanying panchromatic imaging from the far-UV to the radio; the images are presented at full-resolution in an appendix (Appendix B) along with the corresponding multiwavelength images and information on all cataloged galaxies in the images (Appendix C); these images are also available as a single full-resolution atlas at DOI: 10.5281/zenodo.16778896. We also derive point-source photometry, which we use to characterize the stars and stellar clusters that are effective tracers of the on-going star formation which is pervasive in the imaged systems. In Section 3 we present a guide to interpreting the new HST images, whose morphology can reveal the star formation rate intensity (and/or degree of quiescence), gas content, and interaction history, without requiring additional observations. In Section 4, we provide collections of valuable subsamples that are likely of interest for follow-up analysis, including collections of particularly well-resolved galaxies (Section 4.1), low surface brightness galaxies (Section 4.2), systems with rings (Section 4.3) or strong dust features (Section 4.4), strongly asymmetric spirals (Section 4.5), Whirlpool Galaxy analogs (Section 4.6), and galaxies with shells (Section 4.7). We also use morphologies alone to construct a "Toomre sequence" of increasingly advanced merger stages (Section 4.8), which shows interesting evolution in central activity during the merger as revealed in the mid-infrared (MIR) and radio continuum. We use the observed correlations to also select subsamples based upon the strength of their central activity from starbursts and/or active galactic nuclei (AGN) (Section 4.9).

## 2. DATA

The systems in this survey were observed as part of SNAP-15446 (PI: Dalcanton), a "snapshot" program selected as part of HST's "gap-filler" call for proposals in 2017. These proposals were solicited to allow Hubble to observe during orbits that were otherwise unable to accommodate any standard scientific observations, typically due to short orbit durations after large slews. For most of HST's history, no observations were attempted during these orbits, but after a pilot program targeting NGC and IC galaxies (SNAP-14840; A. Bellini et al. 2017) demonstrated that useful imaging[6] could be obtained with a suitably large and flexible target list, a more general call was opened to the community. The call requested proposals for large numbers of targets distributed across the sky, which could be observed with single-band ACS/WFC imaging executed in Hubble's "SNAP" mode. The proposals also were asked to meet more stringent requirements than typical SNAP programs to maximize schedulability, including shorter orbits and no roll-angle requirements, with the understanding that the observations would be scheduled at the lowest priority.

[5] Of the ∼130 systems that had been observed when this program began, ∼75% of the images were of the much smaller subset of 23 galaxies with 40 or more images, the majority of which were observed in the earliest cycles of HST.

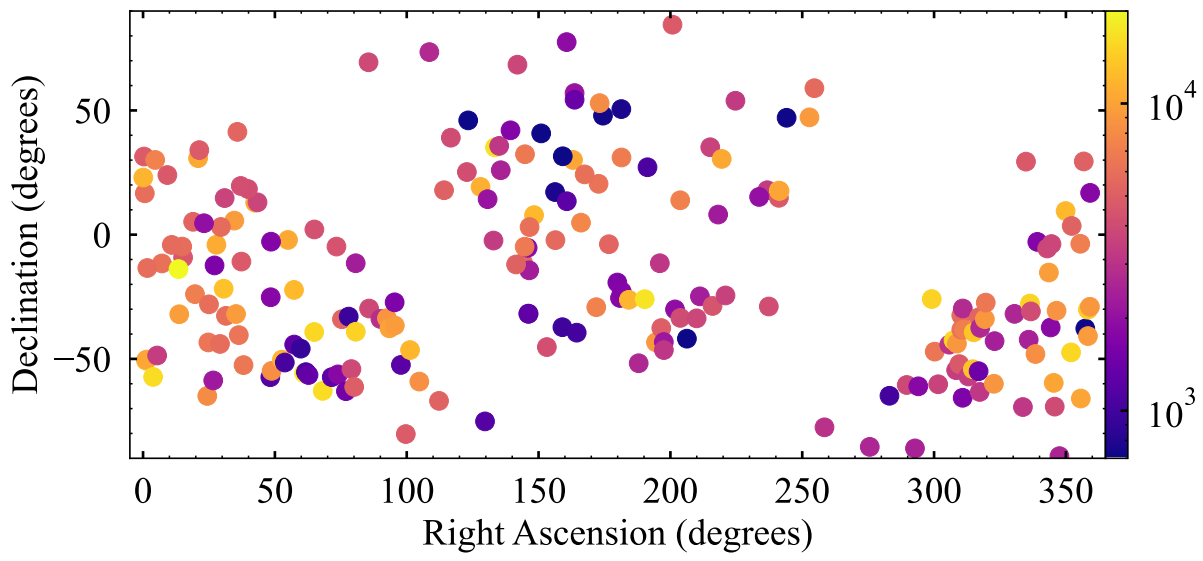

**Figure 1.** Distribution of observed systems on the sky (in J2000 epoch coordinates), color-coded by recessional velocity. The lack of observations at low Galactic latitude is a byproduct of the source catalogs, rather than target selection for the HST observations.

To meet the program's requirements, all observations for SNAP-15446 were identical short two pointing exposures, dithered to cover the ACS/WFC chip gap (ACS-WFC-DITHER-LINE pattern). Exposures were 390 seconds each, for a total exposure time of 780 seconds away from the chip gap. All observations were through the F606W filter, which offered the highest depth in bandpasses that were bluer than what would be available with JWST. This choice also made the program's observations closely analogous to those in the NGC/IC pilot program, allowing SNAP-14840 to serve as a baseline of more "normal" galaxies than those presented here.

We submitted 350 targets to the pool of potential SNAP observations, after several rounds of visual inspection of all candidates in the Arp and Arp-Madore catalogs that were well-matched to the field-of-view (FOV) of ACS/WFC. Within 3 bins of angular diameter ($0.5' - 1'$, $1' - 2'$, and $> 2'$), all coauthors independently ranked candidate targets in priority from 1 to 4. The lead author then inspected the HST archive to assess the quality and scientific utility of any existing HST imaging, and rejected any potential targets that had imaging of comparable depth in similar bandpasses that covered the majority of the most interesting features of the system. The final target list is weighted heavily towards systems with evidence for on-going or recent interactions (tails, shells, irregular or backlit dust features, unusual stellar morphology), with a bias towards closer systems that would be most highly resolved and thus particularly suitable for follow-up. Our selection also included some galaxies that were extremely low surface brightness and nearby, which were likely to be resolved into stars with HST, making them potentially useful targets for understanding the extremes of star formation in galaxies that are less likely to be self-regulated, steady state disks.

[6] We note that the precursor HST imaging of NGC 4993, which contains the first localized gravitational wave source GW170817, was obtained by this program in April 2017, only 4 months before the gravitational wave detection.

**Table 1**. Observational Data

| Target Name | Target RA | Target Dec | Aperture | PA | Start Time | Exp. Time | Flag | ID | $V_r$ | Category |
|---|---|---|---|---|---|---|---|---|---|---|
| | J2000.0 | J2000.0 | | (°) | | sec | | | km s$^{-1}$ | (original) |
| (1) | (2) | (3) | (4) | (5) | (6) | (7) | (8) | (9) | (10) | (11) |
| ARP2 | 16:16:18.203 | +47:02:45.14 | WFC | 60.31 | 2018-10-01 01:17 | 780.00 | NORMAL | UGC 10310 | 712 | A:a |
| ARP3 | 22:36:36.571 | -02:54:23.44 | WFC | -144.88 | 2020-08-23 07:54 | 780.00 | NORMAL | MCG-01-57-016 | 1694 | A:a |
| ARP4 | 01:48:27.593 | -12:22:50.65 | WFC | -61.35 | 2020-10-02 06:06 | 780.00 | NORMAL | MCG-02-05-050 | 1614 | A:a |
| ARP6 | 08:13:16.524 | +45:59:30.60 | WFC | -90.65 | 2018-11-14 09:17 | 780.00 | NORMAL | NGC 2537 | 447 | A:a |
| ARP10 | 02:18:27.297 | +05:39:17.34 | WFC | -110.44 | 2023-08-06 23:14 | 780.00 | NORMAL | UGC 1775 | 9116 | A:b |
| ARP15 | 22:51:36.368 | -05:33:36.61 | WFC | 73.15 | 2021-09-15 16:29 | 780.00 | NORMAL | NGC 7393 | 3744 | A:c |
| ARP18 | 12:05:34.175 | +50:32:21.67 | WFC | -20.35 | 2021-10-07 20:03 | 780.00 | NORMAL | NGC 4088 | 749 | A:c |
| ARP20 | 04:19:53.293 | +02:05:41.99 | WFC1 | -92.91 | 2021-09-17 11:49 | 780.00 | NORMAL | UGC 3014 | 4209 | A:d |
| ARP22 | 11:59:30.600 | -19:15:47.88 | WFC | -79.82 | 2021-11-24 23:01 | 780.00 | NORMAL | NGC 4027 | 1662 | A:e |
| ARP24 | 10:54:37.698 | +56:58:53.04 | WFC | -75.61 | 2019-11-26 23:31 | 780.00 | NORMAL | APG 24 | 2053 | A:e |
| ARP44 | 10:25:34.323 | -02:12:50.91 | WFC | 80.19 | 2021-03-19 00:29 | 780.00 | NORMAL | IC 609 | 5505 | B:a |
| ARP49 | 14:32:26.316 | +08:05:17.73 | WFC | -167.58 | 2023-04-25 11:06 | 780.00 | NORMAL | NGC 5665 | 2218 | B:b |
| ARP58 | 08:31:58.481 | +19:12:31.59 | WFC1 | 101.77 | 2021-03-13 14:05 | 780.00 | NORMAL | UGC 4457 | 10954 | B:b |
| ARP59 | 01:00:45.941 | -09:11:07.52 | WFC1 | -3.62 | 2019-10-14 04:52 | 780.00 | NORMAL | NGC 341 | 4549 | B:b |
| ARP69 | 14:20:27.384 | +35:10:49.09 | WFC | 134.50 | 2023-06-12 10:08 | 780.00 | NORMAL | NGC 5579 | 3573 | B:b |
| ARP70 | 01:23:29.705 | +30:47:14.48 | WFC | -113.50 | 2019-07-17 16:28 | 780.00 | NORMAL | UGC 934 | 10494 | B:b |
| ARP72 | 15:46:56.865 | +17:53:09.42 | WFC | -57.95 | 2022-01-09 13:34 | 780.00 | NORMAL | UGC 10033 | 3306 | B:b |
| ARP75 | 01:51:18.265 | -04:03:37.95 | WFC | 29.27 | 2021-11-11 14:41 | 780.00 | NORMAL | NGC 702 | 10668 | B:b |
| ARP82 | 08:11:14.529 | +25:12:01.11 | WFC | 100.75 | 2019-03-30 14:12 | 780.00 | NORMAL | APG 82 | 4067 | B:c |
| ARP86 | 23:47:03.609 | +29:28:32.99 | WFC | 89.55 | 2019-12-01 11:05 | 780.00 | NORMAL | APG 86 | 4996 | B:c |
| ARP89 | 08:42:41.786 | +14:17:02.78 | WFC | 101.07 | 2019-04-07 20:51 | 780.00 | NORMAL | NGC 2648 | 2064 | B:c |
| ARP91 | 15:34:34.344 | +15:11:06.44 | WFC | 168.43 | 2020-05-22 02:23 | 780.00 | NORMAL | APG 91 | 1980 | B:c |
| ARP97 | 12:05:48.613 | +31:03:39.80 | WFC | 120.92 | 2021-05-26 00:36 | 780.00 | NORMAL | UGC 7085 A | 6830 | B:d |
| ARP100 | 00:28:36.727 | -11:34:50.71 | WFC | -61.77 | 2019-09-14 14:32 | 780.00 | NORMAL | IC 18 | 6071 | B:d |
| ARP101 | 16:04:33.469 | +14:49:48.18 | WFC | 168.55 | 2020-05-16 22:33 | 780.00 | NORMAL | APG 101 | 4599 | B:d |
| ARP107 | 10:52:16.299 | +30:04:04.91 | WFC | -147.17 | 2023-02-17 15:42 | 780.00 | NORMAL | UGC 5984 | 10167 | C:a |
| ARP110 | 22:54:09.000 | -15:14:12.12 | WFC1 | -117.65 | 2023-06-04 23:14 | 780.00 | NORMAL | APG 110 | 9384 | C:b |
| ARP112 | 00:01:29.897 | +31:26:29.19 | WFC | 70.85 | 2019-12-08 13:03 | 780.00 | NORMAL | NGC 7806 | 4789 | C:b |
| ARP113 | 00:18:22.709 | +30:04:09.06 | WFC | -137.95 | 2021-08-15 19:05 | 780.00 | NORMAL | NGC 70 | 7201 | C:c |
| ARP121 | 00:59:23.670 | -04:48:11.37 | WFC | -112.01 | 2023-07-23 23:16 | 780.00 | NORMAL | APG 121 | 5715 | C:c |
| ARP122 | 16:04:26.930 | +17:44:58.99 | WFC1 | -90.32 | 2023-03-04 08:36 | 780.00 | NORMAL | NGC 6039 | 12082 | C:c |
| ARP123 | 05:22:34.708 | -11:30:15.55 | WFC | -93.10 | 2023-09-19 03:26 | 780.00 | NORMAL | NGC 1888 | 2300 | C:c |
| ARP126 | 01:58:06.773 | +03:05:24.30 | WFC-FIX | -120.97 | 2023-06-24 19:09 | 780.00 | NORMAL | APG 126 | 5686 | C:c |
| ARP129 | 09:39:25.275 | +32:21:52.92 | WFC1 | 156.98 | 2021-02-22 07:55 | 780.00 | NORMAL | UGC 5146 | 6511 | C:c |
| ARP130 | 00:02:37.730 | +16:38:50.61 | WFC1 | -106.45 | 2019-05-29 20:02 | 780.00 | NORMAL | IC 5378 | 5832 | C:c |
| ARP136 | 14:58:40.505 | +53:53:44.15 | WFC | -177.71 | 2020-05-02 17:24 | 780.00 | NORMAL | NGC 5820 | 3251 | C:d |

**Table 1** *continued*

**Table 1** *(continued)*

| Target Name | Target RA | Target Dec | Aperture | PA | Start Time | Exp. Time | Flag | ID | $V_r$ | Category |
|---|---|---|---|---|---|---|---|---|---|---|
| | J2000.0 | J2000.0 | | (°) | | sec | | | km s$^{-1}$ | (original) |
| (1) | (2) | (3) | (4) | (5) | (6) | (7) | (8) | (9) | (10) | (11) |
| ARP141 | 07:14:19.612 | +73:28:19.32 | WFC | 85.12 | 2019-04-16 14:12 | 780.00 | NORMAL | UGC 3730 | 2684 | C:e |
| ARP143 | 07:46:54.728 | +39:01:18.28 | WFC | 95.78 | 2020-04-19 23:14 | 780.00 | NORMAL | NGC 2444 | 3955 | C:e |
| ARP144 | 00:06:27.176 | -13:25:16.69 | WFC | 32.63 | 2021-10-10 02:40 | 780.00 | NORMAL | NGC 7828 | 5729 | C:e |
| ARP145 | 02:23:08.630 | +41:22:01.30 | WFC | 47.50 | 2021-03-17 00:19 | 780.00 | NORMAL | UGC 1840 | 5326 | C:e |
| ARP150 | 23:19:31.947 | +09:30:15.64 | WFC | -134.45 | 2020-09-03 02:54 | 780.00 | NORMAL | NGC 7609 | 11888 | D:b |
| ARP156 | 10:42:37.800 | +77:29:42.00 | WFC1 | 35.15 | 2018-07-27 06:14 | 780.00 | NORMAL | UGC 5814 | 10778 | D:c |
| ARP158 | 01:25:21.341 | +34:01:32.35 | WFC | 111.00 | 2019-11-15 18:41 | 780.00 | NORMAL | NGC 523 | 4756 | D:c |
| ARP163 | 12:45:16.800 | +27:07:30.00 | WFC1 | 108.28 | 2018-07-02 05:07 | 780.00 | NORMAL | NGC 4670 | 1072 | D:d |
| ARP164 | 01:15:57.647 | +05:10:42.38 | WFC | 63.33 | 2018-12-04 02:36 | 780.00 | NORMAL | NGC 455 | 5231 | D:d |
| ARP165 | 07:36:38.571 | +17:53:04.37 | WFC | -88.45 | 2020-09-14 00:21 | 780.00 | NORMAL | NGC 2418 | 5043 | D:d |
| ARP172 | 16:05:34.574 | +17:35:50.75 | WFC | -65.25 | 2021-01-25 08:45 | 780.00 | NORMAL | APG 172 | 10043 | D:e |
| ARP176 | 13:03:56.722 | -11:29:49.26 | WFC | -58.55 | 2019-03-20 00:16 | 780.00 | NORMAL | MCG-02-33-101 | 3107 | D:f |
| ARP180 | 04:53:24.545 | -04:47:32.69 | WFC | 100.39 | 2019-04-10 12:12 | 780.00 | NORMAL | MCG-01-13-034 | 4289 | D:g |
| ARP184 | 05:42:03.876 | +69:22:42.41 | WFC | 82.52 | 2020-03-21 14:54 | 780.00 | NORMAL | NGC 1961 | 3888 | D:g |
| ARP190 | 02:50:10.167 | +12:53:09.30 | WFC | 65.85 | 2019-02-07 09:32 | 780.00 | NORMAL | UGC 2320 | 10216 | D:g |
| ARP195 | 08:53:53.071 | +35:08:56.76 | WFC | 111.25 | 2020-04-17 20:28 | 780.00 | NORMAL | UGC 4653 | 16696 | D:h |
| ARP197 | 11:31:01.422 | +20:28:17.15 | WFC1 | 168.17 | 2019-03-21 22:16 | 780.00 | NORMAL | IC 701 | 6077 | D:h |
| ARP200 | 02:53:40.299 | +13:00:48.17 | WFC | 48.55 | 2021-01-29 00:23 | 780.00 | NORMAL | NGC 1134 | 3656 | D:h |
| ARP202 | 09:00:15.600 | +35:43:36.12 | WFC1 | -143.07 | 2021-01-20 03:50 | 780.00 | NORMAL | NGC 2719 | 3144 | D:h |
| ARP204 | 13:23:11.882 | +84:29:55.85 | WFC | 78.86 | 2019-07-22 18:57 | 780.00 | NORMAL | UGC 8454 | 6298 | D:h |
| ARP205 | 10:54:38.661 | +54:18:21.64 | WFC | -95.25 | 2019-12-27 01:04 | 780.00 | NORMAL | NGC 3448 | 1378 | D:h |
| ARP208 | 16:50:58.546 | +47:12:41.36 | WFC | 12.81 | 2020-11-21 23:28 | 780.00 | NORMAL | APG 208 | 9025 | D:h |
| ARP216 | 23:28:47.133 | +03:30:54.41 | WFC | -143.34 | 2019-09-05 06:28 | 780.00 | NORMAL | NGC 7679 | 5126 | D:j |
| ARP219 | 03:39:53.448 | -02:06:51.48 | WFC | 74.02 | 2019-02-13 02:14 | 780.00 | NORMAL | UGC 2812 | 10436 | D:j |
| ARP221 | 09:36:27.264 | -11:20:04.56 | WFC | 6.15 | 2019-02-15 16:37 | 780.00 | NORMAL | MCG-02-25-006 | 5495 | D:k |
| ARP231 | 00:43:33.646 | -04:07:12.92 | WFC | 0.23 | 2019-10-07 02:51 | 780.00 | NORMAL | IC 1575 | 5646 | D:l |
| ARP241 | 14:37:51.000 | +30:28:54.63 | WFC1 | -126.12 | 2021-03-14 01:56 | 780.00 | NORMAL | Segner's Wheel | 10431 | D:m |
| ARP245 | 09:45:45.462 | -14:21:16.57 | WFCENTER | 107.81 | 2018-05-22 08:02 | 780.00 | NORMAL | NGC 2992 | 2292 | D:m |
| ARP248 | 11:46:39.403 | -03:50:55.96 | WFC | -57.95 | 2021-02-23 23:43 | 780.00 | NORMAL | Wild's Triplet | 5275 | D:m |
| ARP249 | 00:00:19.497 | +22:59:28.48 | WFC1 | 146.41 | 2021-10-04 23:12 | 780.00 | NORMAL | UGC 12891 | 11472 | D:m |
| ARP251 | 00:53:47.843 | -13:51:25.45 | WFC1 | -34.70 | 2019-09-30 18:19 | 780.00 | NORMAL | APG 251 | 21997 | D:m |
| ARP253 | 09:43:23.965 | -05:16:38.61 | WFC | 121.19 | 2020-06-29 06:44 | 780.00 | NORMAL | APG 253 | 1867 | D:m |
| ARP255 | 09:53:09.433 | +07:52:07.40 | WFC1 | -26.40 | 2019-02-10 03:08 | 780.00 | NORMAL | UGC 5304 | 11997 | D:m |
| ARP257 | 08:51:38.712 | -02:22:04.30 | WFC | -87.51 | 2021-10-07 02:18 | 780.00 | NORMAL | UGC 4638 | 3323 | D:n |
| ARP258 | 02:39:07.128 | +18:22:51.96 | WFC | -114.78 | 2020-10-22 18:40 | 780.00 | NORMAL | UGC 2140 | 4093 | D:n |
| ARP262 | 23:56:45.031 | +16:48:50.68 | WFC | 123.67 | 2021-10-07 21:08 | 780.00 | NORMAL | UGC 12856 | 1772 | D:n |
| ARP263 | 10:25:04.200 | +17:08:58.20 | WFC | -73.13 | 2022-12-11 23:14 | 780.00 | NORMAL | NGC 3239 | 755 | D:n |
| ARP264 | 10:03:59.021 | +40:45:23.51 | WFC | 118.27 | 2018-04-29 13:23 | 780.00 | NORMAL | NGC 3104 | 602 | D:n |

**Table 1** *continued*

**Table 1** *(continued)*

| Target Name | Target RA | Target Dec | Aperture | PA | Start Time | Exp. Time | Flag | ID | $V_r$ | Category |
|---|---|---|---|---|---|---|---|---|---|---|
| | J2000.0 | J2000.0 | | (°) | | sec | | | km s$^{-1}$ | (original) |
| (1) | (2) | (3) | (4) | (5) | (6) | (7) | (8) | (9) | (10) | (11) |
| ARP267 | 10:36:42.799 | +31:32:47.18 | WFC1 | -159.65 | 2021-02-16 13:46 | 780.00 | NORMAL | UGC 5764 | 582 | D:n |
| ARP275 | 09:25:54.309 | -11:59:37.75 | WFC1 | 101.35 | 2021-05-10 01:05 | 780.00 | NORMAL | NGC 2881 | 5057 | E:b |
| ARP276 | 02:28:11.207 | +19:35:49.32 | WFC | -101.42 | 2021-10-10 01:33 | 780.00 | NORMAL | NGC 935 | 4148 | E:b |
| ARP278 | 22:19:26.260 | +29:23:12.64 | WFC | 69.81 | 2019-11-27 16:27 | 780.00 | NORMAL | NGC 7253 | 4577 | E:b |
| ARP279 | 03:14:09.000 | -02:49:23.88 | WFC | -72.45 | 2021-10-04 21:31 | 780.00 | NORMAL | NGC 1253 | 1711 | E:b |
| ARP280 | 11:37:44.598 | +47:53:28.08 | WFC | 175.93 | 2019-03-15 20:10 | 780.00 | NORMAL | NGC 3769 | 727 | E:b |
| ARP282 | 00:36:54.339 | +23:59:12.17 | WFC | 159.15 | 2021-10-09 14:18 | 780.00 | NORMAL | NGC 169 | 4569 | E:c |
| ARP283 | 09:17:26.405 | +41:59:46.96 | WFC | 101.07 | 2018-05-13 09:23 | 780.00 | NORMAL | APG 283 | 1695 | E:c |
| ARP287 | 09:02:38.120 | +25:56:08.12 | WFC | 117.77 | 2021-03-13 07:44 | 780.00 | NORMAL | APG 287 | 2450 | E:d |
| ARP288 | 13:34:54.683 | +13:50:25.43 | WFC | -60.45 | 2020-12-25 05:49 | 780.00 | NORMAL | NGC 5221 | 7049 | E:d |
| ARP290 | 02:03:48.013 | +14:43:51.39 | WFCENTER | -125.34 | 2021-10-05 19:39 | 780.00 | NORMAL | APG 290 | 3614 | E:d |
| ARP291 | 10:42:48.946 | +13:27:33.48 | WFC1 | -81.04 | 2021-12-30 00:33 | 780.00 | NORMAL | UGC 5832 | 1218 | E:d |
| ARP293 | 16:58:27.739 | +58:56:46.12 | WFC | 87.99 | 2019-09-09 12:57 | 780.00 | NORMAL | APG 293 | 5514 | E:d |
| ARP295 | 23:41:48.000 | -03:39:59.40 | WFC | 13.26 | 2021-09-17 03:28 | 780.00 | NORMAL | MCG-01-60-021 | 6846 | E:e |
| ARP300 | 09:27:58.545 | +68:24:33.25 | WFC | 112.09 | 2018-04-22 22:54 | 780.00 | NORMAL | Mrk 111 | 3736 | E:f |
| ARP301 | 11:09:54.798 | +24:15:19.69 | WFC | -66.24 | 2021-11-17 01:35 | 780.00 | NORMAL | APG 301 | 5965 | E:f |
| ARP303 | 09:46:20.707 | +03:03:42.52 | WFC | -20.06 | 2019-02-09 19:18 | 780.00 | NORMAL | IC 563 | 5964 | E:f |
| ARP306 | 01:32:31.077 | +04:35:56.55 | WFC | -94.95 | 2021-10-05 23:01 | 780.00 | NORMAL | UGC 1102 | 1959 | E:f |
| ARP309 | 02:29:12.285 | -10:49:57.66 | WFC | -75.55 | 2023-09-14 06:20 | 763.71 | MULTIPLE | APG 309 | 4658 | E:f |
| ARP314 | 22:58:06.119 | -03:47:02.24 | WFC | 88.95 | 2019-11-12 15:57 | 780.00 | NORMAL | MCG-01-58-009 | 3704 | F:a |
| ARP321 | 09:38:53.296 | -04:51:05.75 | WFC | 8.55 | 2023-02-19 23:13 | 780.00 | NORMAL | APG 321 | 6611 | F:a |
| ARP322 | 11:32:40.604 | +52:56:06.67 | WFC | 125.50 | 2019-05-12 16:55 | 780.00 | NORMAL | HCG 56 | 7985 | F:b |
| ARP335 | 11:04:22.434 | +04:49:58.56 | WFC | 127.71 | 2023-05-20 17:55 | 780.00 | NORMAL | NGC 3509 | 7574 | F:c |
| ARP-MADORE0001-505 | 00:04:21.474 | -50:38:02.21 | WFC | -147.10 | 2019-05-28 19:52 | 780.00 | NORMAL | ESO 193-14 | 11649 | 23 |
| ARP-MADORE0002-503 | 00:05:28.291 | -50:16:16.24 | WFC | -71.56 | 2021-07-17 21:02 | 780.00 | NORMAL | ESO 193-19 | 10195 | 8 |
| ARP-MADORE0012-573 | 00:15:16.831 | -57:14:41.07 | WFC | -144.10 | 2020-05-06 08:59 | 780.00 | NORMAL | ESO 111-22 | 15900 | 6,10 |
| ARP-MADORE0018-485 | 00:21:23.017 | -48:38:02.67 | WFC | -15.96 | 2020-10-03 23:33 | 780.00 | NORMAL | SCG2 0018-4854 | 3361 | 4 |
| ARP-MADORE0052-321 | 00:54:55.013 | -32:02:02.88 | WFC | -23.95 | 2018-10-04 20:37 | 666.69 | MULTIPLE | AM 0052-321 | 9396 | 2,14 |
| ARP-MADORE0116-241 | 01:18:45.847 | -23:56:37.31 | WFC1 | 21.03 | 2020-10-31 04:25 | 780.00 | NORMAL | ESO 475-16 | 6963 | 8,10 |
| ARP-MADORE0135-650 | 01:37:24.579 | -64:54:11.33 | WFC | -45.49 | 2018-09-07 17:30 | 780.00 | NORMAL | NGC 646 | 7927 | 9 |
| ARP-MADORE0136-433 | 01:39:03.958 | -43:22:16.24 | WFC | 35.84 | 2018-11-29 01:50 | 780.00 | NORMAL | AM 0136-433 | 6239 | 2,15 |
| ARP-MADORE0137-281 | 01:40:00.609 | -28:02:15.87 | WFC | -26.36 | 2021-10-05 17:54 | 780.00 | NORMAL | ESO 413-18 | 5815 | 23 |
| ARP-MADORE0144-585 | 01:46:28.322 | -58:40:00.81 | WFC | 142.37 | 2021-03-16 09:14 | 780.00 | NORMAL | ESO 114-7 | 2210 | 20 |
| ARP-MADORE0154-441 | 01:56:43.271 | -43:59:00.34 | WFC | 17.69 | 2019-11-13 16:07 | 780.00 | NORMAL | ESO 245-10 | 5728 | 8,10,13 |
| ARP-MADORE0200-220 | 02:02:16.824 | -21:45:47.52 | WFC1 | 55.57 | 2020-12-26 01:21 | 780.00 | NORMAL | ESO 544-7 | 12872 | 8,16 |
| ARP-MADORE0203-325 | 02:05:46.346 | -32:40:32.35 | WFC1 | 63.32 | 2021-01-01 01:54 | 780.00 | NORMAL | ESO 354-34 | 5775 | 6,8 |
| ARP-MADORE0218-321 | 02:21:02.374 | -31:56:30.68 | WFC | -117.61 | 2018-07-09 23:31 | 780.00 | NORMAL | ESO 415-19 | 9322 | 1,8,9,13,17 |
| ARP-MADORE0223-403 | 02:25:14.169 | -40:25:46.55 | WFC1 | -33.09 | 2021-10-08 00:18 | 780.00 | NORMAL | MCG-07-06-002 | 6307 | 2,15 |

**Table 1** *continued*

**Table 1** *(continued)*

| Target Name | Target RA | Target Dec | Aperture | PA | Start Time | Exp. Time | Flag | ID | $V_r$ | Category |
|---|---|---|---|---|---|---|---|---|---|---|
| | J2000.0 | J2000.0 | | (°) | | sec | | | km s$^{-1}$ | (original) |
| (1) | (2) | (3) | (4) | (5) | (6) | (7) | (8) | (9) | (10) | (11) |
| ARP-MADORE0230-524 | 02:32:25.416 | -52:29:44.52 | WFC | 147.76 | 2019-04-06 09:34 | 780.00 | NORMAL | ESO 154-2 | 6450 | 2,15 |
| ARP-MADORE0311-252 | 03:13:40.362 | -25:11:06.74 | WFC | 8.35 | 2018-11-23 02:56 | 773.70 | MULTIPLE | ESO 481-14 | 1735 | 8,12 |
| ARP-MADORE0311-573 | 03:13:04.631 | -57:21:33.05 | WFC | -76.39 | 2018-09-03 03:52 | 780.00 | NORMAL | ESO 116-12 | 1140 | 16 |
| ARP-MADORE0313-545 | 03:15:04.068 | -54:49:08.59 | WFC1 | -59.67 | 2019-09-21 03:00 | 780.00 | NORMAL | IC 1908 | 8201 | 2,13 |
| ARP-MADORE0329-502 | 03:30:38.891 | -50:19:14.48 | WFC | 113.33 | 2021-03-16 06:52 | 780.00 | NORMAL | NGC 1356 | 11565 | 1,2 |
| ARP-MADORE0333-513 | 03:35:01.499 | -51:27:18.73 | WFC1 | -165.16 | 2021-06-04 04:21 | 780.00 | NORMAL | ESO 200-45 | 1030 | 20 |
| ARP-MADORE0346-222 | 03:49:00.962 | -22:14:24.17 | WFC1 | -107.87 | 2021-08-08 23:24 | 780.00 | NORMAL | ESO 549-24 | 12141 | 7,16 |
| ARP-MADORE0347-442 | 03:49:17.841 | -44:13:39.02 | WFC1 | 74.88 | 2019-02-11 01:31 | 780.00 | NORMAL | ESO 249-21 | 1248 | 20 |
| ARP-MADORE0357-460 | 03:59:13.349 | -45:52:16.66 | WFC | 109.58 | 2021-03-16 08:32 | 780.00 | NORMAL | Horologium Dwarf | 900 | 8,20 |
| ARP-MADORE0403-555 | 04:04:17.266 | -55:45:37.40 | WFC | -147.14 | 2019-06-29 02:31 | 780.00 | NORMAL | ESO 156-37 | 17093 | 2,6 |
| ARP-MADORE0405-552 | 04:07:03.188 | -55:19:48.19 | WFC | 32.15 | 2019-12-31 05:09 | 780.00 | NORMAL | IC 2032 | 1066 | 16,20 |
| ARP-MADORE0409-563 | 04:11:00.112 | -56:28:47.13 | WFC | -163.72 | 2020-06-14 10:32 | 780.00 | NORMAL | NGC 1536 | 1310 | 14,16 |
| ARP-MADORE0417-391 | 04:19:39.147 | -39:10:22.85 | WFC1 | -37.65 | 2021-10-21 08:59 | 780.00 | NORMAL | ESO 303-11 | 15255 | 6,8 |
| ARP-MADORE0432-625 | 04:32:35.857 | -62:52:04.39 | WFC | -36.21 | 2018-10-29 20:33 | 780.00 | NORMAL | 2MASX J04323376-6251479 | 16063 | 6 |
| ARP-MADORE0445-572 | 04:46:13.388 | -57:20:25.43 | WFC | 130.77 | 2020-04-16 22:46 | 780.00 | NORMAL | ESO 158-3 | 1205 | 10,13,16 |
| ARP-MADORE0454-561 | 04:55:41.791 | -56:14:06.24 | WFC1 | -125.05 | 2020-08-05 14:49 | 780.00 | NORMAL | ESO 158-15 | 1779 | 16 |
| ARP-MADORE0459-340 | 05:01:41.274 | -34:01:53.18 | WFC | -175.57 | 2021-06-13 11:10 | 780.00 | NORMAL | ESO 361-25 | 5269 | 1,24 |
| ARP-MADORE0507-630 | 05:07:45.641 | -62:59:12.19 | WFC | -139.45 | 2019-07-24 15:48 | 780.00 | NORMAL | ESO 85-47 | 1464 | 20 |
| ARP-MADORE0510-330 | 05:11:58.005 | -32:58:41.13 | WFC | -152.27 | 2019-07-08 00:47 | 780.00 | NORMAL | ESO 362-9 | 927 | 20 |
| ARP-MADORE0515-540 | 05:16:04.552 | -54:06:19.83 | WFC | 106.44 | 2019-03-31 08:11 | 780.00 | NORMAL | ESO 159-3 | 3877 | 1,2,15,23 |
| ARP-MADORE0519-611 | 05:20:17.681 | -61:16:40.42 | WFC | -90.11 | 2019-09-17 01:35 | 780.00 | NORMAL | ESO 119-54 | 4871 | 2,8 |
| ARP-MADORE0520-390 | 05:22:42.415 | -39:03:46.50 | WFC1 | 121.40 | 2019-04-20 13:49 | 780.00 | NORMAL | ESO 305-21 | 14734 | 1,6,23 |
| ARP-MADORE0541-294 | 05:42:56.189 | -29:43:46.06 | WFC1 | -3.02 | 2021-12-27 05:21 | 780.00 | NORMAL | ESO 424-10 | 3818 | 15,16,24 |
| ARP-MADORE0558-335 | 06:00:05.584 | -33:55:15.75 | WFC | -76.52 | 2021-10-08 03:44 | 780.00 | NORMAL | IC 2153 | 2872 | 2,15 |
| ARP-MADORE0608-333 | 06:09:57.550 | -33:38:36.90 | WFC | -83.11 | 2021-10-06 02:20 | 780.00 | NORMAL | AM 0608-333 | 8652 | 2 |
| ARP-MADORE0612-373 | 06:13:47.484 | -37:41:19.85 | WFC | -84.31 | 2021-10-05 00:52 | 780.00 | NORMAL | ESO 307-25 | 9734 | 2,14,15 |
| ARP-MADORE0619-271 | 06:21:39.561 | -27:14:01.58 | WFC | -47.78 | 2018-11-18 23:14 | 780.00 | NORMAL | NGC 2217 | 1622 | 8,10,13,14 |
| ARP-MADORE0620-363 | 06:21:55.522 | -36:33:09.06 | WFC | 162.12 | 2021-06-12 13:00 | 780.00 | NORMAL | ESO 365-10 | 9327 | 8,10 |
| ARP-MADORE0630-522 | 06:31:09.885 | -52:25:41.50 | WFC | -61.71 | 2018-10-31 20:12 | 780.00 | NORMAL | ESO 206-16 | 1193 | 20 |
| ARP-MADORE0642-801 | 06:38:36.050 | -80:14:48.01 | WFC1 | 73.74 | 2019-03-16 22:28 | 780.00 | NORMAL | ESO 16-16 | 4779 | 6,8,15 |
| ARP-MADORE0643-462 | 06:45:02.031 | -46:26:35.83 | WFC | 132.72 | 2019-05-17 23:30 | 780.00 | NORMAL | ESO 255-18 | 11774 | 1,6 |
| ARP-MADORE0658-590 | 06:59:05.607 | -59:07:44.59 | WFC | -124.45 | 2018-09-03 16:34 | 780.00 | NORMAL | AM 0658-590 | 8270 | 2,15 |
| ARP-MADORE0728-664 | 07:28:59.865 | -66:53:45.84 | WFC | -172.96 | 2018-07-23 07:25 | 780.00 | NORMAL | ESO 88-17 | 5155 | 1,8 |
| ARP-MADORE0839-745 | 08:38:46.766 | -75:09:06.71 | WFC | 156.57 | 2020-07-07 19:34 | 780.00 | NORMAL | ESO 36-6 | 1142 | 8 |
| ARP-MADORE0942-313A | 09:44:45.534 | -31:50:04.52 | WFC | 86.70 | 2019-04-25 14:55 | 780.00 | NORMAL | ESO 434-33 | 964 | 16,23 |
| ARP-MADORE0942-313B | 09:44:31.816 | -31:47:31.26 | WFC | 98.41 | 2019-05-10 23:21 | 780.00 | NORMAL | IC 2507 | 1253 | 16,23 |
| ARP-MADORE1010-445 | 10:12:33.402 | -45:14:12.77 | WFC | -76.04 | 2018-11-29 00:52 | 780.00 | NORMAL | ESO 263-16 | 4113 | 23 |
| ARP-MADORE1033-365 | 10:36:10.641 | -37:14:25.87 | WFC | 159.05 | 2021-08-12 19:56 | 780.00 | NORMAL | ESO 375-71 | 955 | 20 |

**Table 1** *continued*

**Table 1** *(continued)*

| Target Name | Target RA | Target Dec | Aperture | PA | Start Time | Exp. Time | Flag | ID | $V_r$ | Category |
|---|---|---|---|---|---|---|---|---|---|---|
| | J2000.0 | J2000.0 | | (°) | | sec | | | km s$^{-1}$ | (original) |
| (1) | (2) | (3) | (4) | (5) | (6) | (7) | (8) | (9) | (10) | (11) |
| ARP-MADORE1055-391 | 10:57:51.723 | -39:26:24.52 | WFC | 157.72 | 2021-08-13 11:53 | 780.00 | NORMAL | ESO 318-24 | 1002 | 16,20 |
| ARP-MADORE1125-285 | 11:27:32.408 | -29:11:01.82 | WFC | 5.27 | 2021-03-16 05:53 | 780.00 | NORMAL | ESO 439-10 | 7127 | 2,6,12 |
| ARP-MADORE1200-251 | 12:03:30.629 | -25:28:32.20 | WFC | -18.35 | 2021-03-13 06:24 | 780.00 | NORMAL | ESO 505-7 | 1790 | 20,22 |
| ARP-MADORE1203-223 | 12:06:06.989 | -22:51:06.05 | WFC | -4.92 | 2021-03-21 09:48 | 780.00 | NORMAL | ESO 505-13 | 1722 | 8 |
| ARP-MADORE1214-255 | 12:16:58.470 | -26:12:38.10 | WFC1 | 128.71 | 2021-08-14 16:35 | 780.00 | NORMAL | ESO 505-31 | 11465 | 2,15 |
| ARP-MADORE1229-512 | 12:31:53.691 | -51:44:52.43 | WFC | -18.94 | 2019-03-15 03:59 | 780.00 | NORMAL | ESO 218-8 | 2617 | 7 |
| ARP-MADORE1238-254 | 12:41:04.175 | -25:57:56.03 | WFC | 13.93 | 2019-04-07 14:49 | 780.00 | NORMAL | ESO 506-35 | 16941 | 8,12 |
| ARP-MADORE1255-430 | 12:58:08.228 | -43:19:43.92 | WFC | 118.75 | 2019-08-01 05:12 | 780.00 | NORMAL | ESO 269-20 | 8964 | 15 |
| ARP-MADORE1303-371 | 13:06:10.622 | -37:35:30.96 | WFC | 20.61 | 2019-04-20 03:06 | 780.00 | NORMAL | NGC 4953 | 4897 | 8,16,17 |
| ARP-MADORE1307-425 | 13:10:00.603 | -43:12:47.32 | WFC | 126.95 | 2019-08-20 05:04 | 780.00 | NORMAL | ESO 269-56 | 2135 | 2,8,16 |
| ARP-MADORE1307-461 | 13:10:04.496 | -46:26:14.54 | WFC | -13.45 | 2021-04-06 02:42 | 594.40 | MULTIPLE | ESO 269-57 | 3103 | 6,10 |
| ARP-MADORE1324-294 | 13:27:00.765 | -30:04:51.83 | WFC | -5.35 | 2020-04-12 04:14 | 780.00 | NORMAL | ESO 444-37 | 1902 | 20 |
| ARP-MADORE1332-331 | 13:35:07.307 | -33:29:14.96 | WFC | 30.25 | 2019-04-28 01:45 | 780.00 | NORMAL | NGC 5215 | 3963 | 2,17 |
| ARP-MADORE1342-413 | 13:45:01.977 | -41:51:44.44 | WFC | -87.45 | 2020-12-24 18:40 | 780.00 | NORMAL | ESO 325-11 | 545 | 20 |
| ARP-MADORE1356-332 | 13:59:48.679 | -33:41:09.83 | WFC | -46.95 | 2020-03-28 00:23 | 780.00 | NORMAL | ESO 384-25 | 3706 | 16 |
| ARP-MADORE1401-243A | 14:04:46.509 | -24:49:38.61 | WFC | 2.25 | 2019-04-24 15:12 | 780.00 | NORMAL | AM 1401-243 | 2332 | 12,23 |
| ARP-MADORE1421-282 | 14:23:56.988 | -28:41:22.69 | WFC | -82.24 | 2021-12-30 10:28 | 780.00 | NORMAL | NGC 5592 | 4340 | 24 |
| ARP-MADORE1440-241 | 14:43:36.365 | -24:27:57.01 | WFC | -69.28 | 2021-04-14 21:43 | 780.00 | NORMAL | ESO 512-18 | 3528 | 2 |
| ARP-MADORE1546-284 | 15:49:17.422 | -28:54:33.09 | WFC | 97.55 | 2020-08-22 05:58 | 780.00 | NORMAL | ESO 450-18 | 4045 | 16 |
| ARP-MADORE1705-773 | 17:13:45.986 | -77:32:13.51 | WFC | -83.95 | 2019-03-14 00:28 | 780.00 | NORMAL | IC 4633 | 2953 | 8,20,23 |
| ARP-MADORE1806-852 | 18:22:27.057 | -85:24:27.31 | WFC | 91.94 | 2019-10-05 02:12 | 780.00 | NORMAL | NGC 6438 | 2541 | 2 |
| ARP-MADORE1847-645 | 18:52:19.803 | -64:50:22.27 | WFC | 58.43 | 2018-08-29 14:47 | 780.00 | NORMAL | ESO 104-19 | 1006 | 20 |
| ARP-MADORE1912-860 | 19:31:02.315 | -86:01:08.52 | WFC | -12.46 | 2020-07-02 04:33 | 780.00 | NORMAL | ESO 10-4 | 2444 | 8 |
| ARP-MADORE1914-603 | 19:18:28.714 | -60:30:04.80 | WFC | -93.79 | 2018-04-15 22:24 | 780.00 | NORMAL | IRAS 19140-6035 | 3817 | 3,8 |
| ARP-MADORE1931-610 | 19:36:05.666 | -61:01:53.69 | WFC | -97.50 | 2018-04-15 22:57 | 780.00 | NORMAL | IC 4869 | 1796 | 8,20 |
| ARP-MADORE1953-260 | 19:56:28.654 | -25:54:59.08 | WFC1 | -88.63 | 2019-06-30 15:45 | 780.00 | NORMAL | ESO 526-18 | 14619 | 6 |
| ARP-MADORE1957-471 | 20:00:59.266 | -47:04:26.37 | WFC | -54.39 | 2019-06-25 00:41 | 780.00 | NORMAL | NGC 6845 | 6359 | 4 |
| ARP-MADORE2001-602 | 20:06:14.515 | -60:12:40.87 | WFC | -98.82 | 2019-04-24 15:59 | 780.00 | NORMAL | IC 4938 | 3587 | 6,12 |
| ARP-MADORE2019-442 | 20:22:59.699 | -44:16:13.70 | WFC | -104.95 | 2020-05-10 06:03 | 780.00 | NORMAL | ESO 285-4 | 2963 | 13,15 |
| ARP-MADORE2026-424 | 20:29:32.620 | -42:30:14.86 | WFC1 | -70.65 | 2019-06-19 06:20 | 780.00 | NORMAL | ESO 285-19 | 15124 | 1,6 |
| ARP-MADORE2029-544 | 20:33:28.315 | -54:31:32.77 | WFC | -66.64 | 2019-06-16 15:11 | 780.00 | NORMAL | AM 2029-5441 | 3403 | 23 |
| ARP-MADORE2031-440 | 20:34:40.932 | -43:50:21.22 | WFC | -119.13 | 2019-04-21 21:18 | 780.00 | NORMAL | ESO 285-35 | 8853 | 15 |
| ARP-MADORE2034-521 | 20:38:19.169 | -52:06:42.47 | WFC | 59.16 | 2021-10-05 03:20 | 780.00 | NORMAL | NGC 6935 | 4581 | 23 |
| ARP-MADORE2038-323 | 20:41:16.140 | -32:29:15.88 | WFC | -19.70 | 2020-07-24 06:29 | 780.00 | NORMAL | NGC 6947 | 5551 | 1,6,8 |
| ARP-MADORE2038-382 | 20:41:18.195 | -38:11:38.96 | WFC | 68.65 | 2021-10-07 01:23 | 780.00 | NORMAL | ESO 341-4 | 5996 | 15,16 |
| ARP-MADORE2038-654 | 20:43:21.192 | -65:38:49.92 | WFC1 | -59.05 | 2019-06-20 21:13 | 780.00 | NORMAL | IC 5028 | 1626 | 20 |
| ARP-MADORE2040-295 | 20:43:34.415 | -29:42:23.76 | WFC | -33.02 | 2020-07-25 06:18 | 780.00 | NORMAL | IC 5041 | 2709 | 8,23 |
| ARP-MADORE2042-382 | 20:45:30.016 | -38:10:44.11 | WFC | -27.32 | 2020-07-23 06:39 | 780.00 | NORMAL | ESO 341-11 | 6932 | 1,8 |

**Table 1** *continued*

**Table 1** *(continued)*

| Target Name | Target RA | Target Dec | Aperture | PA | Start Time | Exp. Time | Flag | ID | $V_r$ | Category |
|---|---|---|---|---|---|---|---|---|---|---|
| | J2000.0 | J2000.0 | | (°) | | sec | | | km s$^{-1}$ | (original) |
| (1) | (2) | (3) | (4) | (5) | (6) | (7) | (8) | (9) | (10) | (11) |
| ARP-MADORE2048-571 | 20:52:01.846 | -57:04:17.40 | WFC | -25.82 | 2021-07-18 17:10 | 780.00 | NORMAL | IC 5063 | 3359 | 8,14 |
| ARP-MADORE2055-541 | 20:58:47.300 | -54:09:35.09 | WFC | -131.94 | 2020-03-22 22:38 | 780.00 | NORMAL | AM 2055-541 | 12864 | 2,10 |
| ARP-MADORE2056-392 | 21:00:07.415 | -39:18:00.53 | WFC1 | -5.15 | 2021-08-03 15:54 | 780.00 | NORMAL | AM 2056-392 | (13500) | 6,13,24 |
| ARP-MADORE2103-550 | 21:07:33.451 | -54:57:11.24 | WFC | -34.84 | 2020-07-05 17:39 | 780.00 | NORMAL | ESO 187-51 | 1402 | 20 |
| ARP-MADORE2105-332 | 21:08:04.477 | -33:14:05.79 | WFC | -46.15 | 2018-07-26 18:53 | 780.00 | NORMAL | LEDA 678973 | 5305 | 8,9 |
| ARP-MADORE2105-632 | 21:09:12.481 | -63:17:37.87 | WFC | 50.56 | 2019-10-04 19:46 | 780.00 | NORMAL | IC 5084 | 3142 | 14 |
| ARP-MADORE2106-374 | 21:10:00.548 | -37:30:11.94 | WFC1 | -82.28 | 2021-07-19 02:29 | 780.00 | NORMAL | ESO 342-13 | 2632 | 12,17 |
| ARP-MADORE2113-341 | 21:17:02.827 | -33:59:12.18 | WFC1 | -64.17 | 2021-07-18 09:00 | 780.00 | NORMAL | ESO 402-21 | 8798 | 14,18 |
| ARP-MADORE2115-273 | 21:18:21.503 | -27:20:55.25 | WFC1 | -80.76 | 2021-07-17 20:17 | 780.00 | NORMAL | ESO 464-31 | 6458 | 2,15 |
| ARP-MADORE2126-601 | 21:30:42.615 | -60:00:23.45 | WFC | 17.95 | 2021-08-29 00:33 | 780.00 | NORMAL | IC 5110 | 8660 | 10 |
| ARP-MADORE2128-430 | 21:31:47.944 | -42:50:58.21 | WFC | -82.34 | 2019-06-26 07:02 | 780.00 | NORMAL | ESO 287-34 | 2362 | 10,14,22 |
| ARP-MADORE2159-320 | 22:02:01.274 | -31:52:31.90 | WFC | 0.60 | 2019-08-23 05:23 | 780.00 | NORMAL | NGC 7172 | 2541 | 3,14 |
| ARP-MADORE2210-693 | 22:14:46.283 | -69:22:16.03 | WFC | -46.89 | 2020-07-02 03:50 | 780.00 | NORMAL | IC 5173 | 3155 | 2 |
| ARP-MADORE2220-423 | 22:23:32.719 | -42:16:36.87 | WFC | 42.69 | 2021-10-08 02:53 | 780.00 | NORMAL | ESO 289-26 | 2420 | 20 |
| ARP-MADORE2222-275 | 22:25:27.765 | -27:42:20.34 | WFC | -125.55 | 2019-04-16 15:24 | 780.00 | NORMAL | AM 2222-275 | 14872 | 1,2,9,24 |
| ARP-MADORE2224-310 | 22:26:53.425 | -30:53:13.57 | WFC1 | 29.57 | 2018-09-03 15:45 | 780.00 | NORMAL | ESO 467-62 | 3974 | 1 |
| ARP-MADORE2230-481 | 22:33:46.752 | -48:01:33.46 | WFC1 | -149.78 | 2021-03-26 01:41 | 780.00 | NORMAL | ESO 238-16 | 8185 | 1,6 |
| ARP-MADORE2240-892 | 23:10:12.366 | -89:08:03.09 | WFC | -156.21 | 2019-04-05 08:58 | 780.00 | NORMAL | ESO 1-8 | 2525 | 2,10 |
| ARP-MADORE2254-373 | 22:56:54.973 | -37:20:56.47 | WFC | 43.05 | 2021-10-08 01:20 | 780.00 | NORMAL | NGC 7421 | 1801 | 12,16 |
| ARP-MADORE2258-595 | 23:01:31.271 | -59:39:06.90 | WFC | 10.67 | 2018-09-19 13:52 | 780.00 | NORMAL | ESO 147-19 | 10108 | 2 |
| ARP-MADORE2259-692 | 23:03:02.908 | -69:12:37.80 | WFC | -163.24 | 2021-03-25 01:42 | 780.00 | NORMAL | IC 5279 | 3901 | 15 |
| ARP-MADORE2303-305 | 23:05:48.276 | -30:36:50.43 | WFC | 39.15 | 2021-10-05 01:53 | 780.00 | NORMAL | ESO 469-11 | 8463 | 8,15 |
| ARP-MADORE2325-473 | 23:27:50.184 | -47:22:47.47 | WFC | -153.43 | 2019-04-08 09:12 | 780.00 | NORMAL | ESO 240-3 | 15238 | 1,10 |
| ARP-MADORE2339-661 | 23:42:36.346 | -65:56:55.96 | WFC | 17.40 | 2021-10-05 21:20 | 780.00 | NORMAL | NGC 7733 | 10013 | 2 |
| ARP-MADORE2346-380 | 23:49:23.737 | -37:46:45.05 | WFC | 23.71 | 2021-10-09 17:13 | 780.00 | NORMAL | ESO 348-9 | 647 | 20 |
| ARP-MADORE2350-302 | 23:53:22.645 | -30:09:31.82 | WFC | -27.36 | 2021-09-18 09:53 | 780.00 | NORMAL | ESO 471-37 | 14143 | 8,10,24 |
| ARP-MADORE2350-410 | 23:53:24.912 | -40:48:40.40 | WFC | -90.99 | 2020-06-14 11:37 | 780.00 | NORMAL | ESO 293-8 | 9162 | 3 |
| ARP-MADORE2353-291 | 23:56:24.780 | -29:01:23.91 | WFC1 | -122.35 | 2019-05-28 22:54 | 780.00 | NORMAL | AM 2353-291 | 8828 | 2,6 |

Specific target positions were chosen to optimize the coverage of each system. Given that the program was not allowed to use roll-angle constraints, we took care to select WFC apertures and centers such that the majority of the system would remain within the FOV at any orientation, while minimizing the chance that interesting features would fall within the chip gap. This strategy was largely successful, and only rarely was an observation executed at a particularly unfortunate orientation. A very small number of systems were considered interesting enough that we included two separate positions in the target list, each aimed at a different galaxy in an interacting pair (AM0942-313, for which both positions were observed, and AM1401-243, for which only one position was observed). Note that while the target names for the two positions are supplemented with "A" or "B", these do not necessarily correlate with historical assignments of "A" or "B" to the galaxies in the system.

Observations began in Fall of 2018, and were on-going through September 2023. Table 1 compiles the final list of 216 well-observed targets (column 1), ACS/WFC positions (columns 2 & 3) at the ACS Aperture (column 4), and the position angle (column 5), along with the start time of observation[7] (column 6), exposure time (column 7), and exposure flag (column 8).

Roughly 13 observations were affected by spacecraft issues, and as a SNAP program, lost targets were not eligible for repeat observations. Some of the compromised observations still produced informative imaging, and are included in this atlas (with `FLAG=MULTIPLE` in Table 1, and somewhat shortened exposure times), but 10 were significantly degraded or had zero exposure times, and were excluded, leaving 216 targets out of 225 observations in Table 1, which is ∼62% of our original target list[8]. Systems which were included in the target list but not successfully observed are listed in Table A1 in an appendix (Appendix A) along with thumbnail color images from the Digitized Sky Survey (Figure A1); these remain highly interesting targets for future observing programs.

In addition to the observing parameters, we also include in Table 1 general information about each well-observed system. These include the preferred "Main ID" of the system (or of the dominant galaxy in the image) from the SIMBAD Astronomical Database[9] (column 9); the recessional velocity of the system (taken as the mean of all likely members in the frame, as described below in Section 2.2; column 10); and the original classification of the system from the Arp or Arp & Madore catalogs (column 11), which can be translated using Tables 2 and 3, respectively, and used to select subtypes of interest. We note that the original classifications are based on plate images of the entire system, and in a few cases may not be applicable to just the portion imaged by ACS, or, may not be correct or comprehensive when viewed at HST resolution.

**Table 2.** Arp Catalog Categories

| ID | Arp # | Category |
|---|---|---|
| | | A. Spiral Galaxies |
| a | 1–6 | Low Surface Brightness |
| b | 7– 12 | Split Arm |
| c | 13–18 | Detached Segments |
| d | 19–21 | Three-Armed |
| e | 22–24 | One-Armed |
| f | 25–30 | One Heavy Arm |
| g | 31–36 | Integral Sign |
| | | B. Spiral Galaxies with Companions on Arms |
| a | 37–48 | Low Surface Brightness Companions |
| b | 49–78 | Small, High Surface Brightness Companions |
| c | 79–91 | Large, High Surface Brightness Companions |
| d | 92–101 | Elliptical Galaxy Companions |
| | | C. E and E-like Galaxies |
| a | 102–108 | Connected to Spirals |
| b | 109–112 | Repelling Spiral Arms |
| c | 113–132 | Close to and Perturbing Spirals |
| d | 133–136 | With Nearby Fragments |
| e | 137–145 | Material Emanating from E Galaxies |
| | | D. Galaxies |
| a | 146–148 | With Associated Rings |
| b | 149–152 | With Jets |
| c | 153–160 | Disturbed with Interior Absorption |
| d | 161–166 | With Diffuse Filaments |
| e | 167–172 | Diffuse Counter-Tails |
| f | 173–178 | Narrow Counter-Tails |
| g | 179–193 | Narrow Filaments |
| h | 194–208 | Material Ejected from Nuclei |
| i | 209–214 | Irregularities, Absorption, and Resolution |
| j | 215–220 | Adjacent Loops |
| k | 221–226 | Amorphous Spiral Arms |
| l | 227–231 | Concentric Rings |
| m | 232–256 | Appearance of Fission |
| n | 257–268 | Irregular Clumps |
| | | E. Double Galaxies |
| a | 269–274 | Connected Arms |
| b | 275–280 | Interacting |
| c | 281–286 | Infall and Attraction |
| d | 287–293 | Wind Effects |
| e | 294–297 | Long Filaments |
| f | 298–310 | *Unclassified Double Galaxies* |
| | | F. Miscellaneous |
| a | 311–321 | Groups of Galaxies |
| b | 322–332 | Chains of Galaxies |
| c | 333–338 | *Unclassified Miscellaneous* |

[7] Note that some clusters of observations were associated with times when HST was recovering from being safed, during which the scheduling flexibility offered by the gap-filler programs was particularly valuable.

[8] It is worth remarking that this efficiency is much higher than typical for SNAP-mode observations with HST, demonstrating the efficacy of the gap-filler approach when longer exposures are not needed.

[9] https://simbad.cds.unistra.fr/simbad/

**Table 3.** Arp-Madore Catalog Catagories

| ID | Category |
|---|---|
| 1 | Galaxies with Interacting Companions |
| 2 | Interacting Doubles |
| 3 | Interacting Triples |
| 4 | Interacting Quadruples |
| 5 | Interacting Quintets |
| 6 | Ring Galaxies |
| 7 | Galaxies with Jets |
| 8 | Galaxies with Apparent Companions |
| 9 | M51-types |
| 10 | Galaxies with Peculiar Spiral Arms |
| 11 | Three-Armed Spirals and Multiple-Armed Spirals |
| 12 | Peculiar Spirals |
| 13 | Compact Galaxies |
| 14 | Galaxies with Prominent or Unusual Dust Absorption |
| 15 | Galaxies with Tails, Loops of Material or Debris |
| 16 | Irregular or Disturbed Galaxies |
| 17 | Chains |
| 18 | Groups |
| 19 | Clusters |
| 20 | Dwarf Galaxies (Low Surface Brightness) |
| 21 | Stellar Objects with Associated Nebulosity |
| 22 | Miscellaneous |
| 23 | Close Pairs |
| 24 | Close Triples |
| 25 | Planetary Nebulae |

Figure 1 maps the observed systems across the sky, color coded by their recessional velocities, which range from ∼450 km $s^{-1}$to ∼22,000 km $s^{-1}$. The distribution avoids the Galactic plane, due to biases against low-latitude galaxies obscured by the Milky Way in the original source catalogs from which the HST targets were drawn. In Figure 2, we compare the redshift distribution of the observed systems (discussed below in Section 2) to that of the galaxies from the diameter-limited Uppsala Galaxy Catalog (UGC; P. Nilson 1973), both shown as kernel-density estimator smoothed histograms. The Arp systems in this catalog have a median redshift of ∼4,573 km $s^{-1}$and are biased to lower redshifts than the UGC, due to our selection favoring large, well-resolved systems that were well-matched to ACS/WFC's large field of view. Visually, many of the galaxies in our survey are sufficiently close that they are well-resolved into stars and stellar clusters; we return to this point in Section 2.3 where we analyze photometry from the images.

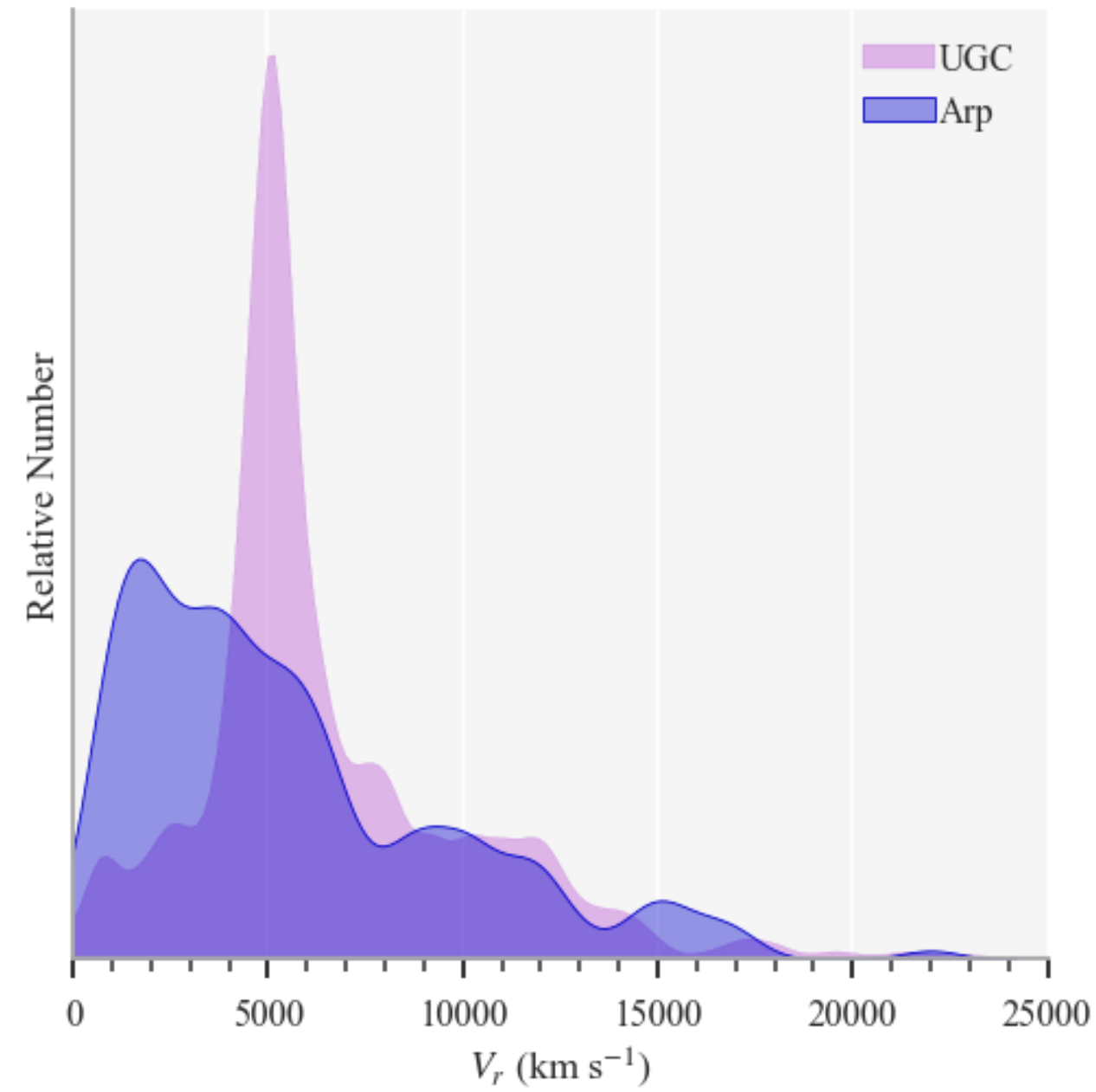

**Figure 2.** Distribution of recessional velocities for the Arp systems in this survey (blue), compared to the UGC catalog (magenta), each normalized by the total number. The galaxies in this atlas are biased towards lower redshifts, since they were chosen to be well-matched to the wide ACS/WFC field of view.

### 2.1. *Images*

The images released as part of this catalog required additional processing to remove cosmic rays, particularly in the chip gaps which were observed by only a single image. HST images were downloaded in Fall 2023 as `*_flc` images. These were then processed through the the machine learning-optimized cosmic-ray identification program `deepCR` (K. Zhang & J. S. Bloom 2020; K. J. Kwon et al. 2021) with the pretrained "`ACS-WFC-F606W-2-32`" model and a threshold of 0.5. All identified cosmic rays in the `*_flc` images were then replaced with values from a local median. The resulting images were then drizzled into a sky-subtracted final image, using medians for the combination step. Each ACS chip was sky subtracted independently during the drizzling process. Seven systems were large and bright enough that the automated sky determination was biased bright, leading to oversubtraction of the sky (Arp 18, Arp 22, Arp 184, Arp 263, Arp 279, AM0619-271, AM1705-773, and AM1914-603). For these few fields, we measured median sky levels by hand in the least contaminated regions, and added an offset to adjust the backgrounds to zero, recording the adjustments in a new header keyword ('SKYTWEAK'). All reduced images are available as high level science products at the Mikulski Archive for Space Telescopes (MAST) via DOI:10.17909/176w-p735.

We note that the final images contain NaNs in the small number of pixels where there was no valid data due to overlapping cosmic rays. When creating plots, these pixels are replaced with the local median (in a 4x4 boxcar), but they are left as NaNs in the distributed catalog `*drz.fits` images. We have also made no attempt to correct for common large scale HST artifacts such as “figure eight” or circular ghosts from bright stars (e.g., Arp 258 in Figure B2 and Arp 221 in Figure B2 as examples, respectively), satellite trails (e.g., Arp 248 in Figure B2), and spikes and bleed trails from saturated stars (e.g., Arp 195 in Figure B2).

In Figure 4 we show a compendium of low-resolution thumbnails of the reduced HST images, plotted in pixel coordinates. The images are ordered following the associated entries in Tables 1. The galaxies from the Arp catalog are plotted first, and because their numbering in the original catalog was based on their category assignment (e.g., Table 2), similar types of systems are typically grouped together. In contrast, the galaxies from the Arp-Madore catalog are sorted by sky position, and are not ordered by their assigned category. To better show the full dynamic range of the data, images are shown in an inverted grayscale, with pixel values $f$ scaled as $[(x - v_{min})/(v_{max} - v_{min})]^{\gamma}$ (“`PowerNorm`” in matplotlib), where $x = \log_{10}(f + 2.5 f_0)$, with $f_0 = 0.05$, $v_{min} = -1$, $v_{max} = 10$, and $\gamma = 0.5$.

Every galaxy name in the caption of Figure 4 links to an associated full-resolution atlas image, shown in Appendix B (Figures B2–B2); an example for Arp 107 is shown in the main body of the text in Figure 3. We stress that these larger images are the preferred presentation of the survey, as they best convey the value of the new HST imaging. Unlike the thumbnails in Figure 4, all atlas images are plotted in sky coordinates, and we include a scale bar in kiloparsecs calculated from the recessional velocity listed in Table 1, assuming a pure Hubble Flow with $H_0 = 70\,\mathrm{km\ s^{-1} Mpc^{-1}}$; the assignment of recessional velocities is discussed in Section 2.2. We note that for the nearest galaxies, expected departures from pure Hubble flow will make the assumed distances and image scales inaccurate. Distances will typically have $\sim$20% scatter for systems with recessional velocities less than than $\sim$1,500$-$2,000 km s$^{-1}$, assuming characteristic peculiar velocities of 300-400 km s$^{-1}$.

#### 2.1.1. *Ancillary Multi-Wavelength Data*

We add further context to the HST images in Figures B2–B2 by presenting matched multi-wavelength imaging from across the electromagnetic spectrum, shown as a row of images below the large grayscale images of the HST data, matched in location and extent to the full plotting area. While the HST images contain far more structural information (e.g., resolving stars, stellar clusters, dust, and background galaxies), the ground-based data offers useful constraints on color and non-thermal emission, both of which aid in interpreting the HST morphologies. In some cases, the ancillary data has more sensitivity to very low surface brightness features that can be detected in the short exposures used in this survey.

From shortest to longest wavelength, the supplementary images are plotted from left to right and include: (1) GALEX FUV+NUV images (4.2″-5.3″resolution) that trace primarily young massive stars, but also hot, old stars with significant mass loss that typically appear red in FUV-NUV color images (D. C. Martin et al. 2005); (2) optical color imaging, primarily from DR9 of the DECaLS Survey (A. Dey et al. 2019), or DR10 when mulitcolor DR9 imaging of a field was not available[10]; (3) NIR ALLWISE (A. M. Meisner et al. 2021) 7-year imaging in W1+W2 bandpasses (6.1″-6.4″resolution, in bandpasses comparable to the Spitzer/IRAC 3.6$\mu$ and 4.5$\mu$ bands) from the NEOWISE mission (A. Mainzer et al. 2014) as a primary tracer of stellar surface density, through the old red giant branch stars which typically dominate the NIR luminosity, along with some minor contributions of PAH emission (at 3.3$\mu$, primarily affecting W1) and hot dust (primarily affecting W2; see S. E. Meidt et al. 2012, for example) and younger, more massive asymptotic giant branch stars; (4) ALLWISE imaging in the W3+W4 filters (6.5″-12.0″, in bandpasses comparable to the IRAS 12$\mu$ and 25$\mu$ bands, or the Spitzer/MIPS 24$\mu$ band for W4) as a tracer of emission from polyaromatic hydrocarbons (PAHs; W3) and warm dust (W4) associated with star-forming regions and active galactic nuclei (AGN); and (5) VLA continuum 1.4 GHz imaging (2.5″resolution) from the VLA Sky Survey (v1.2 M. Lacy et al. 2020, , hereafter VLASS) as a tracer of potential strong AGN activity or concentrated starbursts, through synchrotron or free-free emission, respectively. Not all of the above data is available for every target, with sparser coverage for targets in the southern hemisphere. In addition, not all images for a given bandpass have the same sensitivity, due to variations in survey coverage. All of the imaging above was accessed through `http://www.legacysurvey.org`'s footprint server in August 2024.

[10] The one exception is Arp 158, where SDSS color imaging was the best available.

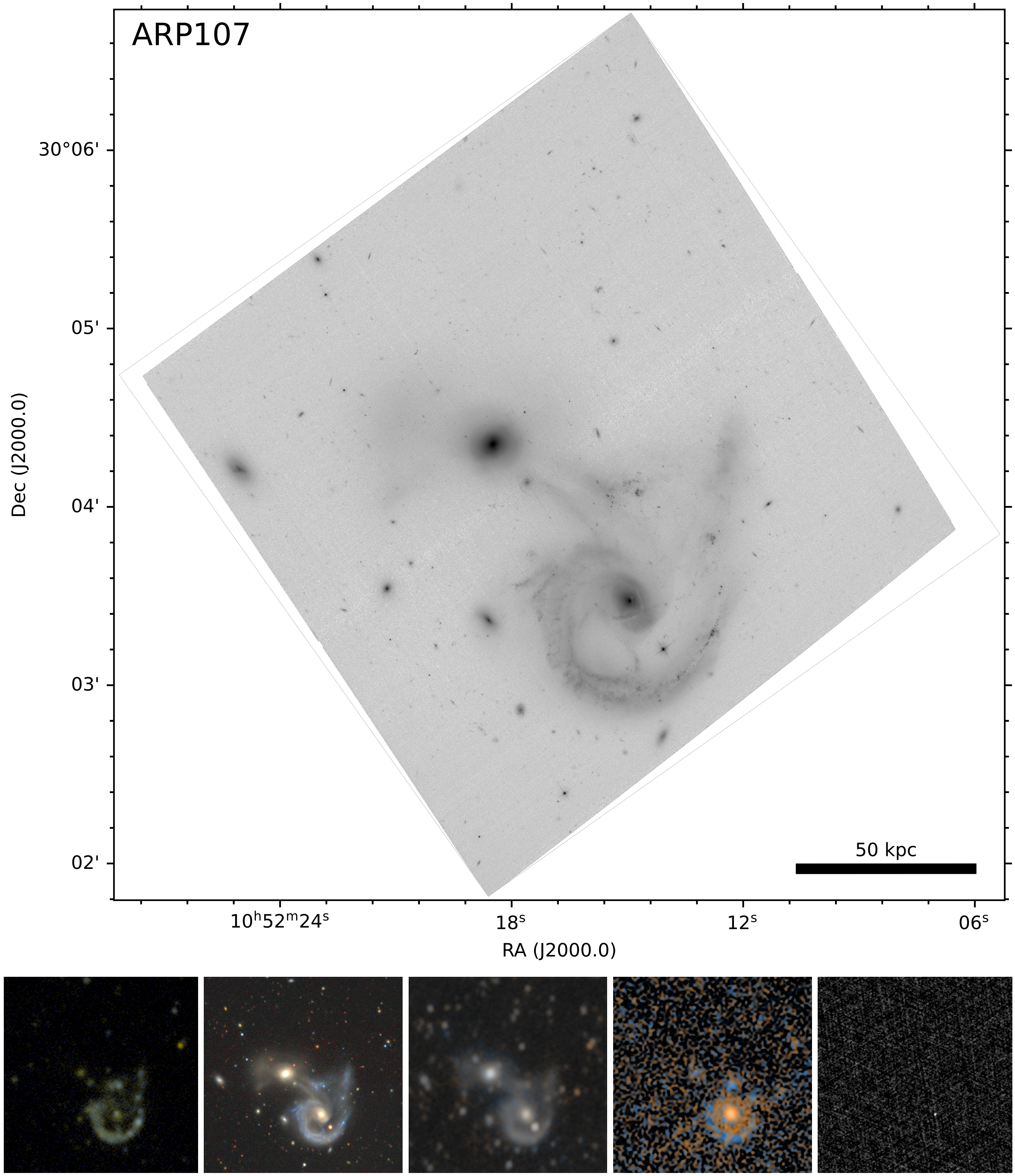

**Figure 3.** Example of a full atlas image, which is available for the entire sample in Appendix B. The large upper panel shows the full-resolution version of Arp 107, as viewed with HST/ACS and plotted in sky coordinates (top; F606W). The scale bar (in kiloparsecs) is calculated from the recessional velocity listed in Table 1, assuming a pure Hubble Flow with $H_0 = 70\,\mathrm{km\,s^{-1}Mpc^{-1}}$; this scale will not necessarily be accurate for nearer galaxies where peculiar velocities are significant, or, for foreground or background interloping galaxies. The bottom row shows a panchromatic view of the galaxy, from short to long wavelengths. From left to right, the panels show: GALEX (NUV+FUV); the Legacy Survey imaging (Optical); NEOWISE (W1+W2 and W3+W4); and the VLA Sky Survey (1.4 GHz continuum), all matched to the field of view of the upper plot.

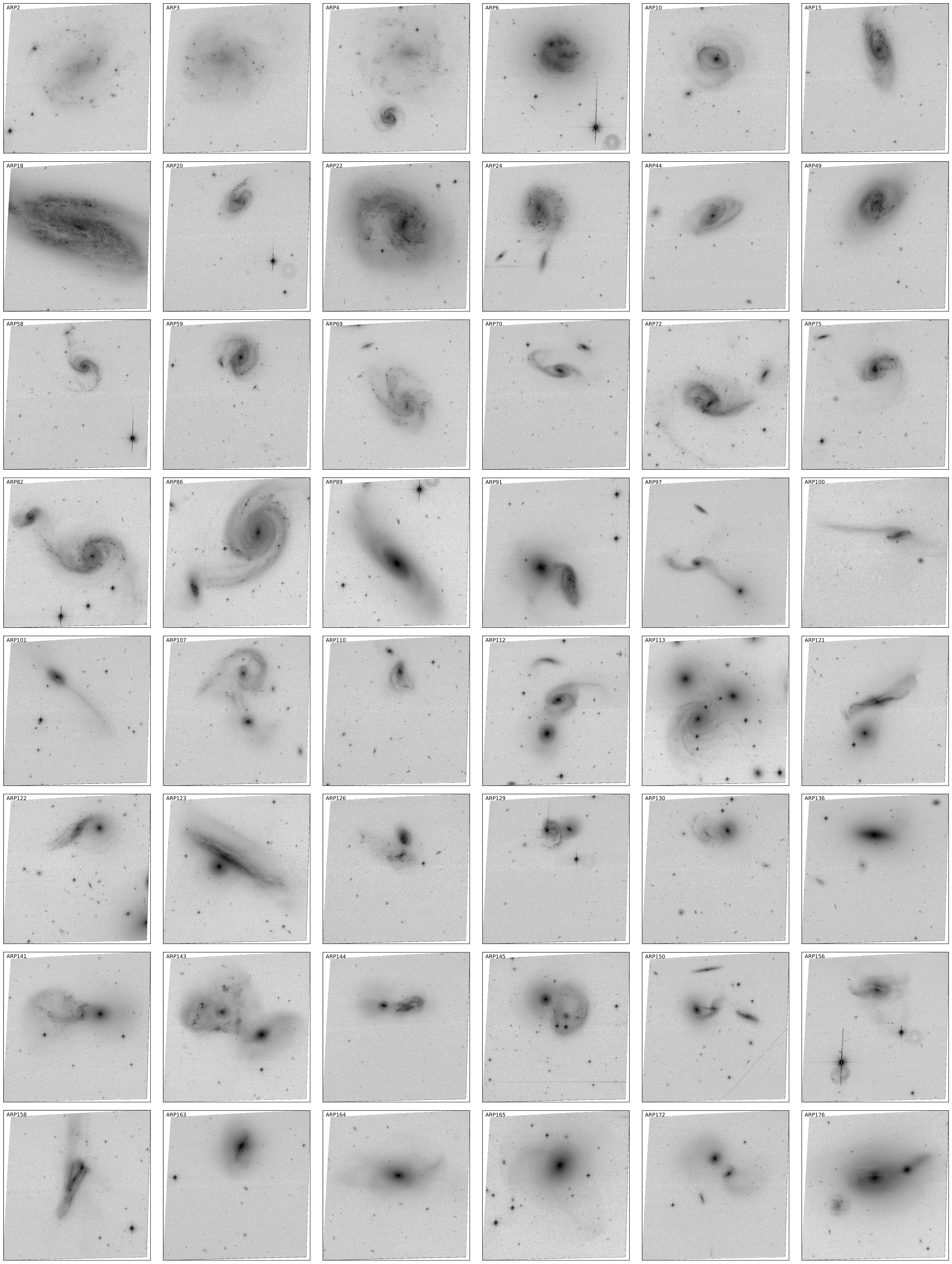

**Figure 4.** Thumbnails of HST images in atlas. Galaxy names in the caption link to full-size HST images with imaging at other wavelengths: ARP2; ARP3; ARP4; ARP6; ARP10; ARP15; ARP18; ARP20; ARP22; ARP24; ARP44; ARP49; ARP58; ARP59; ARP69; ARP70; ARP72; ARP75; ARP82; ARP86; ARP89; ARP91; ARP97; ARP100; ARP101; ARP107; ARP110; ARP112; ARP113; ARP121; ARP122; ARP123; ARP126; ARP129; ARP130; ARP136; ARP141; ARP143; ARP144; ARP145; ARP150; ARP156; ARP158; ARP163; ARP164; ARP165; ARP172; ARP176; [Continued]

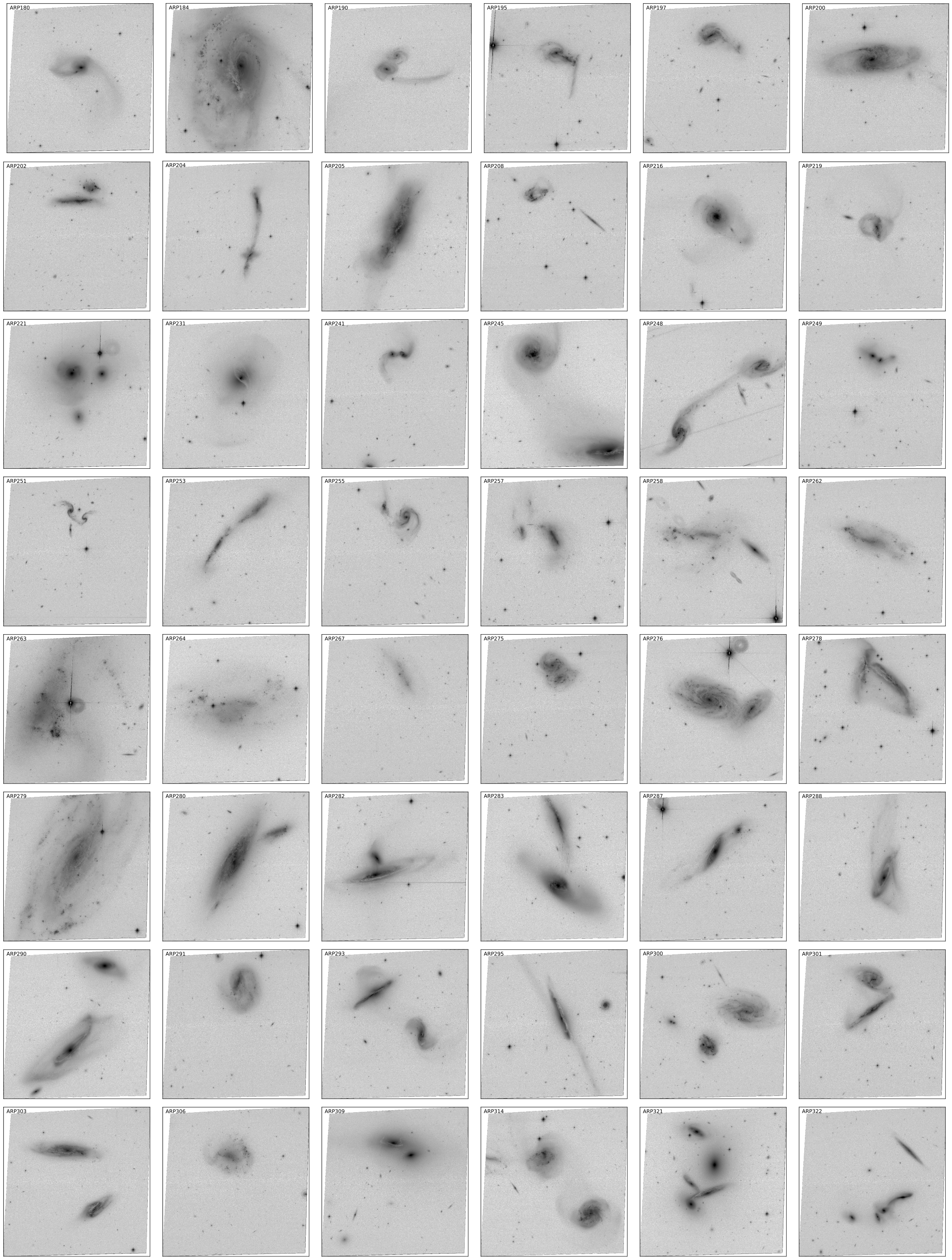

**Figure 4.** [Continued] Thumbnails of HST images in atlas. Galaxy names in the caption link to full-size HST images with imaging at other wavelengths: ARP180; ARP184; ARP190; ARP195; ARP197; ARP200; ARP202; ARP204; ARP205; ARP208; ARP216; ARP219; ARP221; ARP231; ARP241; ARP245; ARP248; ARP249; ARP251; ARP253; ARP255; ARP257; ARP258; ARP262; ARP263; ARP264; ARP267; ARP275; ARP276; ARP278; ARP279; ARP280; ARP282; ARP283; ARP287; ARP288; ARP290; ARP291; ARP293; ARP295; ARP300; ARP301; ARP303; ARP306; ARP309; ARP314; ARP321; ARP322; [Continued]

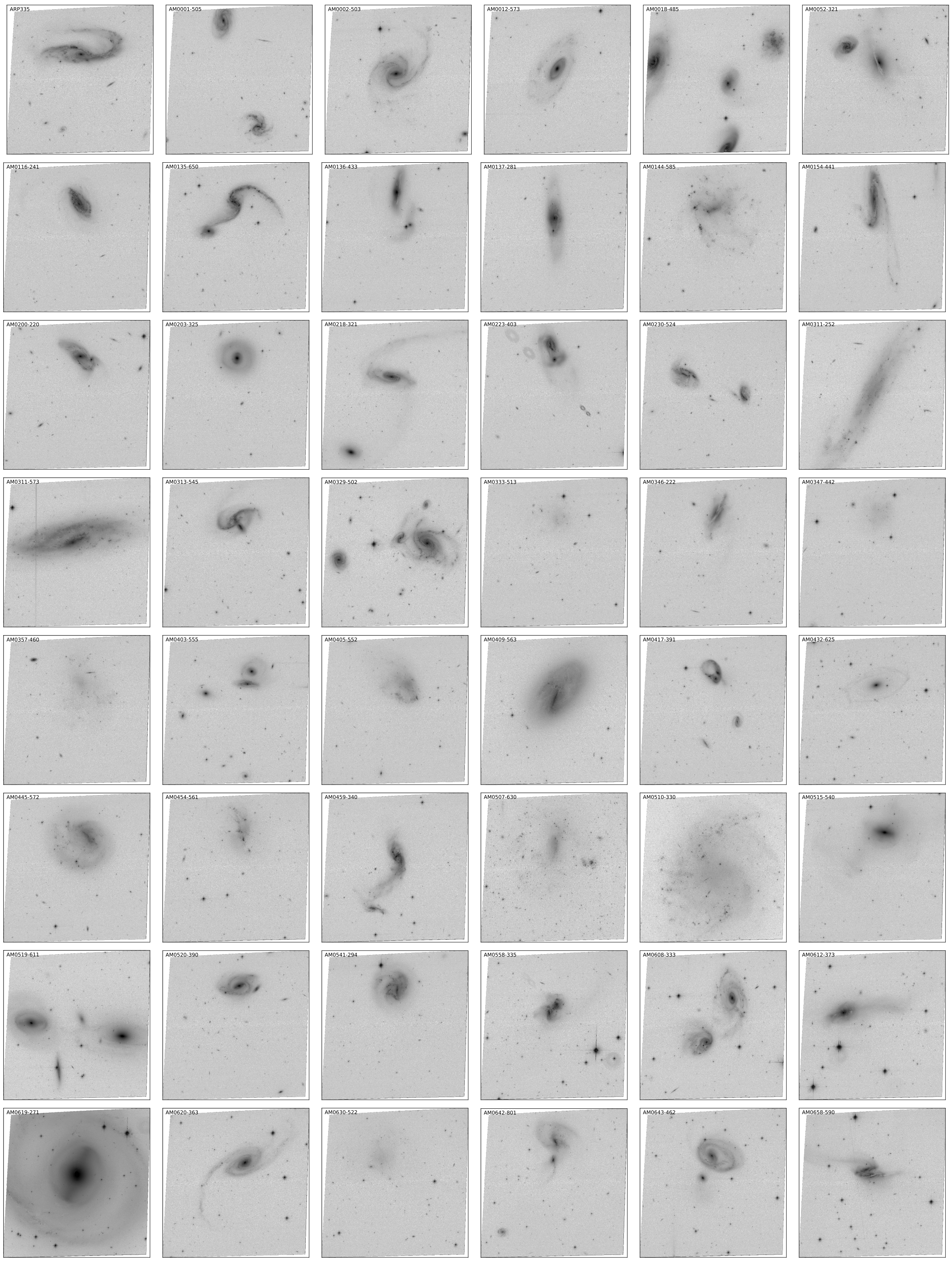

**Figure 4.** [Continued] Thumbnails of HST images in atlas. Galaxy names in the caption link to full-size HST images with imaging at other wavelengths: ARP335; AM0001-505; AM0002-503; AM0012-573; AM0018-485; AM0052-321; AM0116-241; AM0135-650; AM0136-433; AM0137-281; AM0144-585; AM0154-441; AM0200-220; AM0203-325; AM0218-321; AM0223-403; AM0230-524; AM0311-252; AM0311-573; AM0313-545; AM0329-502; AM0333-513; AM0346-222; AM0347-442; AM0357-460; AM0403-555; AM0405-552; AM0409-563; AM0417-391; AM0432-625; AM0445-572; AM0454-561; AM0459-340; AM0507-630; AM0510-330; AM0515-540; AM0519-611; AM0520-390; AM0541-294; AM0558-335; AM0608-333; AM0612-373; AM0619-271; AM0620-363; AM0630-522; AM0642-801; AM0643-462; AM0658-590; [Continued]

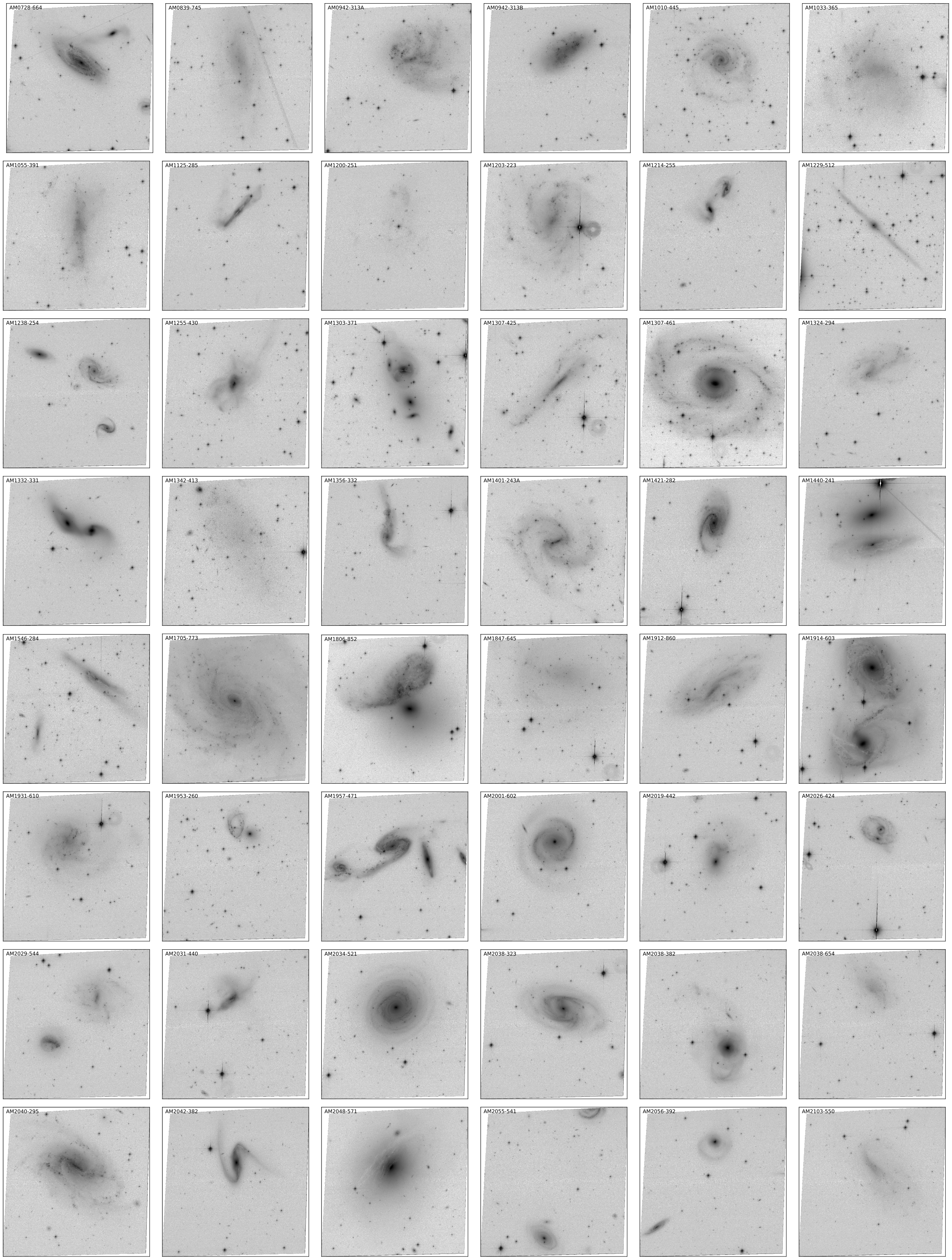

**Figure 4.** [Continued] Thumbnails of HST images in atlas. Galaxy names in the caption link to full-size HST images with imaging at other wavelengths: AM0728-664; AM0839-745; AM0942-313A; AM0942-313B; AM1010-445; AM1033-365; AM1055-391; AM1125-285; AM1200-251; AM1203-223; AM1214-255; AM1229-512; AM1238-254; AM1255-430; AM1303-371; AM1307-425; AM1307-461; AM1324-294; AM1332-331; AM1342-413; AM1356-332; AM1401-243A; AM1421-282; AM1440-241; AM1546-284; AM1705-773; AM1806-852; AM1847-645; AM1912-860; AM1914-603; AM1931-610; AM1953-260; AM1957-471; AM2001-602; AM2019-442; AM2026-424; AM2029-544; AM2031-440; AM2034-521; AM2038-323; AM2038-382; AM2038-654; AM2040-295; AM2042-382; AM2048-571; AM2055-541; AM2056-392; AM2103-550; [Continued]

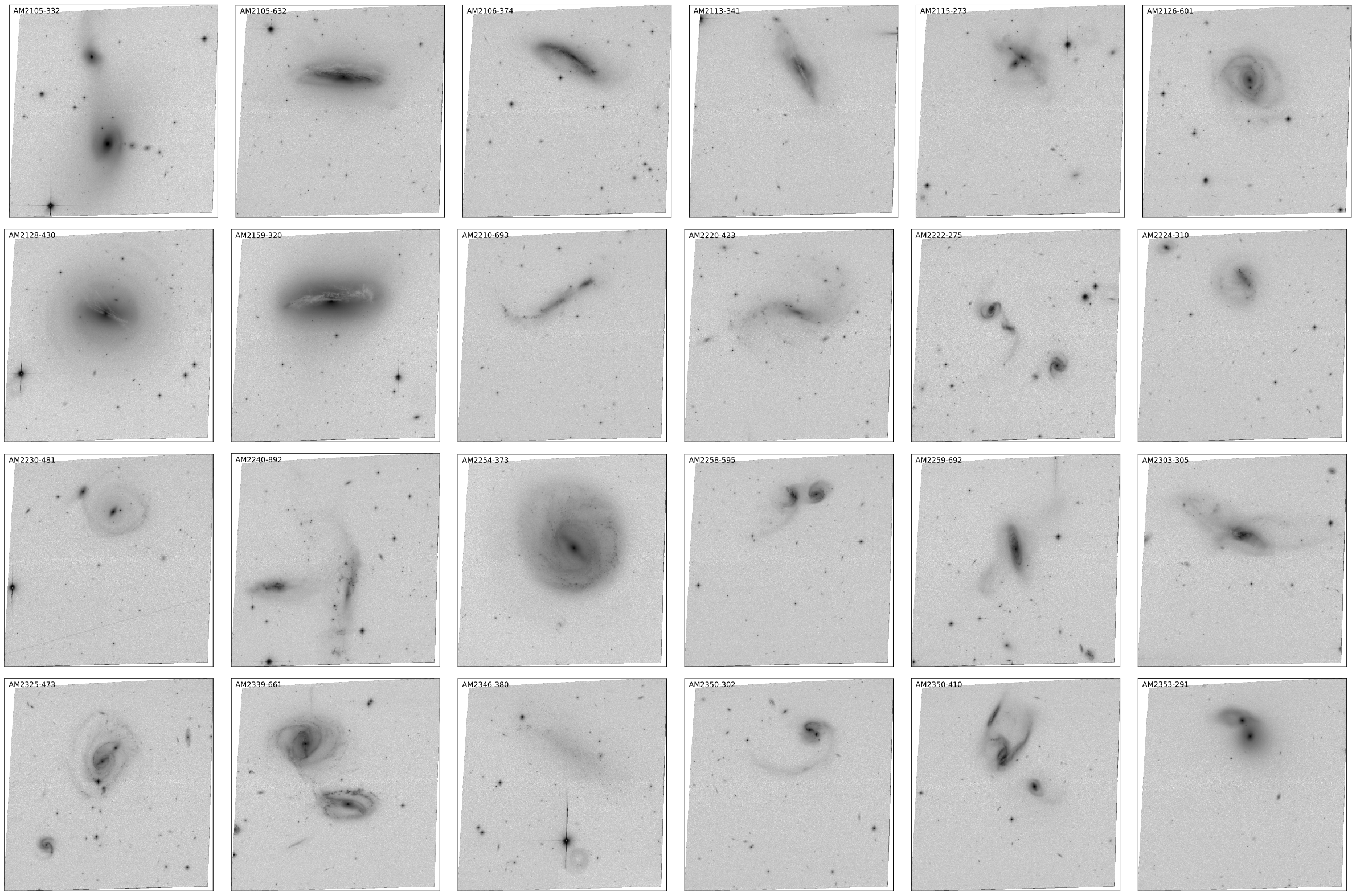

**Figure 4.** [Continued] Thumbnails of HST images in atlas. Galaxy names in the caption link to full-size HST images with imaging at other wavelengths: AM2105-332; AM2105-632; AM2106-374; AM2113-341; AM2115-273; AM2126-601; AM2128-430; AM2159-320; AM2210-693; AM2220-423; AM2222-275; AM2224-310; AM2230-481; AM2240-892; AM2254-373; AM2258-595; AM2259-692; AM2303-305; AM2325-473; AM2339-661; AM2346-380; AM2350-302; AM2350-410; AM2353-291

#### 2.1.2. *Additional Multicolor Images*

One limitation of the atlas images in Figures B2–B2 is that the HST imaging is only available in one bandpass, and thus does not offer color imaging at the full HST resolution. However, combining the ground-based color imaging with the high-resolution HST images from this survey has proven to be a useful strategy for producing informative color images with only a single HST bandpass. When the HST imaging drives intensity, and the ground-based images drive color, the resulting images are extremely useful for interpreting the observed HST morphology (such as identifying dust, young stellar clusters, and foreground stars), although they are not sufficient for more quantitative projects that require accurate colors (say, as one might need for age-dating clusters, or calculating the optical depth of extinction features). Examples of such images can be seen in Figure 5, and are drawn from a number of ESA and NASA releases as summarized in Table 4, as well as the publicly released work of noted astronomical processing wizard Judy Schmidt (see https://flic.kr/s/aHsmAPT18o).

In addition to "hybrid" images constructed from a mix of HST and ground-based images, a few of the targets were sufficiently interesting that they were selected by HST for additional multi-color imaging in support of press releases, as compiled in Table 4. There are also serendipitous observations, where one system was observed independently in F814W by one of the other gap-filler programs (AM2048-571/IC5063; SNAP-15444, PI: Barth), and at least one other was observed subsequently to the submission of the target list for this program (Arp 263, which hosted SN2012A; GO-16179, PI-Filippenko), but before our target was scheduled. Examples of the resulting full-resolution color images are shown in Figure 5, and demonstrate what is possible when the images in this catalog are used as the foundations for larger space-based imaging campaigns.

### 2.2. *Galaxy Identification, Membership, & Redshifts*

The Arp systems targeted by this survey are quite diverse. Some contain a single unusual galaxy, and others contain multiple galaxies. In some cases, our imaging could only include a subregion of the full system (say, for interacting but widely-separated galaxies). Many of these galaxies have been cataloged and characterized in other surveys, and have measured photometry and/or spectroscopic redshifts. To increase the utility of this Atlas, and to improve the completeness and member-ship information for the targets, we have taken advantage of large existing databases to assemble information about the galaxies that fall in each of our images. We forego independent photometry of the galaxies in the HST images, which must typically be matched to the specific purpose for galaxies of this size and resolution, and instead report standardized photometry reported by major databases.

**Table 4.** Color Press Release Images

| Arp Name | Image Credit |
|---|---|
| | HST + Ground |
| Arp 72 | ESA/Hubble & NASA, L. Galbany, J. Dalcanton, Dark Energy Survey/DOE/FNAL/DECam/CTIO/NOIRLab/NSF/AURA |
| Arp 86 | ESA/Hubble & NASA, Dark Energy Survey, J. Dalcanton |
| Arp 91 | ESA/Hubble & NASA, J. Dalcanton; Acknowledgement: J. Schmidt |
| Arp 107 | ESA/Hubble & NASA, J. Dalcanton, Dark Energy Survey/DOE/FNAL/NOIRLab/NSF/AURA, SDSS |
| Arp 122 | ESA/Hubble & NASA, J. Dalcanton, Dark Energy Survey/DOE/FNAL/DECam/CTIO/NOIRLab/NSF/AURA; Acknowledgement: L. Shatz |
| Arp 195 | ESA/Hubble & NASA, J. Dalcanton |
| Arp 248 | ESA/Hubble & NASA, Dark Energy Survey/DOE/FNAL/DECam/CTIO/NOIRLab/NSF/AURA, J. Dalcanton |
| Arp 282 | ESA/Hubble & NASA, J. Dalcanton, Dark Energy Survey, DOE, FNAL/DECam, CTIO/NOIRLab/NSF/AURA, SDSS; Acknowledgement: J. Schmidt |
| Arp 283 | ESA/Hubble & NASA, SDSS, J. Dalcanton; Acknowledgement: Judy Schmidt (Geckzilla) |
| AM 0203-325 | ESA/Hubble & NASA, J. Dalcanton, Dark Energy Survey/DOE/FNAL/DECam/CTIO/NOIRLab/NSF/AURA |
| AM 0329-502 | ESA/Hubble & NASA, J. Dalcanton, Dark Energy Survey/DOE/FNAL/NOIRLab/NSF/AURA; Acknowledgement: L. Shatz |
| AM 0417-391 | ESA/Hubble & NASA, Dark Energy Survey/DOE/FNAL/DECam/CTIO/NOIRLab/NSF/AURA, J. Dalcanton |
| AM 0608-333 | ESA/Hubble & NASA, Dark Energy Survey/DOE/FNAL/DECam/CTIO/NOIRLab/NSF/AURA, J. Dalcanton |
| AM 0619-271 | ESA/Hubble & NASA, J. Dalcanton; Acknowledgement: Judy Schmidt (Geckzilla) |
| AM 1705-773 | ESA/Hubble & NASA, J. Dalcanton, Dark Energy Survey/DOE/FNAL/DECam/CTIO/NOIRLab/NSF/AURA; Acknowledgement: L. Shatz |
| AM 2105-332 | ESA/Hubble & NASA, J. Dalcanton, Dark Energy Survey/DOE/FNAL/NOIRLab/NSF/AURA; Acknowledgement: L. Shatz |
| AM 2339-661 | ESA/Hubble & NASA, J. Dalcanton, Dark Energy Survey/DOE/FNAL/NOIRLab/NSF/AURA; Acknowledgement: L. Shatz |
| AM 2350-410 | ESA/Hubble & NASA, J. Dalcanton, Dark Energy Survey, DOE, FNAL, DECam, CTIO, NOIRLab/NSF/AURA, ESO; Acknowledgement: J. Schmidt |
| | HST Multi-filter |
| Arp 143 | NASA, ESA, STScI, Julianne Dalcanton (Center for Computational Astrophysics/Flatiron Inst., UWashington) |
| Arp 184 | ESA/Hubble & NASA, J. Dalcanton, R. J. Foley (UC Santa Cruz), C. Kilpatrick |
| Arp 263 | ESA/Hubble & NASA, J. Dalcanton, A. Filippenko |
| Arp 300 | NASA/ESA/J. Dalcanton (University of Washington)/R. Windhorst (Arizona State University)/Processing: Gladys Kober (NASA/Catholic University of America) |
| Arp 303 | NASA, ESA, K. Larson (STScI), and J. Dalcanton (University of Washington); Image Processing: G. Kober (NASA Goddard/Catholic University of America) |
| AM 2026-424 | NASA, ESA, Julianne Dalcanton (UWashington), Benjamin F. Williams (UWashington), Meredith Durbin (UWashington) |

Table C2 lists the properties of all galaxies found in each image, compiled from an API query to the SIMBAD Astronomical Database[11] in Fall 2023. The table lists the name of the HST target (Column 1, after shortening "ARP-MADORE" to "AM" for convenience) and the preferred "Main ID" name of the galaxy in the SIMBAD database (Column 3). In Column 2 we list any names from the Arp or Arp-Madore catalogs that are assigned to individual galaxies in the frame; sometimes the Arp identification refers to a single galaxy, but other times there are multiple 'associated galaxies, some of which have been assigned letters (e.g., Arp 293A & Arp 293B) in previous literature, which have then propagated into the official names for the galaxy. Column 2 also includes extended names that have an extra digit for the declination portion of the typical Arp-Madore naming system (e.g., AM1303-3719 is a recognized name in SIMBAD but AM1303-371 is not), even though these are not the official names in the original catalog, which always had an "????-???" format. The reader should also note that we have not culled the table for occasional entries that are for a bright region of a larger galaxy that was deblended into a separate object in a modern automated survey like SDSS. A carefully vetted list of cross-identified major galaxies in the original Arp catalog can be found at D. J. Webb (1996), available through Vizier as catalog "VII/192/", but no such work exists for the much larger Arp-Madore catalog. We discuss some special cases below in Section 2.2.1.

For each galaxy in Table C2, we include the primary SIMBAD classification, the recessional velocity, the angular diameter, and the $B$-band and $K$-band magnitudes, noting that these are not available for all galax-ies. We also include links to an interactive image of the galaxy through AladinLite and to the full galaxy data available on SIMBAD, which includes broader photometric information, associated bibliographic references, and "parent" information about whether a source is part of a larger aggregation of galaxies.

[11] https://simbad.cds.unistra.fr/simbad/

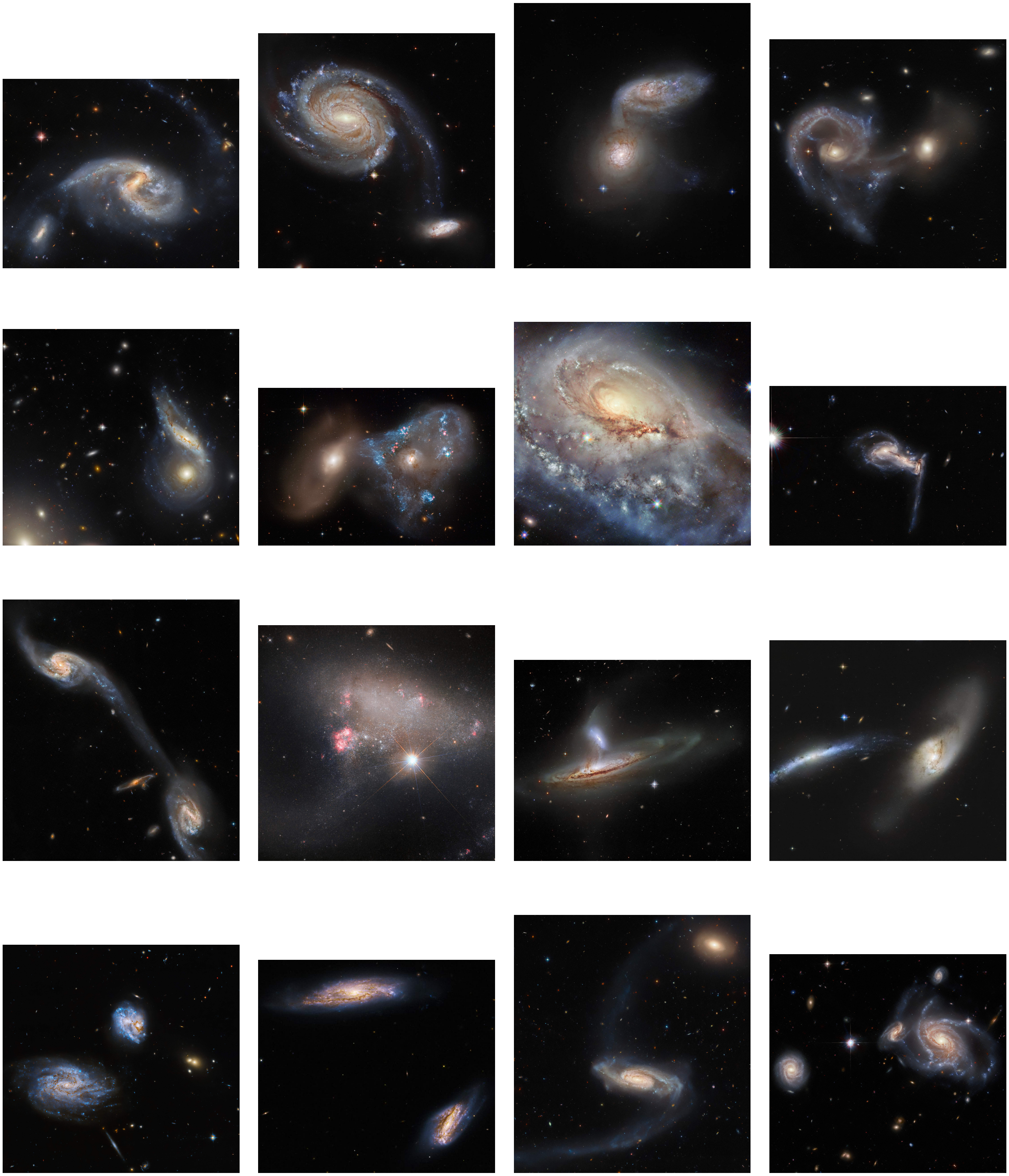

**Figure 5.** Press release images generated from F606W ACS images driving intensity, with color being set by lower-resolution ground-based imaging (Table 4), except for Arp 143, Arp 184, Arp 263, Arp 300 and Arp 303 where HST imaging in other filters was available. From top left to bottom right, systems are: Arp 72; Arp 86; Arp 91; Arp 107; Arp 122; Arp 143; Arp 184; Arp 195; Arp 248; Arp 263; Arp 282; Arp 283; Arp 300; Arp 303; AM0203-325; AM0329-502.

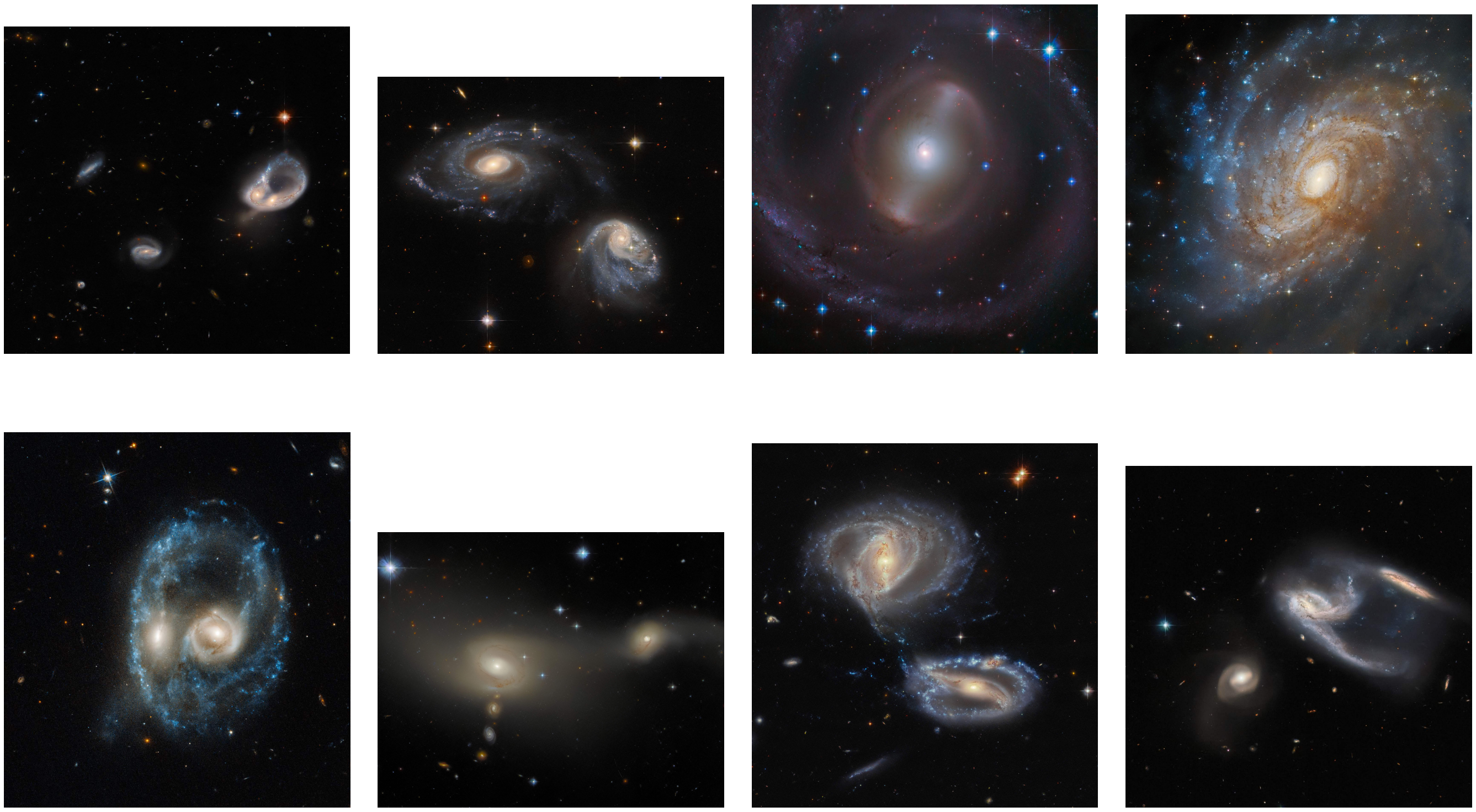

**Figure 5.** Press release images generated from F606W ACS images driving intensity, with color being set by lower-resolution ground-based imaging (Table 4), except for AM2026-424, where HST imaging in other filters was available. From top left to bottom right, systems are: AM0497-391; AM0608-333; AM0619-271; AM1705-773; AM2026-424; AM2105-332; AM2339-661; AM2350-410.

Table C2 also includes our best estimate of whether a galaxy in an image is associated with the Arp system targeted by the observation (Column 4). A flag of "1" in the "Associated?" column indicates a strong likelihood that a table entry is associated with the system, and "-1" indicates the opposite. A flag of "0" indicates that membership is indeterminate. We associate an entry with the Arp system if: (1) there is only one entry; (2) if the primary name in SIMBAD includes the name of the system (see for example, Arp 97 in Table C2); (3) if the recessional velocity is within 500 km $s^{-1}$of the velocity in Table 1. We exclude entries from the system if their recessional velocity is more than 750 km $s^{-1}$away from the velocity in Table 1. Entries without reported recessional velocities or with velocities that differ by 500-750 km $s^{-1}$are considered ambiguous. These velocity criteria are not based on detailed dynamical modeling, but instead use the heuristic that galaxies likely to be associated should have relative velocities that are comparable to the virial velocities of individual galaxies or small groups. We do not use morphological criteria to assign membership, except for a few cases where dramatic mismatches in morphology strongly suggest errors in the catalog data.

#### 2.2.1. *Special Cases*

As is typical with any use of astronomical databases, there are frequently special cases that must be handled individually. While both of the two major databases in widespread use – SIMBAD and the NASA Extragalactic Database (NED) – have largely overlapping information, there are cases where gaps in one are absent in the other. We have relied on SIMBAD for the first pass through the data, but then used NED or the astronomical literature to fill gaps in redshift data. We note several cases here:

*Arp 2* – In addition to being a system in H. Arp (1966), Arp 2 is also the name of a globular cluster from H. C. Arp (1965). Care must be taken to ensure the two are appropriately disaggregated.

*Arp 136* – This edge-on S0 galaxy (NGC 5820) is part of a loose pair with NGC 5821, which falls outside of the ACS image. Arp 136 was included in the H. Arp (1966) catalog because of the nebulosity extending to the south of the galaxy. The HST imaging fully resolves this nebulosity into stars, which, when combined with its large angular size, strongly suggests it is far closer than suggested by the reported recessional velocity of $V_r = 3251\,\mathrm{km\ s^{-1}}$. The nebulosity is therefore most likely to be due to a foreground dwarf galaxy with very low surface brightness, similar to the situation described in J. Yadav et al. (2022) for AM2019-442.

*Arp 156* – We note that Arp 156 appears likely to be a more distant system, based upon its morphology and small angular size, which seems characteristic of a more massive system seen farther away. The velocity of $V_r = 1,888\,\mathrm{km\ s^{-1}}$is based on the 2MASS group catalog of R. B. Tully (2015), but the morphology seems more consistent with the velocity of 10,778 km s$^{-1}$reported by E. E. Falco et al. (1999).

*Arp 204* – The redshift for this system is ambiguous. While SIMBAD did not return recessional velocities for either of the cataloged galaxies, NED has a reported velocity of 4640 km s$^{-1}$for MCG+14-06-025, the larger dwarf irregular from which the tidal stream emerges. It also reports a much larger $V_r = 6298\,\mathrm{km\ s^{-1}}$for the smaller, redder galaxy with the shells (UGC 8454). The galaxies are both highly disturbed, and thus it is far more likely that they are a truly interacting pair (with one discrepant redshift) rather than an extremely fortuitous alignment between a foreground and a background galaxy. We choose to favor the redshift for MCG+14-06-025, which comes from the more recent 2MASS Redshift Survey (M. Bilicki et al. 2014), rather than the older IRAS Redshift Survey redshift (from M. A. Strauss et al. 1992) which targeted the central starburst in the more obscured UGC 8454.

*Arp 290* – The SIMBAD database did not include IC 195, the interacting partner of IC 196. Information from NED was added by hand.

*AM0432-625* – SIMBAD did not report a redshift for 2MASX J04323376-6251479, but NED included one ($V_r$ =16,063 km s$^{-1}$) from the 2MASS redshift survey (M. Bilicki et al. 2014).

*AM2019-442* – SIMBAD reports a recessional velocity of $V_r$ =16,188 km s$^{-1}$for ESO 258-4 (also known as NGC 6902A), centered on the red spheroidal galaxy at the center of the HST image (Figure B2). However, as argued by J. Yadav et al. (2022), this system is a chance superposition between ESO 258-4 and a much closer, diffuse, star-forming dwarf irregular with $V_r = 2963\,\mathrm{km\ s^{-1}}$, which J. Yadav et al. (2022) name "UVIT J202258.73–441623.8", or "UVIT J2022" for short. This interpretation is consistent with the high resolution morphology revealed in the HST image.

*AM2056-392* – This is the only system for which we could not identify a secure redshift. There is second large galaxy in the field ("2MFGC 15905") which does have a recessional velocity of $V_r = 13,539\,\mathrm{km\ s^{-1}}$from the 2MASS Redshift Survey (2MASX J21001518-3919562 J. P. Huchra et al. 2012). Morphologically, there are no signs of interaction to confirm they are neighbors. However, their degree of resolution and comparable sizes and apparent magnitudes are consistent with possibly being members of a single group, so we tentatively assign AM2056-392 a redshift of $V_r = 13,500\,\mathrm{km\ s^{-1}}$in Table 1.

### 2.3. *Photometry*

To measure the depth of the imaging and catalog any observed point sources, all HST data was processed with `hst1pass` (J. Anderson 2022), which is optimized for point spread function (PSF) fitting photometry on undersampled HST images. We do not attempt systematic photometry of extended sources, which, given the complex morphologies of the observed systems, is best left for specific science applications. The point-source optimized photometry, however, is extremely useful for characterizing populations of stars and globular clusters in the galaxies, many of which are sufficiently close for stars and/or clusters be well-resolved.

We adopt fairly permissive detection and fitting parameters for the initial photometry runs, and choose to apply stricter culling criteria at a later step. Our full set of `hst1pass` input parameters are summarized in Table 5. We briefly discuss the most relevant of these here. For in-depth descriptions of all parameters, please refer to J. Anderson (2022). `HMIN` and `FMIN` are finding parameters, which specify the minimum radius at which a pixel can have no brighter neighbors, and the minimum flux in the brightest 2x2 pixels respectively. We set `HMIN` to the minimum allowable value of 2, and `FMIN` to 50. `PSF` specifies either the PSF library file, or else a radius and sky annulus for aperture photometry. The `QMAX`, `CMIN`, and `CMAX` parameters relate to PSF-fitting quality metrics. `Q` is the PSF-weighted absolute sum of the fit residuals, where a perfectly fit source has `Q`=0; we set `QMAX` to 1. `C` is the excess of the central pixel relative to the best-fit PSF, where positive `C` values indicate too-sharp sources, such as cosmic rays, and negative values indicate blurry or extended sources. We choose a range of $-0.2 < \mathtt{C} < 0.1$.

`hst1pass` provides strictly uncalibrated photometry, encouraging users to derive zeropoints for themselves. We apply a Vega magnitude zeropoint of 26.238, corresponding to the 0.25″photometric aperture `hst1pass` uses (86.5% encircled energy in

F606W, R. C. Bohlin et al. 2020) at MJD=59000 (May 31, 2020, the approximate midpoint of the survey). Synthetic photometry was derived under CRDS context hst_synphot_0057.pmap, with `alpha_lyr_stis_011.fits` as the fiducial Vega spectrum.

We cross-match exposure-level catalog pairs for each galaxy using an $0.1''$ search radius, with a maximum magnitude difference of 0.75. We set upper limits on sharpness and quality-of-fit in the combined catalogs as a function of magnitude. Specifically, we require: (1) sharpness <0.0175 for F606W brighter than 23.5, increasing to 0.04 for F606W>25; and (2) quality-of-fit <0.6 for F606W brighter than 21 mag, increasing to 1 for F606W >24.5. We avoid setting a lower limit on sharpness because we do not wish to reject unresolved clusters, which may be less sharp than stars.

### 2.4. *Point Source Luminosities*

In Figure 6, we plot the resulting luminosity function of all sources with reliable measured photometry. The number of sources increases systematically toward fainter magnitudes, but starts to deviate significantly from a power-law by $F606W \approx 25.5$, and is incomplete by more than a factor of two for sources fainter than $F606W \approx 26$, which hereafter we take as our limiting magnitude. We also find increasing issues with saturation for sources brighter than $F606W \approx 18$.

To place these sources in context, in Figure 7 we plot the absolute magnitude of the brightest sources in each image, characterized in two different ways – the magnitude of the third brightest point source in the image (small gray points) and the magnitude that delineates the brightest 2% of the detected point sources (large points color-coded by the total number of point sources in the image). The former is a noisier measure of the brightest luminosity, but may be a more accurate representation of what sources could be available spectroscopically; these and other metrics to characterize the brightest sources are compiled in Table D3. Grey areas indicate regions that are inaccessible, based on the magnitude limits of the data shown in Figure 6. For reference, we include limits indicating the typical $V$-band luminosities of the brightest known stars and the brightest clusters in galaxies (discussed further below). To increase the robustness of the plotted brightnesses, we use only stars that fall within regions of higher than average stellar density; this should reduce contamination from Milky Way foreground stars by eliminating areas that are outside the main body of the targeted galaxies.

Figure 7 shows that the nearest systems in our sample have systematically more detected point sources, due to being nearby enough that (1) numerous but intrinsically fainter stars can still be detected above the magnitude limit and (2) the typical angular separation of sources is large enough that they can be detected individually rather than blending together. In addition, the nearest systems are most likely to be dwarf galaxies (given the effective angular size selection in the Arp catalogs and this survey), which themselves tend to be intrinsically low surface brightness and thus better resolved due to lower stellar blending.

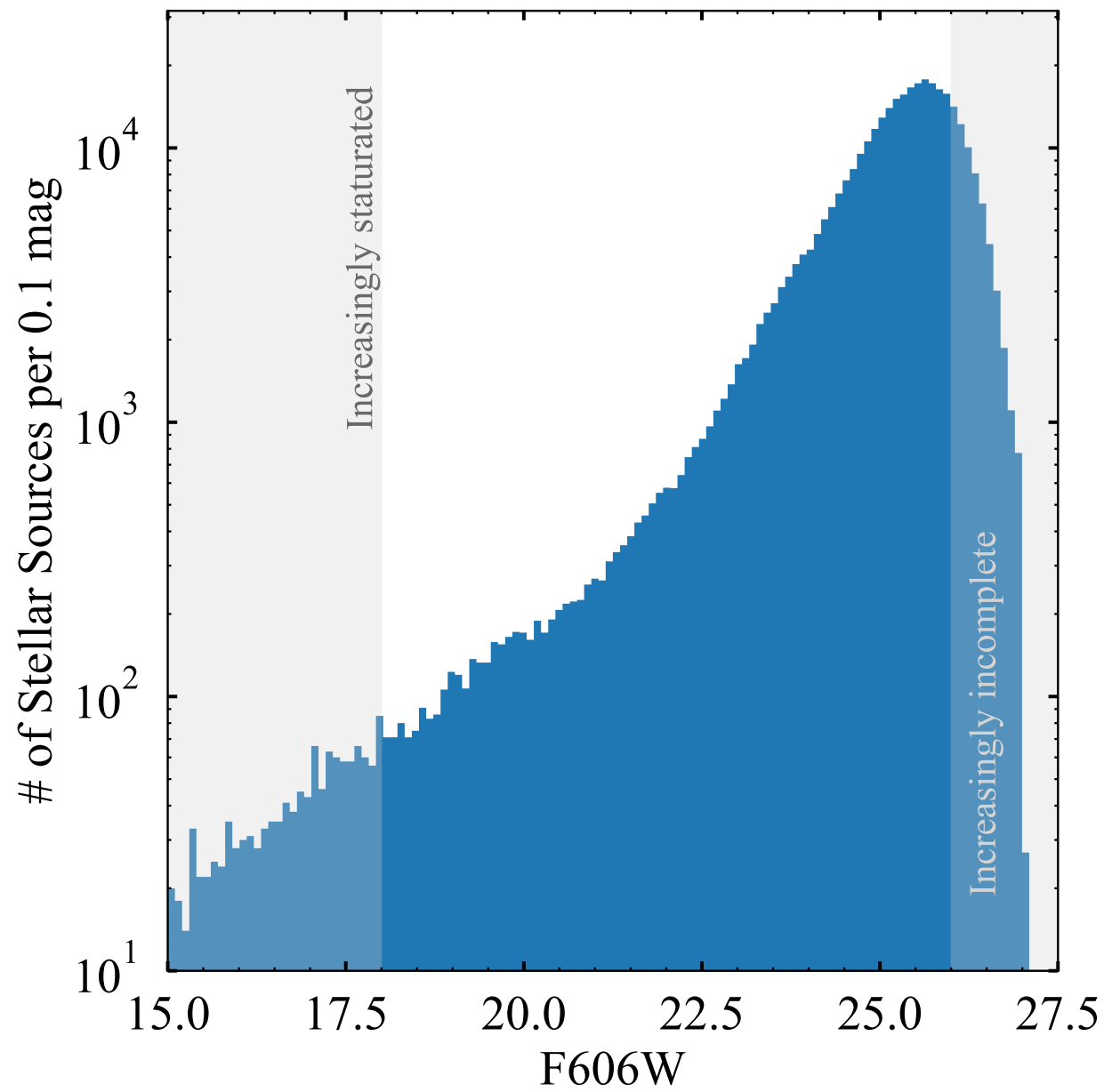

**Figure 6.** Luminosity function of all reliably measured stellar sources in the survey images. Brighter than F606W≈18, stellar sources have an increasing number of saturated pixels, leading to less accuracy at increasingly brighter magnitudes. Fainter than F606W≈26, the luminosity function begins to roll over, indicating increasing incompleteness for dimmer stars. This incompleteness level changes with the local density of stars, leading to deeper photometry in less crowded regions. However, the exposure time limits photometry to brigher than F606W≈27.

Figure 7 is also useful for inferring the likely nature of the detected point sources, based on their absolute magnitudes. We first estimate a characteristic absolute magnitude for the brightest stars. For dwarf galaxies, we do not expect stars to have absolute magnitudes brighter than $M_{F606W} \approx -11$, based on 1 Myr old Padova isochrones at a tenth of solar metallicity, for a 120 $M_\odot$ star (A. Bressan et al. 2012); see also Figure 8, where we show the absolute magnitudes of stars from single stellar populations in the Padova isochrone set, for three metallicities ([Fe/H]=-2,-1,0). At the higher metallicities characteristic of more massive galaxies, the brightest expected magnitudes are typically fainter, with $M_{F606W} \approx -8$ for the youngest stars, which is consistent with measurements of bright blue stars in NGC 253 ($M_{F606W} \approx -8.6$–$-6.8$; M. J. Rodríguez et al. 2018), NGC 247 ($M_{F606W} \approx -7.5$–$-5.8$; M. J. Rodríguez et al. 2019), and 30 Doradus in the LMC ($M_V = -7$; N. R. Walborn & J. C. Blades 1997). Unresolved, non-interacting binaries would tend to be somewhat brighter, but no more than a factor of two in the case of equal mass binaries.

**Table 5.** `hst1pass` parameters

| Name | Value | Description |
|---|---|---|
| `HMIN` | 2 | Minimum distance to brightest neighbor in pixels |
| `FMIN` | 50 | Minimum detection flux |
| `PSF` | `STDPBF_ACSWFC_F606W_SM4` | Library PSF file |
| `GDC` | `STDGDC_OFFICIAL_JFRAME_ACSWFC_F606W` | Distortion model file |
| `DOSATD` | +[a] | Measure saturated star aperture photometry? |
| `PMAX` | 9999999 | Maximum central pixel flux |
| `QMAX` | 1.0[b] | Maximum PSF quality-of-fit |
| `CMIN` | –0.2 | Minimum PSF central excess |
| `CMAX` | 0.1 | Maximum central excess |
| `FOCUS` | –1[c] | Specify or determine telescope focus position |

[a] +/- = True/False

[b] 0 being a perfect fit.

[c] A value of –1 indicates to measure the focus from the image.

Given that some of the point sources will be unresolved clusters, we also use models of stellar clusters to calculate their characteristic absolute magnitudes. Young stellar clusters form with a wide range of stellar masses and thus have luminosities that span several orders of magnitude (see review by M. R. Krumholz et al. 2019) at birth, and fade with age. The most massive, very youngest clusters can be far brighter than any individual star, but lower mass, somewhat older clusters may have integrated luminosities that can be fainter than individual bright massive stars.

We quantify this behavior in Figure 8, where we show expected absolute magnitudes of stellar clusters as a function of age and metallicity for integrated single stellar populations, assuming initial cluster masses of $10^4\,M_\odot$; the upward and downward arrows show the impact of increasing or decreasing the cluster mass by factor of ten. We note that in addition to expected magnitudes from models, there is strong empirical evidence that the brightest stellar clusters in a galaxy have luminosities that scale strongly with the overall star formation rate (SFR), star formation rate intensity, and/or the total number of stellar clusters (e.g., S. S. Larsen 2002; C. Weidner et al. 2004; N. Bastian 2008; T. Vavilkin 2011; B. C. Whitmore et al. 2014). In the most intensely star forming luminous infrared galaxies (LIRGs and ULIRGS), the brightest stellar clusters have $M_V = -14 - -16.5$, and in more typical spirals with $0.1 < \mathrm{SFR} < 10\,M_\odot/\mathrm{yr}$, the brightest clusters have $M_V = -9 - -12$. These dependencies may be due to better sampling of the cluster mass function (i.e., more star formation increasing the chances of forming a very high mass cluster), and/or changes in the mass function itself (i.e., flattening the intrinsic mass function to favor the formation of higher mass clusters). Regardless of its origin, however, we may expect brighter clusters in more intensely star-forming galaxies in the sample.

Based on the above absolute magnitudes for the brightest stars and stellar clusters, the boundaries of the gray "unobservable" regions in Figure 7 suggest that individual stars should be detectable in low-metallicity galaxies with $V_r \lesssim$10,0000 km s$^{-1}$, but only out to $V_r \lesssim$4,000 km s$^{-1}$for more metal rich galaxies. We also show these detectability limits in Figure 8, where the gray regions indicate absolute magnitudes that should be detectable for 5%, 25%, 50%, 75%, or 95% of the systems in the Atlas (from bottom to top), assuming an apparent magnitude limit of 26 mag and pure Hubble flow recessional velocities. For galaxies where stars should be detectable, but are not seen, the absence of such stars may indicate low star formation rates and/or an underpopulated upper IMF. In general, the farther away a galaxy is, less and less of the IMF will be observed for any age population, and fewer stars will be detectable. Detection also will require that stars are sufficiently isolated that they have not been blended together with neighboring sources, favoring lower surface brightness regions of an individual system.

We can perform a similar exercise for the stellar clusters that will be visible in the HST images. Compared to massive stars, young massive stellar clusters are typi-cally more luminous, allowing them to be seen to larger distances. However, they are morphologically indistinguishable from stars for almost all system in this atlas, given that they are physically compact ((with core radii of ∼2±1pc, based on LMC clusters younger than 100 Myr; A. D. Mackey & G. F. Gilmore 2003), and thus are smaller than a ACS/WFC two pixel resolution element for all distances beyond ∼8 Mpc. While morphology is not helpful for separating clusters from stars, one can use luminosity as a discriminant, given that many clusters will be brighter than any plausible stellar model, particularly for the most massive cluster, or for intermediate mass clusters older than ∼10 Myr (Figure 8). The downward and upward facing arrows on Figure 7 indicate regions where the likelihood of being a star or a cluster increases (respectively).

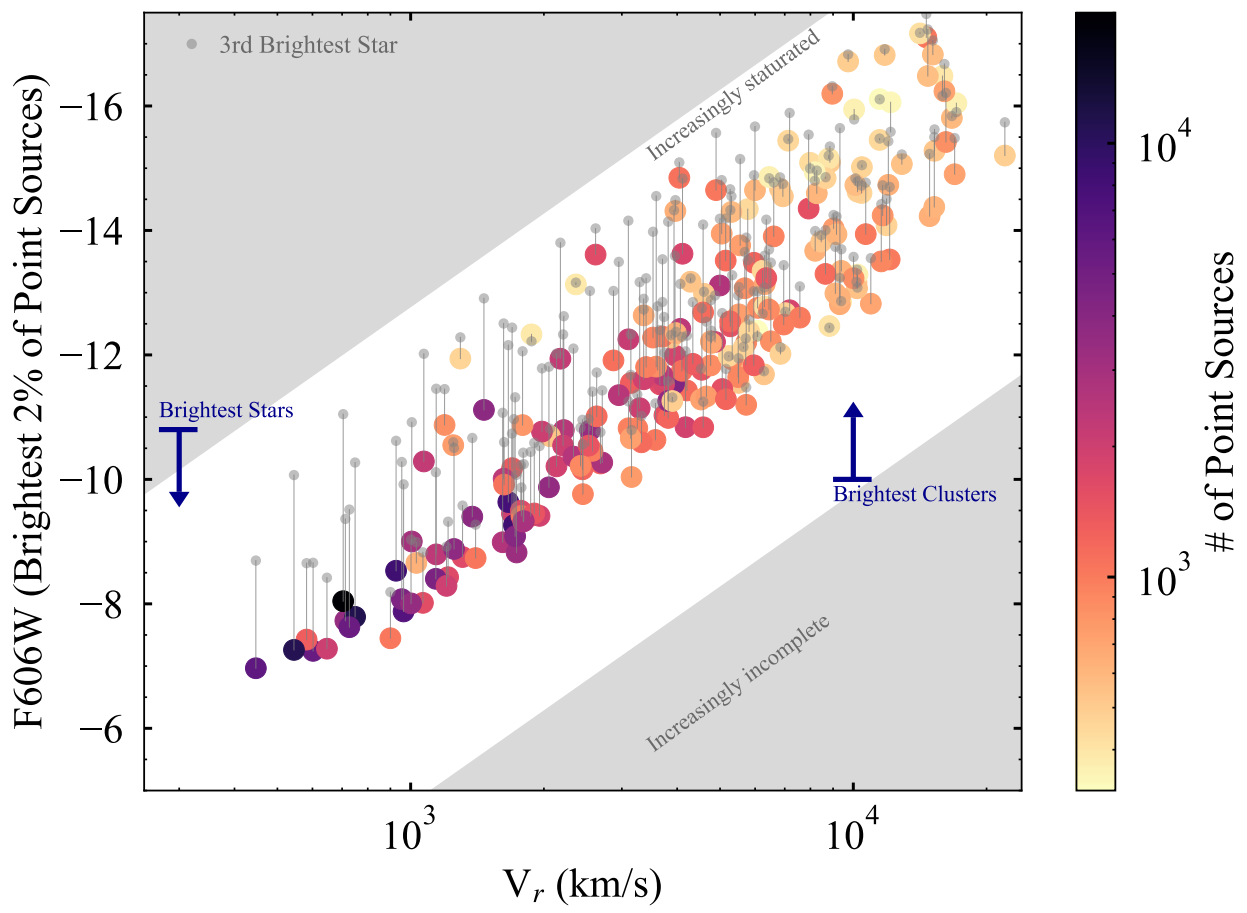

**Figure 7.** The absolute F606W magnitude of the brightest 2% of reliably measured stellar sources in each image, as a function of the recessional velocity of the system. Points are color-coded by the total number of well-measured stellar sources in each system (Table D3), uncorrected for the estimated population of foreground Milky Way field stars. The colored points are connected by thin lines to gray dots, which indicate the magnitude of the third brightest star in the same system, which is a noisier but potentially more accurate indicator of the maximum luminosity of stars or clusters in the system. Generally, there are fewer detected stellar sources in more distant targets, due to the worse resolution at larger distance. The upper and lower limits in blue indicate the typical maximum absolute magnitude of massive stars (at $0.1Z_\odot$, dropping up to two magnitudes fainter for solar metallicities), and the brightest stellar clusters as determined empirically by N. Bastian (2008, , who found that the brightest cluster magnitude was strongly correlated with the overall star formation rate). Systems with recessional velocities beyond 2,000-3,000 km/s are likely to have stellar clusters or blends at their brightest sources, but individual luminous stars can potentially be detected if they are sufficiently isolated. Sources plotted in this figure are restricted to be in regions with the highest 50% of local density among all sources, to minimize contamination from foreground stars.

Figures 7 & 8 suggest that beyond $V_r \sim 2{,}000$ km s$^{-1}$, it becomes increasingly likely that the brightest point sources are stellar clusters, particularly for galaxies with high star formation rate intensities, which are more likely to have recently populated the cluster mass function up to its most massive limits. Even at distances where stars can be detected, the rising cluster luminosity function may increasingly dominate the point source population (i.e., more numerous, lower-mass stellar clusters will have the same luminosity as very rare, high-mass individual stars and binaries). Multicolor photometry (including the UV) and spectroscopy will therefore be useful tools for untangling the recent products of star formation in this sample. Examples of likely cluster-dominated populations can be seen among the more distant systems in Figure 9.

The above discussion focuses on young stellar clusters, but we note that massive old globular clusters will also be detectable in the HST images. Their high mass-to-light ratios make them less detectable than young stellar clusters of the same mass. However, globular clusters typically have higher masses, as is needed for them to remain bound for billions of years. In the Milky Way, the absolute magnitudes of globular clusters can be described as a Gaussian peaked at $M_V \approx -7.4$ mag, with width $\sigma_V = 1.15$ mag (W. E. Harris 2001), and most globulars having luminosities between $-10 < M_V < -4$. The brightest Milky Way globular clusters should be detectable in almost all of these images (out to $V_r \lesssim 10{,}000$ km s$^{-1}$), but globulars at the peak of the distribution would only be detectable to $V_r \lesssim 4{,}000$ km s$^{-1}$. Even in the absence of multicolor imaging, globulars can be distinguished morphologically from their younger, less massive counterparts through their environments, being found preferentially in the halos of massive galaxies, rather than embedded in an obviously star-forming disk. An example of a likely globular cluster system can be seen in Arp 176 (shown in detail in the lower left panel of Figure 9), where there is a population of moderate luminosity point sources clustered around the smooth ellipticals; this population can also be noted in maps of the detected point sources, shown in Figure 12 of Section 2.5 below.

### 2.5. *Characterizing the Distribution of Point Sources*

In addition to photometry catalogs, we have compiled additional summary statistics describing the density of point sources. These can be useful diagnostics for the nature of the point sources, given that the youngest massive stars and stellar clusters tend to be highly clustered, compared to tracers of older stellar populations. These metrics can also help users of this Atlas identify useful systems for follow-up campaigns studying star and cluster formation in the extreme environments present in this sample. The statistics discussed below are all compiled in Table D3.

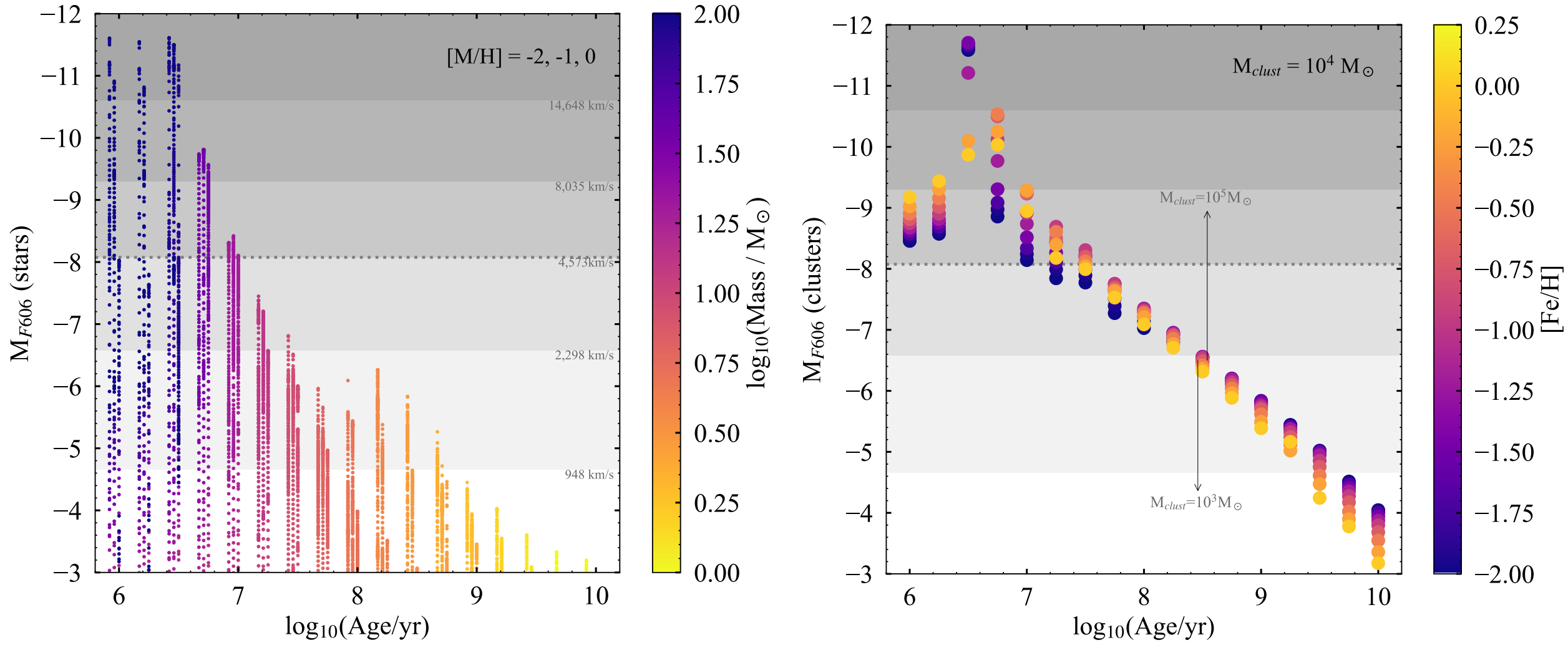

**Figure 8.** The absolute F606W magnitudes of stellar (left) and star clusters (right), as a function their age and metallicity, from the Parsec isochrone sets (A. Bressan et al. 2012) assuming a Kroupa IMF. The stellar magnitudes are shown for 3 metallicities ([Fe/H] = -2, -1, & 0), plotted from left to right at each individual age, and color coded by their initial stellar mass. The cluster magnitudes are shown for a default initial cluster mass of $10^4\,M_\odot$, but can be brighter or fainter at a given age and metallicity for different masses (shown as light gray arrows), with more dispersion expected at low masses due to incomplete sampling of rare massive stars. The gray regions show the absolute magnitudes that can be reached for systems at different distances in the sample. They are drawn assuming pure Hubble flow distances (i.e., $D = V_r/H_0$ with $H_0 = 70\,\mathrm{km\ s^{-1}Mpc^{-1}}$), for values of $V_r$ that correspond to 5%, 25%, 50%, 75%, and 95% of the sample, with the median (i.e., 50%) marked with a dotted line.

#### 2.5.1. *Surface Densities of Point Sources*

We first calculate the local surface density for each point source, $\Sigma_{sources}$, following C. W. Lindberg et al. (2024). We calculate the angular distances to neighboring stars, then weight these distances using a bounded Epanechnikov kernel with an effective radius of $r_e = 10''$ (i.e., half the integrated weight of the 2-dimensional kernel is contained within $r_e$), and take the sum of the weights of all kernels for neighboring stars, and apply the appropriate normalization for kernel area to get the local density in arcsec$^{-2}$. These individual densities are calculated for each star.

We also calculate a kernel density estimated (KDE) interpolated map of the density across the entire image, $\Sigma_{area}$. The resulting maps are similar to, but somewhat sharper than, using an unbounded Gaussian kernel, and better capture local clustering in star-forming regions. Example maps are shown in the left panels of Figure 12, and equivalent maps for all atlas images are in Figure D3, provided in Appendix D. These maps are plotted as a function of position on the image, equivalent to the thumbnail images in Figure 4. The background map shows the interpolated density map, and the individual stellar sources are plotted on top.

For each field, we use the individual local densities for each point source to calculate a number of summary statistics. These include $\Sigma_{max}$, the highest local density of the cataloged point sources, $\Sigma_{sources,50\%}$, the median surface density of the cataloged point sources, and $\Sigma_{area,50\%}$, the median surface density by area, calculated from the KDE-interpolated density maps. The former gives a better estimates of the typical environment of the detected stars and/or stellar clusters, and the latter gives a better estimate of the typical density across the image. A red contour line on the left panels of Figure 12 and Figure D3 is drawn at a density of $\Sigma_{sources,50\%}$. These measures of density are also listed in Table D3.

In addition to the total number of sources per field, we also calculate an estimate of the number of these stars that are foreground Milky Way stars or unresolved background galaxies, which begin to dominate faintward of F606W$\sim$20 mag (M. Postman et al. 1996), explaining the steepening in the slope of number counts in Figure 6. To do so, we take the KDE-interpolated density maps, and calculate $f_{area}(> \Sigma_{area})$, the the fraction of pixels with $> \Sigma_{area}$ as a function of $\Sigma_{area}$ (i.e., such that $f_{area}$ approaches 1 at $\Sigma_{area} = 0$). This function, and the equivalent function for the distribution of densities of individual stars, are shown in the right panels of Fig-ure 12 and Figure D3, as thin red and thick blue lines, respectively. In almost all cases, the red line tracing $f_{area}(> \Sigma_{area})$ shows a sharp rise towards lower densities, when reaching the typical densities of field pixels. We identify the threshold for this upturn by finding the value of $\Sigma_{area,thresh}$ that maximizes the derivative in $f(\Sigma_{area}$ as a function of $\Sigma_{area}$. We take the resulting foreground density $\Sigma_{foreground}$ to be the mean of the interpolated density map pixels that are below this threshold. We then multiply $\Sigma_{foreground}$ by the total area of the image to estimate $N_{foreground}$ the number of field contaminants in the image. This estimate will become less reliable for the few sources that completely fill the ACS field of view.

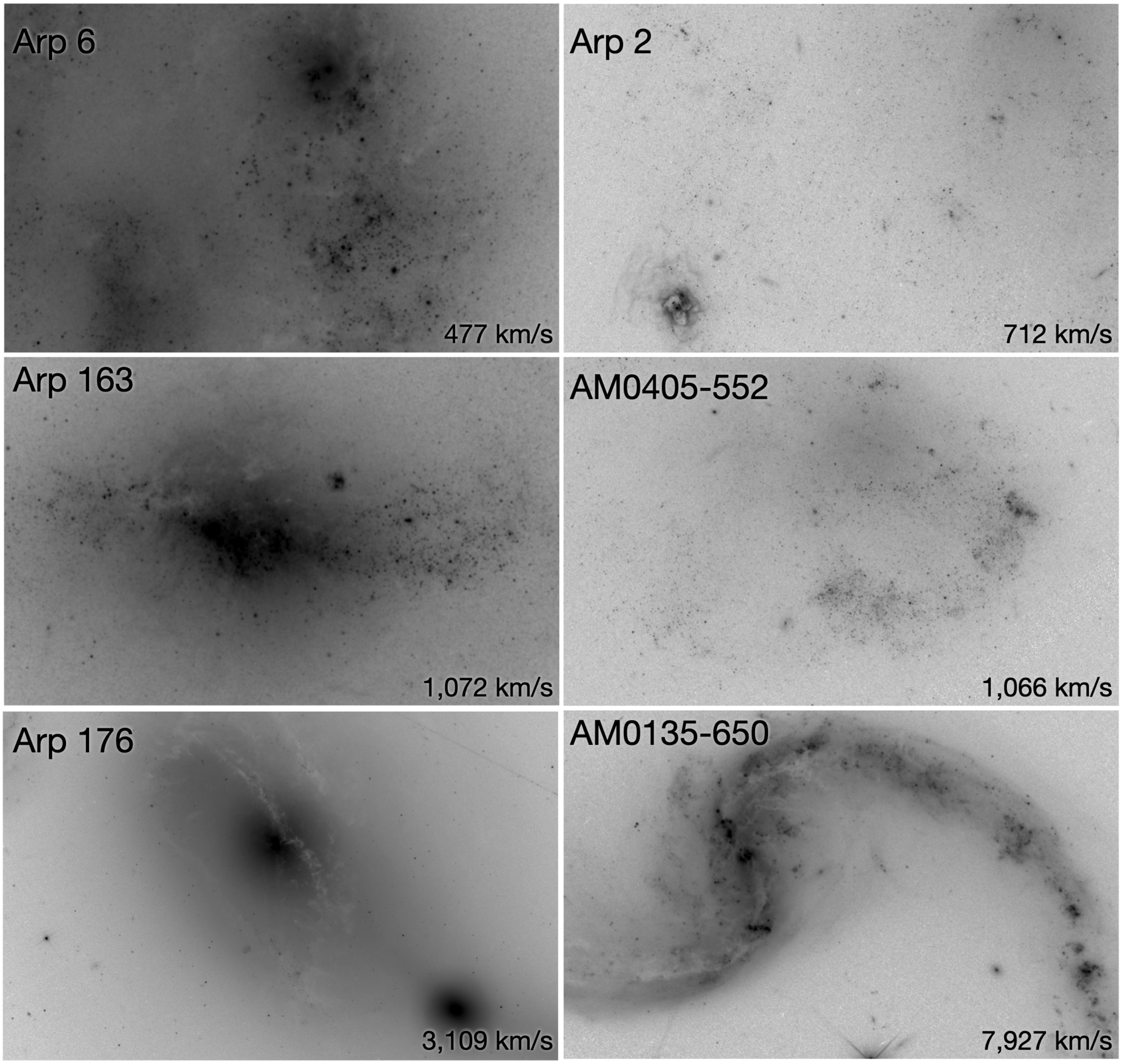

**Figure 9.** Zoomed in F606W images of galaxies from the atlas, showing the variety of morphologies and point source populations at different distances (top left to bottom right), surface densities, and star formation rate intensities, all displayed with a common image scaling, as discussed in Section 2.1. (Top) The top row shows two examples of some of the closest, best-resolved, star-forming systems in the sample, but with very different surface densities and star formation rate intensities. The star formation produces the bright point sources seen in both systems, and drives some of the extended emission from ionized gas, like in the 30 Doradus-like HII region in Arp 2, which also hosts a likely nuclear cluster in the upper right. (Middle) Two star-forming systems at identical distances, but further than those in the top row. The existence of brighter sources in Arp 163 (left) compared to AM0405-552 (right) suggests a higher star formation rate intensity, and/or younger ages, and greater probability that the brightest sources are stellar clusters. (Bottom) The bottom two systems are more distant. The elliptical-dominated system (Arp 176) shows evidence for dense gas, traced by its dust lanes, but does not show a concentration of point sources that would indicate on-going star formation. It does, however, show an extended distribution of point sources, which at this distance is most likely to be a population of globular clusters. The tidally disrupted spiral AM0135-650 (right) shows active star formation which manifests as point sources along the arm, which are likely stellar clusters at this distance; we note that no such clusters are seen on the other arm (not shown, but see Figure B2). Images are oriented such that north is up. See Section 2.4 and Section 3 for more guidance on interpreting these images.

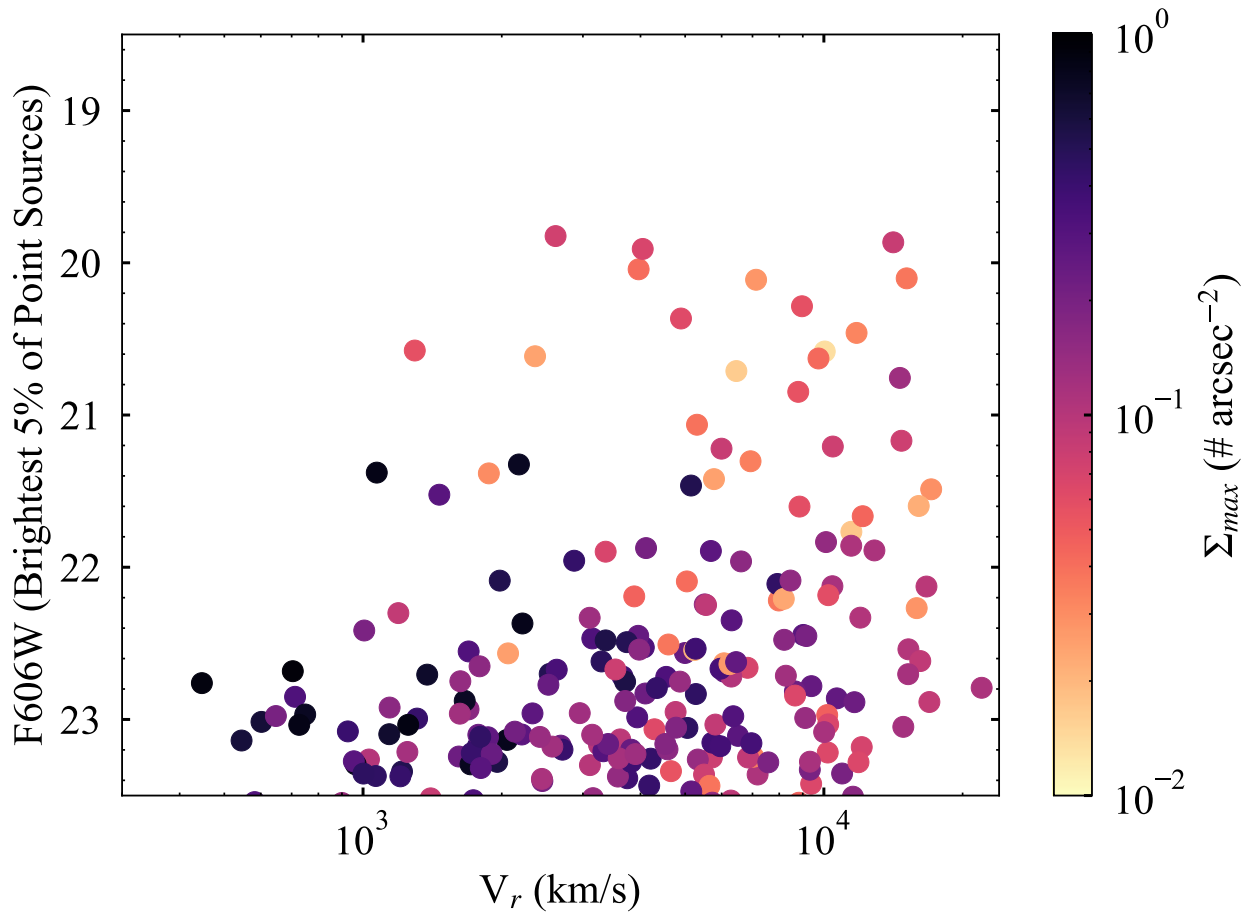

**Figure 10.** The apparent F606W magnitude of the brightest 2% of reliably measured stellar sources in each image, as a function of the recessional velocity of the system. Points are color-coded by the maximum local stellar density of each system (Table D3). Images with low maximum densities are likely to consist of angularly-small and/or poorly-resolved targets, leading their brightest stellar sources to be either stellar clusters or foreground stars.

### 2.5.2. *Properties of Brightest Point Sources*

Finally, we calculate and report statistics for the apparent magnitudes of the brightest point sources in the field. We reduce the effects of field stars by restricting the analysis to stars with higher than the median local density (i.e., $> \Sigma_{sources,50\%}$), which are more likely to be in the system being studied. We report the magnitudes of the 5 brightest sources in this density range, and the magnitude of the brightest 1% of sources, after restricting to sources between 18.5 <F606W< 25.5 to avoid saturation and ensure a uniform magnitude limit unaffected by incompleteness in high density regions. We note, however, that the fluxes of the brightest stars will be rather stochastic, and in systems with intrinsically fainter or less numerous point source populations, there is a greater chance that the brightest sources will be field stars, or in some cases, remnants of bright diffraction spikes that survived various quality cuts. In all cases, direct inspection of the full-resolution atlas images in Appendix B is warranted.

All of the above are presented in Table D3. The complete set of photometric measurements are being released as an associated high level science project on the Mikulski Archive for Space Telescopes (MAST) at DOI:10.17909/176w-p735.

To demonstrate one of the ways the photometric catalogs can be used, in Figure 11 we show close-in views of systems selected to have particularly bright point source populations for their distance (requiring that the top 2% of point sources are brighter than F606W= 22 mag; see Figure 10), high maximum point source densities ($> 0.15\,\text{arcsec}^{-2}$), and at least 3× as many point sources as in the Milky Way foreground. We exclude fields with large numbers of foreground sources (>300 per field), to reduce any issues with contamination from diffraction spikes from bright stars. This combination should naturally identify systems with the youngest and/or most intense star formation. Inspection of the four panels of Figure 11 clearly demonstrates that a selection based on the photometry can highlight a number of systems that seem particularly fruitful for future multiwavelength analyses.

### 2.5.3. *Using Density and Luminosity to Characterize Point Sources*

We can sometimes gain additional insight into the nature of the observed point source populations by looking at their joint distribution in luminosity and local density. In the center panels of Figure 12, we plot the joint distribution of local surface density and apparent magnitude for each image. To facilitate comparisons of the density-dependent luminosity functions, we have normalized the magnitude distributions to have the same values when integrated over all magnitudes brighter F606W=25.5 mag (i.e., where the photometry is highly complete; see Figure 6); histograms of the unnormalized density distributions are shown in the right panels for reference. At each density, we include a red star at the magnitude of the top 2% of the brightest stars, but only if there are more than 50 stars in that density bin. We note that when the red stars are primarily present in the lowest density bins, it is a strong indication that the brightest sources tabulated in Table D3 are either foreground field stars, globular clusters, or possibly nuclear clusters, with AM0630-522, Arp 176, and AM0347-442 being likely examples of each case. We present equivalent figures for all atlas members an Appendix (Figure D3).

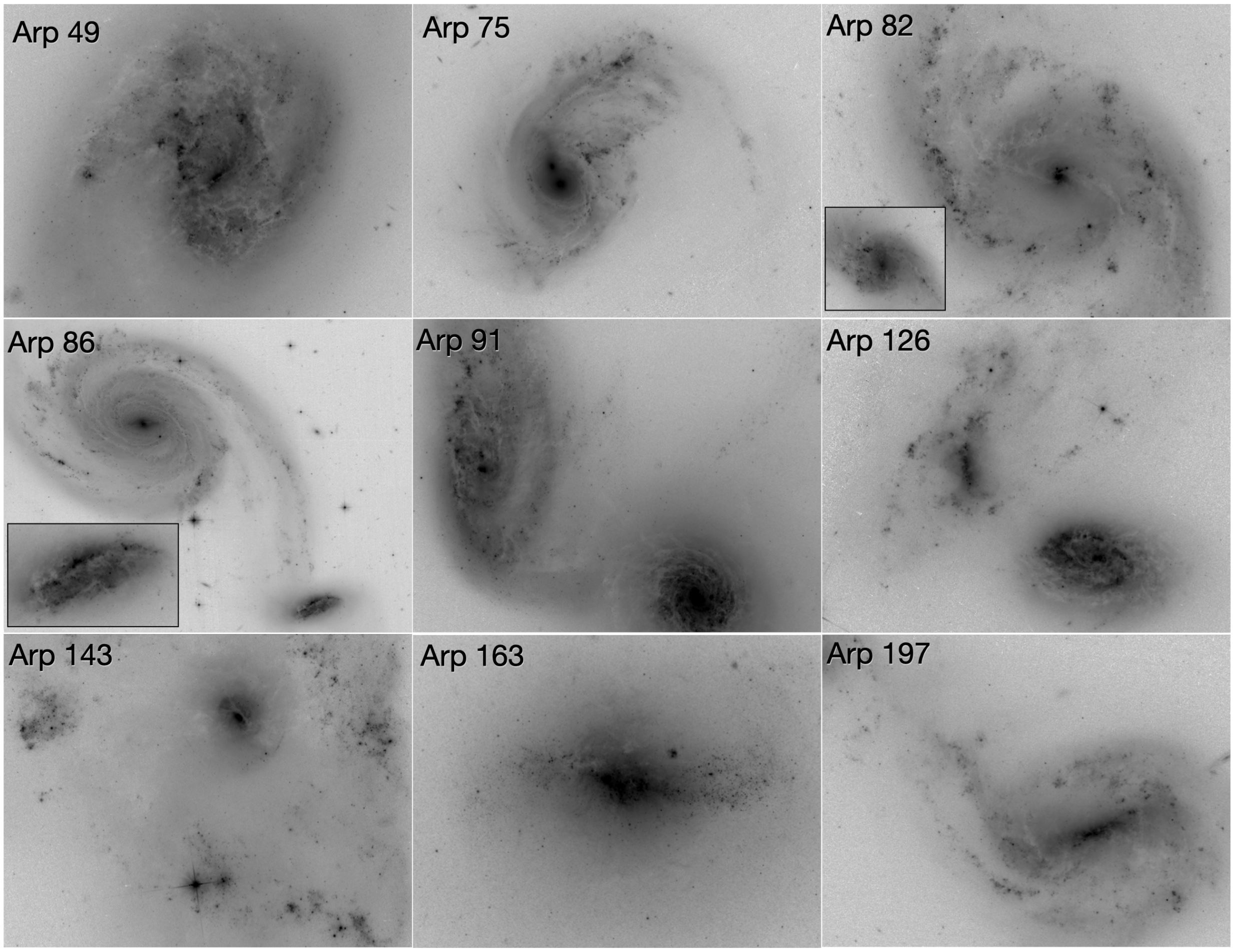

**Figure 11.** Zoom-in views from the atlas images (Appendix B) of systems with notably bright point source populations with high surface density (given their distance) and large numbers of stars above background. The selection favors systems with very recent and/or high-intensity star formation. Fields of view are chosen to highlight areas with the most intense star formation; insets are added when multiple members of the system show evidence for prominent bright star or cluster formation. Systems are ordered by catalog number, and those shown here include: Arp 49; Arp 75; Arp 82; Arp 86; Arp 91; Arp 126; Arp 143; Arp 163; Arp 197. [Continued]

As examples of interpreting these joint distributions, we consider the subset of plots included in Figure 12. At one extreme is Arp 163, a nearby galaxies with morphological evidence for vigorous star formation, also shown in Figure 9. The overall number of point sources detected is very large, with typical stellar densities (blue histogram in the right panels) far above the low density background of Milky Way field stars, and a highly concentrated distribution (left panels). The red stars and the joint distributions in the central panels shows that the typical maximum brightness of the sources increases with density. This behavior is consistent with a picture in which regions of high gas density lead to more vigorous star formation, imprinting a high level of clustering onto the youngest stellar populations. However, as stellar populations age, they tend to diffuse out into the field, occupying regions with lower densities of young stars, at the same time that the brightest stars have evolved, and the characteristic luminosity of the population has faded. Some degree of crowding and blending in regions of high density can also potentially contribute to the increase, particularly in more distant systems.

We can also use Figure 12 to assess the nature of the sources in Arp 163, by considering their luminosity as described above in Section 2.4. If we take the distance modulus (∼31 mag; Table 6) and the apparent magnitude of the brightest 2% of sources in the densest bins (∼20 mag), the brightest sources have absolute magnitudes of $M_{F606W} \approx -11$ mag in Arp 163. Comparing these magnitudes to the expected limits of the brightest stars and clusters, the brightest sources in Arp 163 are too luminous to be probable individual stars, and are therefore likely to be stellar clusters. The fainter sources, however, may well be individual stars.

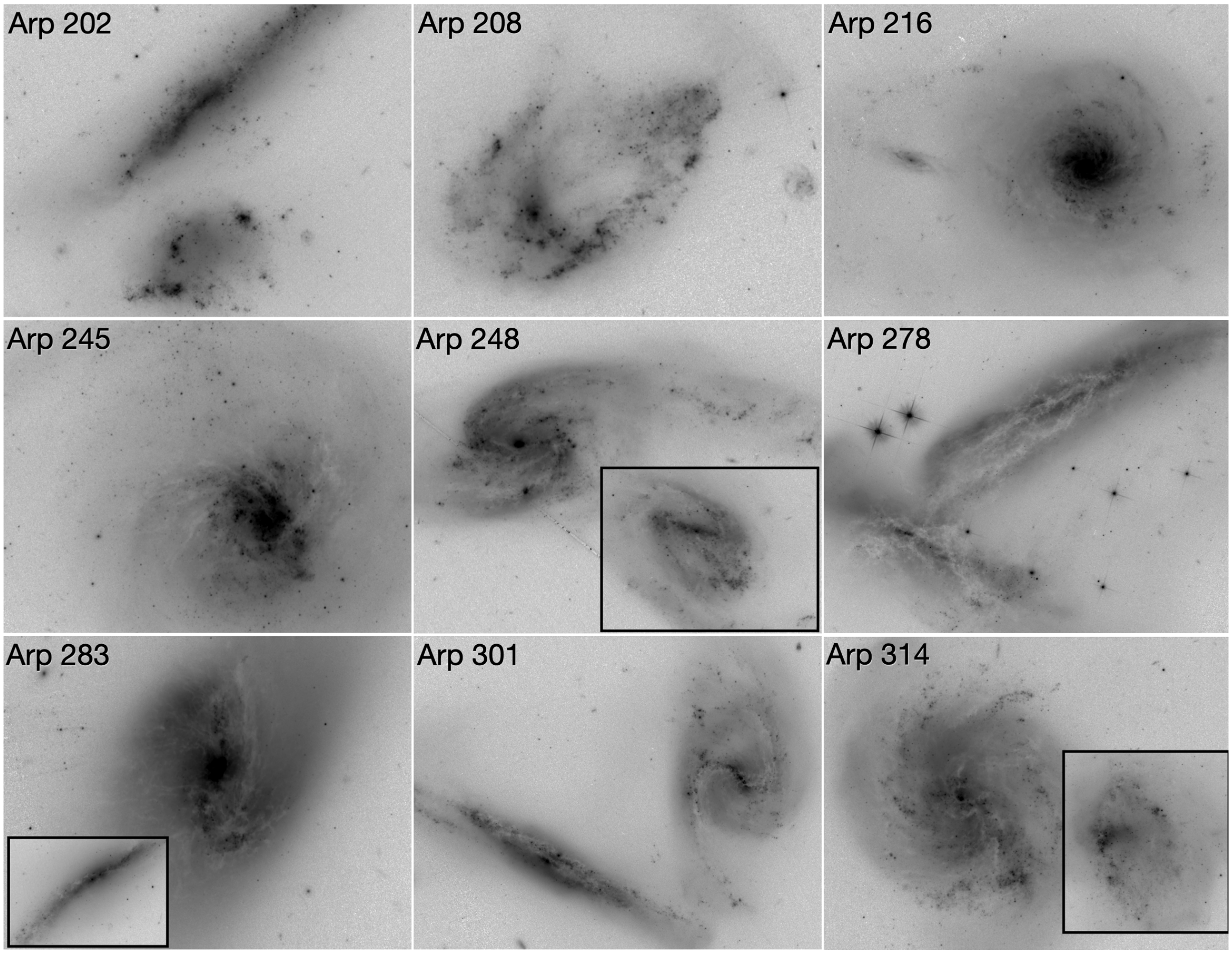

**Figure 11.** [Continued] Zoom-in views from the atlas images (Appendix B) of systems with notably bright point source populations with high surface density (given their distance) and large numbers of stars above background. The selection favors systems with very recent and/or high-intensity star formation. Fields of view are chosen to highlight areas with the most intense star formation; insets are added when multiple members of the system show evidence for prominent bright star or cluster formation. Systems are ordered by catalog number, and those shown here include: Arp 202; Arp 208; Arp 216; Arp 245; Arp 248; Arp 278; Arp 283; Arp 301; Arp 314. [Continued]

## 3. ATLAS IMAGES & THEIR MORPHOLOGICAL INTERPRETATIONS

We present the atlas images for all 216 galaxies in Appendix B (example image at Figure B2, and all images available at `https://doi.org/10.5281/zenodo.16778896` or as electronic figures in the on-line journal). The primary value of the HST atlas images is their exquisite resolution, which offers precision, clarity, and dynamic range. It becomes easy to separate stars from nebulosity (due to gas or background galaxies), to isolate stellar clusters from the background of field stars, and to identify sharp changes in flux due to dust or dynamical effects (e.g., shells, plumes, etc) – none of which are possible when structural features are blurred in lower resolution imaging.

We have therefore reproduced the images at full resolution, to preserve the maximum information contained within. We have chosen a stretch that allows much of the full dynamic range of the images to be visible, from sharp point sources to subtle variations in the unresolved starlight. From these images alone, it is possible to glean a remarkable amount of physical intuition for the likely state of the imaged galaxies, even without spectra, simulations, or physical modeling. The interpretative power of such images increases even further when viewed in tandem with the low-resolution images at other wavelengths, which are included in the figures as well.

In practice, however, we live in an era where astronomers interact with data primarily through catalogs and databases, or in such large quantities that there are far fewer opportunities to simply look at and ponder images of individual well-resolved galaxies. We therefore choose to take a somewhat pedagogical diversion to support interested readers in growing the intuition needed to interpret the images presented here.

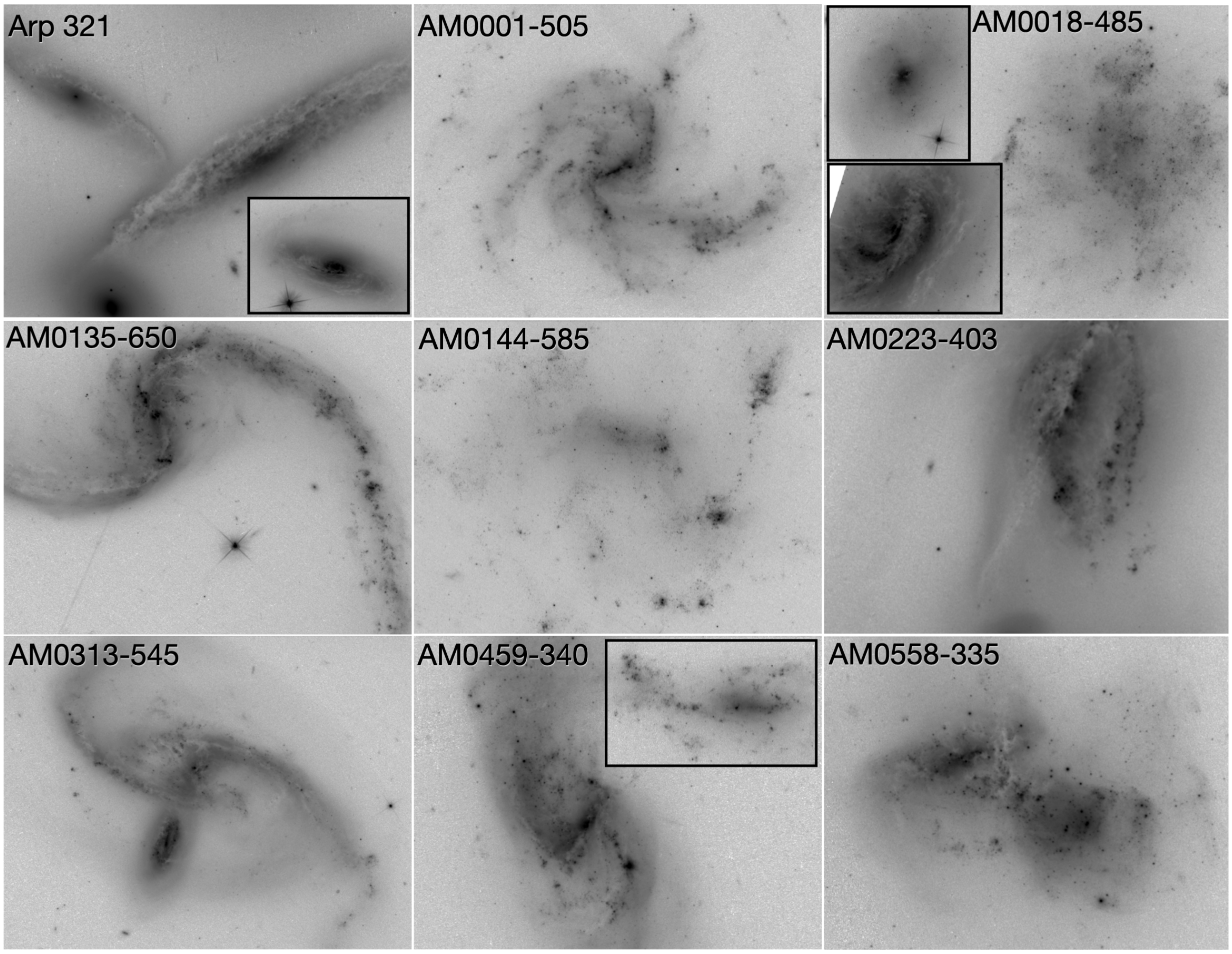

**Figure 11.** [Continued] Zoom-in views from the atlas images (Appendix B) of systems with notably bright point source populations with high surface density (given their distance) and large numbers of stars above background. The selection favors systems with very recent and/or high-intensity star formation. Fields of view are chosen to highlight areas with the most intense star formation; insets are added when multiple members of the system show evidence for prominent bright star or cluster formation. Systems are ordered by catalog number, and those shown here include: Arp 321; AM0001-505; AM0018-485; AM0135-650; AM0144-585; AM0223-403; AM0313-545; AM0459-340; AM0558-335. [Continued]

## 3.1. *How to "Read" the Atlas Images*

### 3.1.1. *What produces the light?*

The first step in interpreting an image is understanding which astrophysical sources will be emitting light in the observed waveband. The ACS F606W filter is quite broad, extending from ∼4760Å to ∼7080Å . For the low redshift ($z < 0.1$) systems imaged here, the majority of light in the filter will be from photons seen at close to their rest wavelength.

*Stars:* In the broad F606W filter, the majority of the detected photons will have originated from stars. Both hot ($T_{eff} > 10,000$K) and cool ($T_{eff} < 4,000$K) stars have significant flux in this bandpass and thus the images will be sensitive to both the younger (hotter) and older (cooler) stellar populations that contribute significantly to the F606W luminosity. As discussed above in Section 2.3 and Figure 7, the starlight will be in the form of: (1) individual bright stars (or binaries) resolved in the image; (2) stellar clusters, which may appear as point sources but not have individually resolved stars within them; (3) unresolved stellar populations, where stars' light is blended together because they are too tightly packed on the image to have any individual star or cluster rise above the background level of a resolution element [12]. Individual stars are most likely to be resolved in closer galaxies, and in regions of lower intrinsic surface densities of stars. The latter favors resolving brighter (i.e., rarer) stars in lower surface brightness regions (see, for example, Arp 163 and AM0405-552 in Figure 9, where bright stars away from the galaxy center are resolved individually, but stars in the denser central regions are blended together).

[12] There are edge cases where the stellar population will be only marginally unresolved. In these cases — such as those used for surface brightness fluctuation distance measurements — no individual star rises above the background, but the brightest stars below the detection limit are rare enough that they add

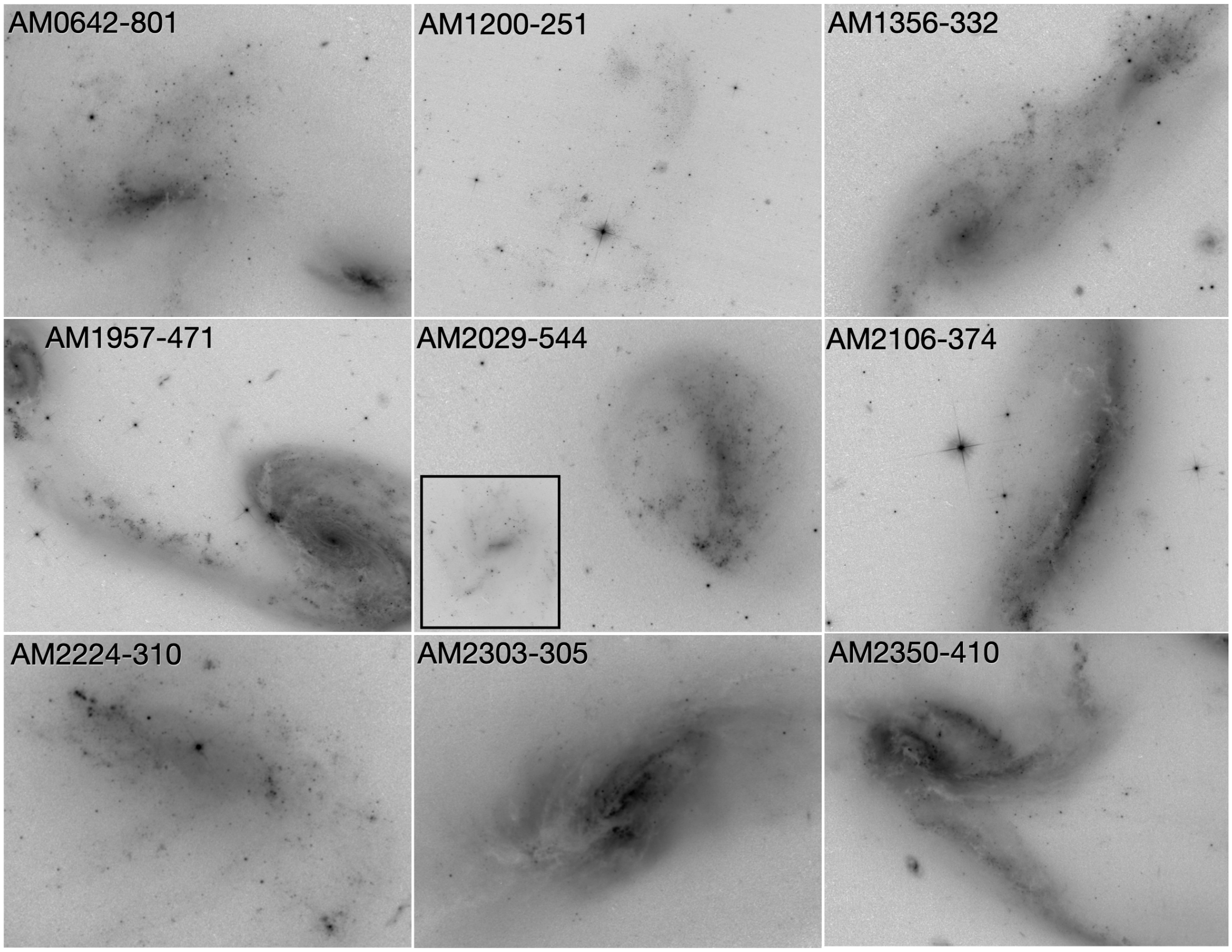

**Figure 11.** [Continued] Zoom-in views from the atlas images (Appendix B) of systems with notably bright point source populations with high surface density (given their distance) and large numbers of stars above background. The selection favors systems with very recent and/or high-intensity star formation. Fields of view are chosen to highlight areas with the most intense star formation; insets are added when multiple members of the system show evidence for prominent bright star or cluster formation. Systems are ordered by catalog number, and those shown here include: AM1200-251; AM1200-251; AM1356-332; AM1957-471; AM2029-544; AM2106-374; AM2224-310; AM2303-305; AM2350-410.

*Gas:* Ionized gas may also appear in F606W images, with the H$\alpha$(6563Å) recombination line and [OIII](5007Å) forbidden lines often having large enough equivalent widths that they contribute significantly to the flux in the F606W bandpass, for recessional velocities of $\lesssim$23,000 km s$^{-1}$. As such, emission lines from warm ($\gtrsim$10,000K) gas (produced by photoionization and/or shocks) may be detectable as nebulosity in young star-forming regions (e.g., HII regions), typically appearing as extended filamentary emission, rather than a point source. Strong emission lines can also be produced by AGN, or by individual massive stars (Be stars, Wolf-Rayet stars, etc). However, unlike ISM emission, such sources will typically need spectroscopy or observations from other wavebands to be cleanly separated from more common point sources (i.e., stars, unresolved stellar clusters, and/or nuclear star clusters).

*Dust:* The F606W filter is sufficiently blue that photons from background stars and gas can be significantly absorbed or scattered by dust along the line of sight. In most conditions, dust is coupled to cool dense gas, significant spatial variance to the background level, above what is expected for Poisson variance. When visible to the eye, these fluctuations indicate a galaxy that is likely closer and/or that has younger unresolved stars, say, from a population of luminous, higher mass AGB stars that one might expect $<$ 1Gyr post-starburst.

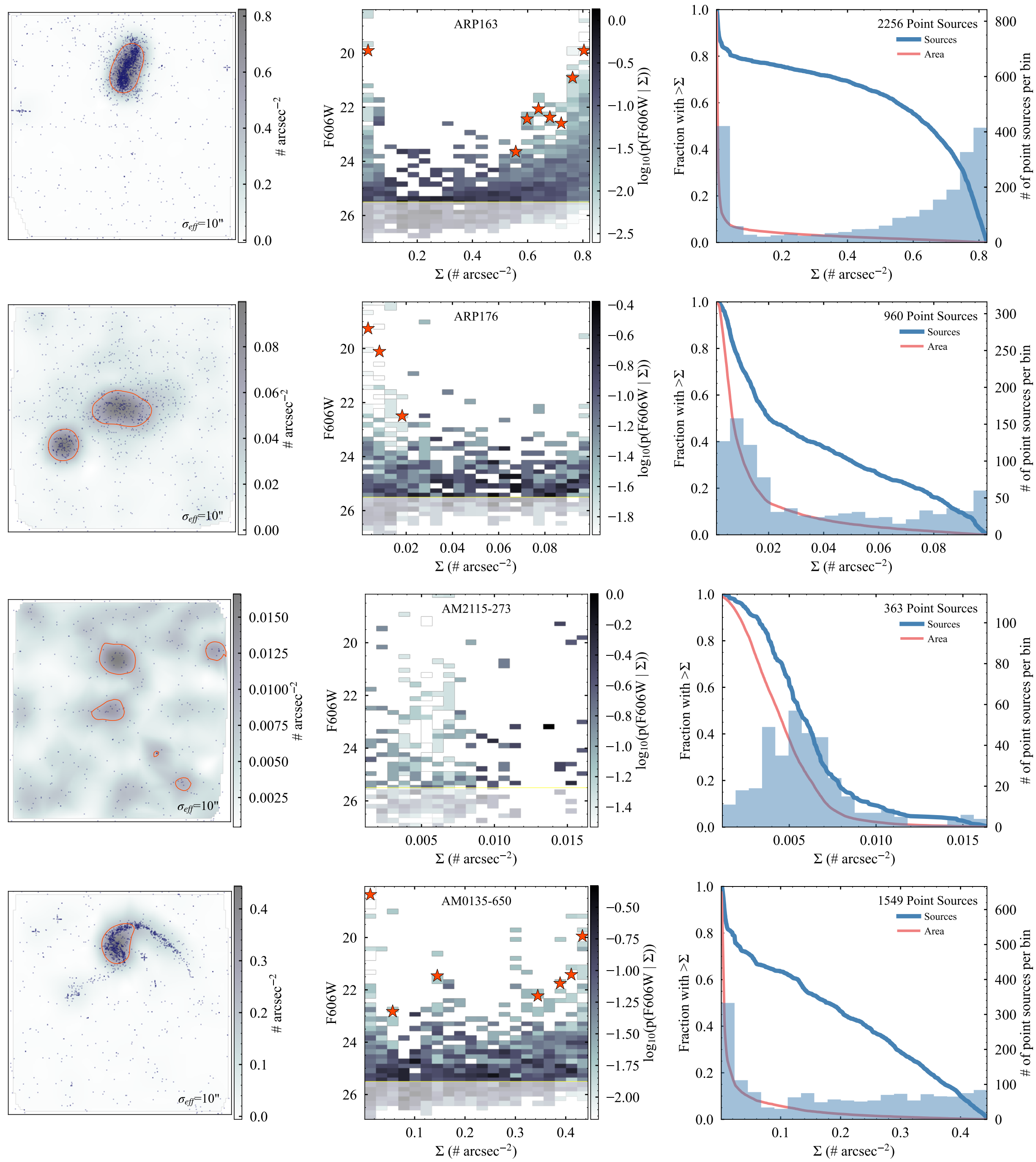

**Figure 12.** Examples of the summary photometry plots found in Appendix D, for a selection of systems with a variety of distances (increasing from top to bottom) and star formation intensities; close-ups of three of these systems are shown in Figure 9, but oriented with north up. The left panels show the location of all point sources (which may be either stars or stellar clusters), plotted on top of a KDE-interpolated density map, with the same orientation as the thumbnails in Figure 4; the red contour is drawn at $\Sigma_{sources,50\%}$, the median surface density of the cataloged point sources. The middle panel shows the density-dependent distribution of point source magnitudes, normalized to have equal probability density brighter than $F606W = 25.5$ mag, in each bin of local surface density; red stars indicate the magnitude of the top 2% of the brightest sources in any density bin with more than 50 stars brighter than $F606W = 25.5$ mag. The right panel shows the histogram (in blue) of local density of all point sources, and its normalized cumulative distribution (thick solid blue line). The thin red line shows the equivalent cumulative distribution of the interpolated density at all pixels in the image; this area-weighted distribution rises sharply at low densities, when the characteristic background level of foreground field stars is reached. Systems with few associated point sources (i.e., no detectable star formation) will have similar cumulative distributions for the red and blue curves (e.g., AM2115-273, and systems with vigorous, highly-clustered star formation will have blue curves that depart dramatically from the red curves, particularly at high densities (e.g., Arp 163 and AM0135-650)). More distant systems will have fewer point sources at the same star formation rate, due to biases against detecting fainter, but more numerous, stars and stellar clusters (Figure 8), and, biases against detecting individual sources when they are highly crowded and blend together.

making dust absorption and/or scattering a useful diagnostic of the presence, location, and/or structure of dense atomic and/or molecular gas. Dust attenuation is a common morphological feature within this atlas, but emission from scattering is typically much less apparent. One notable exception in this atlas is AM2048-571 (i.e., IC 5063), which has strong crepuscular rays from dust-scattered continuum emission from a central AGN – along with dark shadowed "rays" from dust absorption near the nucleus (see W. P. Maksym et al. 2020).

We note, however, that the overall visibility of dust attenuation in the images depends significantly on the geometry of both the dusty ISM and the stars (e.g., A. Natta & N. Panagia 1984; B. G. Elmegreen & D. L. Block 1999). The dust attenuation is most easily seen when: (1) the gas is clumpy or structured on resolvable scales, producing well-defined, dark features across the background light and/or (2) when the background light is smooth, making even subtle attenuation features stand out. At the other extreme, dust attenuation may be harder to detect when: (1) the background emission is heavily structured and/or low surface brightness, or, (2) when the bulk of a stellar population lies in front of, rather than behind, the dust. This latter feature makes dust a useful indicator of the relative location of galaxies (see examples in W. C. Keel et al. 2013; R. E. White et al. 2000), which in interacting galaxies can clarify which galaxy is in front of the other. In all of these cases, the detection of dust structures in images is greatly enhanced by high-resolution, which better isolates dense gas clouds and separates them from other features in the background (e.g., W. C. Keel & R. E. White 2001).

#### 3.1.2. *Interpreting Morphology*

With the above basic grounding of astrophysical sources of flux in the F606W bandpass, we now discuss in more detail how to interpret common morphological features associated with the various sources.

*Identifying recent or on-going star-formation:* There are two widely-used approaches for identifying star formation and/or young stellar populations in galaxies. The first relies on colors, with the simplest implementations using observations in only 2 wavebands (e.g, $B-V$ colors) and more complex approaches using multiple wavebands and fits of the full spectral energy distribution (SED). The second approach to identifying recent star formation relies on measurements in only a single waveband, but one that has a tight, unambiguous connection to star formation, such as the flux in the H$\alpha$ line, in a mid- or far-infrared filter (e.g., 24$\mu$), or in the far UV. Our observations in the single F606W filter do not fit either of these paradigms however. As discussed above, while young, hot, luminous main sequence stars contribute to the flux in F606W, so do much older cool evolving red giant (RGB) and asymptotic giant branch (AGB) stars. There is thus no way to infer the underlying stellar temperatures from our observations alone[13].

However, the high-resolution morphology in the atlas does offer a number of ways to infer the presence of recent or on-going star formation, and some indication of age. The first, and perhaps most direct one, is the presence of resolved point sources, which are individual stars or stellar clusters. As shown in Figure 8, the luminosity of both stars and clusters is a strong function of the age of the stellar population. Generally, stars and clusters younger than 10 Myr old can be an order of magnitude brighter than those that are 100 Myr, and two orders of magnitude brighter than those that are a gigayear old. As such, any galaxy that shows luminous ($M_{F606W} \lesssim -8$) point sources almost certainly has formed stars in the most recent 10-50 Myr. While it requires detailed modeling to convert the observed point-source luminosity function into an age distribution, the presence of the point sources alone is sufficient to identify where recent star formation has occurred, with more luminous point sources suggesting more recent and/or more intense star formation. The only regime where these associations start to break down is for the most distant 5% of systems in this atlas, where the star formation must be both very recent ($\lesssim$ 10 Myr) and very intense, such that large numbers of the most massive stars and/or stellar clusters were formed.

A second indicator of the age of star formation is the degree of spatial clustering among point sources. Stars and clusters are typically formed in the densest regions of the turbulent ISM, and are thus highly clustered when young. However, their velocity dispersions and the expulsion of gas leads to the rapid dispersion of the recently formed stars, leading to increasingly less clustering as the stellar population ages and fades. This process leads to the brightest point sources preferably being found in the most highly clustered regions, as can be seen for Arp 163 in the top row of Figure 12.

A third morphological indication of recent star formation is nebulosity. As noted above, the H$\alpha$ recombination line falls within the F606W bandpass for all galaxies in our sample. In HII regions, this line is produced by ionized gas surrounding luminous hot O-stars, producing a distinctive morphology when the HII region is large enough to be resolved at the distance of a galaxy. A particularly striking example can be found in the nearby galaxy Arp 2 in Figure 9, in the lower left of the image. In more distant galaxies, this nebulosity may appear as barely resolved extended emission around very young stellar clusters. We note that because the timescale for this emission is quite short ($\lesssim$ 5 Myr), H$\alpha$-driven nebulosity will be less common than bright stars and clusters, which are typically detectable for $\sim$10$\times$ longer. This limitation is further increased by the difficulty of detecting the emission-line nebulosity when the flux from the line is dwarfed by the local continuum emission from stars, making weaker or lower surface brightness H$\alpha$ more challenging to detect.

[13] Planned wide-field surveys with Euclid and Roman offer the possibility that the needed high-resolution imaging in other filters will become available for many of these systems. The imaging here will provide opportunities for significantly improved photometric quality and depth for point sources from the upcoming longer-wavelength surveys, following the joint reduction strategies demonstrated in B. F. Williams et al. (2014).

*A demonstration of analyzing recent star formation:* As a demonstration of the above, we can consider the above three indicators of star formation to interpret the close up of AM0405-552 in the center right panel of Figure 9. The image shows point sources across the image, suggesting widespread star formation over the past $\lesssim 50-100$ Myr. However, there are notable variations in the brightness, density, and degree of surrounding nebulosity for the point sources in the image. The brightest region is on the far right, and shows highly clustered, very luminous stars with significant surrounding nebulosity, all indicating that this is a site of significant very recent star formation. There are also somewhat weaker indications of nebulosity in the topmost star forming region, and in some locations in the U-shaped arc of point sources (extending down and too the left of the stronger star forming region). These features also indicate the likely presence of young ionizing O-stars, but, given that the nebulosity and the point source density is weaker than the large star forming region on the right, these are likely intrinsically less massive and/or slightly older star forming regions.

We also see well-resolved point sources on the left half of the image. In contrast to the point sources on the right, however, these sources are typically fainter, and more spread out. This likely indicates a region where star formation has recently slowed down, and initially highly-clustered stars are in the process of dissolving into the field, while the more massive stars have evolved away, leaving a fading stellar population lacking the most luminous stars.

Finally, above the center of the image, there is a broad, largely smooth area of elevated surface brightness. This feature corresponds to the center of the galaxy, and is likely where the dominant stellar mass of the galaxy resides. However, the fact that there are very few detectable stars in the region suggests that there has likely been little significant star formation in the past $\sim$0.5Gyr, based on the distance and the left panel of Figure 8. Given the higher than average surface brightness and the higher stellar age of this region, this central region is where the bulk of the stellar mass is likely to reside, but there must also be a more an extended gaseous reservoir that fuels the observed star formation. We do not see any strong dust absorption, however, which reflects in part the low surface brightness of the background (i.e., such that any dust clumps would be low contrast and hard to detect), but may also indicate an overall low dust-to-gas ratio, consistent with the low metallicities expected for dwarf galaxies (which, morphologically, this system clearly is).

We can take the above insights — gained solely from the F606W image — and compare them to the multiwavelength data shown in the bottom of Figure B2. The optical image strongly confirms what we inferred morphologically, placing the bluest (and thus youngest) star formation in the U-shaped arc, and the most intense star formation on the right. However, the GALEX image, which should be sensitive to $\lesssim$ 100 Myr-old star formation, indicates that there has been pervasive star formation throughout the region on the left, visible as a higher surface brightness arc on the GALEX image. The WISE W1+W2 image also confirms the bulk of the stellar mass as being in the central smooth region we identified. This exercise nicely indicates how just inspection of single HST images can reveal quite nuanced (albeit qualitative) insight into recent star formation across a galaxy.

*Star-forming versus quiescent?* Morphologically, recent star formation produces compact knots of point sources, which can be either young massive stars or stellar clusters. Even when a galaxy in this atlas is too far way for point sources to be detected individually, the clustered star formation (and the associated increase in luminosity) can produce a detectable localized increase in the surface brightness (e.g., see the southern large spiral in Arp 150, shown in Figure B2). Evidence for nearby dust absorption, which is indicative of molecular gas, further increases the certainty that on-going or recent star formation is present. All panels in Figure 9 except Arp 176 show convincing evidence for on-going or very recent star formation.

In contrast, the morphological evidence for quiescence is "smoothness". Recent star formation is invariably highly clustered, but this clustering is transient, as the stars' velocity dispersion and the unbound nature of most stellar clusters leads the aging young stars to disperse rapidly. They become fainter and more distributed, and merge into the background of older stars, becoming largely morphologically indistinguishable, especially as their initially low mass-to-light ratios rapidly increase as the most massive, short-lived stars evolve and disappear. This transition can happen reasonably quickly (empirically, $\lesssim$ 1Gyr in nearby spirals; A. R. Lewis et al. 2015; M. Lazzarini et al. 2022), and if no new star formation takes place, the resulting stellar distribution will be smooth, spread out by the stars' velocity dispersion, which increases with age. In Figure 9, examples include Arp 176, or some of the extended low surface brightness structure in AM0135-650.

*Evidence for Gas:* There are multiple morphological signatures for the presence of gas in the optical images presented here. As discussed above in the context of identifying morphological signatures of star formation, warm ($\gtrsim$10,000K) ionized gas is detectable through the H$\alpha$+[OIII] 5007Å emission lines it produces in the F606W bandpass, typically manifesting as nebulosity in HII regions. More indirectly, the existence of a (presumably atomic-dominated) gas reservoir can be inferred from the morphological evidence for star formation. Empirically, star formation is rare below gas column densities of $3\,M_\odot$ pc$^{-2}$, and almost entirely absent below $1\,M_\odot$ pc$^{-2}$ (F. Bigiel et al. 2008), so the presence of young stars is strong indication of a minimum column density of gas in that region[14]. Moreover, the intensity of star formation goes up as a strong power law of the column density (particularly from 1-10 $M_\odot$pc$^{-2}$), so large numbers of highly-clustered young stars indicate locations with high characteristic gas column densities. These associations do not give a strong indication of the detailed structure of the gas reservoir, but do offer some sense of the extent and likely detectability of 21 cm emission.

An even more revealing morphological indication of gas is dust absorption, which is a direct, high-resolution indicator of cold, dense gas. Dust has been well-established in the Milky Way as a tracer of molecular clouds (e.g., C. J. Lada et al. 1994; L. Cambrésy 1999; M. Lombardi & J. Alves 2001; J. F. Alves et al. 2001; A. A. Goodman et al. 2009; J. L. Pineda et al. 2010), to the point where maps of dust absorption are often preferable to direct detection of CO, which does not always correlate perfectly with $H_2$, particularly at low column densities where CO is not abundant enough to be self-shielding. The presence of structured dust obscuration in an image therefore indicates that dense, presumably molecular, gas exists between the observer and significant background stellar light; Arp 176 in Figure 9 (lower left) shows an obvious example of such obscuration, but patchier dust is also pervasive in Arp 6, Arp 163, and AM0135-650 in the same figure. In all of these cases, the viewer should "read" the dust absorption as a map of molecular gas.

Dust cannot always be detected in this manner, however, if most of the stars are in front of the dust, or, if the stellar background is sufficiently highly structured that it is challenging to separate dust obscuration from spatial variations in the starlight. Optimal conditions for detecting cold gas morphologically are in the inner regions of smooth spheroidal galaxies, but many convincing detections are possible in non-optimal configurations. A morphological suspicion of dust absorption can be confirmed by looking for evidence of localized reddening in the associated optical color images, but must take into account that their lower resolution can smooth out features to make dust less detectable; fully recovering the characteristic reddening typically requires resolving scales of 100 pc or better for individual clouds (W. C. Keel & R. E. White 2001; B. W. Holwerda et al. 2009), though larger coherent structures like dust lanes can still produce notable reddening at low resolutions. Similarly, emission in the W3+W4 NEOWISE images can also indicate emission from PAH's and warm dust emitting at $24\mu$, which is a further sign of a dusty ISM that has been heated by star formation.

*A demonstration of analyzing cold gas structure by interpreting dust absorption:* As an example of how one might use dust absorption to "read" an atlas image, we can return to Arp 176 in the lower left of Figure 9. Because this system contains only smooth spheroidal galaxies, it makes the visual detection and interpretation of dust absorption more straightforward than in a heavily star-forming galaxies; the basics approach, however, can be applied more generally.

The most noticeable dust absorption feature in Arp 176 is the narrow double ring encircling the larger galaxy. The ring is filamentary and has internal structure characteristic of turbulence. There is also a hint of recent star formation in the bottom-most region of the ring, visible as a clustered excess of point sources and some associated nebulosity; this spatially-coincident star formation is confirmation for interpreting the dust absorption as a sign post for molecular gas. Given that the encircled galaxy is a large spheroidal system, and that the double ring does not appear to be aligned in a single disk, it seems likely that this gas is a leftover from an accretion event with sufficient available angular momentum to leave gas in a roughly circular configuration.

[14] We note that given the intrinsic low efficiency of most star formation, it is essentially impossible for gas to be exhausted purely by on-going star formation, and thus there almost certainly is some gas reservoir remaining when star formation is present.

The absorption in this galaxy also offers an instructive case study in the importance of gas+star geometry. There is an global angular asymmetry in the ring, which appears less prominent on the left. While this could be a systematic difference in the characteristic gas density in the ring, it is more plausible that the typical gas column density is actually uniform around the ring, but that the apparent absorption is merely "diluted" by a higher fraction of stars in front of the gas (see arguments in B. G. Elmegreen & D. L. Block 1999; J. J. Dalcanton et al. 2023) on the left side. This visual difference would occur naturally if the left side is the far side of the ring; in contrast, the near side gas lies between us and the bulk of the stars in the galaxy, leading to the more prominent absorption features on the right.

In addition to the rings, the image also shows more subtle radial absorption filaments extending downwards from the center of the larger galaxy; these are lower contrast than the rings, which could indicate either lower intrinsic column densities, or, a greater fraction of starlight in front of the gas, comparable to how star-gas geometry is affecting the far side of the ring on the left. Maintaining a radial gas feature is not obviously compatible with having significant angular momentum — as one would expect is needed to maintain the circular ring — and thus this feature offers a puzzle that likely would place strong constraints on the interaction history of this system with further analysis and simulation.

We also should note that the smaller spheroidal galaxy in the lower right, does not show any evidence for structured dust extinction. This "absence of evidence" is strong indication that the galaxy does not currently have a cold gas reservoir. It is therefore is less likely that the smaller galaxy was the donor of the cold gas, unless the mass difference and interaction pathway was sufficient to entirely strip a pre-existing gas component, which seems unlikely.

*Inferring interaction history:* When interpreting images of interacting systems, it is helpful to keep relevant astrophysical processes in mind. As described above, we see tracers of both gas and stars in the F606W images presented here. Both of these components are subject to gravitational, tidal forces, and when these forces dominate, both gas and stars will have experienced similar forces and wind up in similar morphologies. However, there are significant differences that can play a role in producing more complex morphologies.

First, unlike the gas, stars are collisionless, and have orbits that typically preserve their fine-grained phase space density. These effects lead stars from the original galaxies to have typically smooth features in response to interactions, and well-mixed merger remnants, but with occasional sharp features like shells and shelves at larger radii, associated with "turn around" points in the stellar orbits (see Section 4.7 below).

In contrast, gas is highly dissipative, and is subject to additional magnetohydrodynamic forces that can be much more important than gravitational forces at some points of the interaction. These additional effects can lead to significant differences in the morphologies traced by the gas compared to the stars. These differences can potentially be useful for diagnosing the nature of the past interaction.

It is also important to keep in mind the different timescales of these different classes of forces in mind. Collisionless stellar systems tend to preserve "memory" of their orbital structure long after an interaction, particularly in outer regions where dynamical times are long. In contrast, dissipative gas has the potential to lose energy rapidly, and may rapidly erase a record of its original structure, particularly in denser regions where cooling and interaction timescales are most rapid.

*Identifying and interpreting older massive clusters:* Stellar clusters are generally short-lived features, either dissolving rapidly over a few megayears as their nascent molecular cloud is destroyed, or, or dissolving over hundreds of megayears through tidal shocks and other dynamical processes (M. R. Krumholz et al. 2019). In other words, while Figure 8 shows the brightness of hypothetical stellar clusters over time, the vast majority of the clusters originally formed in a star forming region will be gone well within a gigayear, unless the cluster is massive and tightly bound. As such, we expect the majority of stellar clusters in the F606W images to be young, with the notable exceptions being nuclear clusters (N. Neumayer et al. 2020) and long-lived massive globular clusters (J. P. Brodie & J. Strader 2006). These special cases therefore represent exceptions to the above recommendation that point source populations necessarily indicate recent star formation, and thus we discuss them separately here.

Candidate nuclear clusters can be readily identified in the atlas, particularly in lower surface brightness and nearer systems (compiled below in Table 6 and Table 7, with associated thumbnail images shown in Figure 13 and Figure 14); as an example, Arp 4 (MCG-02-05-050) shows a canonical presentation of a nuclear cluster candidate. While proximity and low contrast backgrounds simplify the by-eye identification of nuclear star clusters, they are typically sufficiently bright ($M_V$ between roughly -8 and -12 for lower mass galaxies) to be de-tectable as luminous point sources throughout this atlas, even when too distant to be spatially resolved with HST resolution. By definition, nuclear clusters are located in the center of the galaxy, but many of the low surface brightness galaxies here are sufficiently diffuse that the "center" is not morphologically well-defined. In addition, galaxies that cover a large area have a non-zero chance of being overlapped by a fortuitously positioned, but unassociated, foreground star from the Milky Way. A full analysis of the nuclear star cluster population — which would involve profile fitting, identification of foreground stars with Gaia, possible color selection using other wavebands, and ideally spectroscopy — is outside the scope of this paper, but we recommend this sample, and the associated photometry (Section 2.3), as being an interesting source of candidates worthy of more detailed study.

The other form of long-lived clusters found in these images are globular clusters, which can be useful tracers of much older star formation and accretion. Taking Omega Cen as being typical of the most luminous globular clusters ($M_V \approx -10$), Figure 7 shows that older globulars are likely only visible for atlas galaxies within $\sim$10,000 km s$^{-1}$, and even closer to see down to the peak of the globular cluster luminosity function. Compared to brighter, larger nuclear clusters, globular clusters can be more easily confused with Milky Way foreground stars. However, they will be spatially clustered, and primarily found around smooth spheroidal galaxies or galaxies with prominent bulges. The larger galaxy in Arp 176 (lower left panel of Figure 9) is an example of a galaxy with a notable globular cluster system. If multicolor data is available, one would expect globular clusters to have colors that mimic those of the main galaxy, which can be helpful to distinguish them from foreground stars in the Milky Way that are too faint to be excluded by Gaia.

## 4. SOME USEFUL SUBSAMPLES

The atlas images presented in thumbnail form in Figure 4 and in full-resolution in Figures B2-B2 are shown in order of their name in the Arp and the Arp-Madore catalogs. While this ordering assists in finding images of a particular system, it does not naturally group "like with like", particularly for the Arp-Madore sources, where names indicate only sky position. The Arp catalog numbers do reflect some level of grouping (see Table 2), but are less than half of the total observed systems. Moreover, many systems could plausibly be placed in multiple categories, as was done in the Arp-Madore catalog.

In this section, we address these limitations of using a single presentation order by showing composites of thumbnails for systems grouped by common properties. These groupings are based both on categorization in the original catalog (as given in Table 1, with keys found in Tables 2 & 3) and on visual inspection. The goal is for these compilations to help readers find compelling candidates for follow-up observations, modeling, or other scientific investigations.

We stress that for any system of interest, the reader should refer to the corresponding full-resolution atlas image, which can be accessed quickly from the link on the galaxy name in the caption.

### 4.1. *Nearest Galaxies*

The closest galaxies offer the highest physical resolution for any given observation, and the greatest sensitivity for detecting intrinsically faint point sources. These properties makes them natural targets for a variety of astrophysical studies, particularly those that rely on detecting individual stars and/or high physical resolution of observations of gas.

Figure 13 and Table 6 show all systems with heliocentric recessional velocities of less than $V_r = 2000$ km s$^{-1}$, which corresponds to $\sim$28 Mpcfor pure Hubble Flow with $H_0 = 70$ km s$^{-1}$Mpc$^{-1}$, in order of increasing $V_r$. This order will not correspond perfectly to increasing distance, due to the galaxies' peculiar velocities and to a lesser extent, our not correcting $V_r$ from heliocentric to Galactocentric velocities[15]. However, this ordering is still sufficient for identifying particularly useful systems for follow-up. Based on Figure 6 and Figure 7, these images should be able to reveal stars with absolute $V$-band magnitudes brighter than -6 for the most distant systems and -3 for the nearest, and have physical resolutions of 3-14 pc per $0.1''$.

The images in Figure 13 are drawn from a much smaller volume of the Universe than the sample at large, and thus are less likely to contain the most extreme systems, which are naturally rare. They are therefore dominated by galaxies with large angular sizes, many of which were cataloged for having irregular internal morphologies or asymmetries, rather than more spectacular interactions. This subsample therefore naturally favors selection of dwarf and/or low surface brightness galaxies, which are numerous in more volume-limited samples.

[15] One concrete example of the effects of peculiar velocity are for the two images of the interacting pair of galaxies in AM0942-313, which have recessional velocities of 964 km s$^{-1}$and 1,252 km s$^{-1}$, but are certainly at a similar distance.

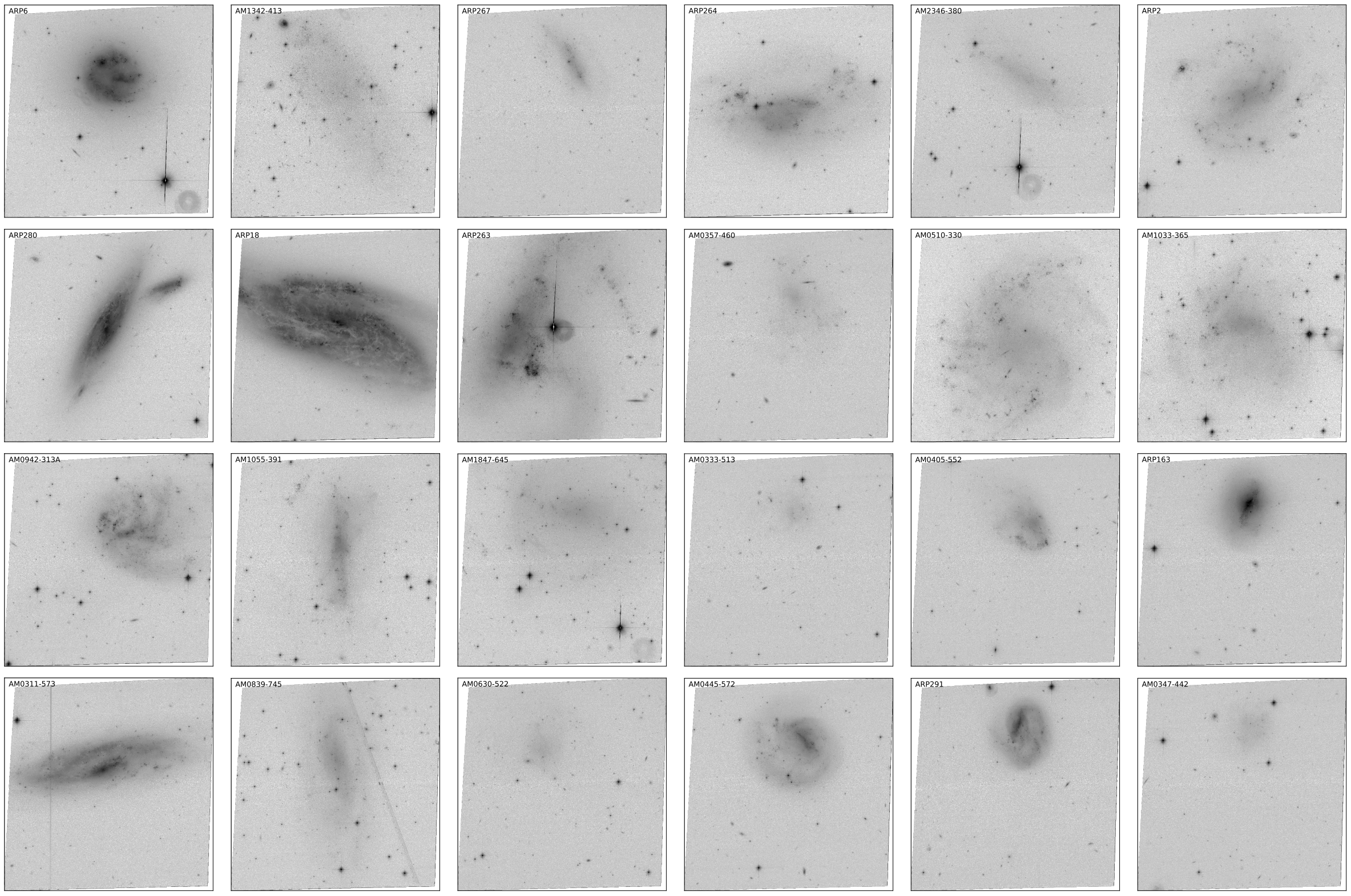

**Figure 13.** Thumbnails of nearest galaxies in order of increasing distance: ARP6; AM1342-413; ARP267; ARP264; AM2346-380; ARP2; ARP280; ARP18; ARP263; AM0357-460; AM0510-330; AM1033-365; AM0942-313A; AM1055-391; AM1847-645; AM0333-513; AM0405-552; ARP163; AM0311-573; AM0839-745; AM0630-522; AM0445-572; ARP291; AM0347-442;

Some systems of notes include: (1) starbursting systems, including the spectacular Arp 6 (the "Bearclaw Galaxy"), Arp 18, Arp 22, Arp 91, Arp 163, Arp 205, and Arp 283; (2) galaxies with particularly notable star forming regions and likely massive star populations, including Arp 262, Arp 263, Arp 264, Arp 280, Arp 306, AM0405-552, AM0454-561, AM0507-630, AM0942-313A, AM0942-313B, and the remarkable 30 Doradus/NGC 604 twin HII region in Arp 2; (3) galaxies likely to be post-merger, including Arp 163 and Arp 205; (4) galaxies with possible nuclear star clusters, including Arp 2, Arp 4, Arp 267, Arp 279, AM0357-460, AM0510-330, AM0942-313A, AM1847-645, AM0347-442, AM0839-745; and (5) galaxies with either extremely thin bars or strongly misaligned outer disks, including AM0445-572, Arp 291, and most notably, AM0409-563.

**Table 6**. Nearest Systems with $V_r \lesssim 2{,}000\,\mathrm{km\ s^{-1}}$

| Arp Target Name | Categories | $V_r$ | $m - M$ | $M_{lim}$ | $M_{F606W}$ | $M_{F606W}$ |
|---|---|---|---|---|---|---|
| | | km s$^{-1}$ | (Approximate) | | (Brightest 5) | (Brightest 1%) |
| (1) | (2) | (3) | (4) | (5) | (6) | (7) |
| ARP6 | A:a | 447 | 29.0 | -3.0 | $-8.7^{-0.73}_{+0.48}$ | -7.5 |
| AM1342-413 | 20 | 545 | 29.5 | -3.5 | $-10.1^{-0.43}_{+0.12}$ | -8.0 |
| ARP267 | D:n | 582 | 29.6 | -3.6 | $-8.6^{-1.74}_{+0.60}$ | -8.6 |
| ARP264 | D:n | 602 | 29.7 | -3.7 | $-8.7^{-0.94}_{+0.29}$ | -7.8 |
| AM2346-380 | 20 | 647 | 29.8 | -3.8 | $-8.4^{-0.15}_{+0.27}$ | -8.2 |
| ARP2 | A:a | 712 | 30.0 | -4.0 | $-9.4^{-1.70}_{+0.68}$ | -8.4 |
| ARP280 | E:b | 727 | 30.1 | -4.1 | $-9.5^{-0.55}_{+0.36}$ | -8.0 |
| ARP18 | A:c | 749 | 30.1 | -4.1 | $-10.3^{-0.77}_{+0.38}$ | -8.4 |

**Table 6** *continued*

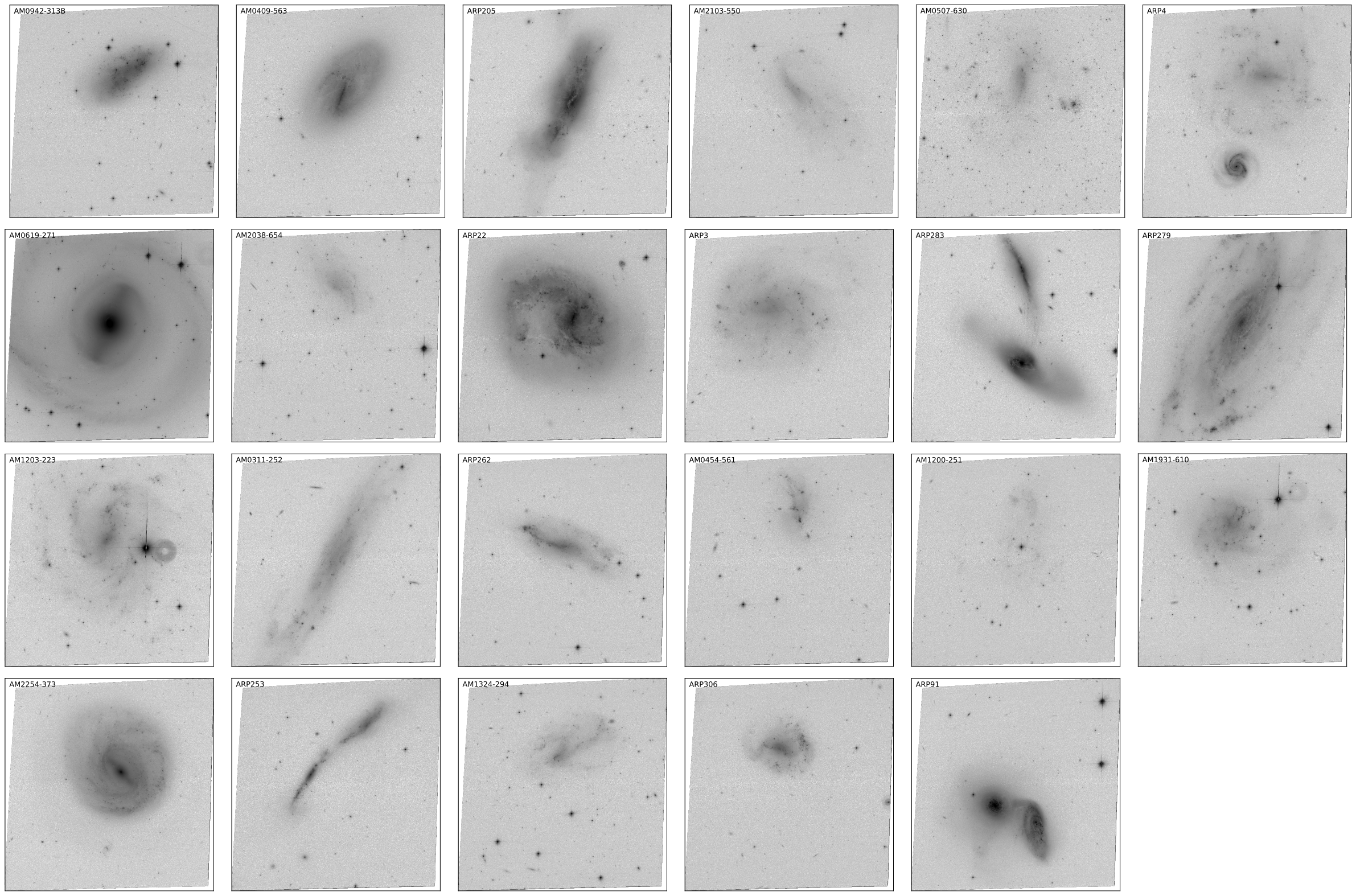

**Figure 13.** [Continued] Thumbnails of nearest galaxies in order of increasing distance: AM0942-313B; AM0409-563; ARP205; AM2103-550; AM0507-630; ARP4; AM0619-271; AM2038-654; ARP22; ARP3; ARP283; ARP279; AM1203-223; AM0311-252; ARP262; AM0454-561; AM1200-251; AM1931-610; AM2254-373; ARP253; AM1324-294; ARP306; ARP91

**Table 6** *(continued)*

| Arp Target Name | Categories | $V_r$ | $m-M$ | $M_{lim}$ | $M_{F606W}$ | $M_{F606W}$ |
|---|---|---|---|---|---|---|
| | | km s$^{-1}$ | (Approximate) | | (Brightest 5) | (Brightest 1%) |
| (1) | (2) | (3) | (4) | (5) | (6) | (7) |
| ARP263 | D:n | 755 | 30.2 | -4.2 | $-11.2^{-0.39}_{+0.56}$ | -8.7 |
| AM0357-460 | 8,20 | 900 | 30.5 | -4.5 | $-8.2^{-3.08}_{+0.71}$ | -8.2 |
| AM0510-330 | 20 | 927 | 30.6 | -4.6 | $-10.6^{-0.68}_{+0.99}$ | -9.0 |
| AM1033-365 | 20 | 955 | 30.7 | -4.7 | $-10.3^{-1.07}_{+0.54}$ | -8.7 |
| AM0942-313A | 16,23 | 964 | 30.7 | -4.7 | $-9.9^{-1.18}_{+0.47}$ | -8.4 |
| AM1055-391 | 16,20 | 1002 | 30.8 | -4.8 | $-9.0^{-0.65}_{+0.16}$ | -8.6 |
| AM1847-645 | 20 | 1006 | 30.8 | -4.8 | $-10.9^{-0.21}_{+0.40}$ | -10.1 |
| AM0333-513 | 20 | 1030 | 30.8 | -4.8 | $-9.0^{-1.81}_{+0.77}$ | -9.5 |
| AM0405-552 | 16,20 | 1066 | 30.9 | -4.9 | $-8.8^{-0.37}_{+0.28}$ | -8.6 |
| ARP163 | D:d | 1072 | 30.9 | -4.9 | $-12.0^{-0.33}_{+0.60}$ | -11.1 |
| AM0311-573 | 16 | 1140 | 31.1 | -5.1 | $-10.1^{-1.78}_{+0.62}$ | -8.9 |
| AM0839-745 | 8 | 1142 | 31.1 | -5.1 | $-11.5^{-0.74}_{+1.09}$ | -9.8 |
| AM0630-522 | 20 | 1193 | 31.2 | -5.2 | $-11.4^{-0.20}_{+0.82}$ | -11.5 |
| AM0445-572 | 10,13,16 | 1205 | 31.2 | -5.2 | $-8.9^{-1.16}_{+0.26}$ | -8.6 |
| ARP291 | E:d | 1218 | 31.2 | -5.2 | $-9.3^{-1.59}_{+0.40}$ | -8.9 |
| AM0347-442 | 20 | 1248 | 31.3 | -5.3 | $-10.6^{-0.77}_{+2.17}$ | -10.6 |
| AM0942-313B | 16,23 | 1253 | 31.3 | -5.3 | $-10.5^{-0.44}_{+0.87}$ | -9.1 |
| AM0409-563 | 14,16 | 1310 | 31.4 | -5.4 | $-9.6^{-1.43}_{+0.36}$ | -9.0 |
| ARP205 | D:h | 1378 | 31.5 | -5.5 | $-10.7^{-0.82}_{+0.58}$ | -9.7 |

**Table 6** *continued*

**Table 6** *(continued)*

| Arp Target Name | Categories | $V_r$ | $m - M$ | $M_{lim}$ | $M_{F606W}$ | $M_{F606W}$ |
|---|---|---|---|---|---|---|
| | | km s$^{-1}$ | (Approximate) | | (Brightest 5) | (Brightest 1%) |
| (1) | (2) | (3) | (4) | (5) | (6) | (7) |
| AM2103-550 | 20 | 1402 | 31.5 | -5.5 | $-9.3^{-2.57}_{+0.43}$ | -9.2 |
| AM0507-630 | 20 | 1464 | 31.6 | -5.6 | $-12.9^{-0.15}_{+0.01}$ | -12.0 |
| ARP4 | A:a | 1614 | 31.8 | -5.8 | $-11.1^{-0.86}_{+0.98}$ | -9.8 |
| AM0619-271 | 8,10,13,14 | 1622 | 31.8 | -5.8 | $-12.5^{-0.33}_{+1.42}$ | -10.8 |
| AM2038-654 | 20 | 1626 | 31.8 | -5.8 | $-10.6^{-0.54}_{+0.57}$ | -10.5 |
| ARP22 | A:e | 1662 | 31.9 | -5.9 | $-12.2^{-0.61}_{+0.76}$ | -10.2 |
| ARP3 | A:a | 1694 | 31.9 | -5.9 | $-10.7^{-1.59}_{+0.52}$ | -10.1 |
| ARP283 | E:c | 1695 | 31.9 | -5.9 | $-12.4^{-0.22}_{+0.40}$ | -12.0 |
| ARP279 | E:b | 1711 | 31.9 | -5.9 | $-11.0^{-1.07}_{+0.24}$ | -9.6 |
| AM1203-223 | 8 | 1722 | 32.0 | -6.0 | $-11.3^{-1.34}_{+1.12}$ | -9.4 |
| AM0311-252 | 8,12 | 1735 | 32.0 | -6.0 | $-10.1^{-1.32}_{+0.66}$ | -9.4 |
| ARP262 | D:n | 1772 | 32.0 | -6.0 | $-10.2^{-1.23}_{+0.20}$ | -9.9 |
| AM0454-561 | 16 | 1779 | 32.0 | -6.0 | $-9.9^{-2.14}_{+0.33}$ | -9.9 |
| AM1200-251 | 20,22 | 1790 | 32.0 | -6.0 | $-12.1^{-0.66}_{+1.56}$ | -12.1 |
| AM1931-610 | 8,20 | 1796 | 32.0 | -6.0 | $-10.4^{-1.80}_{+0.18}$ | -10.0 |
| AM2254-373 | 12,16 | 1801 | 32.1 | -6.1 | $-10.2^{-0.80}_{+0.20}$ | -9.7 |
| ARP253 | D:m | 1867 | 32.1 | -6.1 | $-10.4^{-1.87}_{+0.24}$ | -10.2 |
| AM1324-294 | 20 | 1902 | 32.2 | -6.2 | $-10.6^{-2.17}_{+0.89}$ | -10.0 |
| ARP306 | E:f | 1959 | 32.2 | -6.2 | $-10.5^{-0.78}_{+0.57}$ | -9.8 |
| ARP91 | B:c | 1980 | 32.3 | -6.3 | $-11.8^{-0.28}_{+0.37}$ | -11.0 |

### 4.2. *Low Surface Brightness Galaxies*

Systems from the Arp catalogs are often assumed to be violently interacting galaxies with extreme star formation rates (e.g., Arp 220). However, the Arp catalogs also have ample representation of galaxies occupying the other extreme — star formation that is somehow taking place in extremely tenuous, low density disks, often without obvious triggers. This regime is a particularly interesting test of models of self-regulated disk galaxies, which are typically tuned to the properties of Milky Way-like galaxies.

Figure 14 and Table 7 show all galaxies either classified as low surface brightness ("A:a" or "20" for Arp or Arp-Madore classifications, respectively). We note that Arp 6 is not particularly low surface brightness, due to its active starburst, but it is a dwarf irregular and could return to lower surface brightness as the burst fades. AM1705-773 is also not particularly low surface brightness compared to the other systems here, though it is low surface brightness compared to morphologically-similar spirals; it seems most likely that the low surface brightness of the one half of the galaxy is due to foreground cirrus, as can be seen in the false color press release image in Figure 5.

In addition to the classified systems, we have also included galaxies which we judge to be low surface brightness based upon their HST imaging. For example, AM1229-512 is an unusual edge-on disk which would appear low surface brightness if it were viewed face-on, and AM2019-442 is a chance projection of a diffuse dwarf galaxy seen against a more distant spheroidal galaxy (as initially proposed by J. Yadav et al. 2022, and confirmed by our imaging, as discussed above in Section 2.2.1). We also include Arp 136, which we believe is a similar system to AM2019-442, consisting of a background large lenticular galaxy with an extensive globular cluster system, seen behind a much nearer, extremely diffuse collection of stars to the south east. These stars are clearly visible and well-resolved in the HST imaging, and appear as a blue nebulosity in GALEX imaging (see Figure B2), suggesting the detected stars in the HST imaging are likely "tip of the iceberg" bright young stars, which would be consistent with their degree of clustering. Maps of the resolved stars associated with the foreground dwarf can also be seen in Figure D3. We also show higher-contrast, zoomed-in views of both systems in Figure 15, to better highlight their resolved stars and/or clusters, smooth central low surface brightness component of older stars, and in the case of AM2019-442 (left), possible nuclear cluster and filamentary network of dense dusty gas seen in projection against the smooth background galaxy. Deeper imaging could possibly reveal a comparable network in Arp 136, which morphologically appears to be even lower mass.

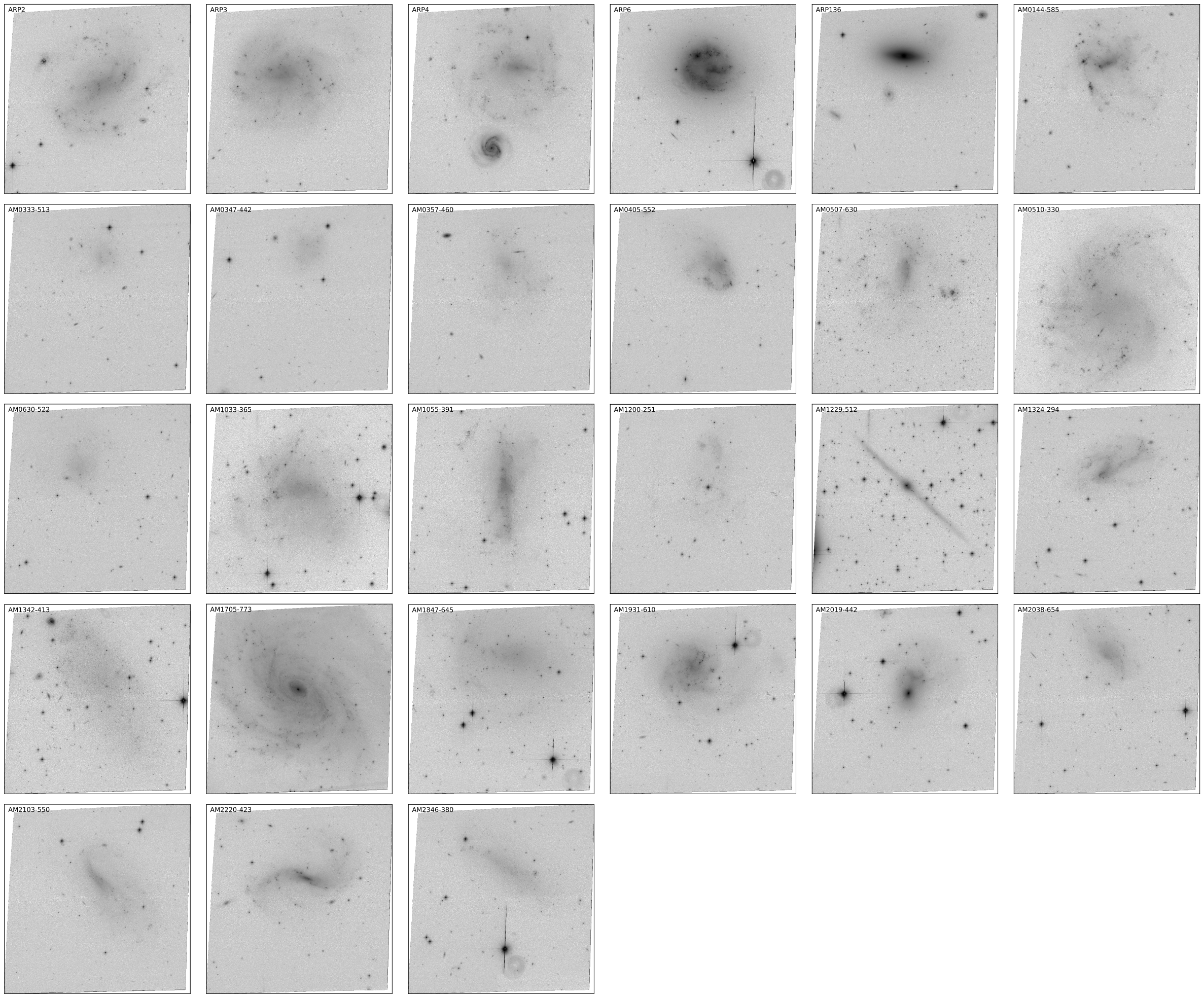

**Figure 14.** Thumbnails of low surface brightness galaxies: ARP2; ARP3; ARP4; ARP6; ARP136; AM0144-585; AM0333-513; AM0347-442; AM0357-460; AM0405-552; AM0507-630; AM0510-330; AM0630-522; AM1033-365; AM1055-391; AM1200-251; AM1229-512; AM1324-294; AM1342-413; AM1705-773; AM1847-645; AM1931-610; AM2019-442; AM2038-654; AM2103-550; AM2220-423; AM2346-380. While the main galaxies in Arp 136 and AM2019-442 are high surface brightness, in both cases there is a foreground low surface brightness galaxy responsible for the neighboring diffuse emission.

In spite of the diffuse nature of these galaxies, all of the low surface brightness systems appear to have young stars, as judged by their GALEX emission (available for ∼3/4 of the sample), blue optical colors (available for ∼2/3 of the sample) and the direct detection of stars or stellar clusters in the HST imaging (which would become undetectable at these distances for older stellar populations, as shown in Figure 8. These systems would be superb targets for follow-up mapping in other filters (to constrain stellar ages and masses), in cold gas tracers (to calculate star formation efficiencies), and with spectroscopy of their likely low metallicity massive stars, which may in many cases be isolated enough for ground-based spectroscopy with large telescopes. Observations of the hot and warm gas would be valuable as well, as it unclear how feedback works in this low density regime, where star formation efficiencies are extremely low, but so are the typical drivers of feedback.

### 4.3. *Galaxies with Rings*

The Arp and Arp-Madore catalogs contain numerous systems with prominent ring structures. Based on their morphologies, these rings have a variety of likely origins (see E. Athanassoula & A. Bosma 1985, for a discussion of principal theories), with the subset of likely "collisional ring" systems from the Arp-Madore catalog previously compiled in (B. F. Madore et al. 2009); such collisional systems are of particular interest, but are quite rare (see R. J. Buta 2017). The Hubble imaging here includes a number of systems from this latter list, including Arp 141, Arp 143, Arp 145, AM0403-555, AM0417-391, AM0520-390, AM0643-462, AM2026-424, and AM2056-392, although not all appear to be colli-

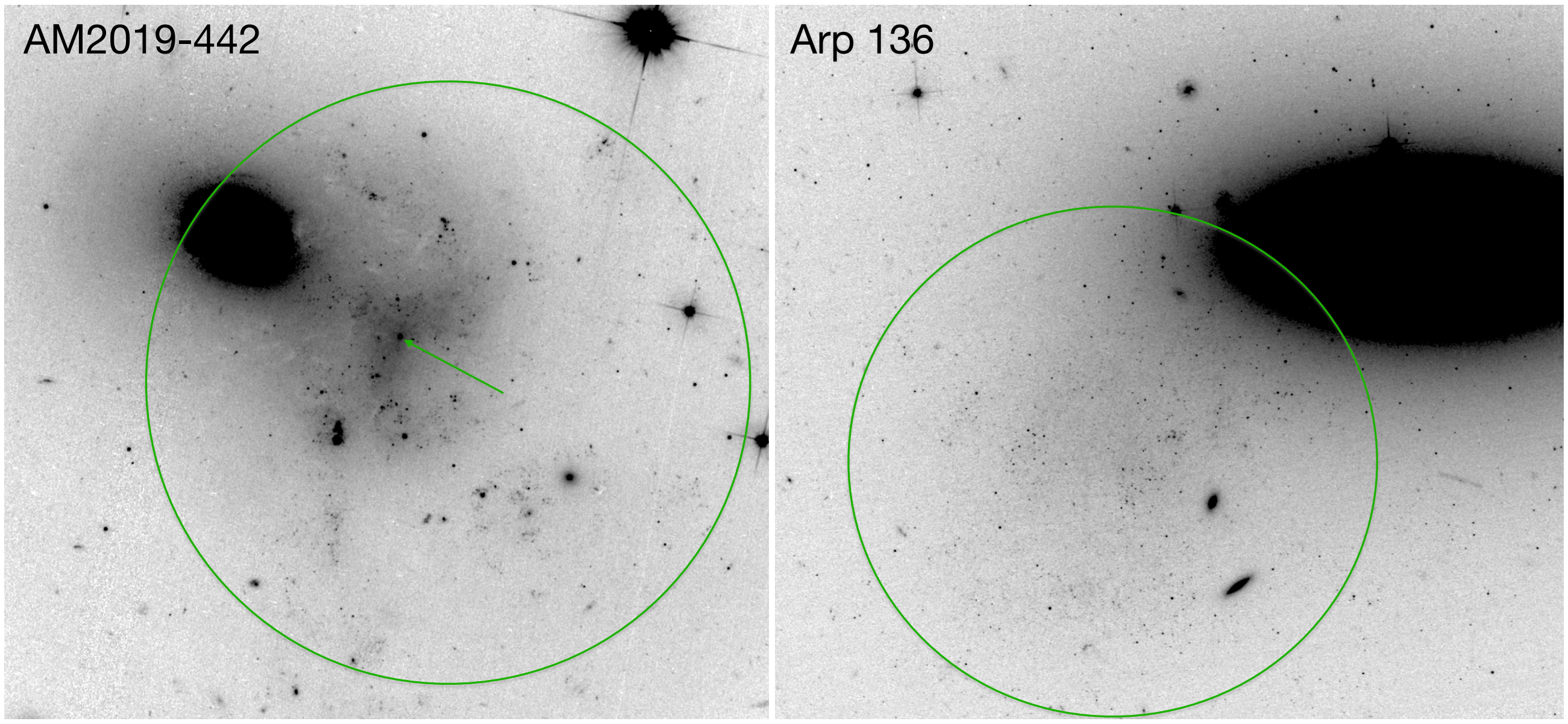

**Figure 15.** Zoomed, high-contrast views of two foreground low surface brightness dwarfs (left: AM2019-442, previously identified by Yadav et al 2022; right: Arp 136) partially superimposed on background early-type galaxies whose morphologies and sizes suggest much larger distances, and thus no true physical association. Both foreground dwarfs have resolved point source populations indicative of recent star formation, centered on a smooth, but low surface brightness, background of older stars. Green circles indicate the approximate extent of the recent star formation in each galaxy, and the green arrow indicates a possible nuclear cluster candidate centered on the smooth background of older stars in AM2019-442's foreground dwarf. The higher density and degree of clustering of point sources in the AM2019-442 dwarf suggests a higher star formation rate intensity than in the Arp 136 system. Notably, there is a rich network of structured dust obscuration against the background galaxy in AM2019-442, which suggests that this might be a favorable system for studying the cold ISM in a low surface brightness galaxy.

sional rings in the new imaging, as discussed immediately below.

In Section 4.3, we show galaxies that had ring classifications in their original catalogs, or, that had ring-like structures from visual inspection; we list the properties of these systems in Table 8. Some appear to be axisymmetric within otherwise unremarkable galaxies, suggesting they are likely to be a fairly stable, long-lived dynamical feature (e.g., AM0619-271, AM1307-461, the inner ring in AM0012-573, and possibly AM2038-323). Some outer, largely axisymmetric rings are tenuous and star-forming, and may potentially be traced to a past gas accretion event (e.g., the outer ring of AM0012-573 and the remarkably wispy ring in AM0432-625). Other similar rings seem to originate primarily from on-going tidal disruption of a satellite, with little obvious star formation (e.g., AM0403-555, AM2056-392, and AM2001-602, which is only a partial outer ring). Other rings are highly non-axisymmetric, and seem likely to be driven by a gravitational disturbance from a non-merging interaction (e.g., Arp 10, AM0203-325, AM0520-390, AM0643-462 (a remarkable twin of AM0644-741, which had earlier been imaged by HST), the inner regions of AM2001-602, and potentially AM2038-323, which is not particularly ring-like when viewed with HST resolution). Similarly to AM2038-323, Arp 231 does not have a clear ring, and appears to be misclassified based on its extensive shells.

The most dramatic ring systems in Section 4.3 are those that appear to be major mergers or head-on collisions (e.g., Arp 141, Arp 145, Arp 150, Arp 208, AM0417-391, AM1953-260, and AM2026-424, which became a popular Halloween press image for NASA, shown in color in Figure 5). When fully modeled, these systems are potentially useful probes of galactic potentials and extended gas reservoirs.

### 4.4. *Galaxies with Strong Dust Features*

The high resolution of HST is superb for revealing the detailed structure of dust absorption. When coherent dust absorption is visible, it is an unambiguous indicator of dense gas (as discussed above in Section 3). However, the absence of dust absorption features does not imply that dense gas is not present, given that the visibility of dust absorption is a strong function of geometry and the nature of the background light. The morphological effects of dust are most clear when the dust is in front of most of the stars, and when the stellar distribution is smooth which favors highly-inclined, early-type systems with a high stellar surface brightness.

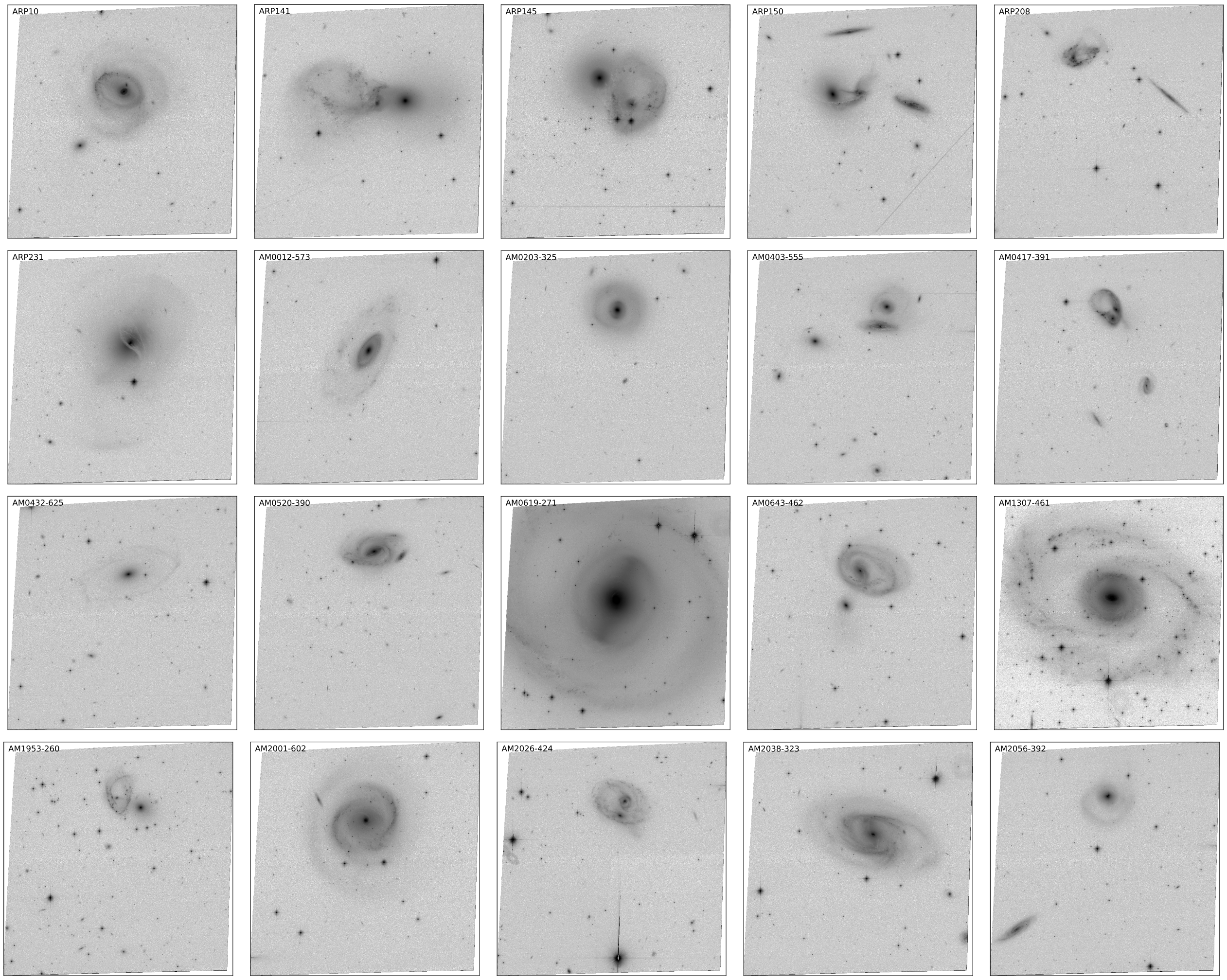

**Figure 16.** Thumbnails of galaxies with ring features: ARP10; ARP141; ARP145; ARP150; ARP208; ARP231; AM0012-573; AM0203-325; AM0403-555; AM0417-391; AM0432-625; AM0520-390; AM0619-271; AM0643-462; AM1307-461; AM1953-260; AM2001-602; AM2026-424; AM2038-323; AM2056-392;

In Section 4.4 we have compiled systems where there are particularly clear, coherent dust absorption features, based on visual inspection of the atlas images. These are systems where it may be easier to infer system geometry (i.e., which features are due to foreground versus background objects), to assess the degree of disturbance in the ISM, and to find relics of past gas accretion in early type systems (e.g., Arp 158, Arp 176, Arp 231, Arp 309, AM2048-571, and AM2128-430, which also has a remarkable shell system). Given the high optical depth of the ISM when seen edge-on, the galaxies in Section 4.4 strongly favor edge-on orientations. We stress that there are ample smaller-scale dust features visible in many star-forming atlas galaxies not included here (for example, the close-ups in Figure 9), and that this list is only an attempt to compile some of the more notable cases.

### 4.5. *Extremely Non-Axisymmetric Spirals*

Canonical spiral disks are often treated as being axisymmetric in their mass distribution. While bars and spiral features add additional structure, that structure is often assumed to have a large degree of rotational symmetry (i.e., with no strong $m = 1$ modes). Underlying this picture is the assumption that the baryonic disk has settled into a principal plane of a relaxed, axisymmetric dark matter distribution.

In practice, spiral galaxies are known to have morphological departures from this expectation. Some non-axisymmetric features are dominated by large-scale differences in local star formation rates (see for example, AM0135-650), typically reflecting transient differences in the gas density. Such features are quite common within this Atlas, given the selection criteria of the original source catalogs.

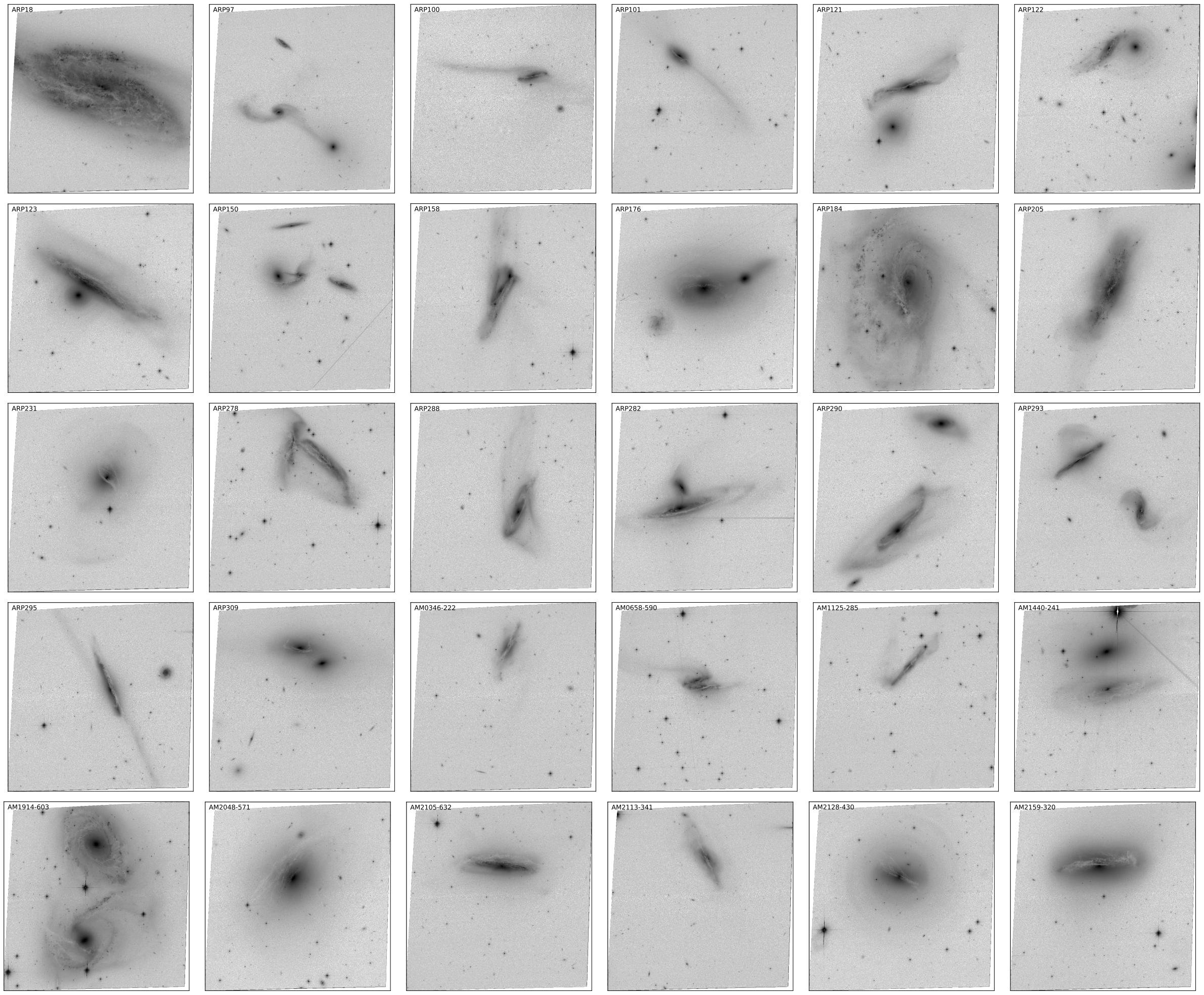

**Figure 17.** Thumbnails of galaxies with notable large-scale dust features, which favors selection of edge-on systems with dense gas, or systems with smooth background light: ARP18; ARP97; ARP100; ARP101; ARP121; ARP122; ARP123; ARP150; ARP158; ARP176; ARP184; ARP205; ARP231; ARP278; ARP288; ARP282; ARP290; ARP293; ARP295; ARP309; AM0346-222; AM0658-590; AM1125-285; AM1440-241; AM1914-603; AM2048-571; AM2105-632; AM2113-341; AM2128-430; AM2159-320;

Of particular interest, however, are features that indicate large-scale, non-axisymmetric departures in the underlying stellar mass distribution. Unlike the gas that drives star formation, stars are collisionless, making their morphological disturbances more long lived and heavily influenced by large-scale features of the gravitational potential. Stellar deviations from axisymmetry are often most visible in the outer regions of galaxies, where the self-gravity of the disk is weak, material is less tightly bound, and the dynamical times are longer. These effects combine to make baryons at large radii more susceptible to visible disturbances from satellite interactions or misaligned (or unrelaxed) dark matter halos, leading to warps or strong $m = 1$ modes.

In Section 4.5 and Table 10 we compile a number of systems with strong $m = 1$ modes, which might be challenging to explain without unrelaxed dark matter halos. In many, there is a likely interaction partner that may have driven the asymmetry (either on or directly off frame), or evidence that the non-axisymmetric galaxy is a member of a larger parent group (listed in Column 4 of Table 10) and may be affected by a larger group potential. A few systems (AM0144-585 and AM0541-294) have no obvious surrounding group or obvious culprits for producing a significant $m = 1$ asymmetry; these systems may be in the process of fully integrating a previously-absorbed companion. Overall, this subsample may offer interesting candidates for investigating relaxation of dark matter matter halos and large scale dynamical sloshing modes, or, as a constraint on the dark matter structure in the larger group environment.

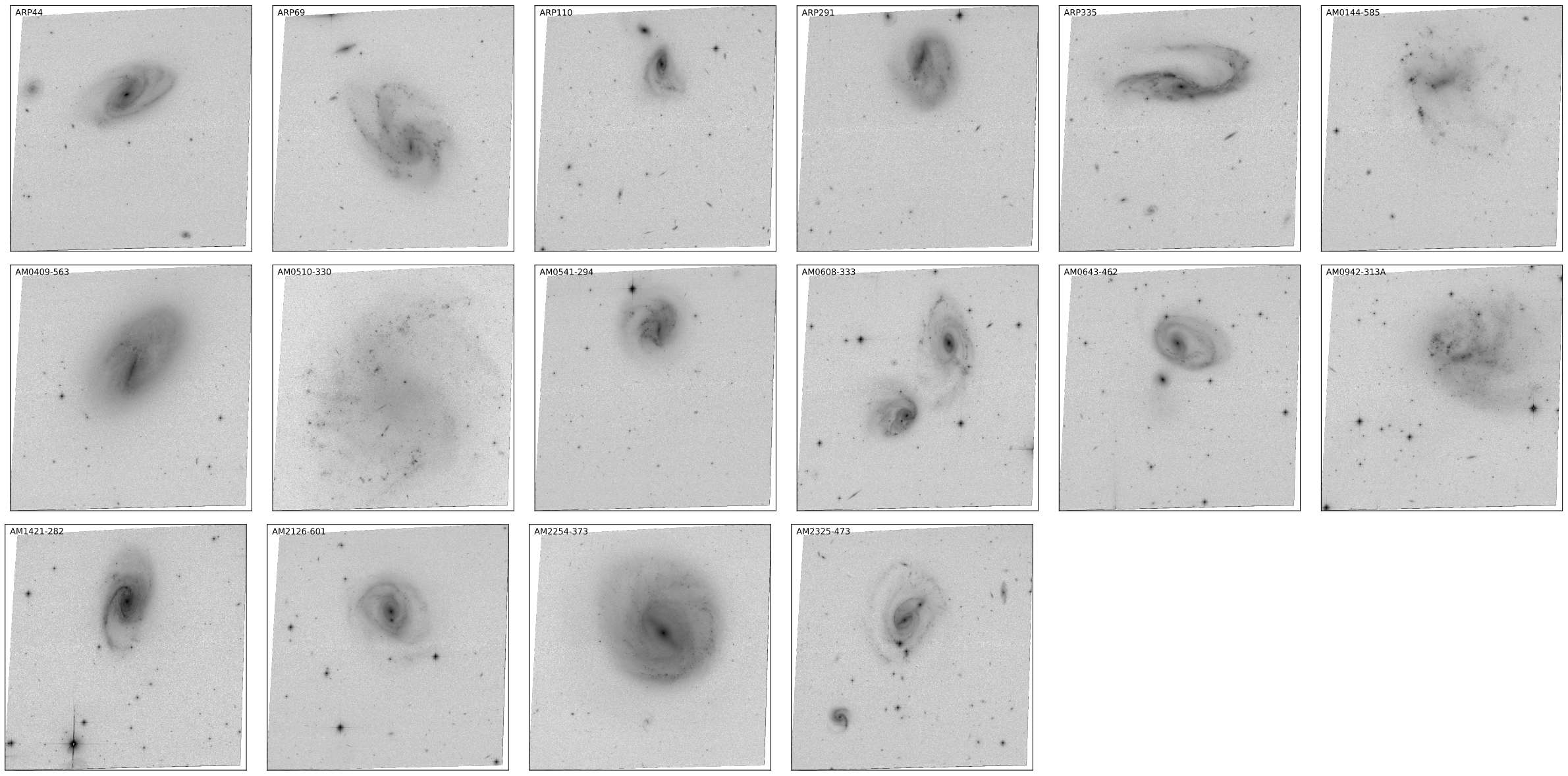

**Figure 18.** Thumbnails of notably asymmetric spirals with strong $m = 1$ asymmetries in their overall stellar distributions: ARP44; ARP69; ARP110; ARP291; ARP335; AM0144-585; AM0409-563; AM0510-330; AM0541-294; AM0608-333; AM0643-462; AM0942-313A; AM1421-282; AM2040-295; AM2126-601; AM2254-373; AM2325-473;

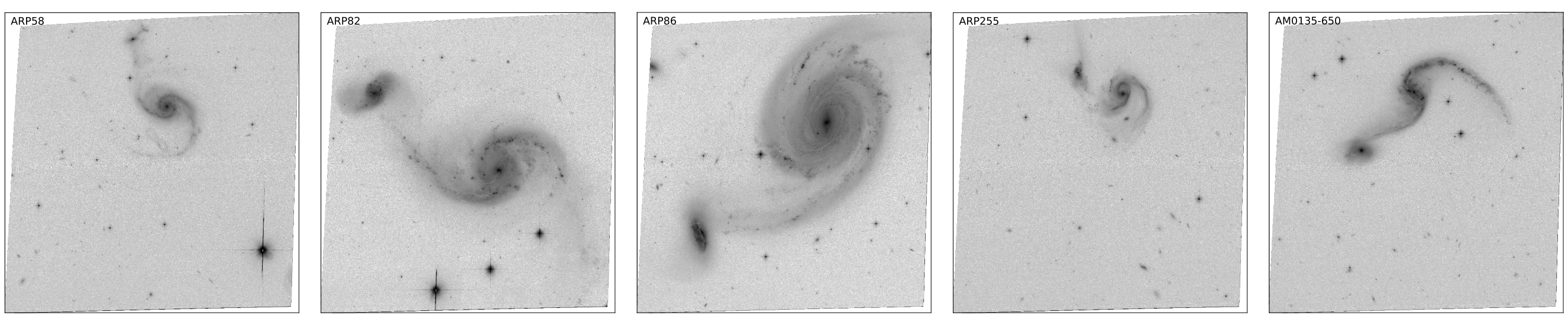

**Figure 19.** Thumbnails of analogs of the Whirlpool galaxy system: ARP58; ARP82; ARP86; ARP255; AM0135-650;

### 4.6. *Whirlpool Galaxy Analogs*

The Whirlpool Galaxy (M51) is a notable example of a close interaction of a low mass companion driving strong $m = 2$ spiral arms and vigorous star formation in a more massive partner. The HST imaging here includes a number of M51-analogs that may be suitable for testing whether models developed to explain M51 are more broadly applicable, albeit at lower resolution given their larger distances ($D_{M51} \approx 7.3\,\mathrm{Mpc}$vs 63–66 Mpcfor Arp 82 and Arp 86). We compile HST images of these systems in Section 4.6, a number of which have already been well-studied at other wavelengths, but have lacked HST imaging until now (e.g., H. Salo & E. Laurikainen 1993; E. Laurikainen et al. 1993; A. V. Zasov et al. 2019, 2022; P. Karera et al. 2022, among others).

### 4.7. *Galaxies with Shells*

Shells in galaxies are a canonical signature of a past merging event. They are associated with the turnaround points of stars on largely radial orbits, with a pile-up resulting from "phase wrapping" of dynamically cold material in either minor (P. J. Quinn 1984) and major (L. Hernquist 1992) mergers; see reviews in J. E. Barnes (1992) and E. Athanassoula & A. Bosma (1985), examples from simulations in A.-R. Pop et al. (2018), and connection to merger orbits in N. C. Amorisco (2015) & D. Hendel & K. V. Johnston (2015). Shells are typically low surface brightness, but they have a large enough angular extent in nearby galaxies that they can be detected with sufficiently good flat-fielding, with the first significant catalogs being based on photographic plates alone (e.g., D. F. Malin & D. Carter 1983).

The images in this atlas include a number of galaxies with clear shells, shown in Section 4.7. The shells are largely found around earlier type galaxies, as was noted in even the earliest studies (e.g., D. F. Malin & D. Carter 1983). There are hints of possible inner shells in AM0612-373, but the complexity of the morphology makes it difficult to separate tidal debris from shells. AM2034-521 was also a candidate for this list, but at HST resolution its candidate shells appear instead to be unusually clean, tight spiral arms. It is also possible that various methods for high-pass filtering (e.g., unsharp masking, subtracting median images, etc) might identify inner shells in additional systems.

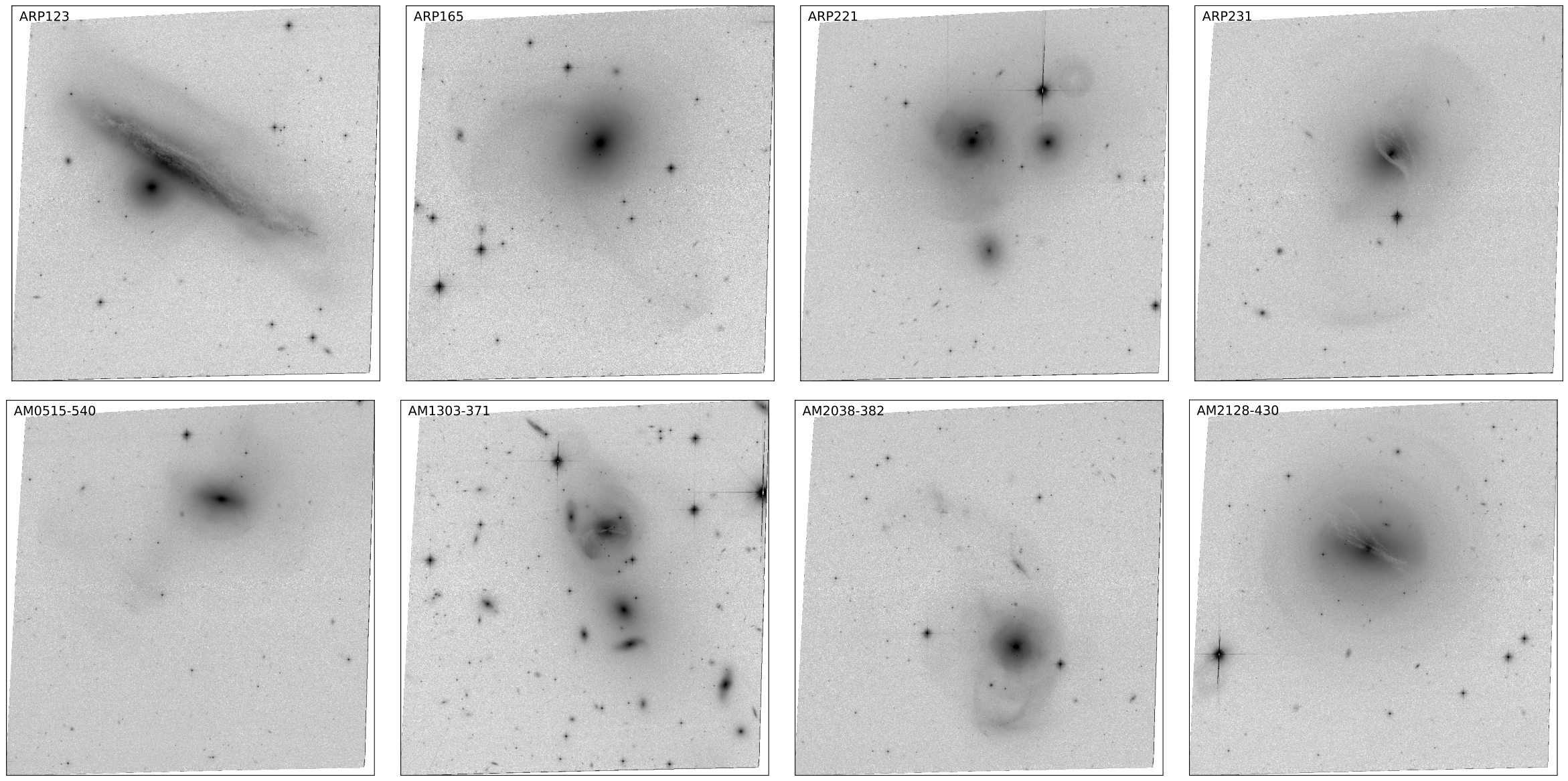

**Figure 20.** Thumbnails of galaxies with shells: ARP123; ARP165; ARP221; ARP231; AM0515-540; AM1303-371; AM2038-382; AM2128-430;

**Table 7.** Low Surface Brightness Galaxies

| Target Name | Type (SIMBAD) | Cat | $V_r$ (km/s) |
|---|---|---|---|
| (1) | (2) | (3) | (4) |
| ARP2 | AGN_Candidate | A:a | 712 |
| ARP3 | AGN_Candidate | A:a | 1694 |
| ARP4 | AGN_Candidate | A:a | 1614 |
| ARP6 | AGN_Candidate | A:a | 447 |
| ARP136 | NGC 5820 | C:d | 3251 |
| AM0144-585 | AGN_Candidate | 20 | 2210 |
| AM0333-513 | EmissionG | 20 | 1030 |
| AM0347-442 | AGN_Candidate | 20 | 1248 |
| AM0357-460 | LowSurfBrghtG | 8,20 | 900 |
| AM0405-552 | EmissionG | 16,20 | 1066 |
| AM0507-630 | AGN_Candidate | 20 | 1464 |
| AM0510-330 | AGN_Candidate | 20 | 927 |
| AM0630-522 | AGN_Candidate | 20 | 1193 |
| AM1033-365 | AGN_Candidate | 20 | 955 |
| AM1055-391 | AGN_Candidate | 16,20 | 1002 |
| AM1200-251 | GinPair | 20,22 | 1790 |
| AM1229-512 | Seyfert2 | 7 | 2617 |
| AM1324-294 | EmissionG | 20 | 1902 |
| AM1342-413 | AGN_Candidate | 20 | 545 |
| AM1705-773 | EmissionG | 8,20,23 | 2953 |
| AM1847-645 | LowSurfBrghtG | 20 | 1006 |
| AM1931-610 | EmissionG | 8,20 | 1796 |
| AM2019-442 | Galaxy | 13,15 | 2963 |
| AM2038-654 | AGN_Candidate | 20 | 1626 |
| AM2103-550 | LowSurfBrghtG | 20 | 1402 |
| AM2220-423 | AGN_Candidate | 20 | 2420 |
| AM2346-380 | AGN_Candidate | 20 | 647 |

**Table 8.** Galaxies with Notable Ring Features

| Target Name | Cat | $V_r$ (km/s) |
|---|---|---|
| ARP10 | A:b | 9116 |
| ARP141 | C:e | 2684 |
| ARP145 | C:e | 5326 |
| ARP150 | D:b | 11888 |
| ARP208 | D:h | 9025 |
| ARP231 | D:l | 5646 |
| AM0012-573 | 6,10 | 15900 |
| AM0203-325 | 6,8 | 5775 |
| AM0403-555 | 2,6 | 17093 |
| AM0417-391 | 6,8 | 15255 |
| AM0432-625 | 6 | 16063 |
| AM0520-390 | 1,6,23 | 14734 |
| AM0619-271 | 8,10,13,14 | 1622 |
| AM0643-462 | 1,6 | 11774 |
| AM1307-461 | 6,10 | 3103 |
| AM1953-260 | 6 | 14619 |
| AM2001-602 | 6,12 | 3587 |
| AM2026-424 | 1,6 | 15124 |
| AM2038-323 | 1,6,8 | 5551 |
| AM2056-392 | 6,13,24 | 13500 |

Given that shells are straightforward to detect with ground-based imaging of nearby galaxies, the strength of the HST imaging here is not so much in detecting whether shells are present or not. Where it does have significant potential is in analyzing the radial structure of shells at high resolution. This structure is set in large part by the physical characteristics of the interacting galaxies (i.e., density profiles, kinematics, etc), and interesting constraints on these quantities can potentially result from the high resolution imaging presented here (see R. E. Sanderson & A. Helmi 2013).

**Table 9.** Systems with Coherent Dust Features

| Target Name | Cat | $V_r$ (km/s) |
|---|---|---|
| ARP18 | A:c | 749 |
| ARP97 | B:d | 6830 |
| ARP100 | B:d | 6071 |
| ARP101 | B:d | 4599 |
| ARP121 | C:c | 5715 |
| ARP122 | C:c | 12082 |
| ARP123 | C:c | 2300 |
| ARP150 | D:b | 11888 |
| ARP158 | D:c | 4756 |
| ARP176 | D:f | 3107 |
| ARP184 | D:g | 3888 |
| ARP205 | D:h | 1378 |
| ARP231 | D:l | 5646 |
| ARP278 | E:b | 4577 |
| ARP288 | E:d | 7049 |
| ARP282 | E:c | 4569 |
| ARP290 | E:d | 3614 |
| ARP293 | E:d | 5514 |
| ARP295 | E:e | 6846 |
| ARP309 | E:f | 4658 |
| AM0346-222 | 7,16 | 12141 |
| AM0658-590 | 2,15 | 8270 |
| AM1125-285 | 2,6,12 | 7127 |
| AM1440-241 | 2 | 3528 |
| AM1914-603 | 3,8 | 3817 |
| AM2048-571 | 8,14 | 3359 |
| AM2105-632 | 14 | 3142 |
| AM2113-341 | 14,18 | 8798 |
| AM2128-430 | 10,14,22 | 2362 |
| AM2159-320 | 3,14 | 2541 |

**Table 10.** Systems Containing Strongly Non-Axisymmetric Spirals

| Target Name | Cat | $V_r$ (km/s) | Parent Group (Simbad) |
|---|---|---|---|
| ARP44 | B:a | 5505 | HDC 580 |
| ARP69 | B:b | 3573 | LGG 30 |
| ARP110 | C:b | 9384 | [T2015] nest 202655 (pair) |
| ARP291 | E:d | 1218 | LGG 214 |
| ARP335 | F:c | 7574 | LDC 775 |
| AM0144-585 | 20 | 2210 | |
| AM0409-563 | 14,16 | 1310 | LGG 114 |
| AM0510-330 | 20 | 927 | LGG 128 |
| AM0541-294 | 15,16,24 | 3818 | |
| AM0608-333 | 2 | 8652 | AM0608-333 |
| AM0643-462 | 1,6 | 11774 | [T2015] nest 201845 (pair) |
| AM0942-313A | 16,23 | 964 | NGC 2997 Group |
| AM1421-282 | 24 | 4340 | [TSK2008] Group 2543 |
| AM2126-601 | 10 | 8660 | |
| AM2254-373 | 12,16 | 1801 | IC 1459 Group |
| AM2325-473 | 1,10 | 15238 | small partners visible off frame |

### 4.8. *Galaxies on the Toomre Sequence*

The classic paper A. Toomre & J. Toomre (1972) was the first to lay out a plausible explanation for galactic bridges and tails. Their early simulations (with rings of particles and no dark matter) showed how tidal forces from interactions between equal mass galaxies naturally produced the features that we now associate with mergers of galaxies with comparable mass (“major” mergers). In subsequent years, a vast literature has explored the simulated properties of such mergers, with ever increasing levels of astrophysical sophistication (e.g., including live dark matter halos, hydrodynamics, star formation, central black holes, stellar and AGN feedback, etc; J. E. Barnes & L. E. Hernquist 1991; J. E. Barnes 1992; J. C. Mihos & L. Hernquist 1996; V. Springel et al. 2005; B. Robertson et al. 2006; T. J. Cox et al. 2008; P. F. Hopkins et al. 2008, 2009, 2013; J. E. Barnes 2016, among many, many others, with earlier results summarized in the review by J. E. Barnes & L. Hernquist (1992)).

In the years since A. Toomre & J. Toomre (1972), several works have taken the approach of using a collection of pairs of interacting galaxies as representative “snapshots” along the canonical merging sequence. The “Toomre sequence” starts with an initial close passage, the development of significant tidal tails and bridges of material as the galaxies move to the apocenter of their orbit, followed by a return to their pericenter and eventual merger. Inspired by these works (e.g., J. E. Hibbard & J. H. van Gorkom 1996; S. Laine et al. 2003; J. Rossa et al. 2007; S. Haan et al. 2011), in Figure 21 we have ordered as series of images from the catalog, based on a plausible sequence in merger stage for large, gas-rich galaxies of comparable mass.

Numerical simulations (such as P. F. Hopkins et al. 2006) have confirmed long-held speculation that gas-rich major mergers should naturally lead to rapid gas inflow into the galaxies’ centers, triggering star formation and AGN activity. This possible activity sequence can be demonstrated by looking at the mutiwavelength counterparts of the optical images in Figure 21, specifically the ALLWISE W3+W4 images, which are sensitive to emission from PAHs and dust heated by star formation or AGN activity.

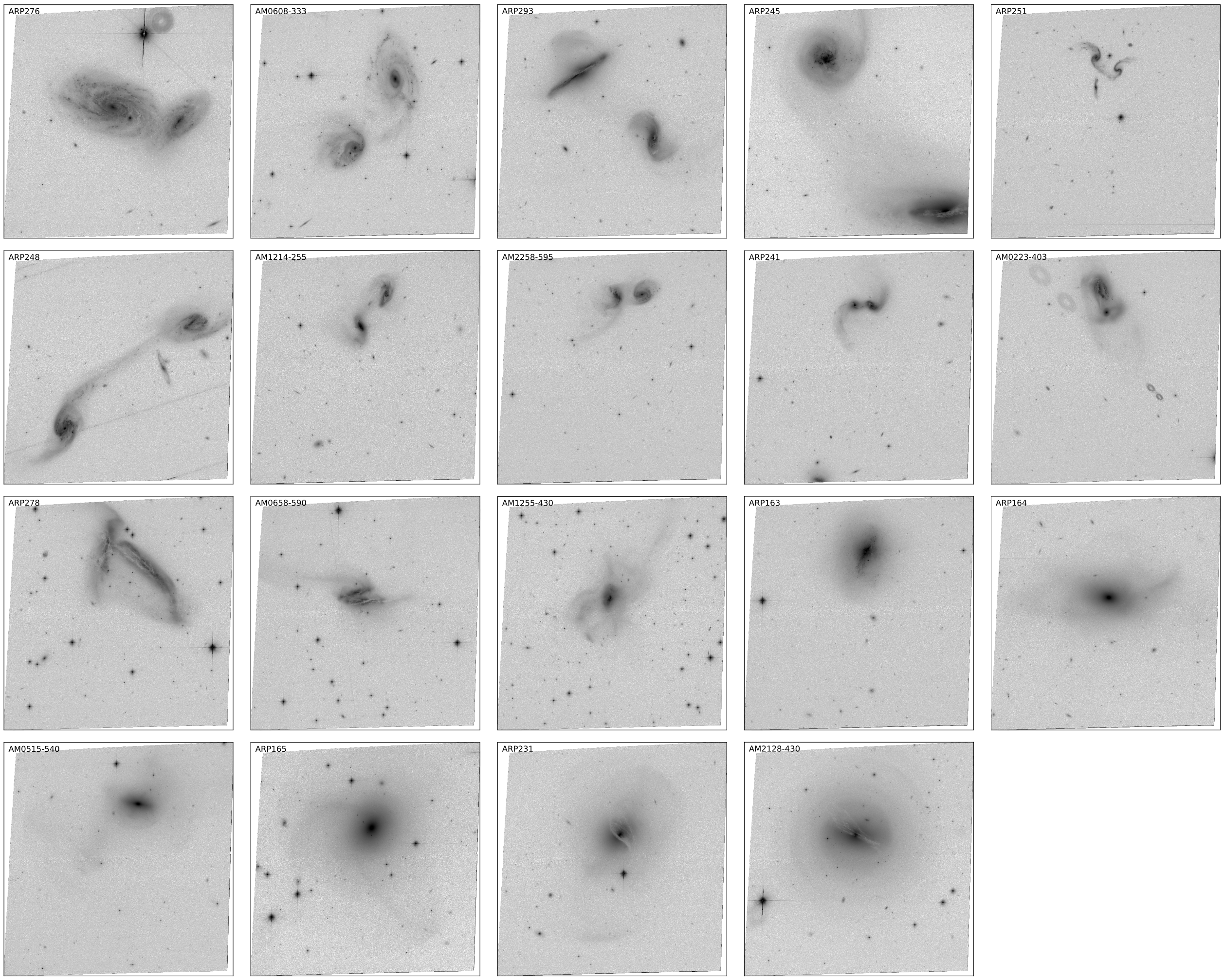

**Figure 21.** Thumbnails of likely major mergers, organized into a plausible analog of the Toomre sequence: ARP276; AM1914-603; ARP293; ARP245; ARP251; ARP248; AM1214-255; AM2258-595; ARP241; AM0223-403; ARP278; AM0658-590; AM1255-430; ARP163; ARP164; AM0515-540; ARP165; ARP231; AM2128-430;

We show the corresponding ALLWISE W3+W4 images for this sequence in Figure 4.8. The first two galaxies in the sequence show signs of distributed emission from galaxy-wide star formation, but show no evidence for a strong central point source, as would be consistent with a central starburst or AGN. This pattern of emission is consistent with their being gas-rich star-forming galaxies on initial approach. At the other end of the sequence, there are 5 galaxies that all appear to be post-merger systems that are now morphologically-disturbed early type systems in the optical. None of these galaxies show significant concentrated central emission, and their W3+W4 is unusually blue.

Compared to the early and late merger stages, the intermediate stages are marked by at least one, if not two, strong central point sources emitting in the mid-infrared, suggestive of central starbursts or AGN. These are always accompanied by a detection in the VLA Sky Survey continuum images, when such images are available. Morphologically, this is confirmation of the general picture of the strongest phases of interaction triggering strong gas inflow in major mergers.

We stress, however, that while the ALLWISE W3+W4 images in Figure 4.8 are strong evidence for central activity that drives concentrated dust emission, further quantitative analysis including other wavebands and/or comparison with spectroscopic classifications (available in Table C2 in Appendix C) would be needed to determine definitively whether this activity is due to central starbursts or AGN. R. Nikutta et al. (2014)'s analysis of the WISE colors of various astronomical objects with SDSS identifications shows that starburst driven emission (from ULIRGS, LIRGS, and other starbursts) has comparable W3−W4 colors as QSOs, Seyferts, ob-scured AGN, and LINERS (see their Figure 4), with mean colors that differ by only ∼0.3 mag (their Table 1). The shorter wavelength W1−W2 colors (or their *Spitzer* equivalents) have proven successful for isolating large samples of extragalactic QSOs out to large redshifts (e.g., T. H. Jarrett et al. 2011; D. Stern et al. 2012; R. J. Assef et al. 2018; S. Guo et al. 2018; K. Storey-Fisher et al. 2024, among others), but again have somewhat overlapping distributions when restricted to nearby samples (R. Nikutta et al. 2014; D. Asmus et al. 2020; F. Z. Zeraatgari et al. 2024), particularly at lower luminosities (see K. N. Hainline et al. 2016; J. A. O'Connor et al. 2016, for examples). This overlap is likely driven by hot dust contamination in W2, which may be particularly strong for high rates of obscured star formation. In Appendix E, we provide paired images in WISE W1+W2 and W3+W4 for all sample galaxies with central emission, to assist the reader. Similarly, the accompanying radio continuum emission is not helpful to distinguish among mechanisms, as it can have contributions from both non-thermal emission from AGN, or the combination of free-free and synchrotron emission associated with ionized gas and supernova remnants in young star forming regions. Instead, dedicated observing programs with more panchromatic coverage into the X-ray and spectroscopy (particularly in the mid-IR) are needed to separate the source powering the mid-IR emission (e.g., see the work in L. Armus et al. 2009; D. Asmus et al. 2020, for examples). Thankfully, many of these systems are close enough that spectroscopic classifications are frequently available, many of which are compiled in Table C2.

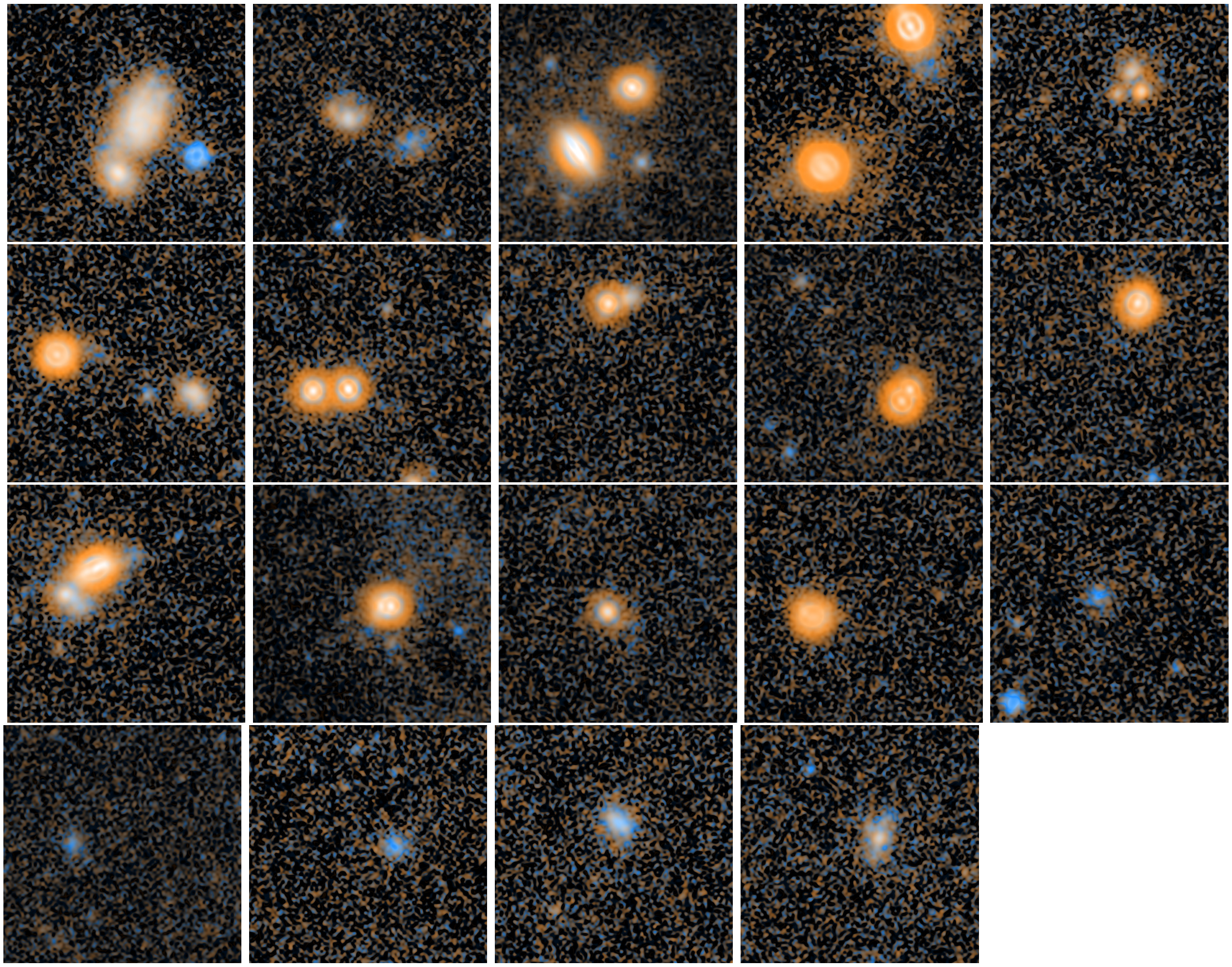

**Figure 22.** WISE W3+W4 thumbnails of likely major mergers, corresponding to the analog Toomre sequence of HST images in Figure 21, but oriented with north up to align with the full atlas images in Appendix B. From top left to bottom right, systems shown are: ARP276; AM1914-603; ARP293; ARP245; ARP251; ARP248; AM1214-255; AM2258-595; ARP241; AM0223-403; ARP278; AM0658-590; AM1255-430; ARP163; ARP164; AM0515-540; ARP165; ARP231; AM2128-430;

### 4.9. *Galaxies with Likely Central Starbursts or AGN*

As could be seen in Figure 21, strong central emission in the mid-infrared is a characteristic feature of strong interactions. This emission is thought to be triggered by the interaction driving a significant amount of a galaxy's ISM into the center, where the high central densities lead to non-linear increases in the star formation rates, while also providing more favorable conditions for feeding gas onto a central black hole (see discussion in Section 4.8).

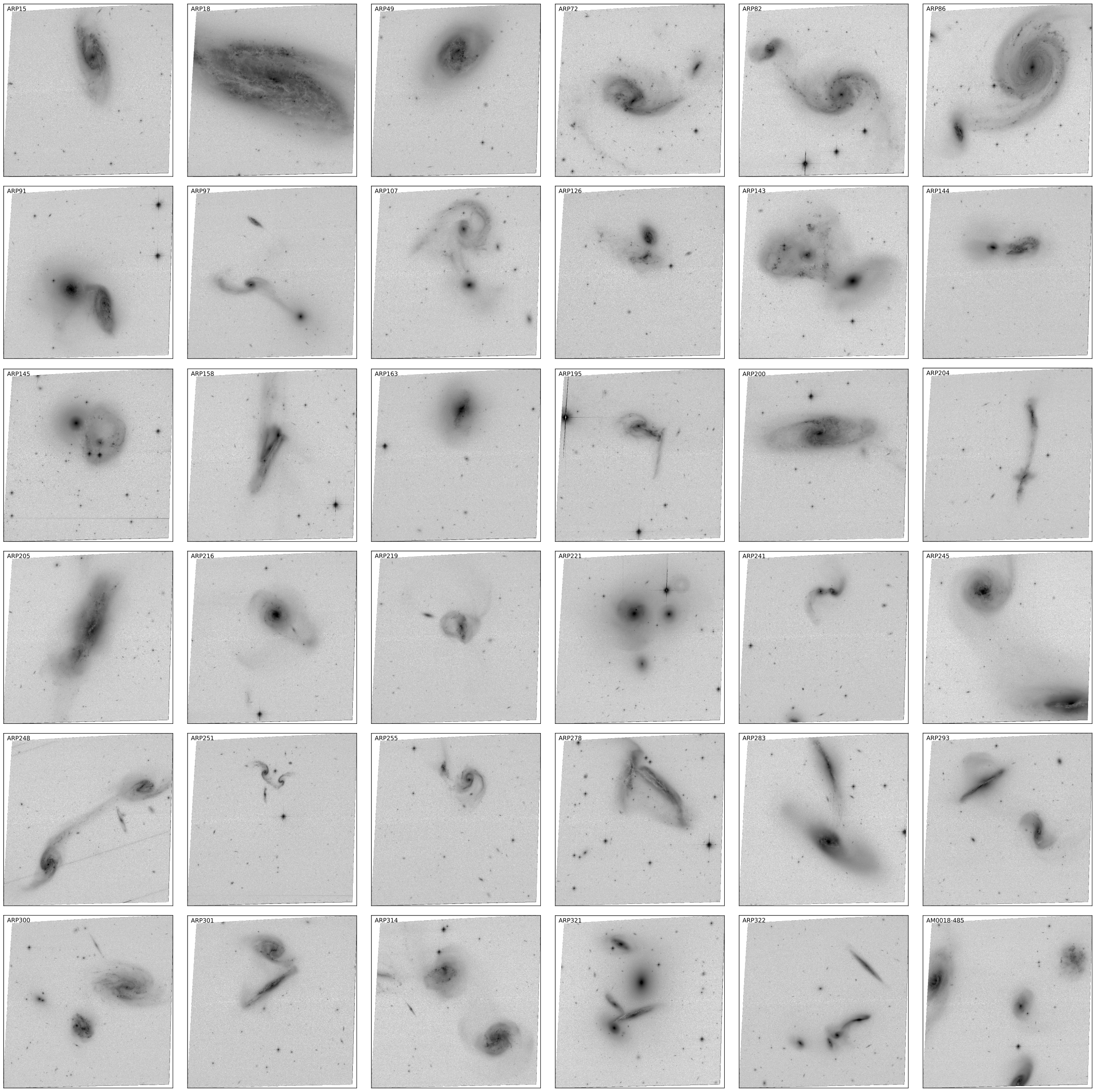

**Figure 23.** Thumbnails of systems containing at least one galaxy with strong, red, central WISE W3+W4 point source emission. All have detectable VLASS emission aligned with the WISE source if observed ($\delta > -40°$), with the exception of Arp 82, AM1421-282, and AM2350-302, which were observed but had no visible radio emission in VLASS. Systems shown are: ARP15; ARP18; ARP49; ARP72; ARP82; ARP86; ARP91; ARP97; ARP107; ARP126; ARP143; ARP144; ARP145; ARP158; ARP163; ARP195; ARP200; ARP204; ARP205; ARP216; ARP219; ARP221; ARP241; ARP245; ARP248; ARP251; ARP255; ARP278; ARP283; ARP293; ARP300; ARP301; ARP314; ARP321; ARP322; AM0018-485; [Continued]

In Figure 23, we essentially reverse the experiment in Figure 21, and select systems that appear to have central, red, point-source emission in the mid-infrared ALLWISE W3+W4 images, comparable to that seen in Figure 4.8, in at least one of the principle galaxies. We then compile the HST images of these systems in Figure 23 and Figure 24, where the latter have somewhat more ambiguous W3+W4 imaging (i.e., central emission that is possibly not a point source, or that may not be as red as the systems in Figure 23); these systems are also tabulated in Table E4. In most cases, the point source completely dominates the W3+W4 flux, although in a few of the galaxies with large angular sizes, some degree of a bluer extended disk is visible in the ALLWISE imaging. Compilations of the corresponding WISE W3+W4 images used for classification are provided in an appendix (Appendix E).

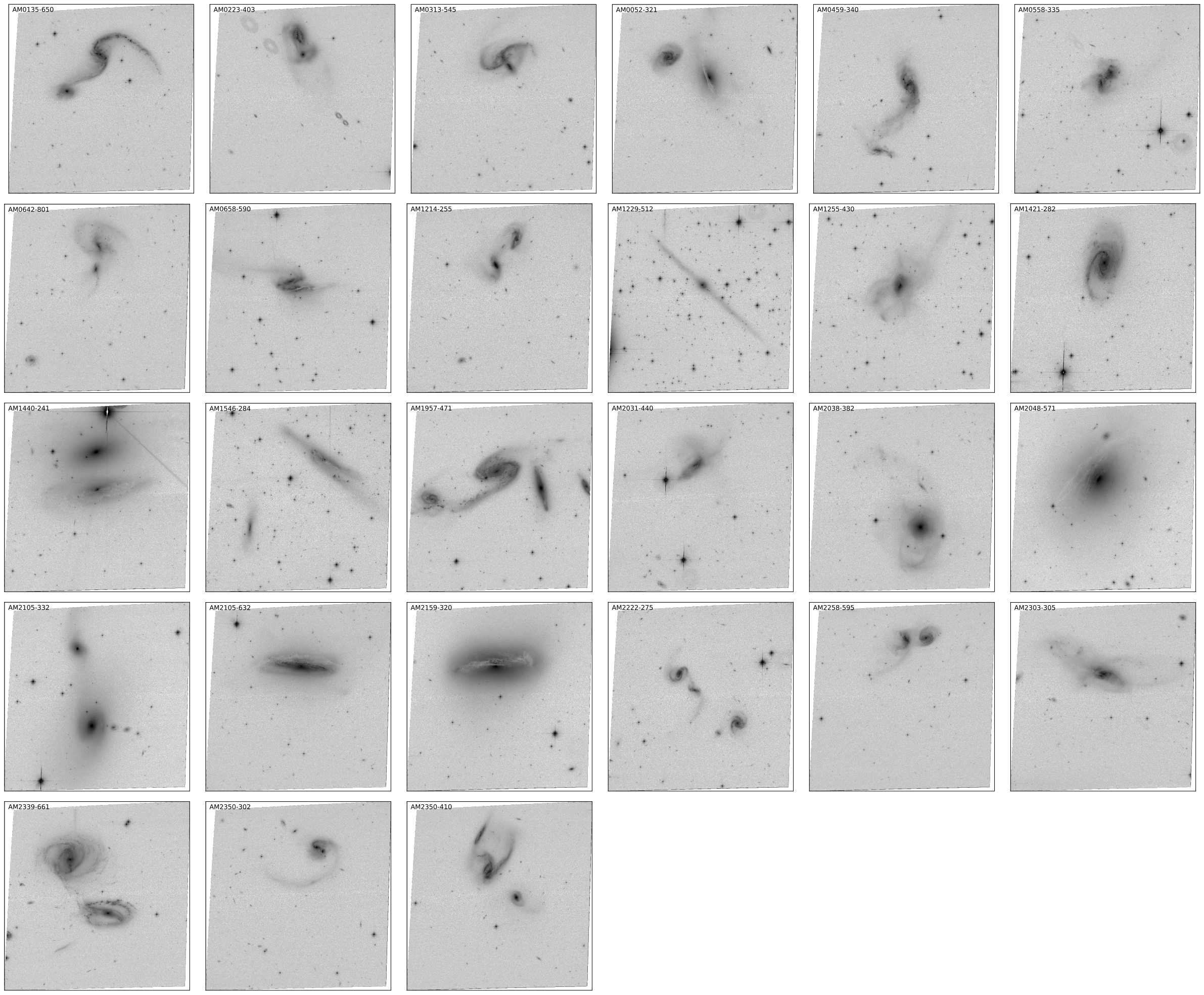

**Figure 23.** [Continued] Thumbnails of systems containing at least one galaxy with strong, red, central WISE W3+W4 point source emission. All have detectable VLASS emission aligned with the WISE source if observed ($\delta > -40°$), with the exception of Arp 82, AM1421-282, and AM2350-302, which were observed but had no visible radio emission in VLASS. Systems shown are: AM0135-650; AM0223-403; AM0313-545; AM0052-321; AM0459-340; AM0558-335; AM0642-801; AM0658-590; AM1214-255; AM1229-512; AM1255-430; AM1421-282; AM1440-241; AM1546-284; AM1957-471; AM2031-440; AM2038-382; AM2048-571; AM2105-332; AM2105-632; AM2159-320; AM2222-275; AM2258-595; AM2303-305; AM2339-661; AM2350-302; AM2350-410

Visual inspection of the HST images in Figure 23 show that the mid-infrared selection naturally identified the subset undergoing (or recovering from) strong interactions. These are likely some of the most compelling targets for imaging and spectroscopy with the James Webb Space Telescope (JWST) or the Atacama Large Millimeter Array (ALMA).

We have inspected the VLA Sky Survey imaging as well, which is available for all sources northward of $-40°$. Nearly all of the galaxies with strong central W3+W4 emission also had visually detectable central VLA sources. The only systems that were in the VLASS imaging but did not have obvious radio continuum counterparts were Arp 82, AM1421-282, AM2042-382, and AM2350-302 in Figure 23 and Arp 6, Arp 208, AM0942-313A, and AM1401-243A in Figure 23.

We used the VLA images to further identify the small subset of galaxies that had significant central radio continuum sources, but did not show up in either of the W3+W4 selected samples. These are shown in Figure 25, and reveal that in almost all cases the radio source is in a smooth, presumably gas-poor galaxy. The associated WISE images are almost always much bluer than for those in Figure 23, as would be expected for systems with little ISM, and thus minimal PAH emission.

For reference, Appendix E shows the associated ALLWISE thumbnails for the HST images in Figures 23–25. We also include in the Appendix a mosaic of HST images for the galaxies that did not show up in either of the mid-IR or radio continuum selected subsamples (Figure E6). These tend to include systems with: (1) individual, non-interacting galaxies; (2) galaxies with evidence for minor mergers; (3) earlier type galaxies; and (4) more distant systems, which are biased against the ALLWISE and VLA selections. However, some of the systems do seem to show major interactions, for which the lack of a strong central mid-IR or continuum source may be surprising, and thus worthy of further investigation.

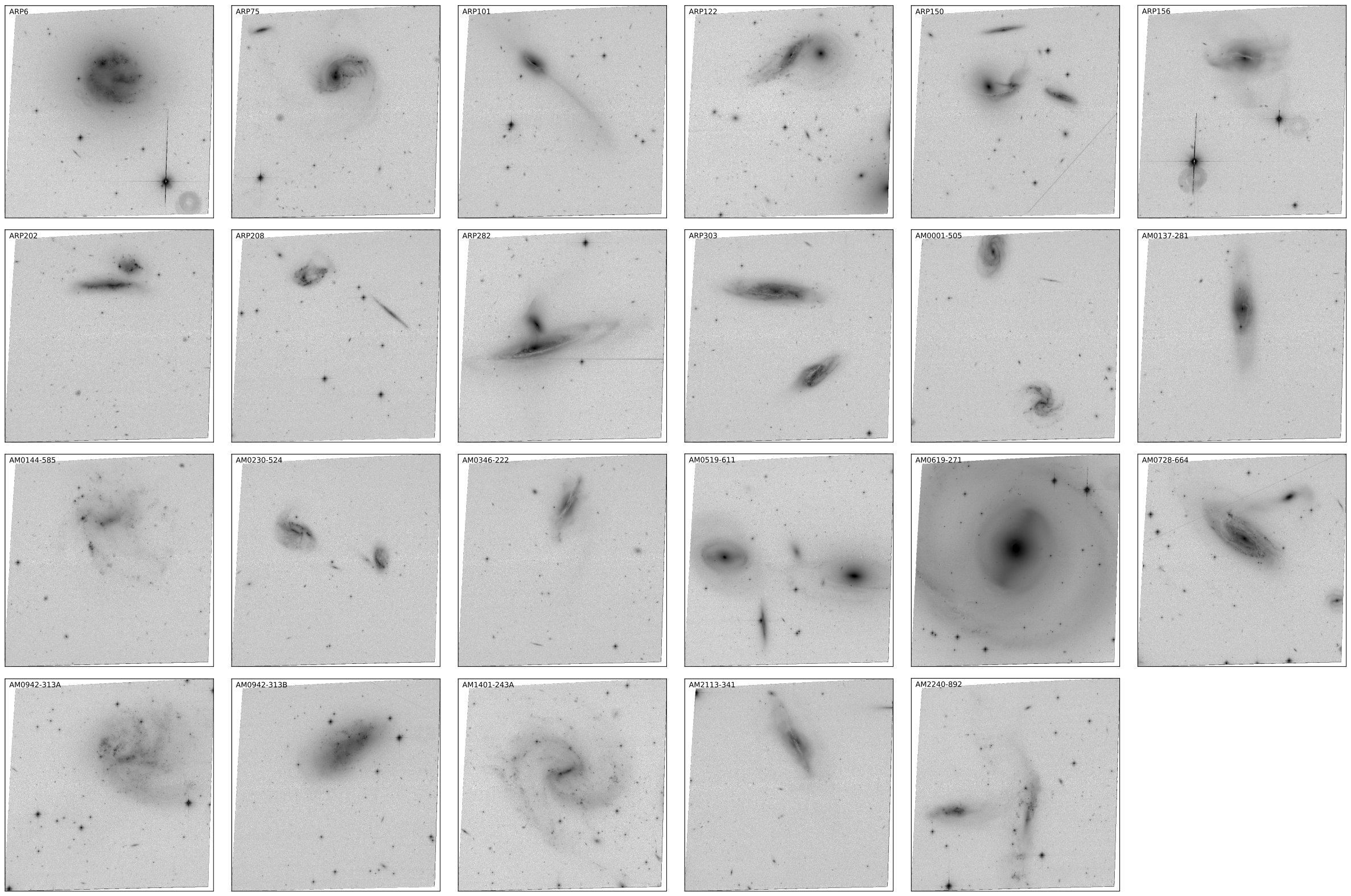

**Figure 24.** Thumbnails of systems containing at least one galaxy with moderate, red, central WISE W3+W4 emission, but weaker than Figure 23 and/or amiguous evidence for being a point source. The galaxies with WISE W3+W4 emission that were observed with VLASS (north of $\delta = -40^{\circ}$) all have detectable VLASS emission aligned with the WISE source, with the exception of Arp 6, Arp 208, AM0942-313A, and AM1401-243A, which were observed but had no visible radio emission in VLASS. Systems shown are: ARP6; ARP75; ARP101; ARP122; ARP150; ARP156; ARP202; ARP208; ARP282; ARP303; AM0001-505; AM0137-281; AM0144-585; AM0230-524; AM0346-222; AM0519-611; AM0619-271; AM0728-664; AM0942-313A; AM0942-313B; AM1401-243A; AM2113-341; AM2240-892

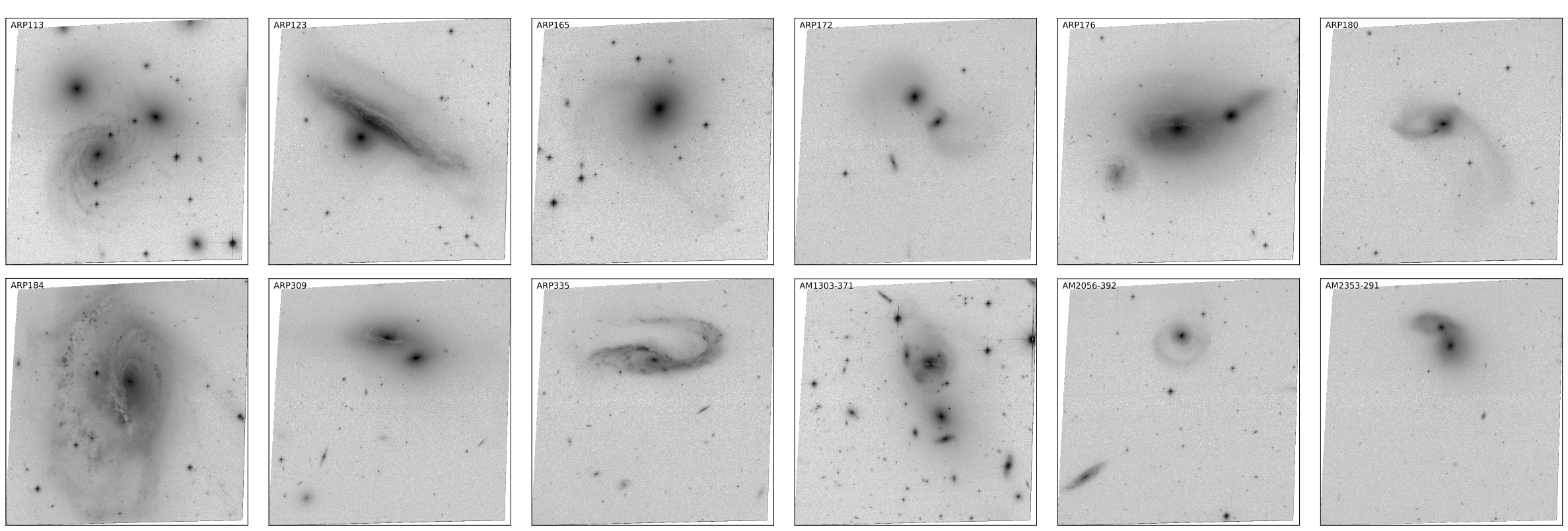

**Figure 25.** Thumbnails of systems containing at least one galaxy with a likely radio detection in VLASS, but lacking red, compact emission in ALLWISE W3+W4. Systems shown are: ARP113; ARP123; ARP165; ARP172; ARP176; ARP180; ARP184; ARP309; ARP335; AM1303-371; AM2056-392; AM2353-291;

## 5. CONCLUSION

The extensive optical HST imaging in this atlas presents hundreds of opportunities for new investigations into the extremes of galaxy formation and evolution. On their own, the images reveal morphologies that strongly constrain the nature of these sources, whose complex structures would easily lead to multiple, conflicting interpretations if only viewed at low resolution. The power of these high-resolution views is multiplied when combined with multiwavelength data (as shown in the main atlas images) and/or spectroscopy. Our hope is that this atlas will be a resource for the astronomical community, and serve as a starting point for new analyses and new proposals for complementary data. Some obvious follow-up strategies include imaging with JWST or ALMA for gas content and embedded sources, wide-field integral field unit (IFU) observations for spatially-resolved spectroscopy across entire systems, adaptive-optics spectroscopy of particularly interesting nuclear activity, and targeted numerical simulations to constrain interaction timescales.

## ACKNOWLEDGMENTS

The authors thank Dustin Lang for assistance with `legacysurvey.org` data products, Bill Keel for backchannel geekery over Galaxies We Love, John O'Meara for general enthusiasm, Judy Schmidt for inspirational image processing (`https://flic.kr/s/aHsmAPT18o`), Christina Lindberg & Zhuo Chen for helpful code snippets, and the HST scheduling team for their service & skill. JJD thanks the Max Planck Institüt for Astronomie for hospitality during some of the writing of this paper. The referee is also warmly thanked for thoughtful comments that helped improve the paper.

This paper made use of the `legacysurvey.org` website to provide non-HST imaging for comparison. The Legacy Surveys consist of three individual and complementary projects: the Dark Energy Camera Legacy Survey (DECaLS; Proposal ID #2014B-0404; PIs: David Schlegel and Arjun Dey), the Beijing-Arizona Sky Survey (BASS; NOAO Prop. ID #2015A-0801; PIs: Zhou Xu and Xiaohui Fan), and the Mayall z-band Legacy Survey (MzLS; Prop. ID #2016A-0453; PI: Arjun Dey). DECaLS, BASS and MzLS together include data obtained, respectively, at the Blanco telescope, Cerro Tololo Inter-American Observatory, NSF's NOIRLab; the Bok telescope, Steward Observatory, University of Arizona; and the Mayall telescope, Kitt Peak National Observatory, NOIRLab. Pipeline processing and analyses of the data were supported by NOIRLab and the Lawrence Berkeley National Laboratory (LBNL). The Legacy Surveys project is honored to be permitted to conduct astronomical research on Iolkam Du'ag (Kitt Peak), a mountain with particular significance to the Tohono O'odham Nation.

NOIRLab is operated by the Association of Universities for Research in Astronomy (AURA) under a cooperative agreement with the National Science Foundation. LBNL is managed by the Regents of the University of California under contract to the U.S. Department of Energy.

This project used data obtained with the Dark Energy Camera (DECam), which was constructed by the Dark Energy Survey (DES) collaboration. Funding for the DES Projects has been provided by the U.S. Department of Energy, the U.S. National Science Foundation, the Ministry of Science and Education of Spain, the Science and Technology Facilities Council of the United Kingdom, the Higher Education Funding Council for England, the National Center for Supercomputing Applications at the University of Illinois at Urbana-Champaign, the Kavli Institute of Cosmological Physics at the University of Chicago, Center for Cosmology and Astro-Particle Physics at the Ohio State University, the Mitchell Institute for Fundamental Physics and Astronomy at Texas A&M University, Financiadora de Estudos e Projetos, Fundacao Carlos Chagas Filho de Amparo, Financiadora de Estudos e Projetos, Fundacao Carlos Chagas Filho de Amparo a Pesquisa do Estado do Rio de Janeiro, Conselho Nacional de Desenvolvimento Cientifico e Tecnologico and the Ministerio da Ciencia, Tecnologia e Inovacao, the Deutsche Forschungsgemeinschaft and the Collaborating Institutions in the Dark Energy Survey. The Collaborating Institutions are Argonne National Laboratory, the University of California at Santa Cruz, the University of Cambridge, Centro de Investigaciones Energeticas, Medioambientales y Tecnologicas-Madrid, the University of Chicago, University College London, the DES-Brazil Consortium, the University of Edinburgh, the Eidgenossische Technische Hochschule (ETH) Zurich,

Fermi National Accelerator Laboratory, the University of Illinois at Urbana-Champaign, the Institut de Ciencies de l'Espai (IEEC/CSIC), the Institut de Fisica d'Altes Energies, Lawrence Berkeley National Laboratory, the Ludwig Maximilians Universitat Munchen and the associated Excellence Cluster Universe, the University of Michigan, NSF's NOIRLab, the University of Nottingham, the Ohio State University, the University of Pennsylvania, the University of Portsmouth, SLAC National Accelerator Laboratory, Stanford University, the University of Sussex, and Texas A&M University.

BASS is a key project of the Telescope Access Program (TAP), which has been funded by the National Astronomical Observatories of China, the Chinese Academy of Sciences (the Strategic Priority Research Program "The Emergence of Cosmological Structures" Grant # XDB09000000), and the Special Fund for Astronomy from the Ministry of Finance. The BASS is also supported by the External Cooperation Program of Chinese Academy of Sciences (Grant # 114A11KYSB20160057), and Chinese National Natural Science Foundation (Grant # 12120101003, # 11433005).

The Legacy Survey team makes use of data products from the Near-Earth Object Wide-field Infrared Survey Explorer (NEOWISE), which is a project of the Jet Propulsion Laboratory/California Institute of Technology. NEOWISE is funded by the National Aeronautics and Space Administration.

The Legacy Surveys imaging of the DESI footprint is supported by the Director, Office of Science, Office of High Energy Physics of the U.S. Department of Energy under Contract No. DE-AC02-05CH1123, by the National Energy Research Scientific Computing Center, a DOE Office of Science User Facility under the same contract; and by the U.S. National Science Foundation, Division of Astronomical Sciences under Contract No. AST-0950945 to NOAO.

This work made use of Astropy:[16] a community-developed core Python package and an ecosystem of tools and resources for astronomy ( Astropy Collaboration et al. 2013, 2018, 2022).

The Flatiron Institute is funded by the Simons Foundation.

*Facilities:* HST(ACS), VLA, WISE, GALEX. All the HST data presented in this paper were obtained from the Mikulski Archive for Space Telescopes (MAST) at the Space Telescope Science Institute. The specific observations analyzed can be accessed via DOI:10.17909/htjd-bj10. Point-source photometry catalogs and reduced HST images are available at MAST as a High Level Science Product via DOI:10.17909/176w-p735.

*Software:* Astropy ( Astropy Collaboration et al. 2013, 2018, 2022), Numpy (C. R. Harris et al. 2020), Seaborn (M. Waskom 2021), Matplotlib (J. D. Hunter 2007) texttthst1pass (J. Anderson 2022)

# APPENDIX

## A. UNOBSERVED TARGETS

As discussed in Section 2, the systems observed in this atlas were part of a larger set of targets that were well-matched to the ACS FOV and the authors found worthy of future study. Because of the nature of the "gap filler" observing program on HST, not all of the systems were actually observed. Here, we provide a list of these targets (Table A1) and corresponding thumbnail images from the Digitized Sky Survey, which may be of interest as potential future targets for high-resolution, space-based observations in the optical or near-infrared.

**Table A1**. Systems without Data

| Unobserved | | | |
|---|---|---|---|
| ARP62 | ARP332 | AM0413-283 | AM1254-321 |

**Table A1** *continued*

[16] http://www.astropy.org

**Table A1** *(continued)*

| Unobserved | | | |
|---|---|---|---|
| ARP83 | ARP334 | AM0421-404 | AM1259-322 |
| ARP84 | AM0003-362 | AM0426-480 | AM1315-350 |
| ARP90 | AM0004-413 | AM0447-595 | AM1316-240 |
| ARP93 | AM0005-301 | AM0448-320 | AM1317-303 |
| ARP106 | AM0028-230 | AM0456-224 | AM1325-292 |
| ARP108 | AM0033-253 | AM0503-380 | AM1349-273 |
| ARP111 | AM0036-660 | AM0507-222 | AM1351-263 |
| ARP119 | AM0038-634 | AM0517-321 | AM1354-250 |
| ARP127 | AM0112-554 | AM0526-393 | AM2035-521 |
| ARP138 | AM0115-444 | AM0537-292 | AM2050-441 |
| ARP159 | AM0117-412 | AM0541-524 | AM2051-410 |
| ARP171 | AM0122-381 | AM0545-453 | AM2100-393 |
| ARP173 | AM0134-373 | AM0545-520 | AM2134-471 |
| ARP192 | AM0136-301 | AM0547-330 | AM2143-464 |
| ARP198 | AM0138-261 | AM0547-474 | AM2145-351 |
| ARP199 | AM0142-435 | AM0949-264 | AM2159-330 |
| ARP203 | AM0155-580 | AM0956-282 | AM2159-541 |
| ARP222 | AM0227-264 | AM1036-272 | AM2203-281 |
| ARP260 | AM0227-484 | AM1122-262 | AM2204-311 |
| ARP270 | AM0239-242 | AM1124-272 | AM2206-280 |
| ARP289 | AM0307-411 | AM1133-374 | AM2209-382 |
| ARP292 | AM0316-235 | AM1157-242 | AM2209-472 |
| ARP294 | AM0322-374 | AM1158-345 | AM2210-262 |
| ARP305 | AM0338-375 | AM1207-292 | AM2213-370 |
| ARP313 | AM0400-420 | AM1223-385 | AM2244-651 |
| ARP320 | AM0410-330 | AM1252-300 | AM2348-405 |

| Failed or Unusable Observations (Not Shown) | | | |
|---|---|---|---|
| AM0003-414 | AM0226-320 | AM2128-302 | AM2146-350 |
| AM0143-781 | AM1213-430 | AM2133-384 | |

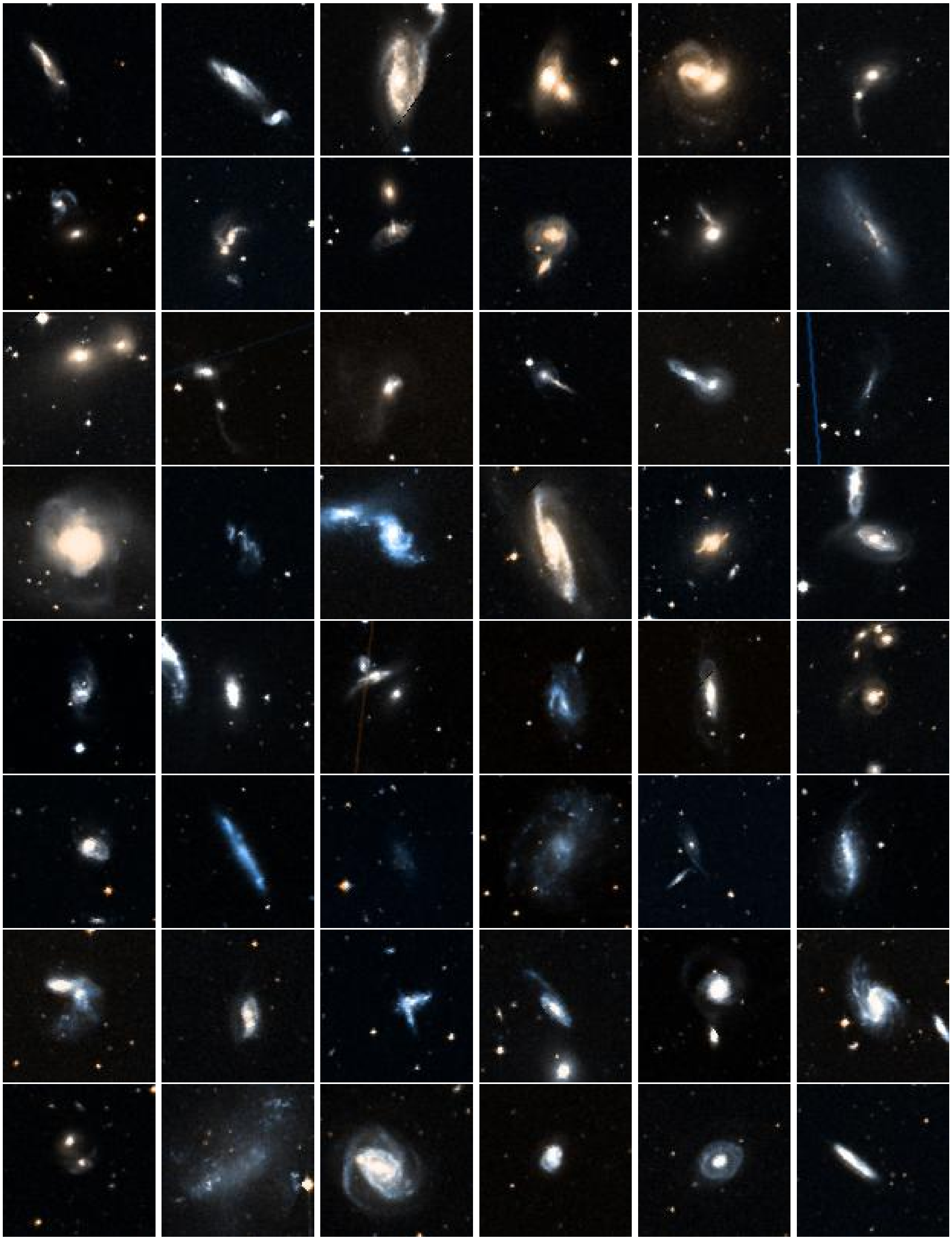

**Figure A1.** Thumbnails from the Digitized Sky Survey showing images selected as snapshot targets, but that were not observed by HST. These are included to provide particularly interesting, well-resolved systems that may be of interest for future studies. Sources are ordered from top left to bottom right, following the order of "Unobserved" section of Table A1. [Continued]

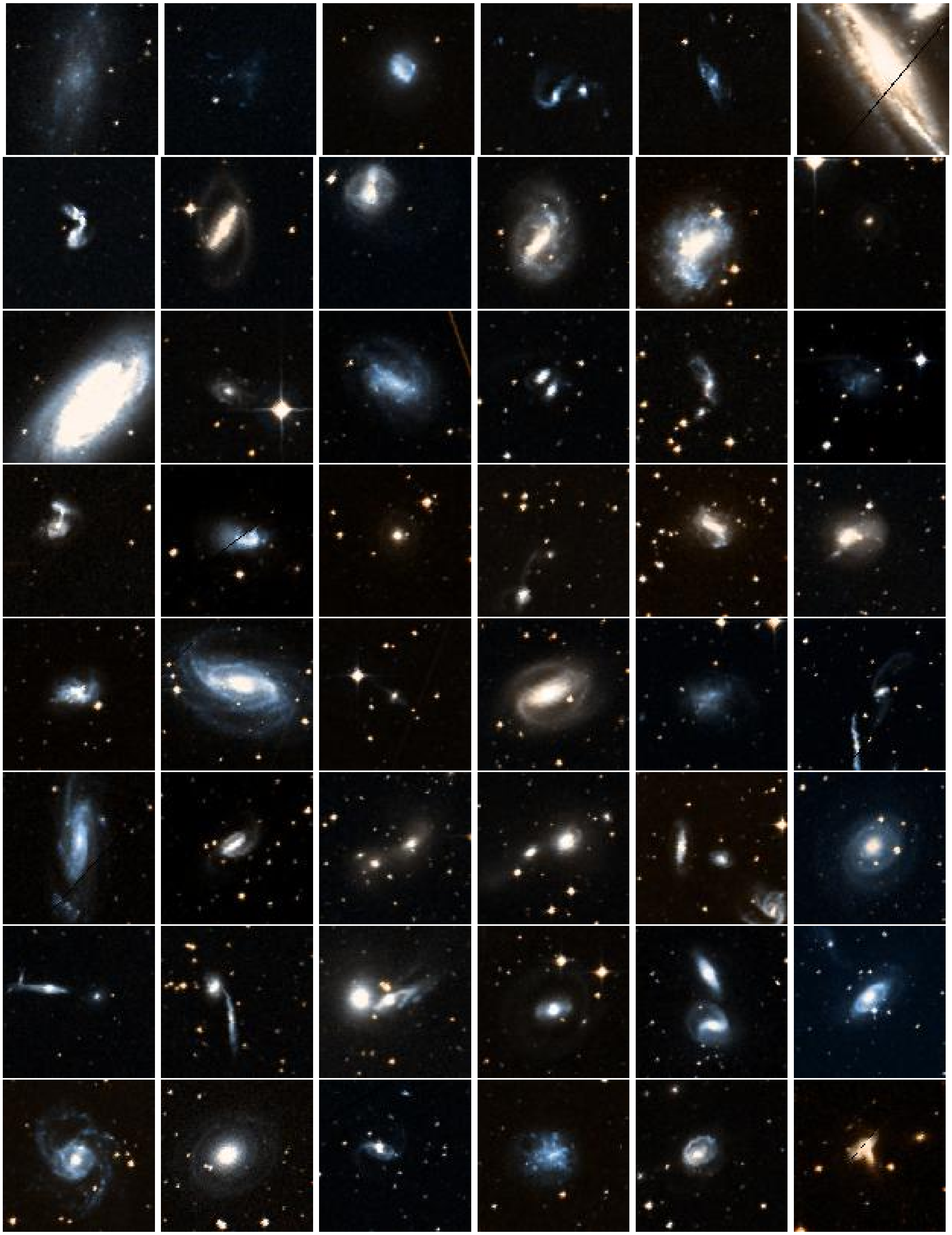

**Figure A1.** Thumbnails from the Digitized Sky Survey showing images selected as snapshot targets, but that were not observed by HST. These are included to provide particularly interesting, well-resolved systems that may be of interest for future studies. Sources are ordered from top left to bottom right, following the order of "Unobserved" section of Table A1. [Continued]

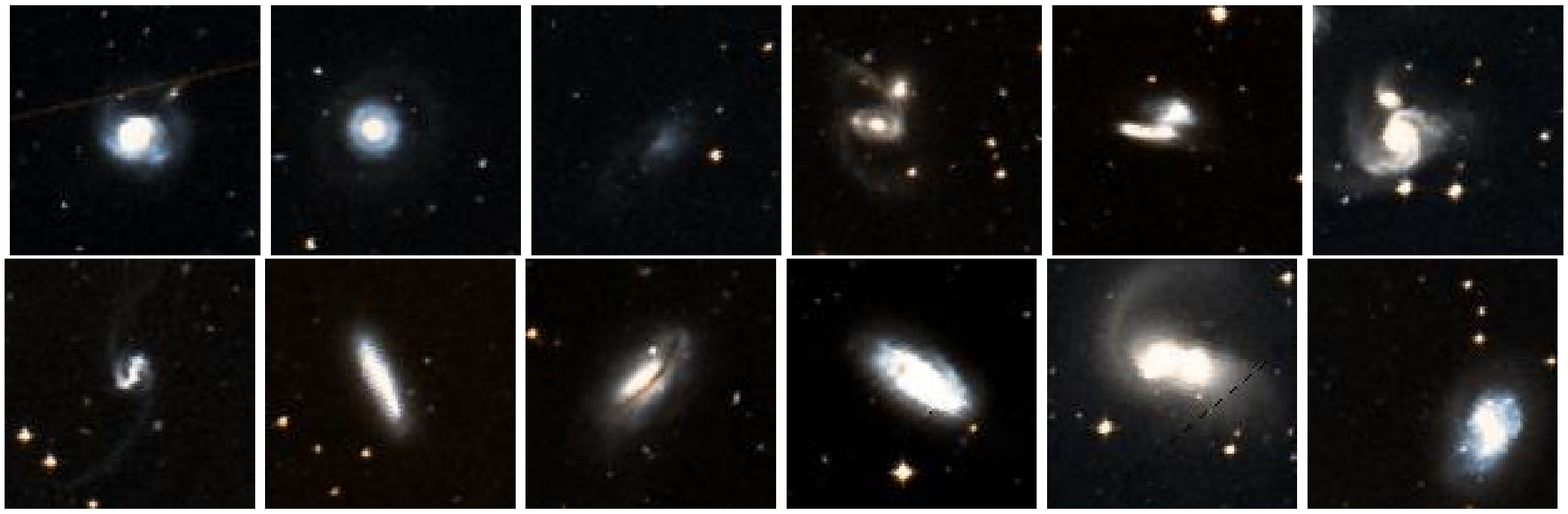

**Figure A1.** Thumbnails from the Digitized Sky Survey showing images selected as snapshot targets, but that were not observed by HST. These are included to provide particularly interesting, well-resolved systems that may be of interest for future studies. Sources are ordered from top left to bottom right, following the order of "Unobserved" section of Table A1.

## B. ATLAS IMAGES

In this appendix we include full page summary images of the systems in this atlas, as described in detail in Section 2.1. The large grayscale images show the new HST F606W imaging at full resolution in sky coordinates, displayed with an inverted grayscale where pixel values $f$ are scaled as $[(x - v_{min})/(v_{max} - v_{min})]^{\gamma}$ ("`PowerNorm`" in matplotlib), where $x = \log_{10}(f + 2.5f_0)$, with $f_0 = 0.05$, $v_{min} = -1$, $v_{max} = 10$, and $\gamma = 0.5$, to preserve detail at both the highest and lowest surface brightnesses. Zooming into the images is highly encouraged. The dark horizontal bar in the lower right of each image indicates an approximate physical scale in kiloparsecs, based on the adopted recessional velocity of the system listed in Table 1, assuming a pure Hubble Flow with $H_0 = 70\,\mathrm{km\ s^{-1}Mpc^{-1}}$; this scale will not necessarily be accurate for nearer galaxies where peculiar velocities are significant, or, for foreground or background interloping galaxies. The bottom row shows a panchromatic view of the galaxy, from short to long wavelengths. From left to right, the panels show: GALEX (NUV+FUV); the Legacy Survey imaging (Optical); NEOWISE (W1+W2 and W3+W4); and the VLA Sky Survey (1.4 GHz continuum), all matched to the field of view of the upper plot. Blank spaces in the lower row indicate cases where there was no suitable multiwavelength data available.

Systems are presented in the same order as Table 1 and as the thumbnails in Figure 4. Please see Section 2.1 for more details. Due to space limitations, only an example image is included in the ArXiv version of the paper; a full PDF including all images is available at `https://doi.org/10.5281/zenodo.16778896`, or as electronic figures in the on-line journal).

**Fig. Set B2. Atlas Images**

## C. GALAXY MEMBERSHIP IN HST ATLAS IMAGES

The Arp and Arp-Madore systems in this Atlas are sufficiently close that many of their constituent galaxies have already been cataloged in other surveys and studied in the literature. Simultaneously, the 4′ field of view of the ACS camera is wide enough that various cataloged, but physically unassociated, background galaxies may serendipitously appear in the frame.

While a detailed discussion of the constituents of every HST image is outside the scope of this paper, we aim to facilitate readers' follow-up analyses of images of interest. We therefore present in Table C2 a complete listing of cataloged galaxies that overlap the image frame. These entries were pulled automatically from the SIMBAD database, and include: the name of the system in this Atlas (column 1); identification name(s) of the system in SIMBAD (column 2); the cataloged galaxies' main identification name (column 3); our automated assessment of whether the galaxy is associated with the system or not (column 4; 1=yes, -1=no, 0=ambiguous; see discussion in Section 2.2); the SIMBAD classification of the galaxy type (column 5); the reported recessional velocity of the galaxy (column 6); the galaxies' angular diameter (column 7), apparent magnitude in $B$ (column 8), and in $K$ (column 9). We also include links that open an AladinLite viewer centered on the entry and the associated SIMBAD entry (columns 10 and 11, respectively), if the reader wishes to make a more detailed inspection of the available data and its literature sources. We note that in a very few cases (discussed in Section 2.2.1) we have overruled some of SIMBAD's assessments of the appropriate recessional velocity, based in part by the morphologies that are now visible in the new HST imaging. We would advise that anyone planning follow-up observations (particularly spectroscopy) double check the provenance of any critical redshift information.

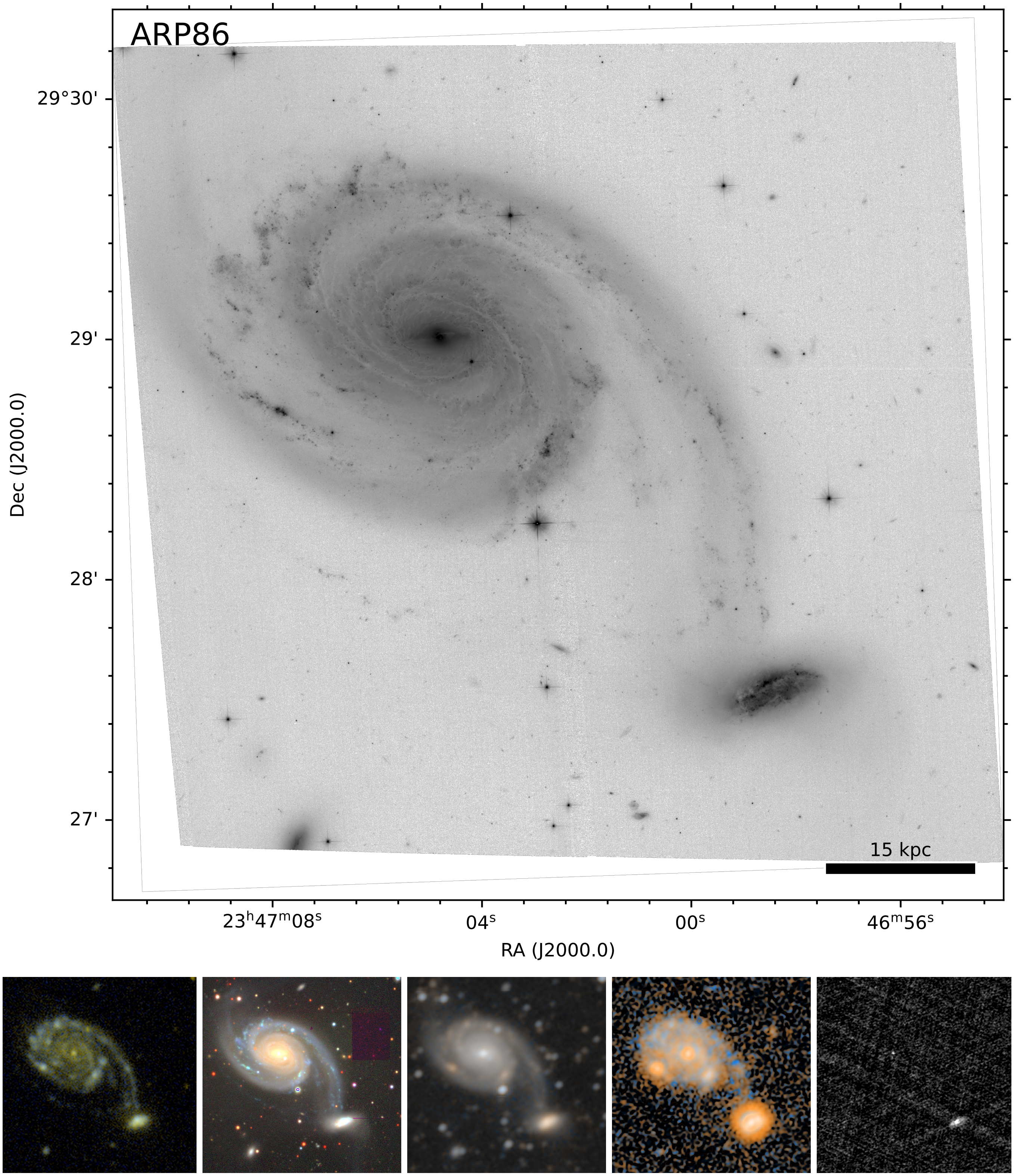

**Figure B2.** Example of multiwavelength imaging for Arp 86; the full set of 216 plots for the catalog are available in the online journal. The large upper panel shows the full-resolution version of Arp 86, as viewed with HST/ACS and plotted in sky coordinates (top; F606W). The scale bar (in kiloparsecs) is calculated from the recessional velocity listed in Table 1, assuming a pure Hubble Flow with $H_0 = 70\,\mathrm{km\ s^{-1}Mpc^{-1}}$; this scale will not necessarily be accurate for nearer galaxies where peculiar velocities are significant, or, for foreground or background interloping galaxies. The bottom row shows a panchromatic view of the galaxy, from short to long wavelengths. From left to right, the panels show: GALEX (NUV+FUV); the Legacy Survey imaging (Optical); NEOWISE (W1+W2 and W3+W4); and the VLA Sky Survey (1.4 GHz continuum), all matched to the field of view of the upper plot.

**Table C2**. Galaxies in Field

| Arp System (1) | Arp ID (2) | Main ID (3) | Associated? (4) | SIMBAD Type (5) | $V_r$ (km/s) (6) | Diam (′) (7) | $m_B$ (8) | $m_K$ (9) | Images (10) | Data (11) |
|---|---|---|---|---|---|---|---|---|---|---|
| ARP2 | ARP2 | UGC 10310 | 1 | InteractingG | 712 | 2.57 | 14.90 | - | Aladin | Simbad |
| | | LEDA 101538 | 1 | HIIG | 710 | 0.35 | 16.90 | - | Aladin | Simbad |
| ARP3 | ARP3 | MCG-01-57-016 | 1 | InteractingG | 1694 | 2.40 | 13.10 | - | Aladin | Simbad |
| ARP4 | ARP4 | MCG-02-05-050 | 1 | InteractingG | 1614 | 2.04 | 13.40 | - | Aladin | Simbad |
| | | 6dFGS gJ014825.7-122253 | 1 | Galaxy | 1617 | - | 14.01 | - | Aladin | Simbad |
| | | MCG-02-05-050a | 1 | EmissionG | 1617 | 0.62 | 14.00 | 11.19 | Aladin | Simbad |
| | | LEDA 953858 | 0 | Galaxy | - | 0.26 | - | - | Aladin | Simbad |
| ARP6 | ARP6 | NGC 2537 | 1 | InteractingG | 447 | - | 12.50 | 9.13 | Aladin | Simbad |
| | | [VV2006c] J081313.1+455926 | 1 | Seyfert1 | 599 | - | 19.18 | - | Aladin | Simbad |
| ARP15 | ARP15 | NGC 7393 | 1 | EmissionG | 3744 | 1.16 | 13.40 | 9.71 | Aladin | Simbad |
| ARP18 | ARP18 | NGC 4088 | 1 | GinPair | 753 | 5.62 | 11.20 | 7.48 | Aladin | Simbad |
| ARP20 | ARP20 | UGC 3014 | 1 | InteractingG | 4190 | 0.68 | 14.70 | 11.67 | Aladin | Simbad |
| ARP22 | ARP22 | NGC 4027 | 1 | InteractingG | 1662 | 1.90 | 11.71 | 8.50 | Aladin | Simbad |
| ARP24 | ARP24 | APG 24 | 1 | Galaxy | - | - | - | - | Aladin | Simbad |
| | | NGC 3445 | 1 | GinPair | 2062 | 0.82 | 12.90 | 10.61 | Aladin | Simbad |
| | | MCG+10-16-024 | 1 | GinPair | 2038 | 0.78 | 15.00 | - | Aladin | Simbad |
| | | NVSS J105436+565921 | 0 | RadioG | - | - | - | - | Aladin | Simbad |
| | | SDSSCGB 51774.3 | 0 | Galaxy | - | - | - | - | Aladin | Simbad |
| | | SDSSCGB 51774.2 | 0 | Galaxy | - | - | - | - | Aladin | Simbad |
| | | SDSSCGB 51774.1 | 0 | Galaxy | - | - | - | - | Aladin | Simbad |
| | | 2MASS J10544551+5659589 | -1 | AGN | 13922 | - | - | 13.98 | Aladin | Simbad |
| ARP44 | ARP44 | IC 609 | 1 | InteractingG | 5505 | 0.74 | 14.40 | 10.34 | Aladin | Simbad |
| | | 2dFGRS TGN222Z290 | 1 | GtowardsGroup | 5524 | - | 14.50 | 10.66 | Aladin | Simbad |
| | | VV 354b | 0 | Galaxy | - | 0.20 | - | - | Aladin | Simbad |
| | | 2dFGRS TGN222Z287 | -1 | Galaxy | 14045 | 0.36 | 17.67 | - | Aladin | Simbad |
| ARP58 | ARP58 | UGC 4457 | 1 | AGN | 10954 | 0.51 | 14.90 | 11.38 | Aladin | Simbad |
| | | 2MASS J08320053+1912058 | 1 | AGN | 11046 | - | - | 15.66 | Aladin | Simbad |
| ARP59 | ARP59 | NGC 341 | 1 | InteractingG | 4549 | 0.77 | 14.10 | 10.17 | Aladin | Simbad |
| | | MCG-02-03-064 | 0 | Galaxy | - | 0.34 | 15.00 | - | Aladin | Simbad |
| ARP70 | ARP70, ARP70B | UGC 934 | 1 | Seyfert2 | 10457 | 0.50 | 14.50 | 10.95 | Aladin | Simbad |
| | ARP70A | APG 70A | 1 | Galaxy | 10121 | 0.28 | 14.77 | 12.06 | Aladin | Simbad |
| ARP72 | ARP72 | UGC 10033 | 1 | InteractingG | 3292 | - | - | - | Aladin | Simbad |
| | ARP72B | NGC 5996 | 1 | HIIG | 3366 | 0.89 | 13.20 | 10.41 | Aladin | Simbad |
| | | SDSSCGB 5972.4 | 0 | Galaxy | - | - | - | - | Aladin | Simbad |
| | | SDSSCGB 5972.2 | 0 | Galaxy | - | - | - | - | Aladin | Simbad |
| | | SDSSCGB 5972.1 | -1 | Galaxy | 54310 | - | - | - | Aladin | Simbad |
| ARP75 | ARP75 | NGC 702 | 1 | InteractingG | 10668 | 1.17 | 14.20 | - | Aladin | Simbad |

**Table C2** *continued*

**Table C2** *(continued)*

| Arp System (1) | Arp ID (2) | Main ID (3) | Associated? (4) | SIMBAD Type (5) | $V_r$ (km/s) (6) | Diam (′) (7) | $m_B$ (8) | $m_K$ (9) | Images (10) | Data (11) |
|---|---|---|---|---|---|---|---|---|---|---|
| | | 2MASX J01511920-0403217 | 1 | EmissionG | 10624 | 0.96 | 17.63 | 10.24 | Aladin | Simbad |
| | | LEDA 144370 | 0 | Galaxy | - | 0.45 | - | - | Aladin | Simbad |
| | | 2MASX J01511833-0402567 | 0 | Galaxy | - | 0.38 | - | 14.44 | Aladin | Simbad |
| ARP82 | ARP82 | APG 82 | 1 | PairG | - | - | - | - | Aladin | Simbad |
| | | LAMOST J081115.12+251225.2 | 1 | Galaxy | 4091 | - | - | - | Aladin | Simbad |
| | ARP82A | NGC 2535 | 1 | RadioG | 4052 | 1.13 | 13.50 | 10.12 | Aladin | Simbad |
| | ARP82B | NGC 2536 | 1 | GinPair | 4082 | 0.56 | 14.80 | 11.71 | Aladin | Simbad |
| | | KPG 156 | 0 | PairG | - | - | - | - | Aladin | Simbad |
| ARP86 | ARP86 | APG 86 | 1 | PairG | - | - | - | - | Aladin | Simbad |
| | | NGC 7753 | 1 | InteractingG | 5153 | 1.47 | 13.20 | 8.61 | Aladin | Simbad |
| | ARP86B | NGC 7752 | 1 | GinPair | 4840 | 0.56 | 14.30 | 11.19 | Aladin | Simbad |
| | | 2MASX J23470758+2926531 | 0 | Galaxy | - | 0.16 | - | 13.80 | Aladin | Simbad |
| ARP89 | ARP89 | NGC 2648 | 1 | GinPair | 2064 | 1.72 | 12.80 | 8.66 | Aladin | Simbad |
| | | KPG 168 | 0 | PairG | - | - | - | - | Aladin | Simbad |
| ARP91 | ARP91 | APG 91 | 1 | InteractingG | - | - | 12.30 | - | Aladin | Simbad |
| | ARP91B | NGC 5953 | 1 | GinPair | 1990 | 1.58 | 13.23 | 9.16 | Aladin | Simbad |
| | ARP91A | NGC 5954 | 1 | GinPair | 1971 | 1.31 | 13.70 | 10.00 | Aladin | Simbad |
| ARP97 | ARP97 | UGC 7085 A | 1 | InteractingG | 6830 | 3.02 | 15.10 | - | Aladin | Simbad |
| | | [BKB95] 1203+3120 | 1 | GtowardsCl | 6910 | - | - | - | Aladin | Simbad |
| | ARP97A | APG 97A | 1 | GtowardsGroup | 6916 | 0.36 | 15.10 | 11.63 | Aladin | Simbad |
| | ARP97B | NVSS B120312+312014 | 1 | RadioG | 6834 | 0.27 | - | 12.16 | Aladin | Simbad |
| | | MCG+05-29-012 | 0 | Galaxy | 7425 | 0.39 | 15.00 | - | Aladin | Simbad |
| ARP100 | ARP100 | IC 18 | 1 | InteractingG | 6071 | 0.62 | 15.00 | 11.04 | Aladin | Simbad |
| | | UGC 10169 | 1 | GtowardsGroup | 4599 | 0.59 | 15.20 | 10.74 | Aladin | Simbad |
| ARP107 | ARP107, ARP107A | UGC 5984 | 1 | Seyfert2 | 10167 | 0.52 | 14.60 | 11.40 | Aladin | Simbad |
| | | MCG+05-26-025 | 1 | GinPair | 10272 | 0.44 | 14.60 | 11.31 | Aladin | Simbad |
| | | LEDA 1889058 | 0 | Galaxy | - | 0.32 | - | - | Aladin | Simbad |
| ARP112 | ARP112 | NGC 7806 | 1 | GinPair | 4767 | 0.69 | 14.40 | 10.35 | Aladin | Simbad |
| | | NGC 7805 | 1 | GinPair | 4791 | 0.61 | 14.30 | 10.13 | Aladin | Simbad |
| | | KUG 2359+311 | 0 | Galaxy | - | - | 16.50 | - | Aladin | Simbad |
| | | MCG+05-01-026 | 0 | Galaxy | - | 0.65 | 17.00 | - | Aladin | Simbad |
| ARP113 | ARP113 | NGC 70 | 1 | Seyfert2 | 7181 | 1.00 | 14.50 | 9.66 | Aladin | Simbad |
| | | NGC 71 | 1 | Seyfert2 | 6623 | 1.80 | 14.80 | 9.54 | Aladin | Simbad |
| | | NGC 69 | 1 | GtowardsGroup | 6607 | 0.33 | 15.70 | 11.65 | Aladin | Simbad |
| | | NGC 67 | 1 | GtowardsGroup | 6636 | 0.21 | 15.70 | 12.59 | Aladin | Simbad |
| | | LEDA 1889390 | 0 | Galaxy | - | 0.40 | - | - | Aladin | Simbad |
| | | NGC 68 | -1 | GtowardsGroup | 5680 | 0.72 | 14.50 | 9.60 | Aladin | Simbad |
| ARP122 | ARP122 | NGC 6039 | 1 | Galaxy | 12082 | 0.89 | 14.60 | - | Aladin | Simbad |
| | | SDSS J160426.50+174431.1 | 1 | LINER | 12008 | 1.03 | 14.60 | 10.64 | Aladin | Simbad |

**Table C2** *continued*

**Table C2** *(continued)*

| Arp System (1) | Arp ID (2) | Main ID (3) | Associated? (4) | SIMBAD Type (5) | $V_r$ (km/s) (6) | Diam (′) (7) | $m_B$ (8) | $m_K$ (9) | Images (10) | Data (11) |
|---|---|---|---|---|---|---|---|---|---|---|
| | | NGC 6040 | 1 | LINER | 12615 | 0.75 | 14.60 | 10.95 | Aladin | Simbad |
| | | [FBD2002] ACO 2151 g6 | 0 | GtowardsCl | - | - | - | - | Aladin | Simbad |
| | | [DKP87] 160216.38+175402.6 | 0 | GtowardsCl | - | - | - | - | Aladin | Simbad |
| | | [BO85] ACO 2151 17 | 0 | GtowardsCl | - | - | - | - | Aladin | Simbad |
| | | 2MASX J16042175+1745431 | -1 | GtowardsCl | 20501 | 0.13 | - | 13.85 | Aladin | Simbad |
| ARP129 | ARP129 | UGC 5146 | 1 | InteractingG | 6511 | 0.62 | 14.30 | - | Aladin | Simbad |
| | | LAMOST J093925.76+322207.8 | 1 | Galaxy | 6919 | - | - | - | Aladin | Simbad |
| | | MCG+06-21-072 | 1 | GinPair | 6859 | 0.87 | 14.00 | 10.86 | Aladin | Simbad |
| ARP130 | ARP130 | IC 5378 | 1 | InteractingG | 6283 | 0.53 | 14.90 | 10.69 | Aladin | Simbad |
| | | MCG+03-01-016 | 1 | GinPair | 5777 | 0.60 | 14.90 | - | Aladin | Simbad |
| ARP136 | ARP136 | NGC 5820 | 1 | GinPair | 3251 | 1.10 | 13.00 | 9.14 | Aladin | Simbad |
| | | [HM2015b] 163 3 | 0 | Galaxy | - | - | - | - | Aladin | Simbad |
| | | LEDA 214353 | 0 | Galaxy | - | 0.30 | - | - | Aladin | Simbad |
| | | 2MASX J14584932+5352159 | -1 | Galaxy | 22140 | 0.17 | - | 13.87 | Aladin | Simbad |
| ARP141 | ARP141 | UGC 3730 | 1 | InteractingG | - | 2.82 | - | - | Aladin | Simbad |
| | | 2MASX J07142032+7328242 | 1 | Galaxy | 2632 | 0.72 | - | 11.07 | Aladin | Simbad |
| | | 2MASX J07142057+7328502 | 1 | GinPair | 2735 | 0.55 | 13.70 | 10.79 | Aladin | Simbad |
| ARP143 | ARP143 | NGC 2444 | 1 | InteractingG | 3955 | 0.41 | 13.10 | 10.14 | Aladin | Simbad |
| | | LEDA 3129223 | 1 | Galaxy | 4063 | - | - | - | Aladin | Simbad |
| | | UZC J074653.4+390019 | 1 | Galaxy | 4091 | - | - | - | Aladin | Simbad |
| | | NGC 2445 | 1 | GinPair | 3950 | 0.61 | 13.10 | 10.26 | Aladin | Simbad |
| | | [CCA99] UGC 4017b | 0 | PartofG | - | - | - | - | Aladin | Simbad |
| | | SDSS J074654.32+390240.7 | -1 | Seyfert1 | 133476 | - | 20.20 | - | Aladin | Simbad |
| ARP144 | ARP144 | NGC 7828 | 1 | InteractingG | 5729 | 0.79 | 14.00 | 10.58 | Aladin | Simbad |
| | | NGC 7829 | 1 | GinPair | 5717 | 0.34 | 15.07 | 10.94 | Aladin | Simbad |
| ARP145 | ARP145 | UGC 1840 | 1 | InteractingG | 5326 | 1.51 | 14.10 | 9.99 | Aladin | Simbad |
| | | LEDA 9060 | 1 | RadioG | 5228 | - | 17.80 | - | Aladin | Simbad |
| ARP150 | ARP150 | NGC 7609 | 1 | InteractingG | 11888 | 0.60 | 15.23 | 10.55 | Aladin | Simbad |
| | | HCG 95 | 1 | Compact_Gr_G | 11637 | 1.50 | - | - | Aladin | Simbad |
| | | Z 406-67 | 1 | Galaxy | 11637 | 0.50 | 16.12 | 11.62 | Aladin | Simbad |
| | | MCG+01-59-046 | 1 | Galaxy | 11562 | 0.48 | 16.38 | 12.25 | Aladin | Simbad |
| | | VV 20b | 0 | Galaxy | 12350 | 0.29 | 17.23 | 11.84 | Aladin | Simbad |
| ARP156 | ARP156 | UGC 5814 | 1 | InteractingG | 1875 | 0.67 | 16.58 | 10.27 | Aladin | Simbad |
| ARP158 | ARP158 | NGC 523 | 1 | InteractingG | 4761 | 1.20 | 13.50 | 9.71 | Aladin | Simbad |
| | | [BDG98] J012519.5+340121 | 1 | GtowardsCl | 4824 | - | - | - | Aladin | Simbad |
| | | [BDG98] J012522.4+340133 | 1 | GtowardsCl | 4824 | - | - | - | Aladin | Simbad |
| ARP163 | ARP163 | NGC 4670 | 1 | InteractingG | 1071 | 0.71 | 12.60 | 10.41 | Aladin | Simbad |
| | | [BKD2008] WR 523 | 1 | HIIG | 1044 | - | - | - | Aladin | Simbad |
| | | SDSS J124517.09+270730.7 | 0 | Galaxy | - | - | - | - | Aladin | Simbad |

**Table C2** *continued*

**Table C2** *(continued)*

| Arp System (1) | Arp ID (2) | Main ID (3) | Associated? (4) | SIMBAD Type (5) | $V_r$ (km/s) (6) | Diam (′) (7) | $m_B$ (8) | $m_K$ (9) | Images (10) | Data (11) |
|---|---|---|---|---|---|---|---|---|---|---|
| | | SDSS J124521.89+270755.3 | -1 | QSO | 129694 | - | - | - | Aladin | Simbad |
| ARP164 | ARP164 | NGC 455 | 1 | InteractingG | 5231 | 0.93 | 13.90 | 9.60 | Aladin | Simbad |
| ARP165 | ARP165 | NGC 2418 | 1 | InteractingG | 5043 | 0.97 | 13.70 | 8.94 | Aladin | Simbad |
| ARP172 | ARP172 | APG 172 | 1 | InteractingG | - | - | - | - | Aladin | Simbad |
| | | IC 1178 | 1 | GtowardsGroup | 10072 | 0.57 | 15.00 | 10.43 | Aladin | Simbad |
| | | IC 1181 | 1 | GtowardsGroup | 10014 | 0.45 | 15.00 | 11.17 | Aladin | Simbad |
| | | 2MASX J16053730+1735558 | 1 | Galaxy | 9732 | 0.22 | - | 14.02 | Aladin | Simbad |
| ARP176 | ARP176, ARP176A | MCG-02-33-101 | 1 | InteractingG | 3107 | 3.47 | 15.00 | - | Aladin | Simbad |
| | | 6dFGS gJ130357.0-112948 | 1 | Galaxy | 3221 | - | 12.67 | - | Aladin | Simbad |
| | | 6dFGS gJ130356.4-112955 | 1 | Galaxy | 3221 | - | 12.50 | - | Aladin | Simbad |
| | ARP176B | IC 4176 | 1 | GtowardsGroup | 3238 | 1.56 | 13.00 | 8.43 | Aladin | Simbad |
| | | MCG-02-33-103 | 1 | GtowardsGroup | 3188 | 0.66 | 17.00 | - | Aladin | Simbad |
| | | NGC 4933 | 0 | Galaxy | - | - | 13.18 | - | Aladin | Simbad |
| ARP180 | ARP180 | MCG-01-13-034 | 1 | InteractingG | 4289 | 0.47 | 16.00 | 11.17 | Aladin | Simbad |
| ARP184 | ARP184 | NGC 1961 | 1 | LINER | 3889 | 4.17 | 11.73 | 7.73 | Aladin | Simbad |
| ARP190 | ARP190 | UGC 2320 | 1 | InteractingG | 10216 | 1.20 | 15.20 | - | Aladin | Simbad |
| | | LEDA 200206 | 0 | Galaxy | - | 0.48 | - | - | Aladin | Simbad |
| ARP195 | ARP195 | UGC 4653 | 1 | InteractingG | 16696 | 0.98 | 15.00 | 10.65 | Aladin | Simbad |
| | | VV 243c | 1 | GtowardsGroup | 16690 | - | - | - | Aladin | Simbad |
| | | NVSS J085354+350856 | 1 | RadioG | 16520 | 0.79 | - | - | Aladin | Simbad |
| | | SDSS J085354.42+350906.7 | 0 | RadioG | - | - | - | - | Aladin | Simbad |
| | | [WZX98] 08507+3520C | 0 | Galaxy | - | - | - | - | Aladin | Simbad |
| | | [WZX98] 08507+3520B | 0 | Galaxy | - | - | - | - | Aladin | Simbad |
| ARP197 | ARP197 | IC 701 | 1 | InteractingG | 6077 | 0.50 | 14.70 | 11.54 | Aladin | Simbad |
| ARP200 | ARP200 | NGC 1134 | 1 | InteractingG | 3611 | 1.45 | 13.20 | 8.81 | Aladin | Simbad |
| ARP202 | ARP202B | MCG+06-20-018 | 1 | GinPair | 3070 | 1.07 | 14.00 | 12.98 | Aladin | Simbad |
| ARP202 | ARP202A | NGC 2719 | 1 | LowSurfBrghtG | 3144 | 0.87 | 13.70 | 11.83 | Aladin | Simbad |
| | | 2MFGC 7027 | 0 | LowSurfBrghtG | - | 0.63 | - | 13.28 | Aladin | Simbad |
| ARP204 | ARP204 | UGC 8454 | 1 | InteractingG | - | 0.30 | 15.60 | 11.72 | Aladin | Simbad |
| | | MCG+14-06-025 | 0 | GinPair | - | 0.42 | 16.00 | 12.57 | Aladin | Simbad |
| ARP205 | ARP205 | NGC 3448 | 1 | InteractingG | 1367 | - | 12.20 | 9.47 | Aladin | Simbad |
| | | DES J105429.71+541858.8 | 0 | GtowardsCl | - | - | - | - | Aladin | Simbad |
| ARP208 | ARP208 | APG 208 | 1 | InteractingG | 9025 | - | - | - | Aladin | Simbad |
| | | MCG+08-31-009 | 1 | Galaxy | 9231 | - | 17.00 | - | Aladin | Simbad |
| | | MCG+08-31-010 | 1 | Galaxy | 9025 | 0.30 | 16.00 | 12.64 | Aladin | Simbad |
| | | MCG+08-31-011 | 0 | GinPair | - | 0.51 | 16.00 | - | Aladin | Simbad |
| | | SDSS J165106.62+471243.8 | -1 | QSO | 228099 | - | - | - | Aladin | Simbad |
| | | SDSS J165103.44+471234.7 | -1 | QSO | 224435 | - | - | - | Aladin | Simbad |
| ARP216 | ARP216 | NGC 7679 | 1 | Seyfert2 | 5149 | 0.61 | 13.20 | 10.02 | Aladin | Simbad |

**Table C2** *continued*

**Table C2** *(continued)*

| Arp System (1) | Arp ID (2) | Main ID (3) | Associated? (4) | SIMBAD Type (5) | $V_r$ (km/s) (6) | Diam (′) (7) | $m_B$ (8) | $m_K$ (9) | Images (10) | Data (11) |
|---|---|---|---|---|---|---|---|---|---|---|
| | | RASSCALS SRGb037.046 | 1 | GtowardsGroup | 5050 | - | - | - | Aladin | Simbad |
| | | SDSS J232848.57+033042.2 | 0 | Galaxy | - | - | - | - | Aladin | Simbad |
| ARP219 | ARP219 | UGC 2812 | 1 | InteractingG | 10436 | 0.57 | 15.26 | 11.31 | Aladin | Simbad |
| | | 2MASX J03395491-0207241 | 0 | Galaxy | - | 0.18 | - | 14.00 | Aladin | Simbad |
| ARP221 | ARP221 | MCG-02-25-006 | 1 | InteractingG | 5495 | 0.96 | 13.91 | 9.32 | Aladin | Simbad |
| | | IRAS 09340-1106 | 1 | Galaxy | 5546 | - | - | - | Aladin | Simbad |
| | | 2MASX J09362516-1119441 | 1 | GtowardsGroup | 5359 | 0.40 | - | 10.63 | Aladin | Simbad |
| | | LEDA 3098245 | 0 | Galaxy | - | 0.30 | - | - | Aladin | Simbad |
| ARP231 | ARP231 | IC 1575 | 1 | InteractingG | 5646 | 0.81 | 16.22 | 9.77 | Aladin | Simbad |
| | | IC 1575B | 1 | GtowardsGroup | 5748 | - | 14.00 | - | Aladin | Simbad |
| | | IC 1575A | 1 | GtowardsGroup | 5671 | 0.98 | 14.00 | - | Aladin | Simbad |
| ARP241 | ARP241 | NAME Segner's Wheel | 1 | InteractingG | 10222 | 0.77 | 15.00 | 11.54 | Aladin | Simbad |
| | | VV 264b | 1 | GinPair | 10228 | 0.65 | - | - | Aladin | Simbad |
| ARP245 | ARP245 | NGC 2992 | 1 | Seyfert2 | 2179 | 3.24 | 13.14 | 8.60 | Aladin | Simbad |
| | | AT 2018ikn | 1 | AGN | 2302 | - | - | - | Aladin | Simbad |
| | | NGC 2993 | 1 | EmissionG | 2430 | 0.61 | 13.11 | 10.13 | Aladin | Simbad |
| ARP248 | ARP248, ARP248B | NAME Wild's Triplet | 1 | EmissionG | 5275 | 0.50 | 14.06 | 10.97 | Aladin | Simbad |
| | | 2dFGRS TGN174Z146 | 1 | Galaxy | 5289 | - | 14.45 | - | Aladin | Simbad |
| | | MCG-01-30-032 | 1 | GtowardsGroup | 5064 | 0.48 | 14.56 | 11.84 | Aladin | Simbad |
| | | LEDA 1065954 | 0 | Galaxy | - | 0.59 | - | - | Aladin | Simbad |
| ARP249 | ARP249 | UGC 12891 | 1 | InteractingG | 11472 | 0.62 | 15.10 | 10.99 | Aladin | Simbad |
| | | [CCA99] UGC 12891b | 0 | Galaxy | - | - | - | - | Aladin | Simbad |
| | | VV 186a | 0 | GinPair | - | 0.26 | - | - | Aladin | Simbad |
| | | SDSS J000021.45+230041.3 | -1 | QSO | 206052 | - | - | - | Aladin | Simbad |
| ARP251 | ARP251 | APG 251 | 1 | GroupG | 21997 | - | - | - | Aladin | Simbad |
| | | 6dFGS gJ005348.4-135115 | 1 | InteractingG | 21997 | 0.26 | 15.00 | 12.66 | Aladin | Simbad |
| | | 2MASX J00534765-1351358 | 1 | Galaxy | 22013 | 0.18 | 17.06 | 13.52 | Aladin | Simbad |
| ARP253 | ARP253 | APG 253 | 1 | Galaxy | - | - | - | - | Aladin | Simbad |
| | | 6dFGS gJ094327.7-051641 | 1 | Galaxy | 1861 | - | 15.37 | - | Aladin | Simbad |
| | | 6dFGS gJ094322.6-051659 | 1 | Galaxy | 1861 | - | 15.86 | - | Aladin | Simbad |
| | | UGCA 173 | 1 | GinPair | 1878 | 1.23 | 15.00 | - | Aladin | Simbad |
| | | EQ 0940-0504 | 0 | Galaxy | - | - | - | - | Aladin | Simbad |
| | | UGCA 174 | 0 | Galaxy | - | 1.26 | 15.00 | - | Aladin | Simbad |
| ARP255 | ARP255 | UGC 5304 | 1 | InteractingG | 11997 | 0.65 | 14.80 | - | Aladin | Simbad |
| | | FIRST J095310.3+075224 | 1 | RadioG | 11953 | 0.21 | - | 13.22 | Aladin | Simbad |
| | | [CCA99] UGC 5304a | 0 | Galaxy | - | - | - | - | Aladin | Simbad |
| | | LAMOST J095309.13+075140.8 | -1 | Galaxy | 17769 | - | - | - | Aladin | Simbad |
| ARP257 | ARP257 | UGC 4638 | 1 | EmissionG | 3323 | 1.02 | 13.80 | 11.69 | Aladin | Simbad |
| | | MCG+00-23-006 | 1 | GinPair | 3315 | 0.65 | 17.00 | - | Aladin | Simbad |

**Table C2** *continued*

**Table C2** *(continued)*

| Arp System (1) | Arp ID (2) | Main ID (3) | Associated? (4) | SIMBAD Type (5) | $V_r$ (km/s) (6) | Diam (′) (7) | $m_B$ (8) | $m_K$ (9) | Images (10) | Data (11) |
|---|---|---|---|---|---|---|---|---|---|---|
| ARP258 | ARP258 | UGC 2140 | 1 | InteractingG | 4099 | 2.00 | 14.60 | - | Aladin | Simbad |
| | | UGC 2140 B | 1 | Galaxy | 4143 | 1.90 | 16.24 | - | Aladin | Simbad |
| | | UGC 2140 A | 1 | GtowardsGroup | 4105 | 1.67 | 15.57 | - | Aladin | Simbad |
| | | UGC 2140 C | 1 | Galaxy | 4067 | 0.72 | 15.73 | - | Aladin | Simbad |
| | | MCG+03-07-038 | -1 | GtowardsGroup | 9911 | 0.58 | 15.62 | 11.48 | Aladin | Simbad |
| ARP262 | ARP262 | UGC 12856 | 1 | InteractingG | 1774 | 2.60 | - | - | Aladin | Simbad |
| | | VV 255b | 1 | Galaxy | 1783 | 2.60 | - | - | Aladin | Simbad |
| | | VV 255a | 1 | Galaxy | 2098 | 0.62 | - | - | Aladin | Simbad |
| ARP263 | ARP263 | NGC 3239 | 1 | InteractingG | 704 | 0.71 | 13.50 | 10.72 | Aladin | Simbad |
| | | LEDA 1525170 | 0 | Galaxy | - | 0.33 | - | - | Aladin | Simbad |
| ARP264 | ARP264 | NGC 3104 | 1 | InteractingG | 602 | 3.02 | 14.20 | - | Aladin | Simbad |
| ARP267 | ARP267 | UGC 5764 | 1 | InteractingG | 582 | 1.00 | 15.60 | - | Aladin | Simbad |
| ARP275 | ARP275 | NGC 2881 | 1 | EmissionG | 5057 | 0.73 | 14.00 | 11.31 | Aladin | Simbad |
| ARP276 | ARP276 | NGC 935 | 1 | GinPair | 4178 | - | 13.56 | 9.32 | Aladin | Simbad |
| | | IC 1801 | 1 | GinPair | 4023 | 1.20 | 14.73 | 10.52 | Aladin | Simbad |
| ARP278 | ARP278 | NGC 7253 | 1 | InteractingG | 4578 | 1.57 | 14.40 | 9.86 | Aladin | Simbad |
| | | [WGB2006] 221712+29080 a | 1 | Galaxy | 4937 | - | - | - | Aladin | Simbad |
| | | UGC 11985 | 1 | GinPair | 4807 | 1.70 | 14.40 | - | Aladin | Simbad |
| | | Gaia DR3 1894245538640380160 | 0 | Galaxy | - | 0.07 | - | - | Aladin | Simbad |
| ARP279 | ARP279 | NGC 1253 | 1 | GinPair | 1711 | 4.47 | 11.80 | 9.26 | Aladin | Simbad |
| | | 6dFGS gJ031415.6-024943 | 1 | Galaxy | 1800 | - | 16.19 | - | Aladin | Simbad |
| ARP280 | ARP280 | NGC 3769 | 1 | GinPair | 704 | 1.60 | 11.70 | 9.20 | Aladin | Simbad |
| | | LAMOST J113744.77+475338.1 | 1 | Galaxy | 708 | - | - | - | Aladin | Simbad |
| | | MCG+08-21-077 | 1 | GinPair | 823 | 0.56 | 14.70 | 12.22 | Aladin | Simbad |
| ARP282 | ARP282 | NGC 169 | 1 | Seyfert | 4528 | 0.99 | 13.70 | 9.24 | Aladin | Simbad |
| | | IC 1559 | 1 | AGN | 4674 | 1.00 | 14.00 | - | Aladin | Simbad |
| ARP283 | ARP283 | APG 283 | 1 | PairG | 1695 | - | 12.20 | - | Aladin | Simbad |
| | | LAMOST J091725.05+420013.2 | 1 | Galaxy | 1732 | - | - | - | Aladin | Simbad |
| | | SDSS J091722.70+415953.9 | 1 | GinPair | 1808 | - | - | - | Aladin | Simbad |
| | | NGC 2798 | 1 | GinPair | 1758 | 2.57 | 13.04 | 9.03 | Aladin | Simbad |
| | | NGC 2799 | 1 | GinPair | 1856 | 0.90 | 14.32 | 11.15 | Aladin | Simbad |
| ARP287 | ARP287 | APG 287 | 1 | InteractingG | 2450 | - | - | - | Aladin | Simbad |
| | ARP287A | NGC 2735 | 1 | GinPair | 2321 | 0.94 | 14.20 | 9.81 | Aladin | Simbad |
| | | MCG+04-22-003 | 1 | GinPair | 2469 | 0.30 | 14.20 | 12.61 | Aladin | Simbad |
| | | SDSS J090236.60+255639.7 | -1 | QSO | 112393 | - | 19.53 | - | Aladin | Simbad |
| ARP288 | ARP288 | NGC 5221 | 1 | InteractingG | 7025 | 1.26 | 14.50 | 10.05 | Aladin | Simbad |
| | | SDSSCGB 12123.2 | 1 | Galaxy | 6763 | - | - | - | Aladin | Simbad |
| | | SDSSCGB 12123.1 | 1 | Galaxy | 7161 | - | - | - | Aladin | Simbad |
| | | SDSSCGB 12123.4 | 0 | Galaxy | - | - | - | - | Aladin | Simbad |

**Table C2** *continued*

**Table C2** *(continued)*

| Arp System (1) | Arp ID (2) | Main ID (3) | Associated? (4) | SIMBAD Type (5) | $V_r$ (km/s) (6) | Diam (′) (7) | $m_B$ (8) | $m_K$ (9) | Images (10) | Data (11) |
|---|---|---|---|---|---|---|---|---|---|---|
| | | SDSSCGB 12123.3 | 0 | Galaxy | - | - | - | - | Aladin | Simbad |
| ARP290 | ARP290 | APG 290 | 1 | GroupG | - | - | - | - | Aladin | Simbad |
| | ARP290B | IC 196 | -1 | InteractingG | 3603 | 1.16 | 14.20 | 9.77 | Aladin | Simbad |
| | | LEDA 212903 | -1 | Galaxy | 9463 | 0.35 | - | 12.97 | Aladin | Simbad |
| ARP291 | ARP291 | UGC 5832 | 1 | GinPair | 1215 | 1.00 | 13.80 | - | Aladin | Simbad |
| | | SDSS J104248.72+132710.8 | 1 | Galaxy | 1230 | - | - | - | Aladin | Simbad |
| | | 2MASXI J1042485+132736 | 1 | LowSurfBrghtG | 1226 | 0.58 | - | 11.92 | Aladin | Simbad |
| ARP293 | ARP293 | APG 293 | 1 | Galaxy | - | - | - | - | Aladin | Simbad |
| | ARP293B | NGC 6286 | 1 | GtowardsGroup | 5500 | 1.02 | 14.20 | 9.76 | Aladin | Simbad |
| | ARP293A | NGC 6285 | 1 | GtowardsGroup | 5528 | 0.64 | 15.30 | 11.05 | Aladin | Simbad |
| ARP295 | ARP295, ARP295A | MCG-01-60-021 | 1 | LINER | 6846 | 1.20 | 14.50 | 10.02 | Aladin | Simbad |
| | | Mrk 933 | 1 | Galaxy | 6537 | 0.23 | 16.50 | 12.90 | Aladin | Simbad |
| ARP300 | ARP300 | Mrk 111 | 1 | InteractingG | 3736 | 0.56 | 13.90 | 11.21 | Aladin | Simbad |
| | | UGC 5029 | 1 | GinPair | 3849 | 0.67 | 14.30 | 10.91 | Aladin | Simbad |
| | | LEDA 2717364 | 0 | Galaxy | - | 0.29 | - | - | Aladin | Simbad |
| ARP301 | ARP301 | APG 301 | 1 | InteractingG | 5965 | 0.40 | - | - | Aladin | Simbad |
| | ARP301B | UGC 6207 | 1 | EmissionG | 6244 | 0.80 | 14.60 | 10.77 | Aladin | Simbad |
| | ARP301A | UGC 6204 | 1 | EmissionG | 5968 | 0.62 | 14.50 | 11.10 | Aladin | Simbad |
| ARP303 | ARP303 | IC 563 | 1 | Galaxy | 5959 | 0.53 | 14.70 | 10.78 | Aladin | Simbad |
| | | LAMOST J094622.51+030421.1 | 1 | Galaxy | 5766 | - | - | - | Aladin | Simbad |
| | | IC 564 | 1 | RadioG | 5930 | 1.07 | 14.10 | 10.03 | Aladin | Simbad |
| ARP306 | ARP306, ARP306A | UGC 1102 | 1 | InteractingG | 1955 | 1.20 | 14.60 | - | Aladin | Simbad |
| | | CAIRNS J013229.58+043606.5 | 1 | GtowardsCl | 1905 | - | - | - | Aladin | Simbad |
| | | VV 173b | 1 | Unknown | 1781 | - | - | - | Aladin | Simbad |
| ARP314 | ARP314, ARP314A | MCG-01-58-009 | 1 | EmissionG | 3704 | 0.63 | 13.70 | 10.45 | Aladin | Simbad |
| | | MCG-01-58-011 | 1 | Galaxy | 3701 | 1.10 | 16.00 | - | Aladin | Simbad |
| ARP321 | ARP321 | APG 321 | 1 | InteractingG | 6611 | 1.74 | 15.00 | - | Aladin | Simbad |
| | | MCG-01-25-008 | 1 | GtowardsGroup | 6628 | - | 14.09 | 9.46 | Aladin | Simbad |
| | | MCG-01-25-012 | 1 | GtowardsGroup | 6842 | - | 15.19 | 10.52 | Aladin | Simbad |
| | | LEDA 27517 | 1 | GtowardsGroup | 6230 | - | - | - | Aladin | Simbad |
| | | MCG-01-25-010 | 1 | GtowardsGroup | 6492 | - | 15.20 | 10.67 | Aladin | Simbad |
| | | MCG-01-25-009 | 1 | GtowardsGroup | 6890 | - | 15.82 | 10.12 | Aladin | Simbad |
| | | MCG-01-25-011 | 1 | GtowardsGroup | 6625 | 0.27 | 17.36 | 12.18 | Aladin | Simbad |
| ARP322 | ARP322 | HCG 56 | 1 | Compact_Gr_G | 7985 | 2.10 | - | - | Aladin | Simbad |
| | | Mrk 176 | 1 | Seyfert2 | 7828 | 1.21 | 15.50 | 10.01 | Aladin | Simbad |
| | | MCG+09-19-113 | 1 | Galaxy | 8128 | 0.79 | 16.36 | 10.48 | Aladin | Simbad |
| | | VV 150b | 1 | Galaxy | 8110 | 0.61 | 15.87 | - | Aladin | Simbad |
| | | VV 150c | 1 | Galaxy | 7930 | 0.42 | 17.01 | - | Aladin | Simbad |
| | | LEDA 35609 | 1 | Galaxy | 7918 | 0.26 | 16.54 | 12.97 | Aladin | Simbad |

**Table C2** *continued*

**Table C2** *(continued)*

| Arp System (1) | Arp ID (2) | Main ID (3) | Associated? (4) | SIMBAD Type (5) | $V_r$ (km/s) (6) | Diam (′) (7) | $m_B$ (8) | $m_K$ (9) | Images (10) | Data (11) |
|---|---|---|---|---|---|---|---|---|---|---|
| | | Gaia DR3 840299051405322112 | 0 | Galaxy | - | 0.07 | - | - | Aladin | Simbad |
| AM0001-505 | AM0001-5055 | ESO 193-14 | 1 | Galaxy | 11649 | 0.60 | 15.13 | 11.31 | Aladin | Simbad |
| AM0002-503 | AM0002-503 | ESO 193-19 | 1 | Galaxy | 10195 | 0.58 | 14.24 | 10.38 | Aladin | Simbad |
| | | 2MASX J00053380-5018171 | 1 | GtowardsCl | 10476 | 0.15 | 15.00 | 13.94 | Aladin | Simbad |
| AM0012-573 | AM0012-573 | ESO 111-22 | 1 | HIIG | 15900 | 0.61 | 15.15 | 11.02 | Aladin | Simbad |
| AM0018-485 | AM0018-485 | SCG2 0018-4854 | 1 | Compact_Gr_G | 3361 | - | 13.91 | - | Aladin | Simbad |
| | | NGC 92 | 1 | GtowardsCl | 3362 | 0.72 | 13.71 | 9.38 | Aladin | Simbad |
| | | NGC 87 | 1 | GtowardsGroup | 3525 | 0.41 | 14.70 | 13.53 | Aladin | Simbad |
| | | NGC 88 | 1 | GtowardsGroup | 3462 | 0.32 | 15.11 | 12.61 | Aladin | Simbad |
| AM0052-321 | AM0052-321 | AM 0052-321 | 1 | PairG | 9396 | - | - | - | Aladin | Simbad |
| | | ESO 411-29 | 1 | Seyfert2 | 9468 | 0.57 | 14.98 | 10.57 | Aladin | Simbad |
| | | ESO 411-30 | 1 | Galaxy | 9427 | 0.43 | 14.77 | 11.42 | Aladin | Simbad |
| AM0116-241 | AM0116-241 | ESO 475-16 | 1 | Galaxy | 6963 | 0.74 | 14.23 | 10.54 | Aladin | Simbad |
| AM0135-650 | AM0135-650 | NGC 646 | 1 | InteractingG | 7927 | 1.51 | 14.37 | - | Aladin | Simbad |
| | | VV 443a | 1 | GinPair | 8047 | 0.56 | 14.24 | 11.09 | Aladin | Simbad |
| | | VV 443b | 1 | GinPair | 8013 | 0.32 | 15.65 | 11.71 | Aladin | Simbad |
| AM0136-433 | AM0136-433 | AM 0136-433 | 1 | InteractingG | - | - | - | - | Aladin | Simbad |
| | | ESO 244-47 | 1 | GinPair | 6239 | 0.92 | 14.69 | 10.33 | Aladin | Simbad |
| | | 2dFGRS TGS850Z057 | 0 | Galaxy | 0 | - | 16.58 | - | Aladin | Simbad |
| | | ESO 244-46 | 0 | GinPair | - | 0.68 | 15.15 | - | Aladin | Simbad |
| AM0137-281 | AM0137-2817 | ESO 413-18 | 1 | GtowardsGroup | 5815 | 0.81 | 14.48 | 10.15 | Aladin | Simbad |
| AM0144-585 | AM0144-585 | ESO 114-7 | 1 | Galaxy | 2210 | 1.58 | 14.22 | - | Aladin | Simbad |
| | | 6dFGS gJ014630.0-584025 | 1 | Galaxy | 2213 | - | 14.05 | - | Aladin | Simbad |
| | | 6dFGS gJ014629.0-584033 | 1 | Galaxy | 2201 | - | 16.36 | - | Aladin | Simbad |
| | | LEDA 6512 | 0 | Galaxy | - | 1.55 | 14.44 | - | Aladin | Simbad |
| | | LEDA 6508 | 0 | Galaxy | - | 0.33 | 17.22 | - | Aladin | Simbad |
| AM0154-441 | AM0154-441 | ESO 245-10 | 1 | Galaxy | 5728 | 1.24 | 14.32 | 10.83 | Aladin | Simbad |
| | | Lu YC 0154-44 | 0 | Galaxy | - | - | - | - | Aladin | Simbad |
| AM0200-220 | AM0200-220 | ESO 544-7 | 1 | Galaxy | 12872 | 0.84 | 14.87 | 10.52 | Aladin | Simbad |
| | | 6dFGS gJ020216.0-214535 | 1 | Galaxy | 13170 | - | 19.13 | - | Aladin | Simbad |
| AM0203-325 | AM0203-325 | ESO 354-34 | 1 | GtowardsGroup | 5775 | 0.42 | 14.61 | 10.45 | Aladin | Simbad |
| AM0218-321 | AM0218-3210 | ESO 415-19 | 1 | EmissionG | 9322 | 0.50 | 14.81 | 11.30 | Aladin | Simbad |
| | | ESO 415-20 | 1 | Galaxy | 9626 | 0.40 | 15.71 | 11.16 | Aladin | Simbad |
| AM0223-403 | AM0223-403 | MCG-07-06-002 | 1 | GinPair | 6307 | 0.97 | 14.40 | 10.87 | Aladin | Simbad |
| | | MCG-07-06-003 | 1 | GinPair | 5318 | 0.88 | 15.63 | 13.28 | Aladin | Simbad |
| | | 6dFGS gJ023223.0-522959 | 1 | Galaxy | 6372 | - | - | - | Aladin | Simbad |
| | | ESO 154-2 | 1 | InteractingG | 6450 | 0.43 | 14.84 | 12.34 | Aladin | Simbad |
| | | LEDA 95390 | 0 | Galaxy | - | 0.57 | 15.76 | - | Aladin | Simbad |
| AM0311-252 | AM0311-252 | ESO 481-14 | 1 | GtowardsGroup | 1735 | 4.07 | 13.82 | - | Aladin | Simbad |

**Table C2** *continued*

**Table C2** *(continued)*

| Arp System (1) | Arp ID (2) | Main ID (3) | Associated? (4) | SIMBAD Type (5) | $V_r$ (km/s) (6) | Diam (′) (7) | $m_B$ (8) | $m_K$ (9) | Images (10) | Data (11) |
|---|---|---|---|---|---|---|---|---|---|---|
| | | APMBGC 481+079+017 | 0 | Galaxy | - | 2.35 | 14.75 | - | Aladin | Simbad |
| AM0311-573 | AM0311-573 | ESO 116-12 | 1 | GtowardsGroup | 1140 | 1.01 | 12.96 | 10.07 | Aladin | Simbad |
| AM0313-545 | AM0313-545 | IC 1908 | 1 | GinPair | 8201 | 0.62 | 14.53 | 11.03 | Aladin | Simbad |
| | | LEDA 12086 | 0 | Galaxy | - | 0.31 | 16.14 | - | Aladin | Simbad |
| AM0329-502 | AM0329-502 | NGC 1356 | 1 | InteractingG | 11565 | 0.77 | 14.05 | 10.19 | Aladin | Simbad |
| | | 6dFGS gJ033038.9-501859 | 1 | Galaxy | 11342 | - | 19.77 | - | Aladin | Simbad |
| | | IC 1947 | 1 | Galaxy | 11323 | 0.40 | 15.50 | 11.77 | Aladin | Simbad |
| | | LEDA 467699 | 0 | Galaxy | - | 0.32 | - | - | Aladin | Simbad |
| | | Gaia DR3 4737365660680635008 | 0 | Galaxy | - | 0.08 | - | - | Aladin | Simbad |
| | | 2MASX J03303982-5019126 | -1 | Galaxy | 17646 | 0.31 | 16.77 | 12.02 | Aladin | Simbad |
| AM0333-513 | AM0333-513 | ESO 200-45 | 1 | EmissionG | 1030 | 0.72 | 16.24 | - | Aladin | Simbad |
| | | LEDA 456200 | 0 | Galaxy | - | 0.30 | - | - | Aladin | Simbad |
| | | LEDA 456178 | 0 | Galaxy | - | 0.28 | - | - | Aladin | Simbad |
| AM0346-222 | AM0346-222 | ESO 549-24 | 1 | EmissionG | 12141 | 0.57 | 15.31 | 10.98 | Aladin | Simbad |
| AM0347-442 | AM0347-442 | ESO 249-21 | 1 | Galaxy | 1248 | 0.89 | 16.24 | - | Aladin | Simbad |
| AM0357-460 | AM0357-460 | NAME Horologium Dwarf | 1 | LowSurfBrghtG | 900 | 1.26 | 15.19 | - | Aladin | Simbad |
| | | HIPASS J0359-45 | 1 | Galaxy | 898 | - | - | - | Aladin | Simbad |
| AM0403-555 | AM0403-555 | ESO 156-37 | 1 | Galaxy | 17093 | 0.26 | 15.96 | 12.25 | Aladin | Simbad |
| | | 2MASX J04041706-5546047 | 1 | Galaxy | 17296 | 0.42 | 15.96 | 11.86 | Aladin | Simbad |
| | | ESO-LV 156-0371 | 0 | Galaxy | - | 0.70 | 16.29 | - | Aladin | Simbad |
| | | LEDA 406284 | 0 | Galaxy | - | 0.38 | - | - | Aladin | Simbad |
| | | LEDA 406135 | 0 | Galaxy | - | 0.28 | - | - | Aladin | Simbad |
| | | LEDA 406404 | 0 | Galaxy | - | 0.24 | - | - | Aladin | Simbad |
| | | Gaia DR3 4683419432629897088 | 0 | Galaxy | - | 0.23 | - | - | Aladin | Simbad |
| AM0405-552 | AM0405-552 | IC 2032 | 1 | EmissionG | 1066 | 1.23 | 14.78 | - | Aladin | Simbad |
| AM0409-563 | AM0409-563 | NGC 1536 | 1 | InteractingG | 1310 | 0.98 | 13.29 | 9.86 | Aladin | Simbad |
| AM0417-391 | AM0417-391 | ESO 303-11 | 1 | LINER | 15255 | 0.54 | 15.07 | 11.31 | Aladin | Simbad |
| | | ESO-LV 303-0110 | 1 | Galaxy | 14867 | 0.95 | 16.11 | - | Aladin | Simbad |
| | | LEDA 14884 | 0 | Galaxy | - | 0.51 | 16.90 | - | Aladin | Simbad |
| | | ESO-LV 303-0111 | 0 | Galaxy | - | 0.33 | 17.97 | - | Aladin | Simbad |
| | | LEDA 601375 | 0 | Galaxy | 14454 | 0.32 | - | - | Aladin | Simbad |
| | | 2MASX J04194052-3911390 | 0 | Galaxy | - | 0.19 | - | 13.83 | Aladin | Simbad |
| AM0432-625 | AM0432-625 | 2MASX J04323376-6251479 | 1 | Galaxy | - | 0.26 | - | 12.55 | Aladin | Simbad |
| AM0445-572 | AM0445-572 | ESO 158-3 | 1 | Galaxy | 1205 | 0.79 | 13.96 | - | Aladin | Simbad |
| | | 6dFGS gJ044617.5-572037 | 1 | Galaxy | 1208 | - | 14.26 | - | Aladin | Simbad |
| AM0454-561 | AM0454-561 | ESO 158-15 | 1 | Galaxy | 1779 | 1.38 | 15.34 | - | Aladin | Simbad |
| | | 6dFGS gJ045542.4-561408 | 1 | Galaxy | 1770 | - | 15.01 | - | Aladin | Simbad |
| | | LEDA 16342 | 0 | Galaxy | - | 1.38 | 14.64 | - | Aladin | Simbad |
| | | ESO-LV 158-0151 | 0 | Galaxy | - | 0.50 | 16.92 | - | Aladin | Simbad |

**Table C2** *continued*

**Table C2** *(continued)*

| Arp System (1) | Arp ID (2) | Main ID (3) | Associated? (4) | SIMBAD Type (5) | $V_r$ (km/s) (6) | Diam (′) (7) | $m_B$ (8) | $m_K$ (9) | Images (10) | Data (11) |
|---|---|---|---|---|---|---|---|---|---|---|
| AM0459-340 | AM0459-340 | ESO 361-25 | 1 | InteractingG | 5269 | 1.62 | 14.26 | 11.93 | Aladin | Simbad |
| AM0507-630 | AM0507-630 | ESO 85-47 | 1 | Galaxy | 1464 | 1.35 | 14.57 | - | Aladin | Simbad |
| | | 6dFGS gJ050743.6-625923 | 1 | Galaxy | 1465 | - | 15.85 | - | Aladin | Simbad |
| | | 6dFGS gJ050743.2-625940 | 1 | Galaxy | 1465 | - | 18.14 | - | Aladin | Simbad |
| AM0510-330 | AM0510-330 | ESO 362-9 | 1 | Galaxy | 927 | 1.77 | 13.05 | - | Aladin | Simbad |
| | | 6dFGS gJ051200.2-325756 | 1 | Galaxy | 931 | - | 18.95 | - | Aladin | Simbad |
| AM0515-540 | AM0515-540 | ESO 159-3 | 1 | Galaxy | 3877 | 0.62 | 13.97 | 10.42 | Aladin | Simbad |
| AM0519-611 | AM0519-611 | ESO 119-54 | 1 | GtowardsGroup | 4871 | 0.81 | 14.33 | 10.42 | Aladin | Simbad |
| | | ESO 119-55 | 1 | InteractingG | 4746 | 0.65 | 13.72 | 9.97 | Aladin | Simbad |
| | | LEDA 356638 | 0 | Galaxy | - | 0.72 | - | - | Aladin | Simbad |
| | | Gaia DR3 4760921017975791872 | 0 | Galaxy | - | 0.03 | - | - | Aladin | Simbad |
| AM0520-390 | AM0520-390 | ESO 305-21 | 1 | Galaxy | 14734 | 0.74 | 14.62 | 10.72 | Aladin | Simbad |
| AM0541-294 | AM0541-294 | ESO 424-10 | 1 | EmissionG | 3818 | 0.53 | 14.15 | 11.10 | Aladin | Simbad |
| AM0558-335 | AM0558-335 | IC 2153 | 1 | InteractingG | 2872 | 1.23 | 14.16 | - | Aladin | Simbad |
| | | LEDA 18213 | 1 | Galaxy | 2852 | 1.07 | - | - | Aladin | Simbad |
| | | 6dFGS gJ060005.4-335506 | 1 | EmissionG | 2866 | 0.63 | 14.67 | 11.61 | Aladin | Simbad |
| | | ESO-LV 364-0221 | 0 | Galaxy | - | 1.06 | 14.27 | - | Aladin | Simbad |
| AM0608-333 | AM0608-333 | AM 0608-333 | 1 | PairG | - | - | - | - | Aladin | Simbad |
| | | ESO 364-36 | 1 | GinPair | 8551 | 0.48 | 14.39 | 11.90 | Aladin | Simbad |
| | | ESO 364-35 | 1 | GinPair | 8753 | 0.40 | 14.59 | 11.39 | Aladin | Simbad |
| AM0612-373 | AM0612-373 | ESO 307-25 | 1 | InteractingG | 9734 | 0.76 | 14.76 | 10.30 | Aladin | Simbad |
| AM0619-271 | AM0619-271 | NGC 2217 | 1 | LINER | 1622 | 4.90 | 11.04 | 7.09 | Aladin | Simbad |
| AM0620-363 | AM0620-363 | ESO 365-10 | 1 | GtowardsGroup | 9164 | 0.51 | 14.74 | 10.90 | Aladin | Simbad |
| AM0630-522 | AM0630-522 | ESO 206-16 | 1 | Galaxy | 1193 | 0.56 | 17.14 | - | Aladin | Simbad |
| AM0642-801 | AM0642-801 | ESO 16-16 | 1 | Galaxy | 4779 | 0.38 | 15.44 | 12.01 | Aladin | Simbad |
| | | 2MASX J06382833-8014582 | -1 | Galaxy | 7518 | 0.20 | 17.16 | 13.03 | Aladin | Simbad |
| AM0643-462 | AM0643-462 | ESO 255-18 | 1 | Galaxy | 11774 | 0.58 | 14.52 | 11.40 | Aladin | Simbad |
| | | 2MASX J06450041-4626432 | 1 | Galaxy | 11873 | 0.22 | 17.16 | 12.31 | Aladin | Simbad |
| | | Lu YC 0643-46 | 0 | Galaxy | - | - | - | - | Aladin | Simbad |
| AM0658-590 | AM0658-590 | AM 0658-590 | 1 | Galaxy | - | - | - | - | Aladin | Simbad |
| | NAMEAM0658-590NE | NAME AM 0658-590 NE | 1 | InteractingG | 8148 | 2.57 | 14.73 | - | Aladin | Simbad |
| | | 6dFGS gJ065905.5-590738 | 1 | GtowardsGroup | 8392 | 0.64 | 17.11 | 10.35 | Aladin | Simbad |
| AM0728-664 | AM0728-664 | ESO 88-17 | 1 | Galaxy | 5155 | 1.14 | 13.69 | 9.36 | Aladin | Simbad |
| | | ESO 88-18 | 1 | Galaxy | 5275 | 0.31 | 14.91 | 11.72 | Aladin | Simbad |
| | | LEDA 301093 | 0 | Galaxy | - | 0.41 | - | - | Aladin | Simbad |
| AM0839-745 | AM0839-745 | ESO 36-6 | 1 | Galaxy | 1142 | 1.95 | 14.19 | - | Aladin | Simbad |
| | | ESO 434-33 | 1 | EmissionG | 964 | 2.14 | 13.19 | - | Aladin | Simbad |
| | | ZSFW N2997 b | 0 | Galaxy | - | - | 13.20 | - | Aladin | Simbad |
| | | IC 2507 | 1 | EmissionG | 1253 | 1.63 | 13.33 | 11.05 | Aladin | Simbad |

**Table C2** *continued*

**Table C2** *(continued)*

| Arp System (1) | Arp ID (2) | Main ID (3) | Associated? (4) | SIMBAD Type (5) | $V_r$ (km/s) (6) | Diam (′) (7) | $m_B$ (8) | $m_K$ (9) | Images (10) | Data (11) |
|---|---|---|---|---|---|---|---|---|---|---|
| AM1010-445 | AM1010-445 | ESO 263-16 | 1 | InteractingG | 4113 | 0.46 | 13.85 | 11.02 | Aladin | Simbad |
| AM1033-365 | AM1033-365 | ESO 375-71 | 1 | Galaxy | 955 | 2.95 | 13.10 | - | Aladin | Simbad |
| AM1055-391 | AM1055-391 | ESO 318-24 | 1 | Galaxy | 1002 | 2.19 | 14.00 | - | Aladin | Simbad |
| | | 6dFGS gJ105752.0-392626 | 1 | Galaxy | 1014 | - | 13.98 | - | Aladin | Simbad |
| AM1125-285 | AM1125-285 | ESO 439-10 | 1 | Galaxy | 7127 | 0.76 | 15.73 | 11.26 | Aladin | Simbad |
| | | 6dFGS gJ112731.7-291025 | 1 | Galaxy | 7099 | - | 16.88 | - | Aladin | Simbad |
| AM1200-251 | AM1200-251 | ESO 505-7 | 1 | GinPair | 1790 | 2.14 | 17.64 | - | Aladin | Simbad |
| | | LEDA 778683 | 0 | Galaxy | - | 0.32 | - | - | Aladin | Simbad |
| AM1203-223 | AM1203-223 | ESO 505-13 | 1 | GtowardsGroup | 1722 | 2.40 | 13.04 | - | Aladin | Simbad |
| | | 6dFGS gJ120607.2-225054 | 1 | Galaxy | 1776 | - | 17.79 | - | Aladin | Simbad |
| | | VV 49a | 0 | GinPair | - | - | - | - | Aladin | Simbad |
| AM1214-255 | AM1214-255 | ESO 505-31 | 1 | Seyfert2 | 11465 | 0.35 | 15.13 | 11.89 | Aladin | Simbad |
| | | ESO 505-30 | 1 | Seyfert2 | 11576 | 0.39 | 14.88 | 10.95 | Aladin | Simbad |
| AM1229-512 | AM1229-512 | ESO 218-8 | 1 | Galaxy | 2617 | 0.20 | 16.20 | 12.84 | Aladin | Simbad |
| AM1238-254 | AM1238-254 | ESO 506-35 | 1 | InteractingG | 16941 | 0.27 | 15.44 | 12.10 | Aladin | Simbad |
| | | LEDA 3094720 | 0 | Galaxy | - | 0.68 | - | - | Aladin | Simbad |
| | | 2MASX J12410900-2557312 | -1 | Galaxy | 13185 | 0.79 | - | - | Aladin | Simbad |
| AM1255-430 | AM1255-430 | ESO 269-20 | 1 | GtowardsGroup | 8964 | 0.48 | 14.53 | 10.73 | Aladin | Simbad |
| AM1303-371 | AM1303-3719 | NGC 4953 | 1 | GtowardsGroup | 4897 | 0.73 | 14.12 | 9.98 | Aladin | Simbad |
| | | 6dFGS gJ130607.3-373603 | 1 | GtowardsCl | 14836 | - | 16.44 | - | Aladin | Simbad |
| | | LEDA 45343 | 1 | Galaxy | 14946 | 0.83 | 15.49 | 10.45 | Aladin | Simbad |
| | | 2MASX J13060231-3736036 | 1 | GtowardsCl | 14867 | 0.26 | 16.85 | 12.03 | Aladin | Simbad |
| | | 2MASX J13061435-3736146 | 1 | GtowardsCl | 14957 | 0.15 | 17.93 | 13.85 | Aladin | Simbad |
| | | 6dFGS gJ130612.3-373510 | -1 | GtowardsCl | 16074 | - | 17.02 | - | Aladin | Simbad |
| AM1307-425 | AM1307-425 | ESO 269-56 | 1 | InteractingG | 2135 | 0.56 | 14.04 | 14.53 | Aladin | Simbad |
| AM1307-461 | AM1307-461 | ESO 269-57 | 1 | GtowardsGroup | 3103 | 0.96 | 12.49 | 8.78 | Aladin | Simbad |
| AM1324-294 | AM1324-294 | ESO 444-37 | 1 | GtowardsCl | 1902 | 1.74 | 14.87 | - | Aladin | Simbad |
| AM1332-331 | AM1332-331 | NGC 5215 | 1 | PairG | - | - | - | - | Aladin | Simbad |
| | | ESO 383-28 | 1 | InteractingG | 3913 | 1.57 | 14.16 | 9.98 | Aladin | Simbad |
| | | ESO 383-29 | 1 | InteractingG | 4013 | 1.29 | 13.93 | - | Aladin | Simbad |
| | | 2MASX J13351155-3328576 | 1 | Galaxy | 3986 | 0.81 | - | 12.78 | Aladin | Simbad |
| | | ShaSS 401036410 | -1 | Galaxy | 79576 | - | - | - | Aladin | Simbad |
| AM1342-413 | AM1342-413 | ESO 325-11 | 1 | Galaxy | 545 | 2.75 | 14.02 | - | Aladin | Simbad |
| AM1356-332 | AM1356-332 | ESO 384-25 | 1 | InteractingG | 3706 | 1.29 | 15.11 | - | Aladin | Simbad |
| | | 6dFGS gJ135948.7-334052 | 1 | Galaxy | 3662 | - | 15.80 | - | Aladin | Simbad |
| | | ESO-LV 384-0251 | 1 | GinPair | 3954 | 1.19 | 15.59 | - | Aladin | Simbad |
| | | 6dFGS gJ140444.7-245006 | 1 | Galaxy | 2279 | - | 17.98 | - | Aladin | Simbad |
| | | [PCM2000] 12 | 1 | PairG | 2117 | - | - | - | Aladin | Simbad |
| | | ESO 510-59 | 1 | GinPair | 2332 | 0.49 | 13.49 | 12.17 | Aladin | Simbad |

**Table C2** *continued*

**Table C2** *(continued)*

| Arp System (1) | Arp ID (2) | Main ID (3) | Associated? (4) | SIMBAD Type (5) | $V_r$ (km/s) (6) | Diam (′) (7) | $m_B$ (8) | $m_K$ (9) | Images (10) | Data (11) |
|---|---|---|---|---|---|---|---|---|---|---|
| AM1421-282 | AM1421-282 | NGC 5592 | 1 | EmissionG | 4315 | 0.93 | 13.62 | 9.39 | Aladin | Simbad |
| AM1440-241 | AM1440-241 | ESO 512-18 | 1 | GtowardsGroup | 3528 | 1.66 | 12.96 | 8.86 | Aladin | Simbad |
| | | 6dFGS gJ144336.5-242758 | 1 | Galaxy | 3682 | - | 14.87 | - | Aladin | Simbad |
| | | ESO 512-19 | 1 | GinPair | 3700 | 2.40 | 13.05 | - | Aladin | Simbad |
| AM1546-284 | AM1546-284 | ESO 450-18 | 1 | EmissionG | 4045 | 0.92 | 15.62 | 10.04 | Aladin | Simbad |
| | | ESO 450-17 | 0 | Galaxy | - | 0.30 | 17.15 | 13.36 | Aladin | Simbad |
| AM1705-773 | AM1705-773 | IC 4633 | 1 | GtowardsGroup | 2953 | 1.85 | 12.40 | 8.47 | Aladin | Simbad |
| AM1806-852 | AM1806-852 | NGC 6438 | 1 | InteractingG | 2541 | 1.91 | 13.37 | 8.28 | Aladin | Simbad |
| | | 6dFGS gJ182259.7-852417 | 1 | Galaxy | 2504 | - | 19.11 | - | Aladin | Simbad |
| | | 6dFGS gJ182230.1-852416 | 1 | Galaxy | 2504 | - | 12.46 | - | Aladin | Simbad |
| | | ESO 10-2 | 1 | InteractingG | 2282 | 3.16 | 12.57 | - | Aladin | Simbad |
| AM1847-645 | AM1847-645 | ESO 104-19 | 1 | LowSurfBrghtG | 1006 | 2.29 | 13.21 | - | Aladin | Simbad |
| | | 6dFGS gJ185224.0-644955 | 1 | Galaxy | 1001 | - | 17.68 | - | Aladin | Simbad |
| AM1912-860 | AM1912-860 | ESO 10-4 | 1 | Galaxy | 2444 | 2.40 | 14.00 | - | Aladin | Simbad |
| | | 6dFGS gJ193102.1-860056 | 1 | Galaxy | 2429 | - | 13.82 | - | Aladin | Simbad |
| AM1914-603 | AM1914-603 | IRAS 19140-6035 | 1 | Infrared | - | - | - | - | Aladin | Simbad |
| | | NGC 6769 | 1 | GtowardsGroup | 3783 | 2.57 | 12.33 | 8.48 | Aladin | Simbad |
| | | NGC 6770 | 1 | GtowardsGroup | 3851 | 2.34 | 12.44 | 8.86 | Aladin | Simbad |
| AM1931-610 | AM1931-610 | IC 4869 | 1 | Galaxy | 1796 | 0.45 | 13.63 | 12.19 | Aladin | Simbad |
| AM1953-260 | AM1953-260 | ESO 526-18 | 1 | Galaxy | - | 0.33 | - | 13.67 | Aladin | Simbad |
| | | LEDA 63839 | 1 | Galaxy | 14802 | 0.52 | - | - | Aladin | Simbad |
| | | 2MASX J19562912-2555158 | 1 | Galaxy | 14436 | 0.36 | 16.90 | 11.49 | Aladin | Simbad |
| AM1957-471 | AM1957-471 | NGC 6845 | 1 | InteractingG | 6359 | 0.87 | 15.81 | 9.92 | Aladin | Simbad |
| | | NAME NGC 6845 Quartet | 1 | Compact_Gr_G | 6804 | 2.92 | - | - | Aladin | Simbad |
| | | 6dFGS gJ200056.8-470504 | 1 | GtowardsGroup | 6755 | 0.85 | 14.52 | 10.42 | Aladin | Simbad |
| | | 2MASX J20010526-4703340 | 1 | GtowardsCl | 6677 | 0.48 | 17.18 | 11.56 | Aladin | Simbad |
| | | 6dFGS gJ200053.6-470543 | 1 | GtowardsGroup | 7070 | 0.38 | 15.50 | 11.87 | Aladin | Simbad |
| AM2001-602 | AM2001-602 | IC 4938 | 1 | GtowardsGroup | 3587 | 0.95 | 13.33 | 9.92 | Aladin | Simbad |
| | | [A81] 200157-6023 | 0 | Galaxy | - | - | - | - | Aladin | Simbad |
| | | FRL 66 | 0 | Galaxy | - | - | 15.00 | - | Aladin | Simbad |
| AM2019-442 | AM2019-442 | ESO 285-4 | 1 | Galaxy | 16188 | 0.79 | 14.42 | 10.88 | Aladin | Simbad |
| | | LEDA 64576 | 0 | Galaxy | - | 1.32 | - | - | Aladin | Simbad |
| AM2026-424 | AM2026-424 | ESO 285-19 | 1 | Galaxy | - | 1.00 | 14.20 | - | Aladin | Simbad |
| | | 6dFGS gJ202933.6-423024 | 1 | Galaxy | 15087 | - | 16.43 | - | Aladin | Simbad |
| | | 2MASX J20293243-4230230 | 1 | GinPair | 15161 | 0.63 | 15.08 | 12.08 | Aladin | Simbad |
| | | ESO-LV 285-0191 | 0 | Galaxy | - | 1.09 | 15.16 | - | Aladin | Simbad |
| | | ESO-LV 285-0190 | 0 | GinPair | - | 0.92 | 15.57 | - | Aladin | Simbad |
| AM2029-544 | AM2029-5441 | AM 2029-5441 | 1 | Galaxy | 3403 | - | 15.02 | - | Aladin | Simbad |
| AM2029-544 | AM2029-5442 | ESO 186-60 | 1 | Galaxy | 3621 | 1.10 | 14.79 | - | Aladin | Simbad |

**Table C2** *continued*

**Table C2** *(continued)*

| Arp System (1) | Arp ID (2) | Main ID (3) | Associated? (4) | SIMBAD Type (5) | $V_r$ (km/s) (6) | Diam (′) (7) | $m_B$ (8) | $m_K$ (9) | Images (10) | Data (11) |
|---|---|---|---|---|---|---|---|---|---|---|
| | | IC 5021 | 0 | InteractingG | - | 0.35 | 15.02 | 12.93 | Aladin | Simbad |
| AM2031-440 | AM2031-440 | ESO 285-35 | 1 | InteractingG | 8853 | 0.69 | 14.86 | 11.03 | Aladin | Simbad |
| AM2034-521 | AM2034-521 | NGC 6935 | 1 | GinPair | 4581 | 1.27 | 12.77 | 8.88 | Aladin | Simbad |
| | | MCL 20 203439.0-521706 | 0 | Galaxy | - | - | - | - | Aladin | Simbad |
| AM2038-323 | AM2038-323 | NGC 6947 | 1 | Galaxy | 5551 | 1.04 | 14.84 | 9.99 | Aladin | Simbad |
| | | Gaia DR3 6792390349407758720 | 0 | Galaxy | - | 0.02 | - | - | Aladin | Simbad |
| | | [A81] 203813-3242A | -1 | Galaxy | 37163 | - | - | - | Aladin | Simbad |
| | | [A81] 203813-3242B | -1 | Galaxy | 40732 | - | - | - | Aladin | Simbad |
| | | 2XMM J204118.8-322711 | -1 | EmissionG | 37762 | 0.27 | 19.27 | - | Aladin | Simbad |
| AM2038-382 | AM2038-382 | ESO 341-4 | 1 | InteractingG | 5996 | 0.61 | 13.40 | 10.01 | Aladin | Simbad |
| | | LEDA 611786 | 0 | Galaxy | - | 1.38 | - | - | Aladin | Simbad |
| AM2038-654 | AM2038-654 | IC 5028 | 1 | Galaxy | 1626 | 1.29 | 15.24 | - | Aladin | Simbad |
| | | 6dFGS gJ204322.0-653852 | 1 | Galaxy | 1597 | - | 15.23 | - | Aladin | Simbad |
| AM2040-295 | AM2040-295 | IC 5041 | 1 | GinPair | 2709 | 1.03 | 13.11 | 10.11 | Aladin | Simbad |
| AM2042-382 | AM2042-3821 | ESO 341-11 | 1 | GtowardsGroup | 6932 | 1.02 | 14.29 | 10.23 | Aladin | Simbad |
| AM2048-571 | AM2048-571 | IC 5063 | 1 | Seyfert2 | 3361 | 1.24 | 12.92 | 8.75 | Aladin | Simbad |
| AM2055-541 | AM2055-541 | AM 2055-541 | 1 | Galaxy | - | - | - | - | Aladin | Simbad |
| | | ESO 187-38 | 1 | Galaxy | 12950 | 0.48 | 14.90 | 11.11 | Aladin | Simbad |
| | | ESO 187-40 | 1 | Galaxy | 12779 | 0.35 | 15.43 | 12.12 | Aladin | Simbad |
| | | LEDA 422408 | 0 | Galaxy | - | 0.35 | - | - | Aladin | Simbad |
| AM2056-392 | AM2056-392 | AM 2056-392 | 1 | Galaxy | - | 0.80 | 15.90 | - | Aladin | Simbad |
| | | FRL 1169 | 0 | GroupG | - | 0.83 | - | - | Aladin | Simbad |
| AM2103-550 | AM2103-550 | ESO 187-51 | 1 | LowSurfBrghtG | 1402 | 1.78 | 14.90 | - | Aladin | Simbad |
| | | 6dFGS gJ210733.1-545705 | 1 | Galaxy | 1357 | - | 16.11 | - | Aladin | Simbad |
| AM2105-332 | AM2105-332 | LEDA 678973 | 1 | Galaxy | - | 2.34 | - | - | Aladin | Simbad |
| | | ESO 402-10 | 1 | InteractingG | 5305 | 1.05 | 14.74 | 9.36 | Aladin | Simbad |
| | | ESO 402-9 | 1 | EmissionG | 5429 | 0.42 | 16.01 | 11.25 | Aladin | Simbad |
| AM2105-632 | AM2105-632 | IC 5084 | 1 | GtowardsGroup | 3142 | 1.36 | 13.53 | 9.25 | Aladin | Simbad |
| | | MCL 8 210509.0-632936 | 0 | Galaxy | - | - | - | - | Aladin | Simbad |
| AM2106-374 | AM2106-374 | ESO 342-13 | 1 | GinPair | 2632 | 1.06 | 13.78 | 10.85 | Aladin | Simbad |
| AM2113-341 | AM2113-341 | ESO 402-21 | 1 | EmissionG | 8798 | 0.71 | 14.97 | 10.79 | Aladin | Simbad |
| AM2115-273 | AM2115-273 | ESO 464-31 | 1 | PairG | 6458 | 0.68 | 15.64 | 10.66 | Aladin | Simbad |
| | | SGC 211526-2733.6 | 0 | Galaxy | - | - | - | - | Aladin | Simbad |
| | | LEDA 66529 | -1 | GinPair | 10879 | 1.17 | 15.02 | - | Aladin | Simbad |
| AM2126-601 | AM2126-601 | IC 5110 | 1 | InteractingG | 8660 | 0.72 | 14.26 | 10.63 | Aladin | Simbad |
| AM2128-430 | AM2128-430 | ESO 287-34 | 1 | GtowardsGroup | 2362 | 1.22 | 13.01 | 9.31 | Aladin | Simbad |
| | | MCL 34 212836.0-430400 | 0 | Galaxy | - | - | - | - | Aladin | Simbad |
| AM2159-320 | AM2159-320 | NGC 7172 | 1 | Seyfert2 | 2575 | 2.57 | 12.72 | 8.32 | Aladin | Simbad |
| | | 2dFGRS TGS407Z097 | 1 | Galaxy | 2384 | - | 15.56 | 13.50 | Aladin | Simbad |

**Table C2** *continued*

**Table C2** *(continued)*

| Arp System (1) | Arp ID (2) | Main ID (3) | Associated? (4) | SIMBAD Type (5) | $V_r$ (km/s) (6) | Diam (′) (7) | $m_B$ (8) | $m_K$ (9) | Images (10) | Data (11) |
|---|---|---|---|---|---|---|---|---|---|---|
| | | MCL 59 215907.0-320635 | 0 | Galaxy | - | - | - | - | Aladin | Simbad |
| | | APMBGC 466-093+108 | 0 | Galaxy | - | 1.00 | 14.99 | - | Aladin | Simbad |
| | | 2dFGRS TGS407Z103 | -1 | Galaxy | -30 | - | 19.33 | - | Aladin | Simbad |
| | | 2dFGRS TGS407Z106 | -1 | Galaxy | 36746 | 0.05 | 18.75 | - | Aladin | Simbad |
| AM2210-693 | AM2210-693 | IC 5173 | 1 | InteractingG | 3155 | 1.55 | 15.94 | - | Aladin | Simbad |
| | | 6dFGS gJ221444.7-692157 | 1 | Galaxy | 3155 | - | 14.88 | - | Aladin | Simbad |
| | | ESO-LV 76-0082 | 1 | Galaxy | 3121 | 1.50 | 15.72 | - | Aladin | Simbad |
| | | 6dFGS gJ221438.7-692204 | 1 | Galaxy | 3167 | 0.78 | 16.31 | - | Aladin | Simbad |
| | | IC 5173B | 0 | GinPair | - | 0.84 | 16.07 | - | Aladin | Simbad |
| | | LEDA 279703 | 0 | Galaxy | - | 0.45 | - | - | Aladin | Simbad |
| AM2220-423 | AM2220-423 | ESO 289-26 | 1 | Galaxy | 2420 | 1.91 | 14.37 | - | Aladin | Simbad |
| | | LEDA 565710 | 0 | Galaxy | - | 0.34 | - | - | Aladin | Simbad |
| AM2222-275 | AM2222-275 | AM 2222-275 | 1 | Compact_Gr_G | 14872 | - | - | - | Aladin | Simbad |
| | | 2MASX J22253211-2742295 | 1 | EmissionG | 14615 | 0.31 | 16.15 | 12.31 | Aladin | Simbad |
| | | ESO 467-56 | 1 | Galaxy | 15259 | 0.29 | 15.58 | 12.08 | Aladin | Simbad |
| | | 2MASX J22252774-2742091 | 1 | Galaxy | 14945 | 0.19 | 16.22 | 13.40 | Aladin | Simbad |
| | | LEDA 751081 | -1 | Galaxy | 50666 | 0.26 | - | - | Aladin | Simbad |
| AM2224-310 | AM2224-310 | ESO 467-62 | 1 | Galaxy | 3974 | 0.19 | 14.94 | 13.85 | Aladin | Simbad |
| | | 2dFGRS TGS337Z148 | -1 | Galaxy | 16574 | 0.36 | 17.40 | - | Aladin | Simbad |
| AM2230-481 | AM2230-481 | ESO 238-16 | 1 | Galaxy | 8185 | 0.29 | 15.10 | 12.07 | Aladin | Simbad |
| | | 2MASX J22334341-4801300 | -1 | Galaxy | 10142 | 0.21 | 17.08 | 12.51 | Aladin | Simbad |
| AM2240-892 | AM2240-892 | ESO 1-8 | 1 | InteractingG | 2525 | 0.38 | 15.32 | 12.11 | Aladin | Simbad |
| | | 6dFGS gJ231157.3-890708 | 1 | Galaxy | 2547 | - | 17.76 | - | Aladin | Simbad |
| | | IRAS 22443-8923 | 1 | Galaxy | 2578 | - | - | - | Aladin | Simbad |
| | | ESO 1-9 | 0 | InteractingG | - | 0.37 | 14.88 | 12.65 | Aladin | Simbad |
| AM2254-373 | AM2254-373 | NGC 7421 | 1 | GtowardsGroup | 1801 | 1.28 | 12.64 | 9.25 | Aladin | Simbad |
| AM2258-595 | AM2258-595 | ESO 147-19 | 1 | InteractingG | 10108 | 0.33 | 15.84 | 12.23 | Aladin | Simbad |
| | | 2MASX J23012961-5938090 | 1 | Galaxy | 10043 | 0.33 | 15.51 | 11.85 | Aladin | Simbad |
| | | LEDA 70283 | 0 | Galaxy | - | 0.68 | - | - | Aladin | Simbad |
| | | ESO-LV 147-0191 | 0 | Galaxy | - | 0.65 | 15.44 | - | Aladin | Simbad |
| AM2259-692 | AM2259-692 | IC 5279 | 1 | GtowardsGroup | 3901 | 0.77 | 14.31 | 10.76 | Aladin | Simbad |
| AM2303-305 | AM2303-305 | ESO 469-11 | 1 | Seyfert2 | 8463 | 0.72 | 14.39 | 10.61 | Aladin | Simbad |
| | | LEDA 714883 | 0 | Galaxy | - | 0.31 | - | - | Aladin | Simbad |
| | | 2dFGRS TGS345Z059 | -1 | Galaxy | 35463 | - | 18.55 | - | Aladin | Simbad |
| AM2325-473 | AM2325-473 | ESO 240-3 | 1 | Galaxy | 15238 | 0.51 | 14.83 | 11.17 | Aladin | Simbad |
| | | 6dFGS gJ232749.7-472235 | 1 | Galaxy | 14818 | - | 16.16 | - | Aladin | Simbad |
| | | LEDA 502448 | 0 | Galaxy | - | 0.40 | - | - | Aladin | Simbad |
| | | LEDA 502014 | 0 | Galaxy | - | 0.39 | - | - | Aladin | Simbad |
| AM2339-661 | AM2339-661, AM2339-6613 | NGC 7733 | 1 | Seyfert2 | 10013 | 0.85 | 14.49 | 10.93 | Aladin | Simbad |

**Table C2** *continued*

**Table C2** *(continued)*

| Arp System (1) | Arp ID (2) | Main ID (3) | Associated? (4) | SIMBAD Type (5) | $V_r$ (km/s) (6) | Diam (′) (7) | $m_B$ (8) | $m_K$ (9) | Images (10) | Data (11) |
|---|---|---|---|---|---|---|---|---|---|---|
| | | NGC 7733N | 1 | Galaxy | 10653 | - | - | - | Aladin | Simbad |
| | | NGC 7734 | 1 | InteractingG | 10383 | 0.79 | 13.95 | 10.05 | Aladin | Simbad |
| AM2346-380 | AM2346-380 | ESO 348-9 | 1 | Galaxy | 647 | 2.09 | 14.60 | - | Aladin | Simbad |
| | | APMBGC 348+032-109 | 0 | Galaxy | - | 1.92 | 15.97 | - | Aladin | Simbad |
| AM2350-302 | AM2350-302 | ESO 471-37 | 1 | InteractingG | 14143 | 1.74 | 17.84 | - | Aladin | Simbad |
| | | 2dFGRS TGS356Z090 | 1 | Galaxy | 36981 | - | 19.18 | - | Aladin | Simbad |
| | | 2dFGRS TGS355Z012 | 1 | Galaxy | 37189 | - | 19.29 | - | Aladin | Simbad |
| | | [VV2006] J235323.1-301057 | -1 | QSO | 209178 | - | 20.10 | - | Aladin | Simbad |
| | | 6dFGS gJ235320.0-300902 | -1 | Seyfert2 | 14660 | 0.33 | 16.53 | 11.51 | Aladin | Simbad |
| AM2350-410 | AM2350-410 | ESO 293-8 | 1 | InteractingG | 9162 | 0.55 | 15.35 | 10.86 | Aladin | Simbad |
| | | 2dFGRS TGS883Z423 | 1 | Galaxy | 8539 | - | 14.25 | - | Aladin | Simbad |
| | | LEDA 72763 | 1 | Galaxy | 8832 | 0.78 | - | - | Aladin | Simbad |
| | | 2MFGC 17933 | 1 | Galaxy | 8726 | 0.44 | 16.12 | 12.15 | Aladin | Simbad |
| | | 2dFGRS TGS883Z419 | 1 | GtowardsGroup | 8713 | 0.27 | 16.45 | 11.67 | Aladin | Simbad |
| | | [KK2000] 76A | 0 | GtowardsGroup | - | - | - | - | Aladin | Simbad |
| | | APMBGC 293+092+054 | 0 | Galaxy | - | 1.13 | 14.42 | - | Aladin | Simbad |
| | | ESO-LV 293-0082 | 0 | Galaxy | - | 1.08 | 15.22 | - | Aladin | Simbad |
| AM2353-291 | AM2353-291 | AM 2353-291 | 1 | GinPair | - | 0.94 | 14.69 | - | Aladin | Simbad |
| | | LEDA 72950 | 1 | Galaxy | 8828 | - | - | - | Aladin | Simbad |
| | | MCG-05-01-007 | 1 | GtowardsGroup | 8931 | 1.29 | 15.00 | - | Aladin | Simbad |
| | | IC 5364 | 1 | InteractingG | 8931 | 1.23 | 15.16 | 9.60 | Aladin | Simbad |
| | | 2dFGRS TGS275Z164 | -1 | Galaxy | 21255 | - | 19.23 | - | Aladin | Simbad |

## D. POINT-SOURCE PHOTOMETRY DIAGNOSTICS

In Section 2.5, we presented examples of diagnostics that are useful for interpreting the point source photometry discussed in Section 2.3. Here, we provide a table summarizing the available photometry for each system at a high level (Table D3), including: the target name (column 1); the number of detected point sources (column 2); the approximate number of these sources that are likely foreground stars (column 3); the peak surface density of point sources in the KDE density map, calculated with an Epanechnikov smoothing kernel with a 10″ radius (column 4); the surface density contour level containing 50% of the sources (column 5); the median surface density of all the pixels (column 6); and the apparent F606W magnitudes of the brightest sources (top 5 by number in column 7, and top 1% in column 8) that fall within the surface density contour from column 5.

In addition, we supplement the representative example plots from Figure 12 with the full set of diagnostic plots (Figure D3) showing the spatial distribution of point sources (left), the conditional luminosity function in bins of spatial density (middle), and the histogram and cumulative distributions of surface density by star and by pixel (right). A fuller description of these plots can be found in Section 2.5 and in the caption to Figure 12. While in most cases the plots are dominated by point sources in the target galaxies, there are images that are entirely dominated by the Milky Way foreground (with AM0432-625 being a typical example), leading to more uniform density maps (left panel), and nearly identical area and source density cumulative distributions with very low median densities. Such cases are more common for quiescent, more distant systems.

We note that, in spite of our best attempts to limit photometry to only bona fide, well-measured point sources, there is occasional contamination from deblended diffraction spikes from bright Milky Way foreground stars (with Arp 263 and Arp 276 being relatively clear examples). Readers may wish to undergo a more detailed culling of spurious photometry for some applications.

**Table D3**. Point Source Photometry Data

| Target (1) | # Point Sources (Total) (2) | # Point Sources (Foreground) (3) | $\Sigma_{max}$ (# arcsec$^{-2}$) (4) | $\Sigma_{sources,50\%}$ (# arcsec$^{-2}$) (5) | $\Sigma_{area,50\%}$ (# arcsec$^{-2}$) (6) | F606W (Brightest 5) (7) | F606W (Brightest 1%) (8) |
|---|---|---|---|---|---|---|---|
| ARP2 | 3508 | ∼230 | 0.320 | 0.110 | 0.023 | $20.67^{-1.70}_{+0.68}$ | 21.61 |
| ARP3 | 1945 | ∼150 | 0.163 | 0.076 | 0.008 | $21.18^{-1.59}_{+0.52}$ | 21.77 |
| ARP4 | 2637 | ∼180 | 0.227 | 0.096 | 0.015 | $20.74^{-0.86}_{+0.98}$ | 22.03 |
| ARP6 | 6124 | ∼200 | 0.875 | 0.685 | 0.012 | $20.33^{-0.73}_{+0.48}$ | 21.54 |
| ARP10 | 616 | ∼90 | 0.139 | 0.026 | 0.003 | $21.84^{-2.66}_{+0.81}$ | 21.42 |
| ARP15 | 1541 | ∼140 | 0.392 | 0.247 | 0.004 | $21.61^{-0.88}_{+0.41}$ | 22.16 |
| ARP18 | 11328 | ∼160 | 0.739 | 0.491 | 0.027 | $19.88^{-0.77}_{+0.38}$ | 21.77 |
| ARP20 | 1160 | ∼130 | 0.426 | 0.218 | 0.004 | $21.59^{-1.69}_{+0.32}$ | 21.73 |
| ARP22 | 7908 | ∼230 | 0.727 | 0.509 | 0.015 | $19.72^{-0.61}_{+0.76}$ | 21.66 |
| ARP24 | 3607 | ∼140 | 0.785 | 0.441 | 0.005 | $20.53^{-0.46}_{+0.69}$ | 21.76 |
| ARP44 | 530 | ∼130 | 0.089 | 0.014 | 0.004 | $22.53^{-3.69}_{+0.75}$ | 22.44 |
| ARP49 | 2489 | ∼150 | 0.792 | 0.513 | 0.005 | $19.88^{-0.22}_{+0.87}$ | 21.19 |
| ARP58 | 709 | ∼170 | 0.201 | 0.019 | 0.004 | $22.42^{-2.08}_{+0.78}$ | 22.26 |
| ARP59 | 1462 | ∼180 | 0.324 | 0.170 | 0.005 | $21.32^{-2.57}_{+0.57}$ | 21.83 |
| ARP69 | 1032 | ∼160 | 0.148 | 0.062 | 0.005 | $22.32^{-0.71}_{+0.48}$ | 22.75 |
| ARP70 | 466 | ∼150 | 0.077 | 0.008 | 0.004 | $20.82^{-1.75}_{+0.32}$ | 20.55 |
| ARP72 | 2039 | ∼190 | 0.472 | 0.225 | 0.006 | $20.19^{-0.69}_{+0.48}$ | 21.43 |
| ARP75 | 1014 | ∼180 | 0.258 | 0.127 | 0.005 | $21.15^{-1.96}_{+0.68}$ | 21.34 |
| ARP82 | 2367 | ∼180 | 0.318 | 0.173 | 0.009 | $20.68^{-0.66}_{+0.09}$ | 20.98 |
| ARP86 | 2663 | ∼130 | 0.246 | 0.124 | 0.010 | $20.08^{-0.78}_{+0.28}$ | 20.69 |
| ARP89 | 539 | ∼190 | 0.025 | 0.008 | 0.005 | $21.53^{-2.45}_{+0.28}$ | 21.31 |
| ARP91 | 2048 | ∼130 | 0.525 | 0.344 | 0.004 | $20.47^{-0.28}_{+0.37}$ | 21.22 |
| ARP97 | 519 | ∼140 | 0.069 | 0.011 | 0.004 | $20.09^{-1.59}_{+2.29}$ | 19.98 |
| ARP100 | 322 | ∼140 | 0.027 | 0.005 | 0.004 | $22.40^{-0.42}_{+0.22}$ | 22.09 |

**Table D3** *continued*

**Table D3** *(continued)*

| Target | # Point Sources (Total) | # Point Sources (Foreground) | $\Sigma_{max}$ (# arcsec$^{-2}$) | $\Sigma_{sources,50\%}$ (# arcsec$^{-2}$) | $\Sigma_{area,50\%}$ (# arcsec$^{-2}$) | F606W (Brightest 5) | F606W (Brightest 1%) |
|---|---|---|---|---|---|---|---|
| (1) | (2) | (3) | (4) | (5) | (6) | (7) | (8) |
| ARP101 | 506 | ∼200 | 0.038 | 0.007 | 0.005 | $21.07^{-1.33}_{+0.59}$ | 20.19 |
| ARP107 | 528 | ∼140 | 0.048 | 0.010 | 0.004 | $22.64^{-3.55}_{+0.26}$ | 22.05 |
| ARP110 | 388 | ∼130 | 0.066 | 0.006 | 0.003 | $22.77^{-2.88}_{+0.38}$ | 21.20 |
| ARP112 | 667 | ∼140 | 0.090 | 0.015 | 0.005 | $21.29^{-1.35}_{+0.77}$ | 21.23 |
| ARP113 | 1811 | ∼550 | 0.111 | 0.030 | 0.018 | $19.17^{-0.36}_{+2.72}$ | 21.89 |
| ARP121 | 631 | ∼190 | 0.079 | 0.010 | 0.005 | $22.30^{-3.21}_{+0.58}$ | 21.63 |
| ARP122 | 1009 | ∼280 | 0.070 | 0.021 | 0.008 | $21.49^{-0.92}_{+1.12}$ | 21.56 |
| ARP123 | 1000 | ∼130 | 0.233 | 0.075 | 0.004 | $21.94^{-0.39}_{+0.48}$ | 22.24 |
| ARP126 | 842 | ∼120 | 0.278 | 0.144 | 0.004 | $20.90^{-1.42}_{+0.38}$ | 20.90 |
| ARP129 | 847 | ∼130 | 0.281 | 0.134 | 0.004 | $21.36^{-0.22}_{+1.05}$ | 21.48 |
| ARP130 | 524 | ∼150 | 0.091 | 0.009 | 0.004 | $23.03^{-0.75}_{+0.33}$ | 22.54 |
| ARP136 | 978 | ∼190 | 0.099 | 0.027 | 0.006 | $21.97^{-0.22}_{+0.40}$ | 22.09 |
| ARP141 | 1273 | ∼160 | 0.232 | 0.107 | 0.005 | $22.16^{-1.21}_{+0.13}$ | 22.26 |
| ARP143 | 1851 | ∼170 | 0.247 | 0.154 | 0.005 | $20.17^{-0.25}_{+0.74}$ | 21.00 |
| ARP144 | 917 | ∼140 | 0.375 | 0.233 | 0.004 | $20.68^{-1.67}_{+1.50}$ | 21.89 |
| ARP145 | 879 | ∼190 | 0.143 | 0.027 | 0.006 | $21.13^{-1.20}_{+0.90}$ | 21.71 |
| ARP150 | 467 | ∼120 | 0.066 | 0.009 | 0.003 | $21.64^{-2.22}_{+1.08}$ | 21.22 |
| ARP156 | 390 | ∼140 | 0.030 | 0.007 | 0.004 | $19.92^{-1.39}_{+1.03}$ | 19.28 |
| ARP158 | 870 | ∼220 | 0.142 | 0.019 | 0.006 | $21.87^{-3.10}_{+0.16}$ | 21.89 |
| ARP163 | 2256 | ∼180 | 0.825 | 0.647 | 0.005 | $18.91^{-0.33}_{+0.60}$ | 19.87 |
| ARP164 | 454 | ∼200 | 0.021 | 0.007 | 0.005 | $22.39^{-0.62}_{+0.15}$ | 22.29 |
| ARP165 | 739 | ∼250 | 0.041 | 0.012 | 0.008 | $19.48^{-0.44}_{+0.15}$ | 19.50 |
| ARP172 | 366 | ∼170 | 0.014 | 0.006 | 0.004 | $20.00^{-0.97}_{+0.57}$ | 19.44 |
| ARP176 | 960 | ∼220 | 0.099 | 0.020 | 0.007 | $21.99^{-2.46}_{+0.23}$ | 22.13 |
| ARP180 | 538 | ∼180 | 0.063 | 0.008 | 0.005 | $20.71^{-1.06}_{+0.32}$ | 20.50 |
| ARP184 | 4338 | ∼340 | 0.311 | 0.126 | 0.028 | $19.42^{-0.77}_{+0.94}$ | 21.60 |
| ARP190 | 362 | ∼120 | 0.066 | 0.006 | 0.003 | $22.74^{-1.84}_{+0.28}$ | 21.80 |
| ARP195 | 587 | ∼140 | 0.095 | 0.013 | 0.004 | $21.05^{-0.37}_{+0.17}$ | 20.97 |
| ARP197 | 862 | ∼160 | 0.268 | 0.069 | 0.005 | $21.67^{-1.43}_{+0.21}$ | 21.71 |
| ARP200 | 1631 | ∼90 | 0.387 | 0.226 | 0.004 | $21.19^{-1.94}_{+0.42}$ | 21.64 |
| ARP202 | 1318 | ∼180 | 0.320 | 0.220 | 0.005 | $20.93^{-0.99}_{+0.18}$ | 21.12 |
| ARP204 | 428 | ∼100 | 0.105 | 0.052 | 0.003 | $21.69^{-1.32}_{+0.79}$ | 21.13 |
| ARP205 | 3863 | ∼120 | 0.692 | 0.496 | 0.006 | $20.81^{-0.82}_{+0.58}$ | 21.78 |
| ARP208 | 758 | ∼130 | 0.282 | 0.147 | 0.004 | $21.30^{-1.22}_{+0.14}$ | 21.32 |
| ARP216 | 1134 | ∼190 | 0.521 | 0.271 | 0.005 | $20.14^{-0.47}_{+0.37}$ | 20.51 |
| ARP219 | 466 | ∼160 | 0.116 | 0.007 | 0.004 | $21.15^{-2.05}_{+0.75}$ | 20.28 |
| ARP221 | 914 | ∼250 | 0.076 | 0.019 | 0.007 | $21.52^{-2.06}_{+1.24}$ | 21.67 |
| ARP231 | 468 | ∼190 | 0.038 | 0.007 | 0.005 | $22.41^{-3.33}_{+0.30}$ | 21.83 |
| ARP241 | 482 | ∼160 | 0.059 | 0.008 | 0.004 | $21.03^{-2.08}_{+0.78}$ | 20.13 |
| ARP245 | 2222 | ∼190 | 0.746 | 0.169 | 0.007 | $18.66^{-0.16}_{+0.60}$ | 19.96 |
| ARP248 | 1500 | ∼130 | 0.446 | 0.290 | 0.005 | $20.06^{-0.86}_{+0.65}$ | 21.14 |
| ARP249 | 328 | ∼130 | 0.017 | 0.006 | 0.003 | $19.97^{-1.23}_{+1.82}$ | 19.69 |
| ARP251 | 485 | ∼130 | 0.111 | 0.010 | 0.004 | $21.75^{-1.06}_{+0.92}$ | 21.56 |
| ARP253 | 1221 | ∼120 | 0.264 | 0.164 | 0.004 | $21.69^{-1.87}_{+0.24}$ | 21.93 |
| ARP255 | 644 | ∼190 | 0.104 | 0.013 | 0.005 | $20.74^{-0.93}_{+1.12}$ | 20.51 |
| ARP257 | 1148 | ∼170 | 0.301 | 0.081 | 0.006 | $22.36^{-0.38}_{+0.12}$ | 22.47 |
| ARP258 | 1122 | ∼130 | 0.228 | 0.105 | 0.005 | $21.55^{-0.53}_{+0.43}$ | 21.93 |
| ARP262 | 1666 | ∼140 | 0.432 | 0.165 | 0.005 | $21.79^{-1.23}_{+0.20}$ | 22.07 |
| ARP263 | 20006 | ∼920 | 0.925 | 0.679 | 0.098 | $18.96^{-0.39}_{+0.56}$ | 21.45 |

**Table D3** *continued*

**Table D3** *(continued)*

| Target | # Point Sources (Total) | # Point Sources (Foreground) | $\Sigma_{max}$ (# arcsec$^{-2}$) | $\Sigma_{sources,50\%}$ (# arcsec$^{-2}$) | $\Sigma_{area,50\%}$ (# arcsec$^{-2}$) | F606W (Brightest 5) | F606W (Brightest 1%) |
|---|---|---|---|---|---|---|---|
| (1) | (2) | (3) | (4) | (5) | (6) | (7) | (8) |
| ARP264 | 5588 | ∼250 | 0.623 | 0.312 | 0.015 | $21.01^{-0.94}_{+0.29}$ | 21.88 |
| ARP267 | 1328 | ∼150 | 0.307 | 0.131 | 0.004 | $20.95^{-1.74}_{+0.60}$ | 21.04 |
| ARP275 | 1550 | ∼180 | 0.455 | 0.279 | 0.005 | $21.62^{-2.29}_{+0.30}$ | 22.06 |
| ARP276 | 1917 | ∼110 | 0.417 | 0.192 | 0.004 | $21.82^{-1.50}_{+0.52}$ | 22.72 |
| ARP278 | 1152 | ∼220 | 0.162 | 0.038 | 0.007 | $19.98^{-0.85}_{+0.83}$ | 20.43 |
| ARP279 | 8384 | ∼90 | 0.772 | 0.242 | 0.067 | $20.97^{-1.07}_{+0.24}$ | 22.33 |
| ARP280 | 5436 | ∼190 | 0.821 | 0.551 | 0.009 | $20.57^{-0.55}_{+0.36}$ | 22.13 |
| ARP282 | 721 | ∼180 | 0.174 | 0.022 | 0.005 | $22.28^{-2.25}_{+0.34}$ | 22.31 |
| ARP283 | 1334 | ∼140 | 0.314 | 0.125 | 0.005 | $19.48^{-0.22}_{+0.40}$ | 19.93 |
| ARP287 | 928 | ∼140 | 0.208 | 0.051 | 0.005 | $22.16^{-0.27}_{+0.71}$ | 22.34 |
| ARP288 | 475 | ∼190 | 0.034 | 0.008 | 0.005 | $22.32^{-0.44}_{+0.49}$ | 22.28 |
| ARP290 | 601 | ∼140 | 0.093 | 0.015 | 0.004 | $20.78^{-2.25}_{+1.14}$ | 20.70 |
| ARP291 | 1691 | ∼140 | 0.521 | 0.296 | 0.004 | $21.88^{-1.59}_{+0.40}$ | 22.30 |
| ARP293 | 743 | ∼180 | 0.148 | 0.020 | 0.005 | $21.39^{-2.13}_{+0.41}$ | 21.46 |
| ARP295 | 478 | ∼150 | 0.089 | 0.008 | 0.004 | $22.83^{-1.79}_{+0.20}$ | 22.82 |
| ARP300 | 1763 | ∼130 | 0.500 | 0.202 | 0.005 | $21.03^{-1.87}_{+0.27}$ | 21.69 |
| ARP301 | 1264 | ∼120 | 0.349 | 0.178 | 0.004 | $19.77^{-0.33}_{+0.82}$ | 20.60 |
| ARP303 | 1326 | ∼130 | 0.362 | 0.220 | 0.004 | $21.63^{-0.62}_{+0.67}$ | 22.33 |
| ARP306 | 1913 | ∼160 | 0.560 | 0.245 | 0.004 | $21.70^{-0.78}_{+0.57}$ | 22.41 |
| ARP309 | 765 | ∼210 | 0.055 | 0.012 | 0.007 | $21.58^{-2.06}_{+1.17}$ | 21.59 |
| ARP314 | 2400 | ∼80 | 0.556 | 0.333 | 0.006 | $20.77^{-0.19}_{+0.46}$ | 21.57 |
| ARP321 | 937 | ∼200 | 0.159 | 0.020 | 0.007 | $20.10^{-1.27}_{+0.32}$ | 20.13 |
| ARP322 | 431 | ∼150 | 0.038 | 0.007 | 0.005 | $20.29^{-0.69}_{+1.78}$ | 19.87 |
| ARP335 | 999 | ∼140 | 0.191 | 0.071 | 0.004 | $22.07^{-2.35}_{+0.26}$ | 22.21 |
| ARP-MADORE0001-505 | 808 | ∼130 | 0.226 | 0.066 | 0.004 | $21.38^{-0.76}_{+0.49}$ | 21.37 |
| ARP-MADORE0002-503 | 545 | ∼150 | 0.064 | 0.012 | 0.004 | $22.45^{-2.33}_{+0.58}$ | 21.48 |
| ARP-MADORE0012-573 | 397 | ∼150 | 0.028 | 0.006 | 0.004 | $20.62^{-1.80}_{+1.74}$ | 19.66 |
| ARP-MADORE0018-485 | 1943 | ∼140 | 0.545 | 0.265 | 0.005 | $20.51^{-0.74}_{+0.43}$ | 21.25 |
| ARP-MADORE0052-321 | 619 | ∼150 | 0.249 | 0.024 | 0.004 | $21.94^{-0.43}_{+0.36}$ | 21.93 |
| ARP-MADORE0116-241 | 982 | ∼150 | 0.358 | 0.216 | 0.004 | $21.55^{-0.48}_{+0.70}$ | 21.90 |
| ARP-MADORE0135-650 | 1549 | ∼180 | 0.444 | 0.179 | 0.006 | $19.73^{-0.60}_{+0.29}$ | 20.06 |
| ARP-MADORE0136-433 | 456 | ∼150 | 0.027 | 0.007 | 0.005 | $22.09^{-2.49}_{+0.53}$ | 19.88 |
| ARP-MADORE0137-281 | 453 | ∼140 | 0.087 | 0.007 | 0.004 | $22.03^{-0.49}_{+0.76}$ | 21.98 |
| ARP-MADORE0144-585 | 2045 | ∼150 | 0.275 | 0.112 | 0.008 | $20.17^{-0.28}_{+0.72}$ | 20.89 |
| ARP-MADORE0154-441 | 910 | ∼140 | 0.153 | 0.097 | 0.004 | $23.09^{-0.61}_{+0.28}$ | 23.06 |
| ARP-MADORE0200-220 | 543 | ∼130 | 0.112 | 0.012 | 0.004 | $21.10^{-1.91}_{+0.56}$ | 20.63 |
| ARP-MADORE0203-325 | 438 | ∼180 | 0.025 | 0.008 | 0.005 | $21.06^{-1.55}_{+0.40}$ | 19.69 |
| ARP-MADORE0218-321 | 660 | ∼190 | 0.197 | 0.010 | 0.005 | $22.39^{-0.71}_{+0.50}$ | 22.38 |
| ARP-MADORE0223-403 | 681 | ∼130 | 0.287 | 0.029 | 0.004 | $21.18^{-1.47}_{+0.46}$ | 21.18 |
| ARP-MADORE0230-524 | 888 | ∼130 | 0.256 | 0.142 | 0.004 | $21.13^{-1.08}_{+0.90}$ | 21.44 |
| ARP-MADORE0311-252 | 3258 | ∼180 | 0.319 | 0.163 | 0.007 | $21.90^{-1.32}_{+0.66}$ | 22.60 |
| ARP-MADORE0311-573 | 4532 | ∼180 | 0.666 | 0.281 | 0.007 | $20.94^{-1.78}_{+0.62}$ | 22.19 |
| ARP-MADORE0313-545 | 633 | ∼140 | 0.165 | 0.016 | 0.004 | $21.35^{-2.01}_{+0.65}$ | 21.29 |
| ARP-MADORE0329-502 | 914 | ∼130 | 0.163 | 0.058 | 0.004 | $21.67^{-1.72}_{+0.15}$ | 21.67 |
| ARP-MADORE0333-513 | 633 | ∼150 | 0.101 | 0.024 | 0.004 | $21.85^{-1.81}_{+0.77}$ | 21.30 |
| ARP-MADORE0346-222 | 338 | ∼140 | 0.044 | 0.005 | 0.004 | $20.61^{-1.91}_{+1.26}$ | 19.45 |
| ARP-MADORE0347-442 | 747 | ∼140 | 0.121 | 0.039 | 0.004 | $20.66^{-0.77}_{+2.17}$ | 20.63 |
| ARP-MADORE0357-460 | 1039 | ∼120 | 0.134 | 0.038 | 0.006 | $22.36^{-3.08}_{+0.71}$ | 22.37 |
| ARP-MADORE0403-555 | 365 | ∼140 | 0.029 | 0.006 | 0.004 | $21.04^{-1.27}_{+0.51}$ | 20.50 |

**Table D3** *continued*

**Table D3** *(continued)*

| Target | # Point Sources (Total) | # Point Sources (Foreground) | $\Sigma_{max}$ (# arcsec$^{-2}$) | $\Sigma_{sources,50\%}$ (# arcsec$^{-2}$) | $\Sigma_{area,50\%}$ (# arcsec$^{-2}$) | F606W (Brightest 5) | F606W (Brightest 1%) |
|---|---|---|---|---|---|---|---|
| (1) | (2) | (3) | (4) | (5) | (6) | (7) | (8) |
| ARP-MADORE0405-552 | 1596 | ∼140 | 0.448 | 0.212 | 0.004 | $22.09^{-0.37}_{+0.28}$ | 22.35 |
| ARP-MADORE0409-563 | 1664 | ∼140 | 0.415 | 0.220 | 0.004 | $21.78^{-1.43}_{+0.36}$ | 22.39 |
| ARP-MADORE0417-391 | 525 | ∼180 | 0.102 | 0.009 | 0.005 | $21.07^{-1.11}_{+0.76}$ | 20.57 |
| ARP-MADORE0432-625 | 719 | ∼350 | 0.022 | 0.010 | 0.008 | $20.13^{-1.15}_{+0.46}$ | 20.11 |
| ARP-MADORE0445-572 | 1999 | ∼180 | 0.418 | 0.178 | 0.005 | $22.26^{-1.16}_{+0.26}$ | 22.53 |
| ARP-MADORE0454-561 | 1016 | ∼140 | 0.232 | 0.129 | 0.004 | $22.16^{-2.14}_{+0.33}$ | 22.17 |
| ARP-MADORE0459-340 | 1194 | ∼180 | 0.385 | 0.118 | 0.006 | $19.72^{-1.14}_{+0.71}$ | 20.33 |
| ARP-MADORE0507-630 | 3857 | ∼1360 | 0.289 | 0.053 | 0.035 | $18.69^{-0.15}_{+0.01}$ | 19.58 |
| ARP-MADORE0510-330 | 8449 | ∼820 | 0.401 | 0.186 | 0.079 | $19.99^{-0.68}_{+0.99}$ | 21.57 |
| ARP-MADORE0515-540 | 526 | ∼160 | 0.046 | 0.008 | 0.005 | $21.40^{-2.58}_{+0.52}$ | 21.35 |
| ARP-MADORE0519-611 | 1910 | ∼780 | 0.138 | 0.024 | 0.019 | $21.26^{-1.86}_{+0.40}$ | 21.77 |
| ARP-MADORE0520-390 | 565 | ∼160 | 0.077 | 0.008 | 0.005 | $19.39^{-0.21}_{+1.26}$ | 19.38 |
| ARP-MADORE0541-294 | 1210 | ∼170 | 0.391 | 0.194 | 0.005 | $22.11^{-0.97}_{+0.46}$ | 22.36 |
| ARP-MADORE0558-335 | 1103 | ∼200 | 0.423 | 0.168 | 0.006 | $20.04^{-0.22}_{+0.18}$ | 20.20 |
| ARP-MADORE0608-333 | 1134 | ∼180 | 0.224 | 0.047 | 0.007 | $21.46^{-1.42}_{+0.08}$ | 21.48 |
| ARP-MADORE0612-373 | 497 | ∼190 | 0.043 | 0.008 | 0.005 | $18.89^{-0.21}_{+0.32}$ | 18.77 |
| ARP-MADORE0619-271 | 2390 | ∼520 | 0.126 | 0.058 | 0.019 | $19.32^{-0.33}_{+1.42}$ | 21.03 |
| ARP-MADORE0620-363 | 624 | ∼150 | 0.116 | 0.011 | 0.005 | $19.98^{-0.30}_{+2.85}$ | 19.93 |
| ARP-MADORE0630-522 | 840 | ∼190 | 0.088 | 0.021 | 0.006 | $19.71^{-0.20}_{+0.82}$ | 19.68 |
| ARP-MADORE0642-801 | 698 | ∼150 | 0.180 | 0.032 | 0.004 | $21.78^{-2.87}_{+0.23}$ | 21.70 |
| ARP-MADORE0643-462 | 509 | ∼200 | 0.032 | 0.007 | 0.005 | $19.22^{-0.48}_{+0.45}$ | 19.08 |
| ARP-MADORE0658-590 | 606 | ∼180 | 0.124 | 0.010 | 0.005 | $20.40^{-1.38}_{+0.48}$ | 20.27 |
| ARP-MADORE0728-664 | 1229 | ∼230 | 0.256 | 0.099 | 0.006 | $22.56^{-3.02}_{+0.06}$ | 22.60 |
| ARP-MADORE0839-745 | 2347 | ∼280 | 0.162 | 0.066 | 0.012 | $19.61^{-0.74}_{+1.09}$ | 21.27 |
| ARP-MADORE0942-313A | 6872 | ∼290 | 0.793 | 0.534 | 0.010 | $20.78^{-1.18}_{+0.47}$ | 22.31 |
| ARP-MADORE0942-313B | 3742 | ∼200 | 0.855 | 0.683 | 0.006 | $20.76^{-0.44}_{+0.87}$ | 22.15 |
| ARP-MADORE1010-445 | 1786 | ∼540 | 0.202 | 0.030 | 0.014 | $19.01^{-0.47}_{+0.31}$ | 19.61 |
| ARP-MADORE1033-365 | 4432 | ∼460 | 0.286 | 0.110 | 0.038 | $20.40^{-1.07}_{+0.54}$ | 22.00 |
| ARP-MADORE1055-391 | 3545 | ∼400 | 0.458 | 0.195 | 0.015 | $21.75^{-0.65}_{+0.16}$ | 22.22 |
| ARP-MADORE1125-285 | 449 | ∼190 | 0.028 | 0.007 | 0.005 | $19.57^{-0.58}_{+0.31}$ | 19.07 |
| ARP-MADORE1200-251 | 806 | ∼190 | 0.159 | 0.023 | 0.005 | $19.98^{-0.66}_{+1.56}$ | 19.94 |
| ARP-MADORE1203-223 | 4077 | ∼360 | 0.410 | 0.097 | 0.035 | $20.64^{-1.34}_{+1.12}$ | 22.55 |
| ARP-MADORE1214-255 | 441 | ∼160 | 0.110 | 0.007 | 0.004 | $20.59^{-0.27}_{+0.05}$ | 20.49 |
| ARP-MADORE1229-512 | 1617 | ∼730 | 0.075 | 0.020 | 0.017 | $18.83^{-0.28}_{+0.16}$ | 19.06 |
| ARP-MADORE1238-254 | 718 | ∼150 | 0.088 | 0.018 | 0.005 | $21.44^{-0.83}_{+0.61}$ | 21.40 |
| ARP-MADORE1255-430 | 818 | ∼340 | 0.057 | 0.011 | 0.009 | $19.22^{-0.23}_{+0.09}$ | 19.27 |
| ARP-MADORE1303-371 | 1044 | ∼370 | 0.063 | 0.015 | 0.011 | $18.66^{-0.11}_{+0.57}$ | 18.79 |
| ARP-MADORE1307-425 | 1947 | ∼250 | 0.210 | 0.108 | 0.009 | $20.40^{-1.24}_{+0.36}$ | 21.32 |
| ARP-MADORE1307-461 | 2149 | ∼490 | 0.130 | 0.042 | 0.022 | $19.08^{-0.24}_{+0.23}$ | 19.75 |
| ARP-MADORE1324-294 | 1466 | ∼190 | 0.241 | 0.107 | 0.006 | $21.58^{-2.17}_{+0.89}$ | 22.20 |
| ARP-MADORE1332-331 | 654 | ∼280 | 0.041 | 0.010 | 0.007 | $19.28^{-0.70}_{+0.20}$ | 19.27 |
| ARP-MADORE1342-413 | 10897 | ∼670 | 0.605 | 0.332 | 0.064 | $19.39^{-0.43}_{+0.12}$ | 21.48 |
| ARP-MADORE1356-332 | 1015 | ∼260 | 0.177 | 0.025 | 0.006 | $20.08^{-1.50}_{+1.06}$ | 20.64 |
| ARP-MADORE1401-243A | 2564 | ∼240 | 0.272 | 0.081 | 0.016 | $20.51^{-1.90}_{+0.99}$ | 21.74 |
| ARP-MADORE1421-282 | 1371 | ∼180 | 0.426 | 0.180 | 0.005 | $21.11^{-0.49}_{+0.26}$ | 21.49 |
| ARP-MADORE1440-241 | 922 | ∼210 | 0.078 | 0.019 | 0.007 | $19.53^{-0.69}_{+1.10}$ | 20.09 |
| ARP-MADORE1546-284 | 1092 | ∼440 | 0.068 | 0.014 | 0.011 | $18.72^{-0.17}_{+0.10}$ | 18.80 |
| ARP-MADORE1705-773 | 2462 | ∼370 | 0.128 | 0.052 | 0.025 | $19.63^{-0.46}_{+0.36}$ | 20.20 |
| ARP-MADORE1806-852 | 3428 | ∼220 | 0.513 | 0.294 | 0.007 | $19.77^{-1.22}_{+0.16}$ | 21.37 |

**Table D3** *continued*

**Table D3** *(continued)*

| Target | # Point Sources (Total) | # Point Sources (Foreground) | $\Sigma_{max}$ (# arcsec$^{-2}$) | $\Sigma_{sources,50\%}$ (# arcsec$^{-2}$) | $\Sigma_{area,50\%}$ (# arcsec$^{-2}$) | F606W (Brightest 5) | F606W (Brightest 1%) |
|---|---|---|---|---|---|---|---|
| (1) | (2) | (3) | (4) | (5) | (6) | (7) | (8) |
| ARP-MADORE1847-645 | 2957 | ∼590 | 0.192 | 0.054 | 0.026 | $19.87^{-0.21}_{+0.40}$ | 20.65 |
| ARP-MADORE1912-860 | 1234 | ∼160 | 0.114 | 0.059 | 0.005 | $21.76^{-2.99}_{+0.65}$ | 22.35 |
| ARP-MADORE1914-603 | 3624 | ∼340 | 0.246 | 0.120 | 0.019 | $19.54^{-0.33}_{+1.14}$ | 21.93 |
| ARP-MADORE1931-610 | 2043 | ∼200 | 0.441 | 0.213 | 0.006 | $21.62^{-1.80}_{+0.18}$ | 22.01 |
| ARP-MADORE1953-260 | 1048 | ∼380 | 0.131 | 0.013 | 0.010 | $19.12^{-0.30}_{+0.15}$ | 19.23 |
| ARP-MADORE1957-471 | 1427 | ∼240 | 0.307 | 0.090 | 0.007 | $20.62^{-1.99}_{+0.33}$ | 21.12 |
| ARP-MADORE2001-602 | 874 | ∼160 | 0.119 | 0.052 | 0.005 | $19.00^{-0.21}_{+1.76}$ | 19.48 |
| ARP-MADORE2019-442 | 837 | ∼240 | 0.090 | 0.015 | 0.007 | $20.61^{-0.47}_{+0.61}$ | 20.64 |
| ARP-MADORE2026-424 | 570 | ∼230 | 0.037 | 0.009 | 0.006 | $19.62^{-0.84}_{+0.27}$ | 19.44 |
| ARP-MADORE2029-544 | 951 | ∼190 | 0.238 | 0.033 | 0.005 | $20.20^{-0.94}_{+1.18}$ | 20.55 |
| ARP-MADORE2031-440 | 507 | ∼190 | 0.058 | 0.008 | 0.005 | $20.16^{-1.33}_{+0.68}$ | 20.07 |
| ARP-MADORE2034-521 | 1514 | ∼260 | 0.183 | 0.106 | 0.008 | $22.83^{-0.16}_{+0.26}$ | 23.13 |
| ARP-MADORE2038-323 | 738 | ∼210 | 0.091 | 0.012 | 0.006 | $19.35^{-0.71}_{+1.29}$ | 19.38 |
| ARP-MADORE2038-382 | 607 | ∼190 | 0.083 | 0.009 | 0.006 | $18.99^{-0.45}_{+1.20}$ | 18.96 |
| ARP-MADORE2038-654 | 1088 | ∼200 | 0.146 | 0.047 | 0.006 | $21.23^{-0.54}_{+0.57}$ | 21.34 |
| ARP-MADORE2040-295 | 2813 | ∼220 | 0.409 | 0.165 | 0.010 | $21.51^{-2.04}_{+0.32}$ | 22.26 |
| ARP-MADORE2042-382 | 520 | ∼210 | 0.034 | 0.008 | 0.005 | $20.23^{-1.38}_{+0.28}$ | 19.59 |
| ARP-MADORE2048-571 | 661 | ∼200 | 0.067 | 0.010 | 0.006 | $20.68^{-1.38}_{+0.10}$ | 20.66 |
| ARP-MADORE2055-541 | 466 | ∼150 | 0.056 | 0.007 | 0.004 | $19.05^{-0.42}_{+0.84}$ | 18.92 |
| ARP-MADORE2056-392 | 433 | ∼190 | 0.031 | 0.006 | 0.005 | $20.04^{-0.88}_{+0.22}$ | 19.69 |
| ARP-MADORE2103-550 | 1061 | ∼140 | 0.083 | 0.034 | 0.007 | $22.24^{-2.57}_{+0.43}$ | 22.35 |
| ARP-MADORE2105-332 | 579 | ∼200 | 0.039 | 0.010 | 0.006 | $19.85^{-0.31}_{+0.32}$ | 19.77 |
| ARP-MADORE2105-632 | 730 | ∼160 | 0.120 | 0.017 | 0.005 | $21.58^{-0.47}_{+0.96}$ | 21.62 |
| ARP-MADORE2106-374 | 1166 | ∼120 | 0.340 | 0.203 | 0.005 | $21.16^{-1.45}_{+0.30}$ | 21.46 |
| ARP-MADORE2113-341 | 398 | ∼130 | 0.056 | 0.006 | 0.003 | $20.29^{-0.74}_{+0.44}$ | 20.24 |
| ARP-MADORE2115-273 | 363 | ∼180 | 0.016 | 0.006 | 0.004 | $19.99^{-0.68}_{+0.65}$ | 19.82 |
| ARP-MADORE2126-601 | 467 | ∼140 | 0.066 | 0.008 | 0.004 | $20.61^{-1.14}_{+0.76}$ | 20.25 |
| ARP-MADORE2128-430 | 390 | ∼160 | 0.024 | 0.006 | 0.004 | $19.48^{-0.11}_{+0.49}$ | 19.39 |
| ARP-MADORE2159-320 | 987 | ∼270 | 0.124 | 0.018 | 0.007 | $21.42^{-1.22}_{+0.49}$ | 21.55 |
| ARP-MADORE2210-693 | 758 | ∼130 | 0.140 | 0.063 | 0.004 | $22.48^{-0.62}_{+0.74}$ | 22.53 |
| ARP-MADORE2220-423 | 892 | ∼140 | 0.137 | 0.032 | 0.005 | $21.86^{-0.75}_{+0.46}$ | 22.00 |
| ARP-MADORE2222-275 | 639 | ∼180 | 0.120 | 0.012 | 0.005 | $21.41^{-2.67}_{+1.00}$ | 21.42 |
| ARP-MADORE2224-310 | 633 | ∼150 | 0.162 | 0.013 | 0.004 | $21.40^{-0.43}_{+0.17}$ | 21.32 |
| ARP-MADORE2230-481 | 354 | ∼140 | 0.024 | 0.005 | 0.004 | $20.54^{-1.34}_{+1.63}$ | 19.64 |
| ARP-MADORE2240-892 | 1446 | ∼220 | 0.239 | 0.125 | 0.006 | $21.81^{-2.39}_{+0.38}$ | 22.18 |
| ARP-MADORE2254-373 | 2908 | ∼140 | 0.287 | 0.172 | 0.005 | $21.81^{-0.80}_{+0.20}$ | 22.39 |
| ARP-MADORE2258-595 | 572 | ∼140 | 0.148 | 0.014 | 0.004 | $20.96^{-0.39}_{+0.11}$ | 20.89 |
| ARP-MADORE2259-692 | 493 | ∼140 | 0.131 | 0.008 | 0.004 | $22.41^{-3.74}_{+0.18}$ | 21.15 |
| ARP-MADORE2303-305 | 526 | ∼160 | 0.153 | 0.009 | 0.004 | $21.49^{-0.47}_{+0.29}$ | 21.39 |
| ARP-MADORE2325-473 | 614 | ∼130 | 0.118 | 0.027 | 0.004 | $21.19^{-2.13}_{+1.16}$ | 20.98 |
| ARP-MADORE2339-661 | 852 | ∼130 | 0.116 | 0.045 | 0.004 | $22.20^{-3.18}_{+0.33}$ | 22.29 |
| ARP-MADORE2346-380 | 2029 | ∼150 | 0.200 | 0.104 | 0.007 | $21.41^{-0.15}_{+0.27}$ | 21.65 |
| ARP-MADORE2350-302 | 421 | ∼140 | 0.083 | 0.007 | 0.004 | $19.35^{-0.13}_{+0.28}$ | 19.35 |
| ARP-MADORE2350-410 | 607 | ∼140 | 0.197 | 0.025 | 0.004 | $21.36^{-0.13}_{+0.35}$ | 21.36 |
| ARP-MADORE2353-291 | 406 | ∼150 | 0.046 | 0.007 | 0.004 | $23.07^{-3.66}_{+0.45}$ | 22.55 |

Note—Column data includes: (1) the target name; (2) the number of well-detected point sources; (3) the approximate number of these sources that are likely foreground stars rather than sources in the target system; (4) the peak surface density of point sources in the KDE density map, calculated with an Epanechnikov smoothing kernel with a 10″ radius; (5) the surface density contour level containing 50% of the sources; (6) the median surface density of all pixels; (7) the apparent F606W magnitude of the 3rd brightest source that falls within the surface density contour from column 5, with the upper and lower limits indicating the magnitude range between the 1st and 5th brightest; (8) the apparent F606W magnitude of the brightest 1% of sources that fall within the surface density contour from column 5.

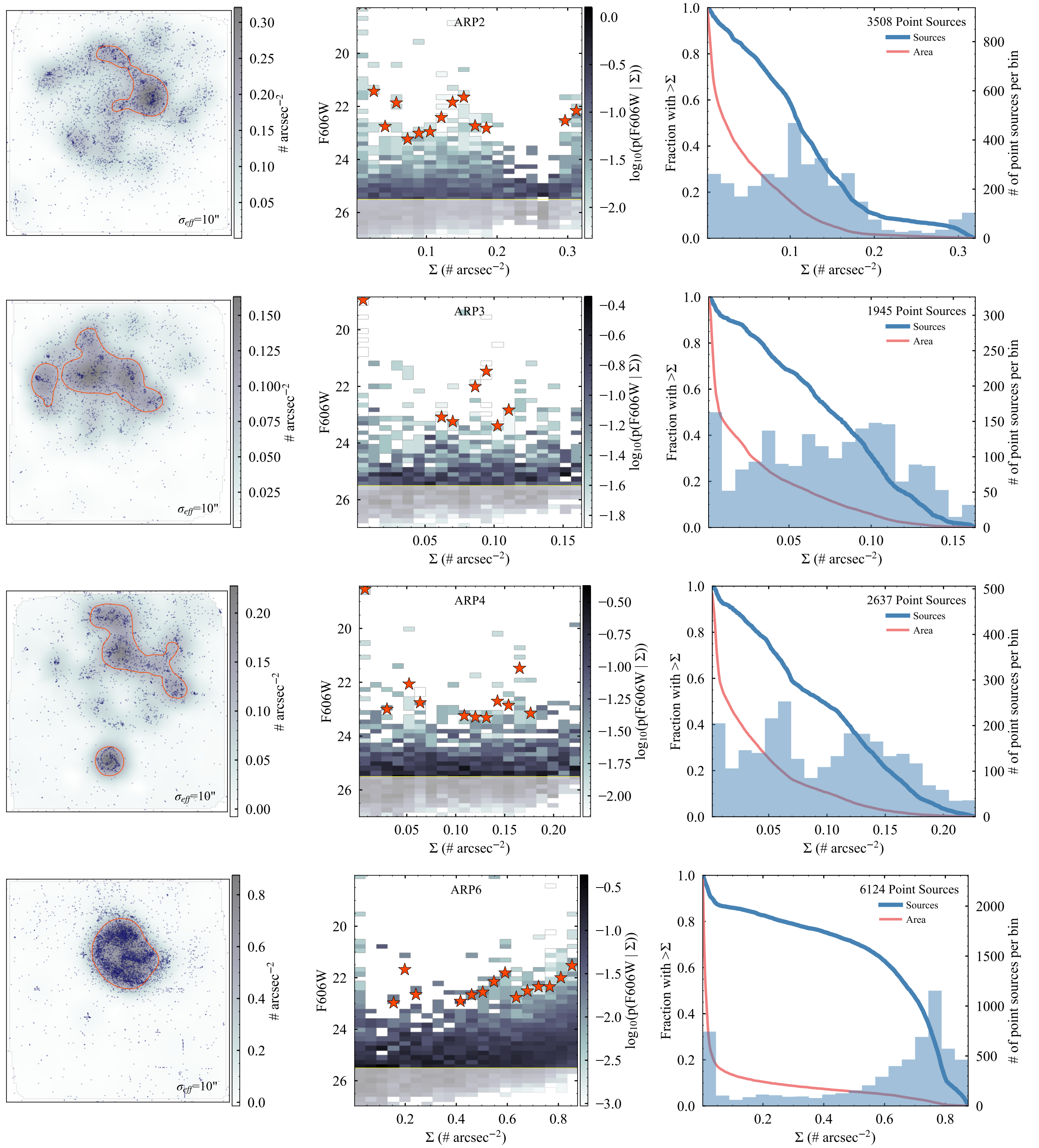

**Figure D3.** ARP2, ARP3, ARP4, and ARP6 (top to bottom). [Left] The spatial distribution of point sources (points) and the smoothed background density field (greyscale); [Middle] the conditional luminosity function (greyscale) in bins of local surface density, with red stars marking the magnitude of the top 2% of the brightest sources in any density bin with more than 50 stars brighter than F606W=25.5; [Right] the histogram (shaded blue; right axis) and cumulative distributions (solid lines) of densities calculated at the location of point sources (blue thick line) or pixels (red thin line). Please see Section 2.5 and Figure 12 for a fuller description. [Continued]

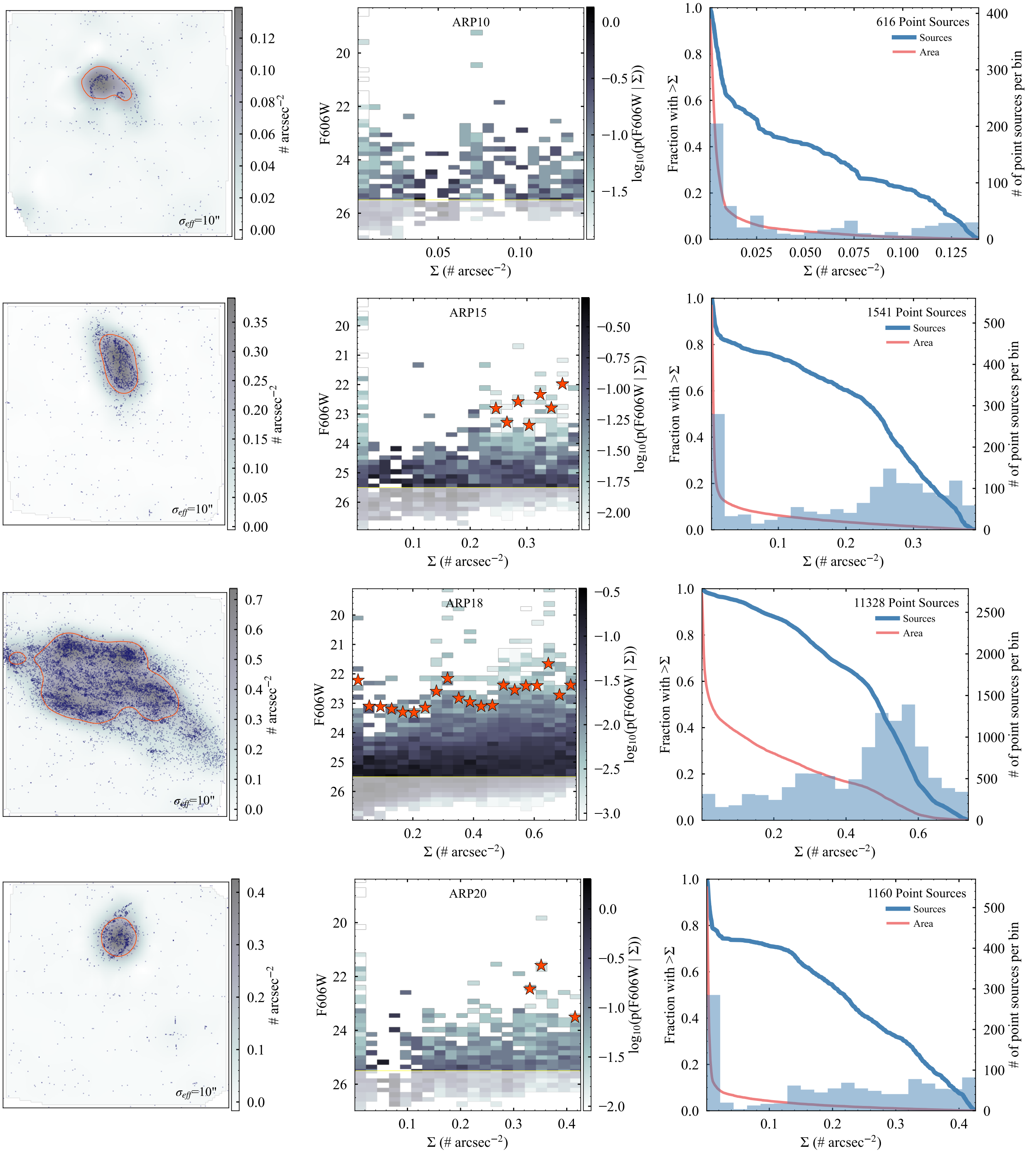

**Figure D3.** [Continued] ARP10, ARP15, ARP18, and ARP20 (top to bottom). [Left] The spatial distribution of point sources (points) and the smoothed background density field (greyscale); [Middle] the conditional luminosity function (greyscale) in bins of local surface density, with red stars marking the magnitude of the top 2% of the brightest sources in any density bin with more than 50 stars brighter than F606W=25.5; [Right] the histogram (shaded blue; right axis) and cumulative distributions (solid lines) of densities calculated at the location of point sources (blue thick line) or pixels (red thin line). Please see Section 2.5 and Figure 12 for a fuller description. [Continued]

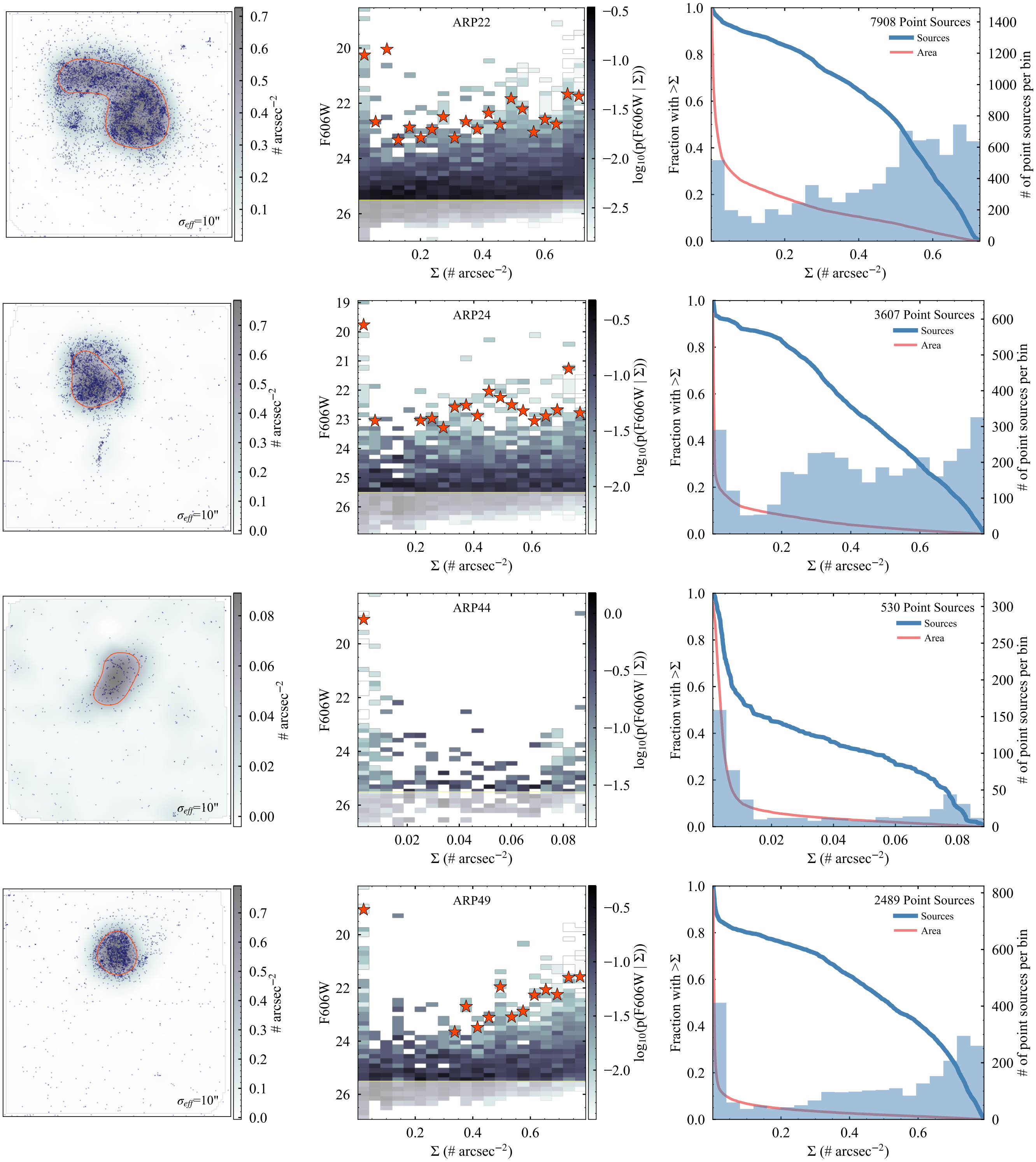

**Figure D3.** [Continued] ARP22, ARP24, ARP44, and ARP49 (top to bottom). [Left] The spatial distribution of point sources (points) and the smoothed background density field (greyscale); [Middle] the conditional luminosity function (greyscale) in bins of local surface density, with red stars marking the magnitude of the top 2% of the brightest sources in any density bin with more than 50 stars brighter than F606W=25.5; [Right] the histogram (shaded blue; right axis) and cumulative distributions (solid lines) of densities calculated at the location of point sources (blue thick line) or pixels (red thin line). Please see Section 2.5 and Figure 12 for a fuller description. [Continued]

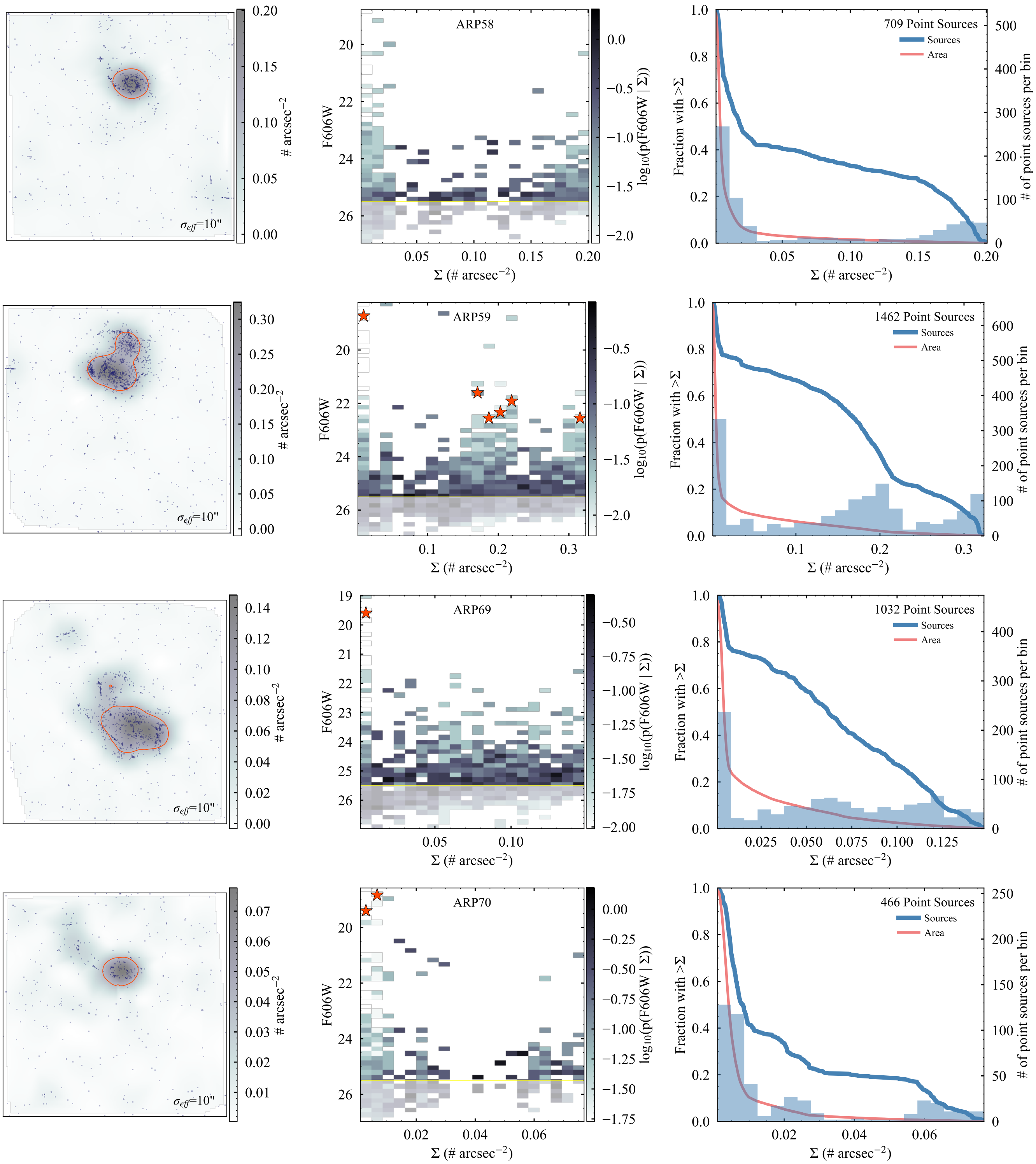

**Figure D3.** [Continued] ARP58, ARP59, ARP69, and ARP70 (top to bottom). [Left] The spatial distribution of point sources (points) and the smoothed background density field (greyscale); [Middle] the conditional luminosity function (greyscale) in bins of local surface density, with red stars marking the magnitude of the top 2% of the brightest sources in any density bin with more than 50 stars brighter than F606W=25.5; [Right] the histogram (shaded blue; right axis) and cumulative distributions (solid lines) of densities calculated at the location of point sources (blue thick line) or pixels (red thin line). Please see Section 2.5 and Figure 12 for a fuller description. [Continued]

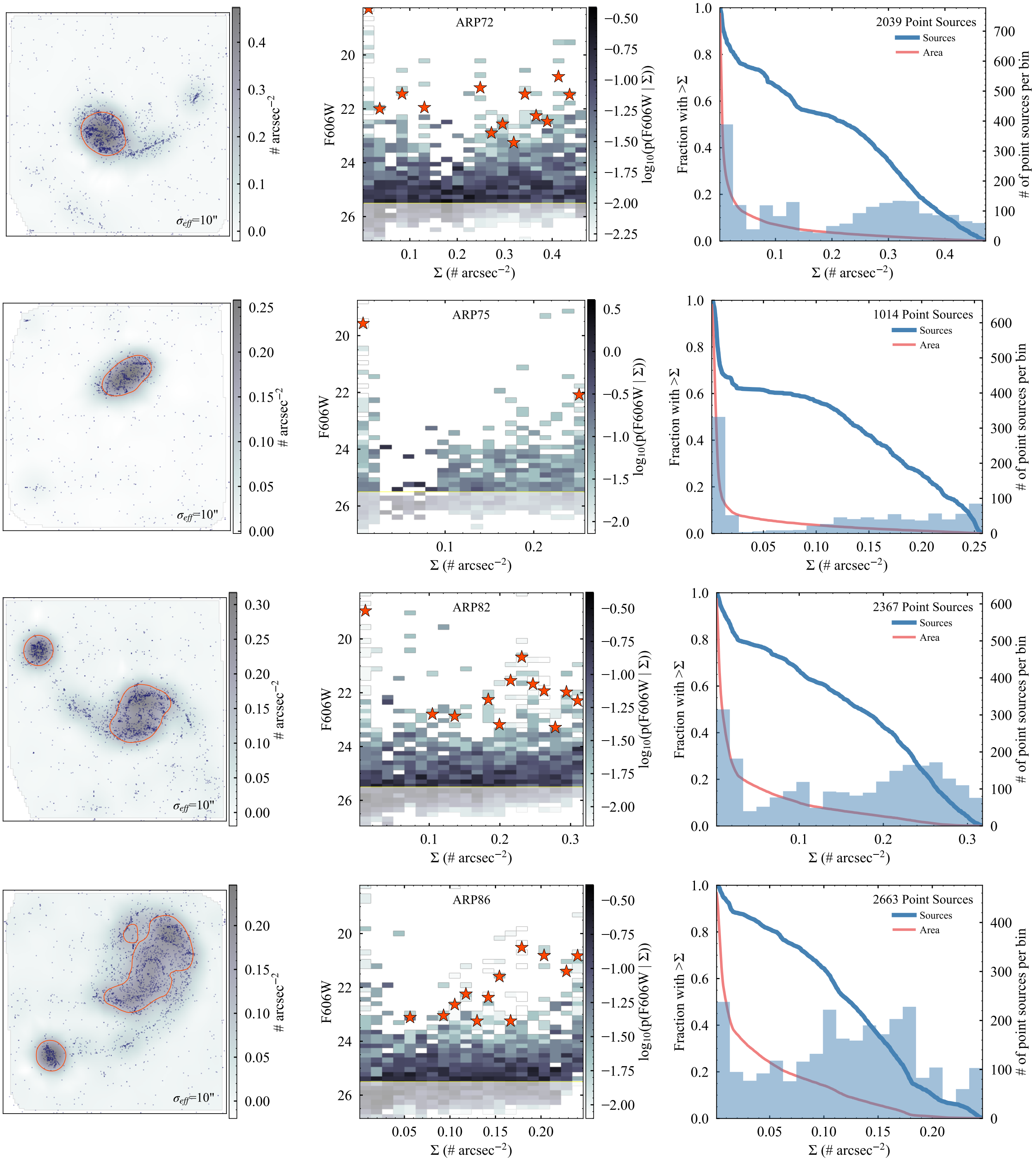

**Figure D3.** [Continued] ARP72, ARP75, ARP82, and ARP86 (top to bottom). [Left] The spatial distribution of point sources (points) and the smoothed background density field (greyscale); [Middle] the conditional luminosity function (greyscale) in bins of local surface density, with red stars marking the magnitude of the top 2% of the brightest sources in any density bin with more than 50 stars brighter than F606W=25.5; [Right] the histogram (shaded blue; right axis) and cumulative distributions (solid lines) of densities calculated at the location of point sources (blue thick line) or pixels (red thin line). Please see Section 2.5 and Figure 12 for a fuller description. [Continued]

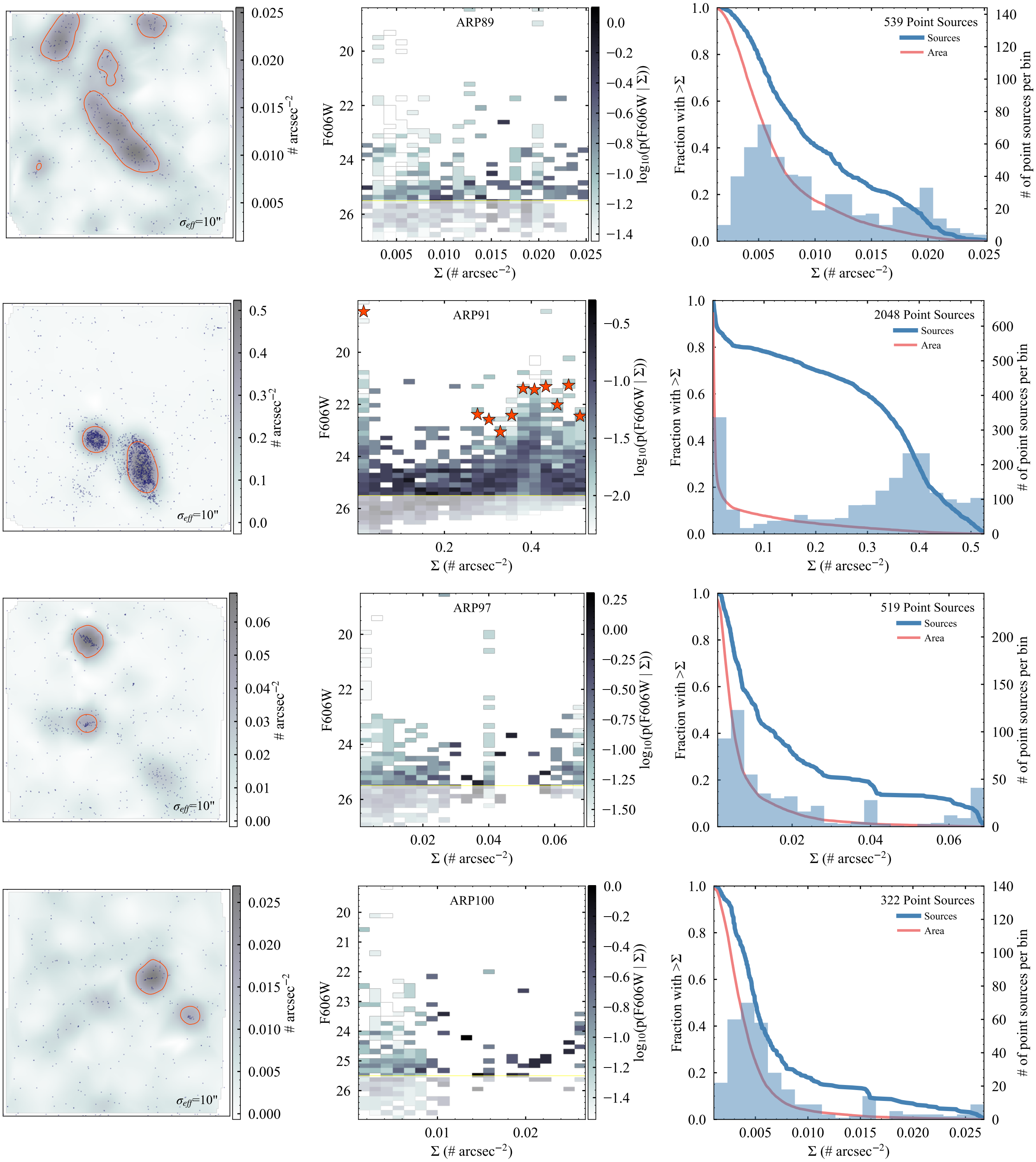

**Figure D3.** [Continued] ARP89, ARP91, ARP97, and ARP100 (top to bottom). [Left] The spatial distribution of point sources (points) and the smoothed background density field (greyscale); [Middle] the conditional luminosity function (greyscale) in bins of local surface density, with red stars marking the magnitude of the top 2% of the brightest sources in any density bin with more than 50 stars brighter than F606W=25.5; [Right] the histogram (shaded blue; right axis) and cumulative distributions (solid lines) of densities calculated at the location of point sources (blue thick line) or pixels (red thin line). Please see Section 2.5 and Figure 12 for a fuller description. [Continued]

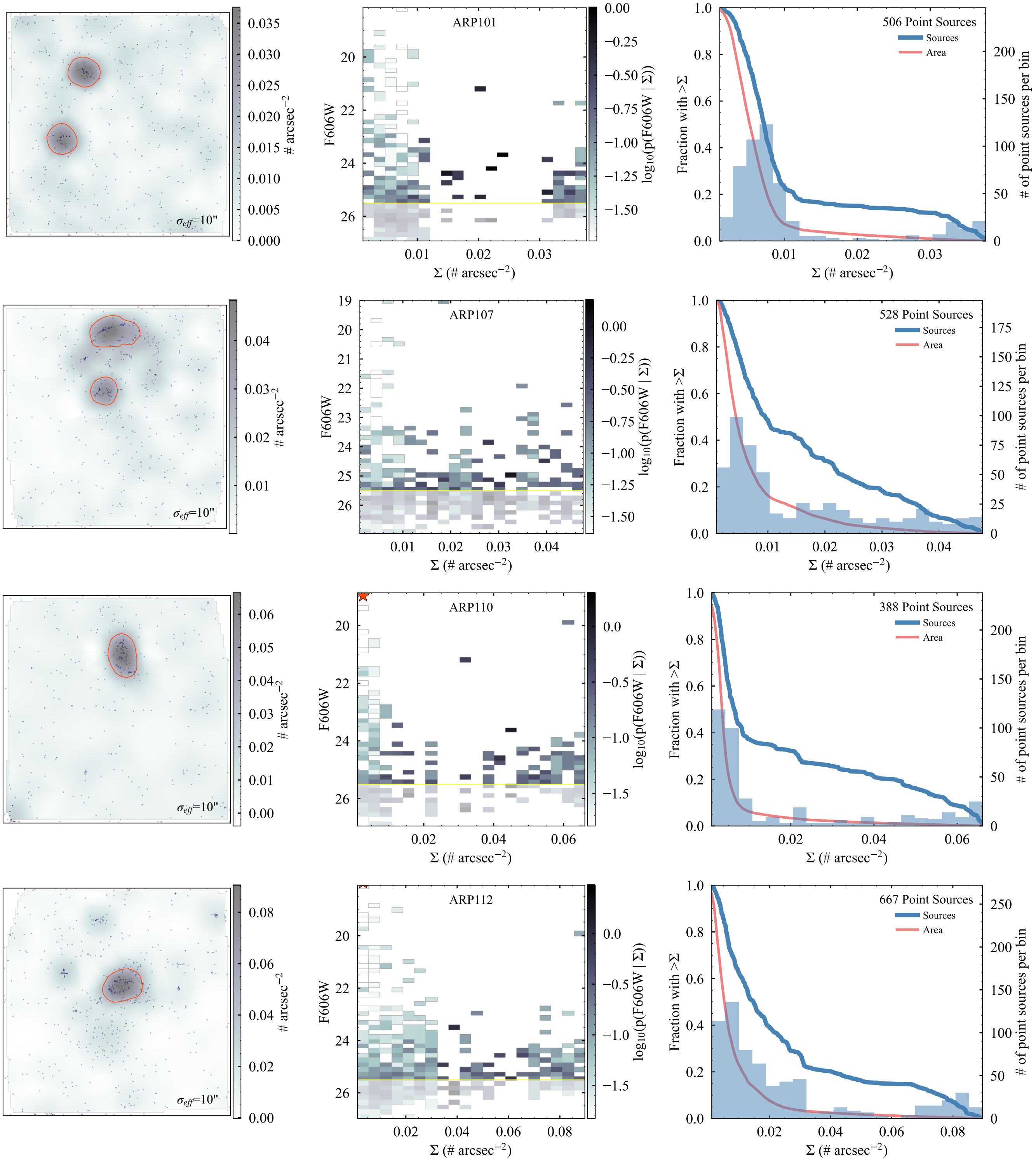

**Figure D3.** [Continued] ARP101, ARP107, ARP110, and ARP112 (top to bottom). [Left] The spatial distribution of point sources (points) and the smoothed background density field (greyscale); [Middle] the conditional luminosity function (greyscale) in bins of local surface density, with red stars marking the magnitude of the top 2% of the brightest sources in any density bin with more than 50 stars brighter than F606W=25.5; [Right] the histogram (shaded blue; right axis) and cumulative distributions (solid lines) of densities calculated at the location of point sources (blue thick line) or pixels (red thin line). Please see Section 2.5 and Figure 12 for a fuller description. [Continued]

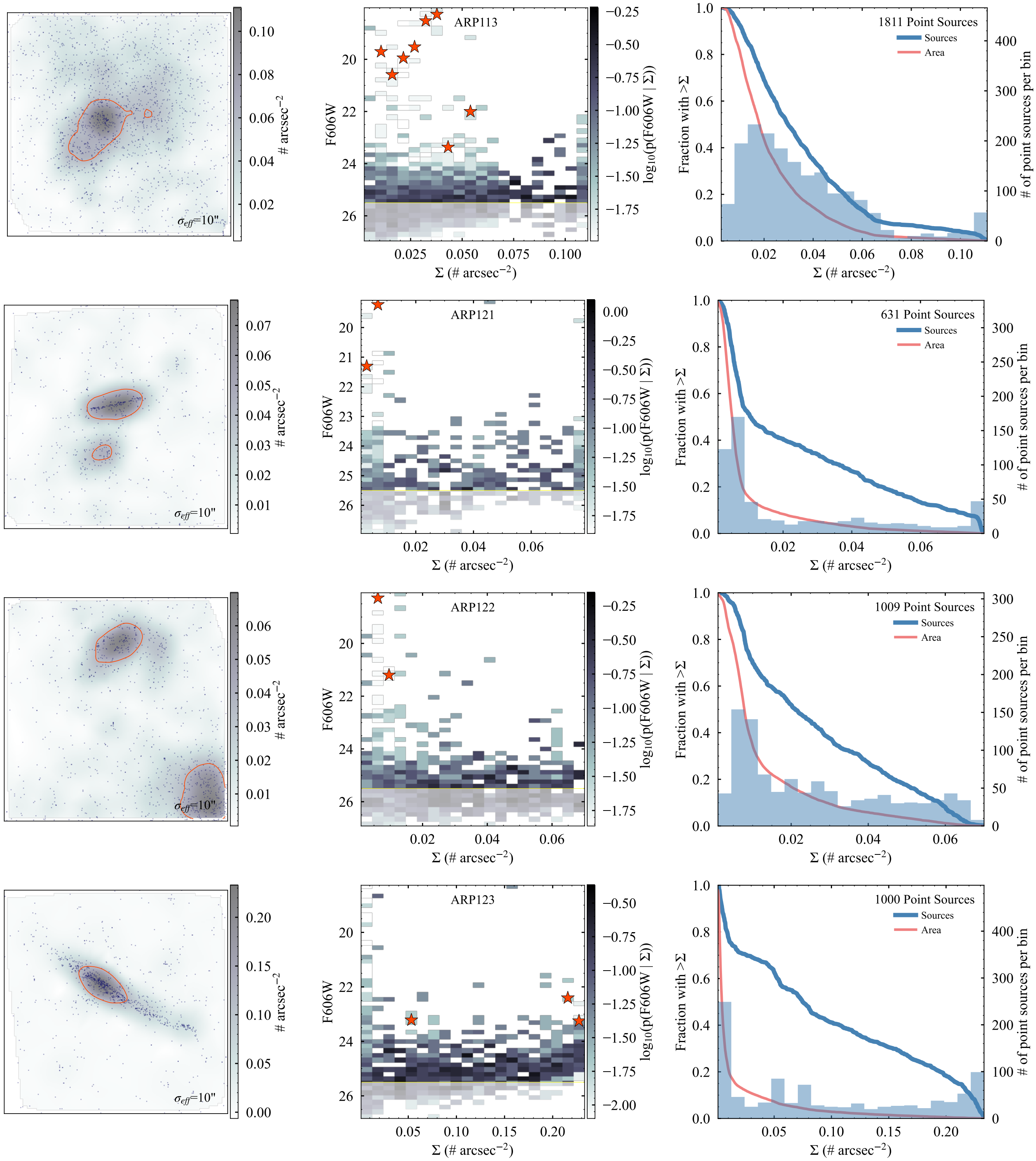

**Figure D3.** [Continued] ARP113, ARP121, ARP122, and ARP123 (top to bottom). [Left] The spatial distribution of point sources (points) and the smoothed background density field (greyscale); [Middle] the conditional luminosity function (greyscale) in bins of local surface density, with red stars marking the magnitude of the top 2% of the brightest sources in any density bin with more than 50 stars brighter than F606W=25.5; [Right] the histogram (shaded blue; right axis) and cumulative distributions (solid lines) of densities calculated at the location of point sources (blue thick line) or pixels (red thin line). Please see Section 2.5 and Figure 12 for a fuller description. [Continued]

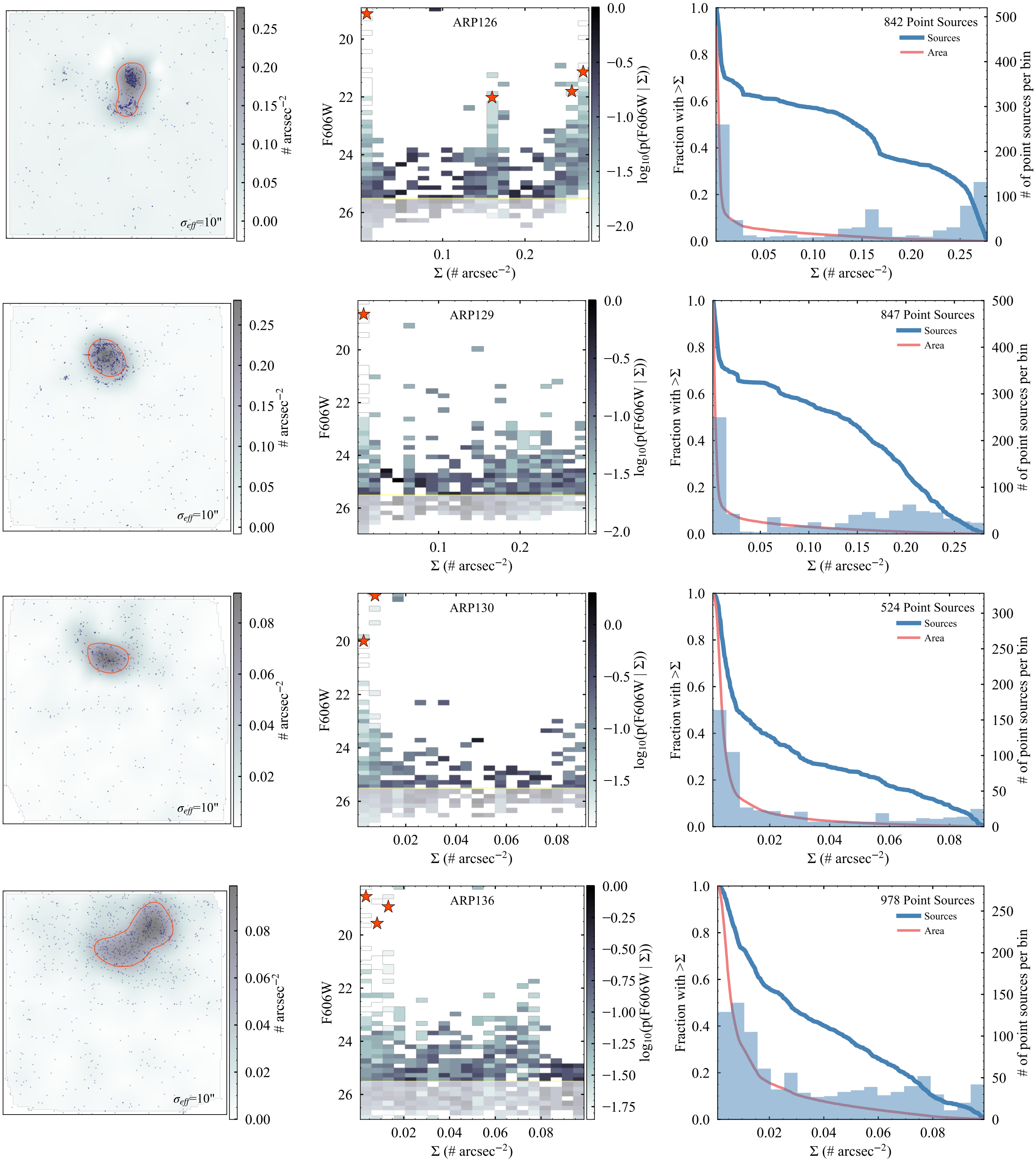

**Figure D3.** [Continued] ARP126, ARP129, ARP130, and ARP136 (top to bottom). [Left] The spatial distribution of point sources (points) and the smoothed background density field (greyscale); [Middle] the conditional luminosity function (greyscale) in bins of local surface density, with red stars marking the magnitude of the top 2% of the brightest sources in any density bin with more than 50 stars brighter than F606W=25.5; [Right] the histogram (shaded blue; right axis) and cumulative distributions (solid lines) of densities calculated at the location of point sources (blue thick line) or pixels (red thin line). Please see Section 2.5 and Figure 12 for a fuller description. [Continued]

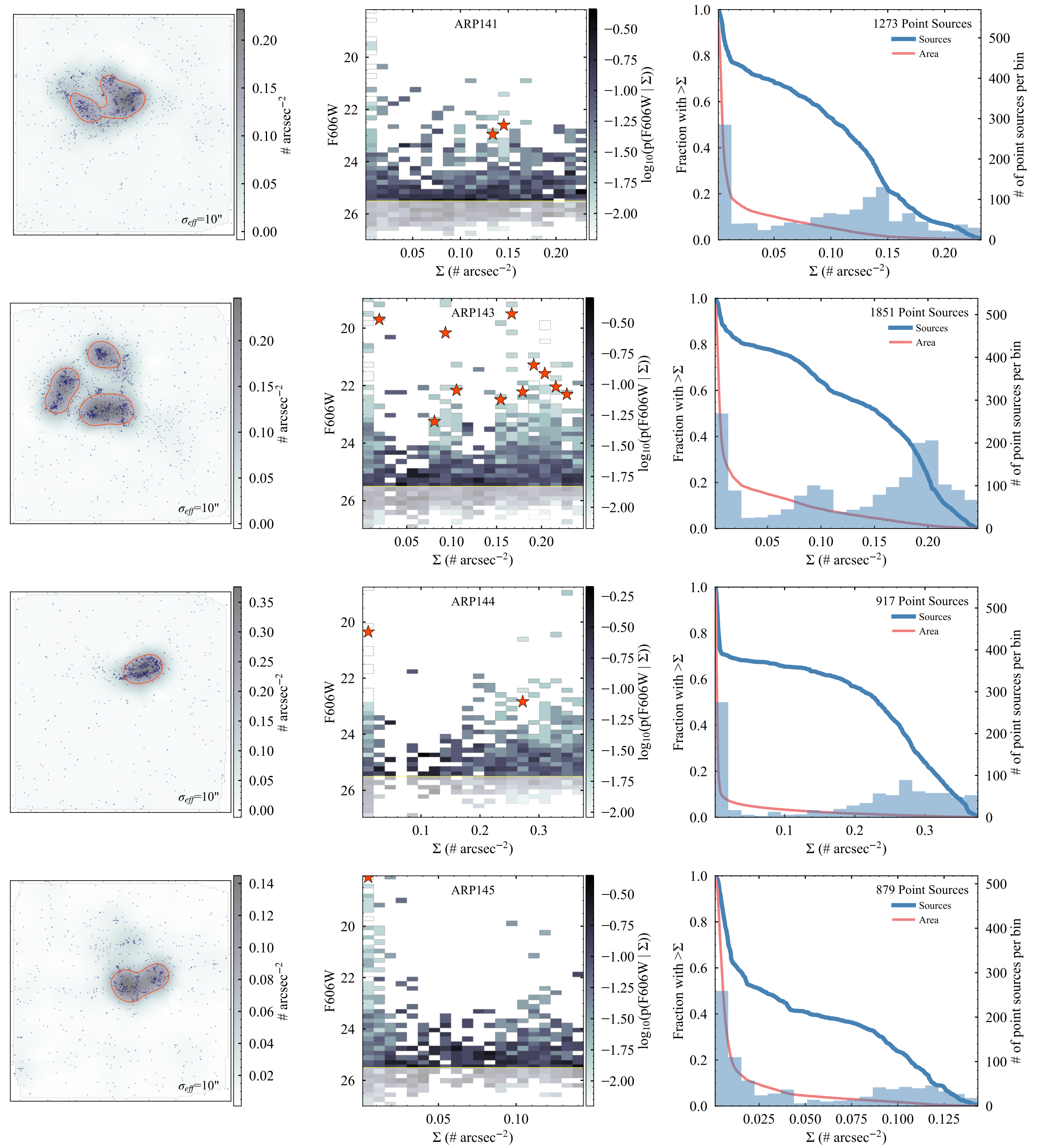

**Figure D3.** [Continued] ARP141, ARP143, ARP144, and ARP145 (top to bottom). [Left] The spatial distribution of point sources (points) and the smoothed background density field (greyscale); [Middle] the conditional luminosity function (greyscale) in bins of local surface density, with red stars marking the magnitude of the top 2% of the brightest sources in any density bin with more than 50 stars brighter than F606W=25.5; [Right] the histogram (shaded blue; right axis) and cumulative distributions (solid lines) of densities calculated at the location of point sources (blue thick line) or pixels (red thin line). Please see Section 2.5 and Figure 12 for a fuller description. [Continued]

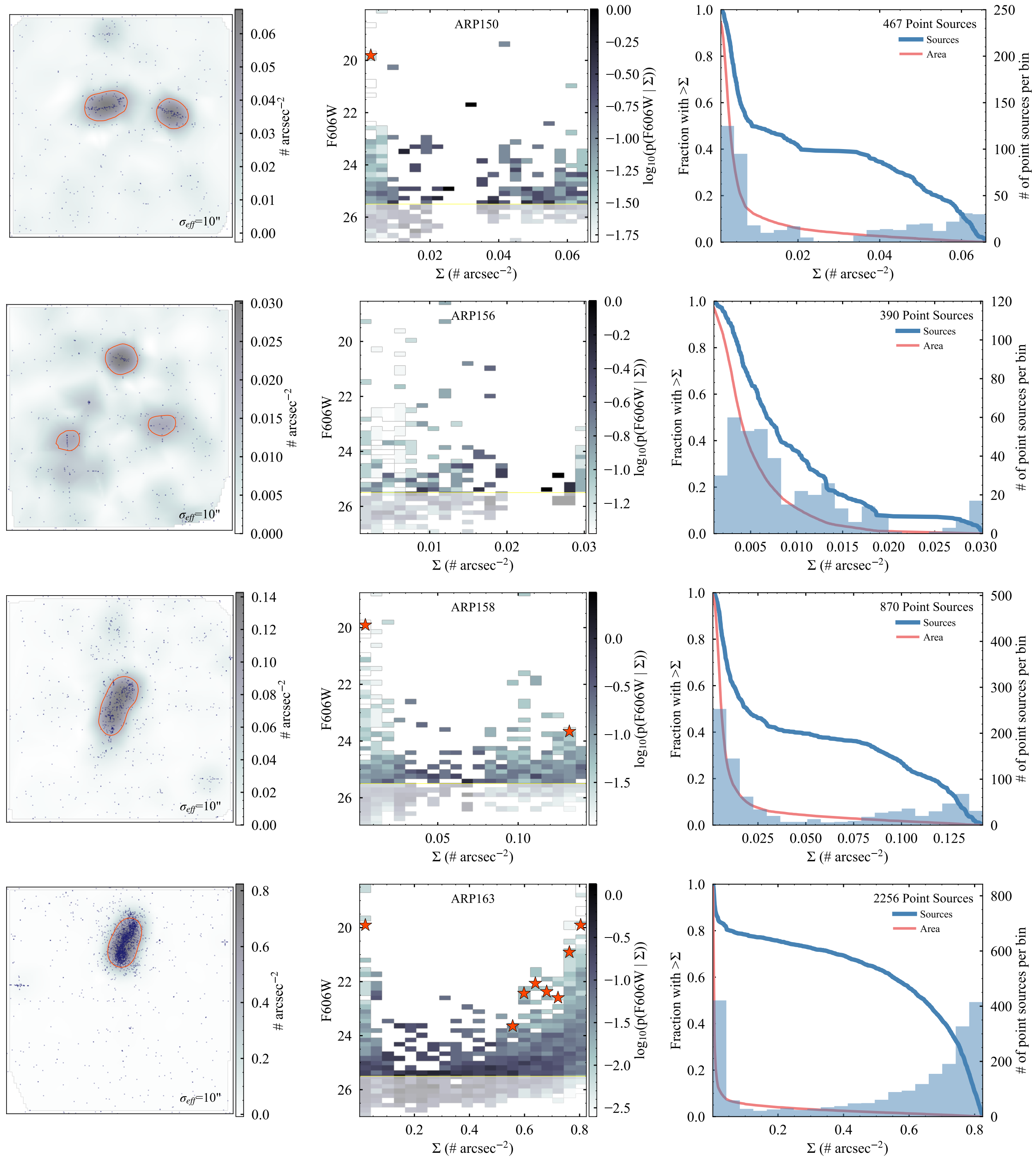

**Figure D3.** [Continued] ARP150, ARP156, ARP158, and ARP163 (top to bottom). [Left] The spatial distribution of point sources (points) and the smoothed background density field (greyscale); [Middle] the conditional luminosity function (greyscale) in bins of local surface density, with red stars marking the magnitude of the top 2% of the brightest sources in any density bin with more than 50 stars brighter than F606W=25.5; [Right] the histogram (shaded blue; right axis) and cumulative distributions (solid lines) of densities calculated at the location of point sources (blue thick line) or pixels (red thin line). Please see Section 2.5 and Figure 12 for a fuller description. [Continued]

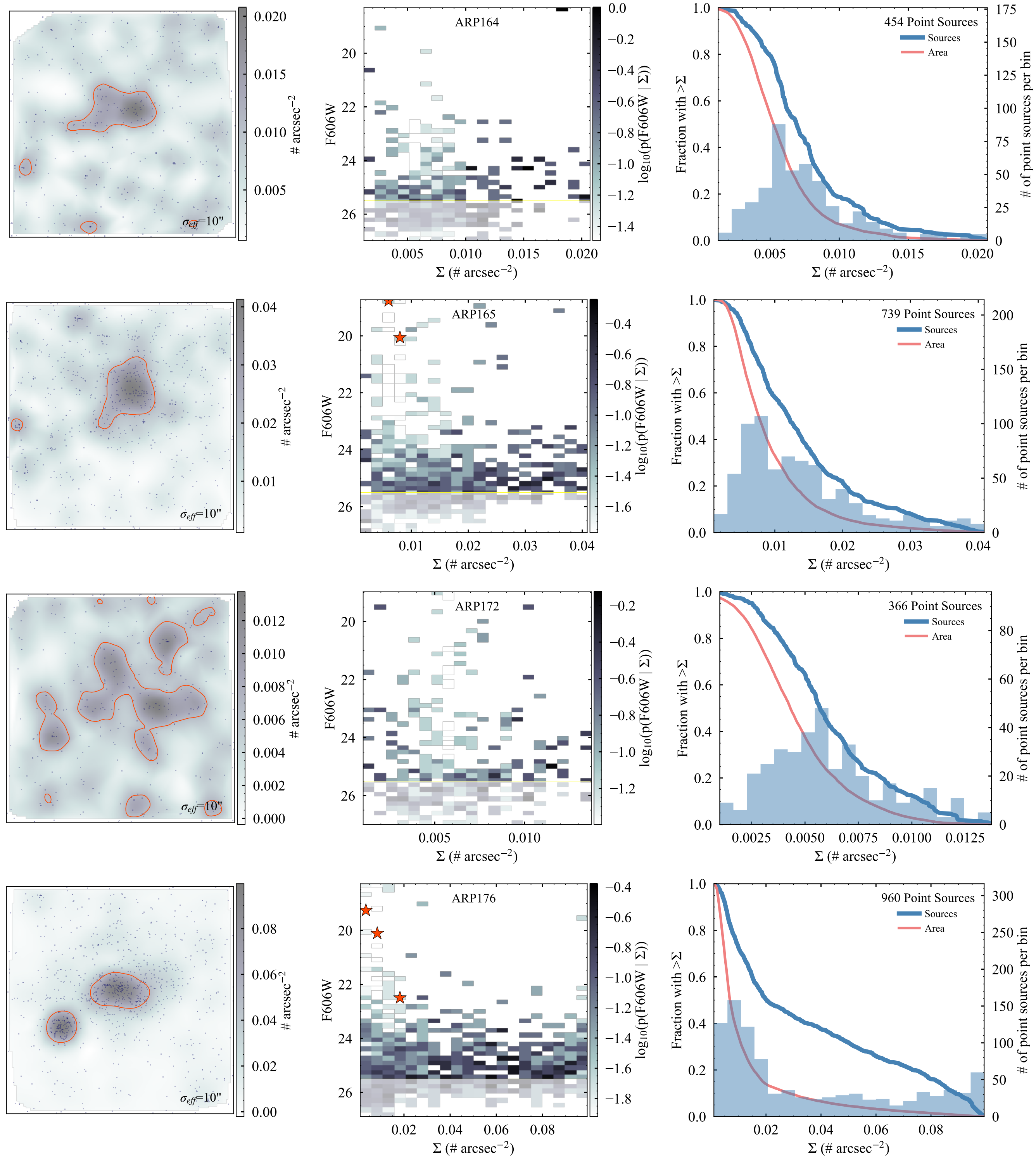

**Figure D3.** [Continued] ARP164, ARP165, ARP172, and ARP176 (top to bottom). [Left] The spatial distribution of point sources (points) and the smoothed background density field (greyscale); [Middle] the conditional luminosity function (greyscale) in bins of local surface density, with red stars marking the magnitude of the top 2% of the brightest sources in any density bin with more than 50 stars brighter than F606W=25.5; [Right] the histogram (shaded blue; right axis) and cumulative distributions (solid lines) of densities calculated at the location of point sources (blue thick line) or pixels (red thin line). Please see Section 2.5 and Figure 12 for a fuller description. [Continued]

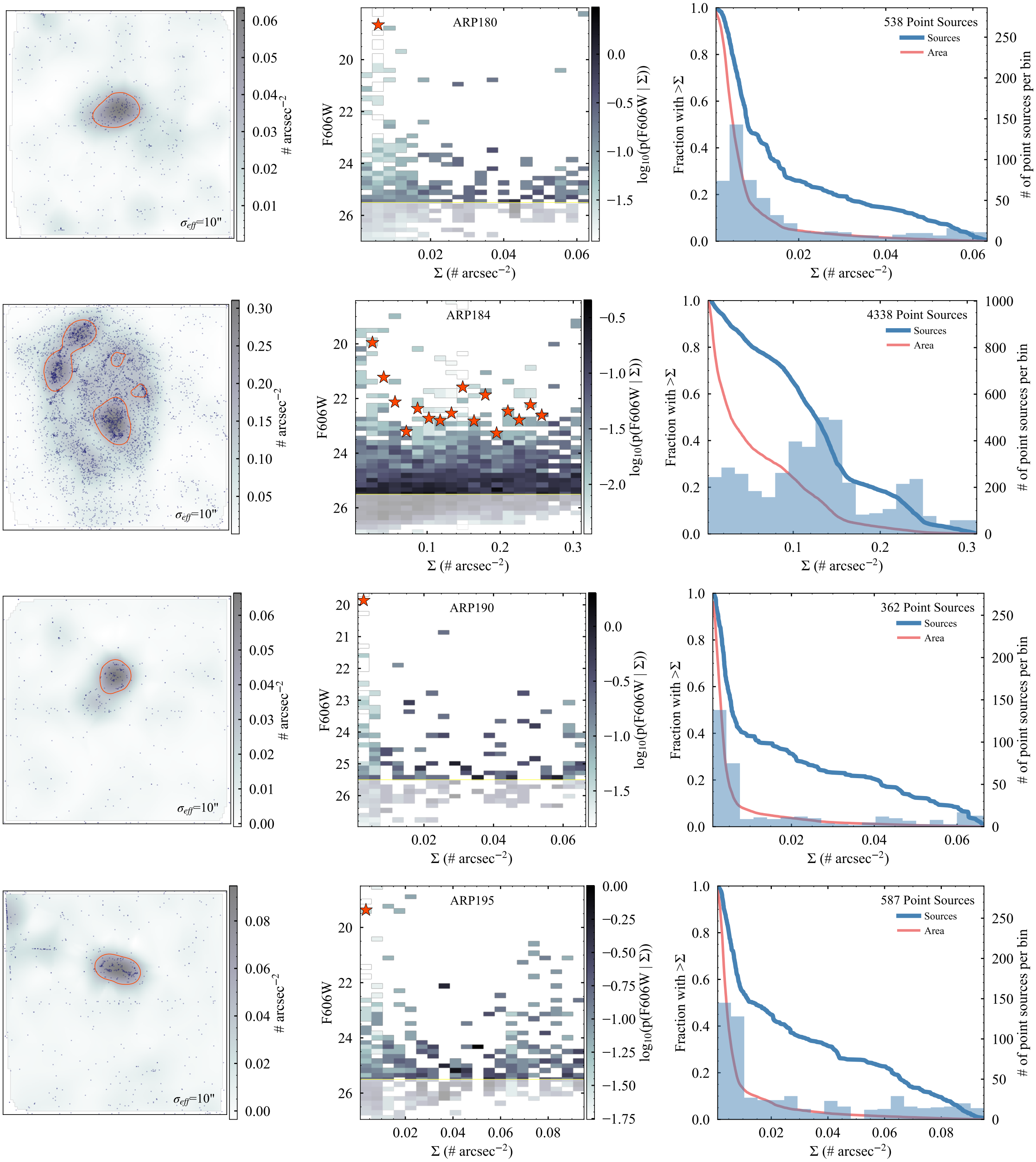

**Figure D3.** [Continued] ARP180, ARP184, ARP190, and ARP195 (top to bottom). [Left] The spatial distribution of point sources (points) and the smoothed background density field (greyscale); [Middle] the conditional luminosity function (greyscale) in bins of local surface density, with red stars marking the magnitude of the top 2% of the brightest sources in any density bin with more than 50 stars brighter than F606W=25.5; [Right] the histogram (shaded blue; right axis) and cumulative distributions (solid lines) of densities calculated at the location of point sources (blue thick line) or pixels (red thin line). Please see Section 2.5 and Figure 12 for a fuller description. [Continued]

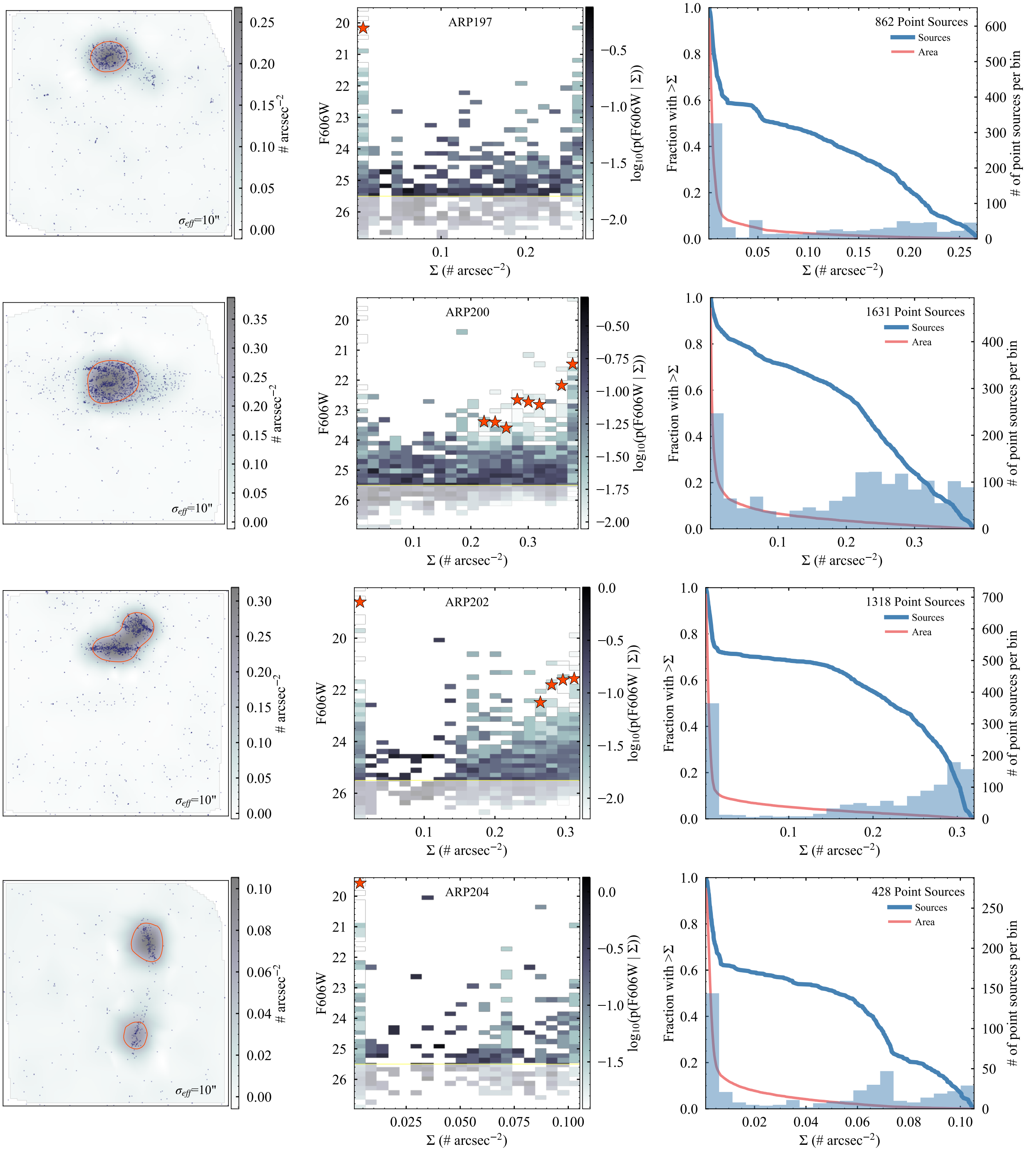

**Figure D3.** [Continued] ARP197, ARP200, ARP202, and ARP204 (top to bottom). [Left] The spatial distribution of point sources (points) and the smoothed background density field (greyscale); [Middle] the conditional luminosity function (greyscale) in bins of local surface density, with red stars marking the magnitude of the top 2% of the brightest sources in any density bin with more than 50 stars brighter than F606W=25.5; [Right] the histogram (shaded blue; right axis) and cumulative distributions (solid lines) of densities calculated at the location of point sources (blue thick line) or pixels (red thin line). Please see Section 2.5 and Figure 12 for a fuller description. [Continued]

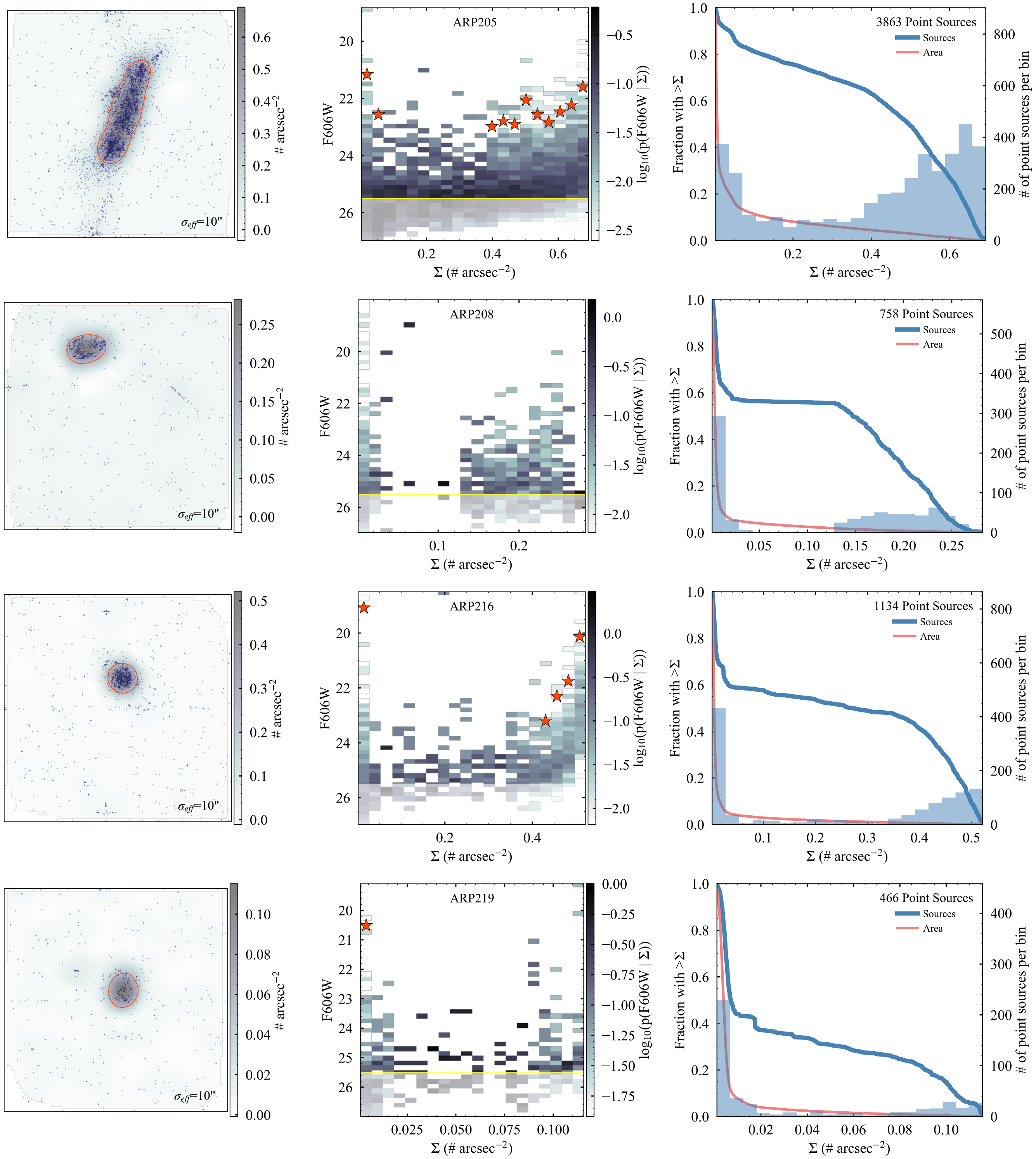

**Figure D3.** [Continued] ARP205, ARP208, ARP216, and ARP219 (top to bottom). [Left] The spatial distribution of point sources (points) and the smoothed background density field (greyscale); [Middle] the conditional luminosity function (greyscale) in bins of local surface density, with red stars marking the magnitude of the top 2% of the brightest sources in any density bin with more than 50 stars brighter than F606W=25.5; [Right] the histogram (shaded blue; right axis) and cumulative distributions (solid lines) of densities calculated at the location of point sources (blue thick line) or pixels (red thin line). Please see Section 2.5 and Figure 12 for a fuller description. [Continued]

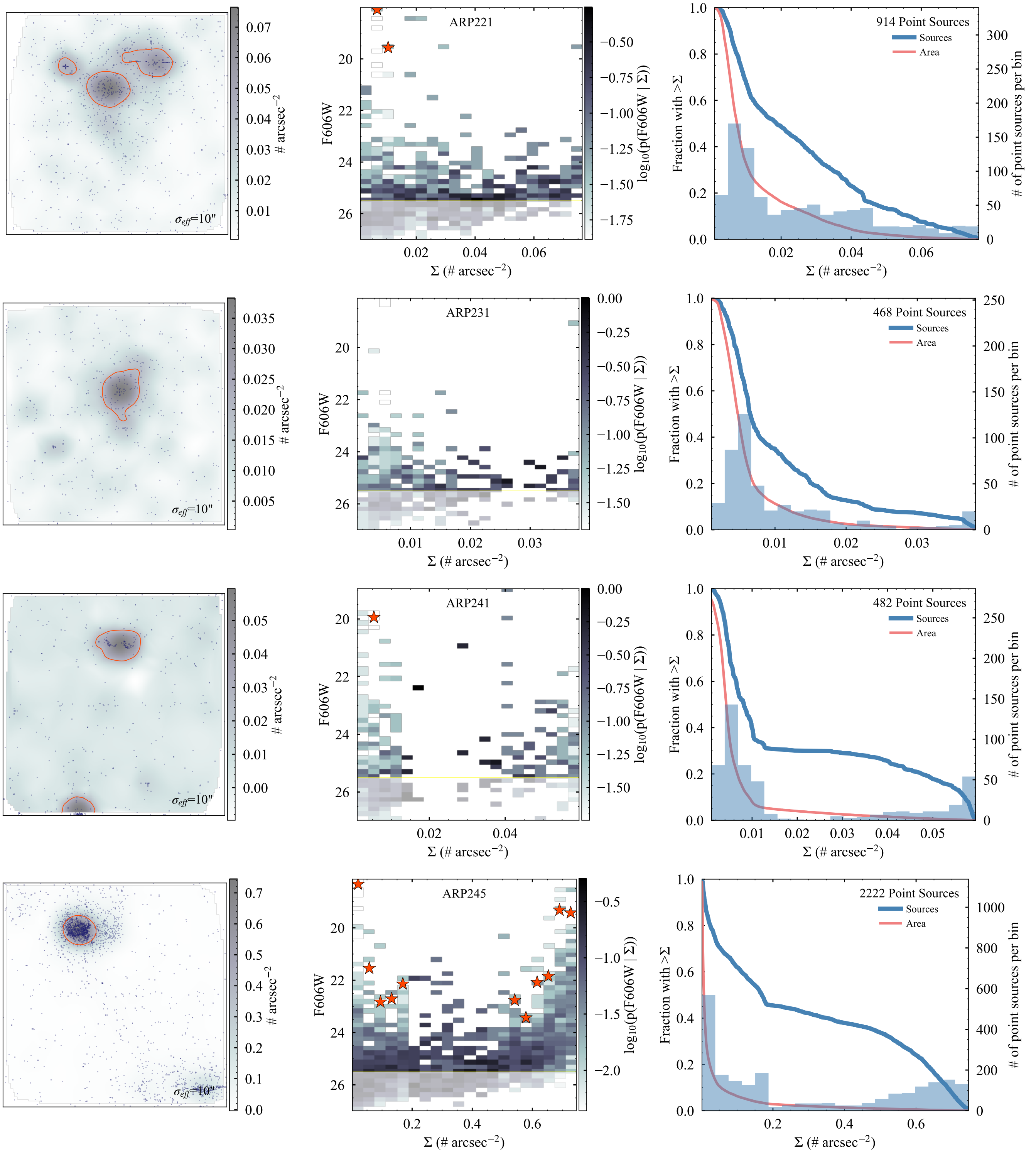

**Figure D3.** [Continued] ARP221, ARP231, ARP241, and ARP245 (top to bottom). [Left] The spatial distribution of point sources (points) and the smoothed background density field (greyscale); [Middle] the conditional luminosity function (greyscale) in bins of local surface density, with red stars marking the magnitude of the top 2% of the brightest sources in any density bin with more than 50 stars brighter than F606W=25.5; [Right] the histogram (shaded blue; right axis) and cumulative distributions (solid lines) of densities calculated at the location of point sources (blue thick line) or pixels (red thin line). Please see Section 2.5 and Figure 12 for a fuller description. [Continued]

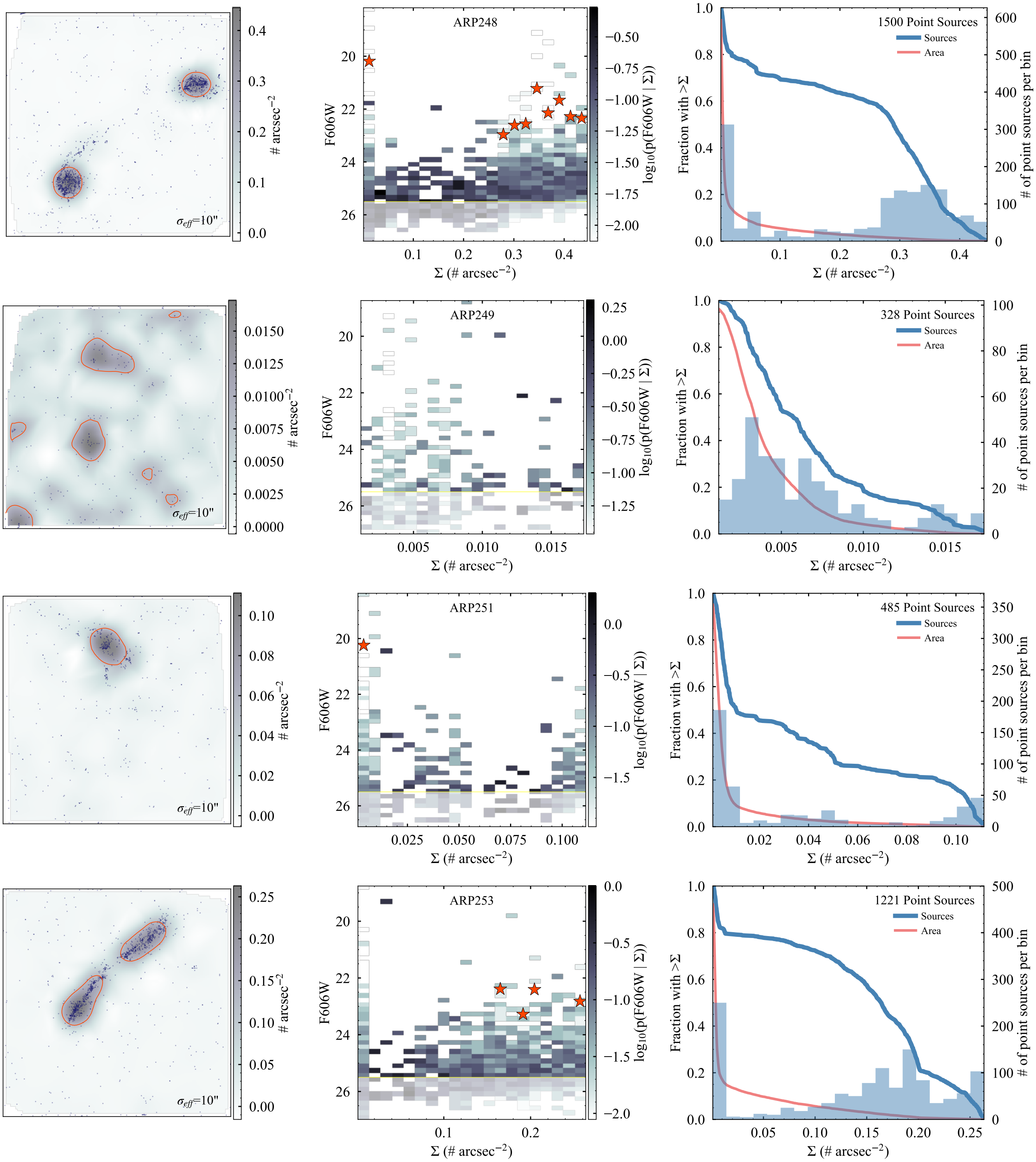

**Figure D3.** [Continued] ARP248, ARP249, ARP251, and ARP253 (top to bottom). [Left] The spatial distribution of point sources (points) and the smoothed background density field (greyscale); [Middle] the conditional luminosity function (greyscale) in bins of local surface density, with red stars marking the magnitude of the top 2% of the brightest sources in any density bin with more than 50 stars brighter than F606W=25.5; [Right] the histogram (shaded blue; right axis) and cumulative distributions (solid lines) of densities calculated at the location of point sources (blue thick line) or pixels (red thin line). Please see Section 2.5 and Figure 12 for a fuller description. [Continued]

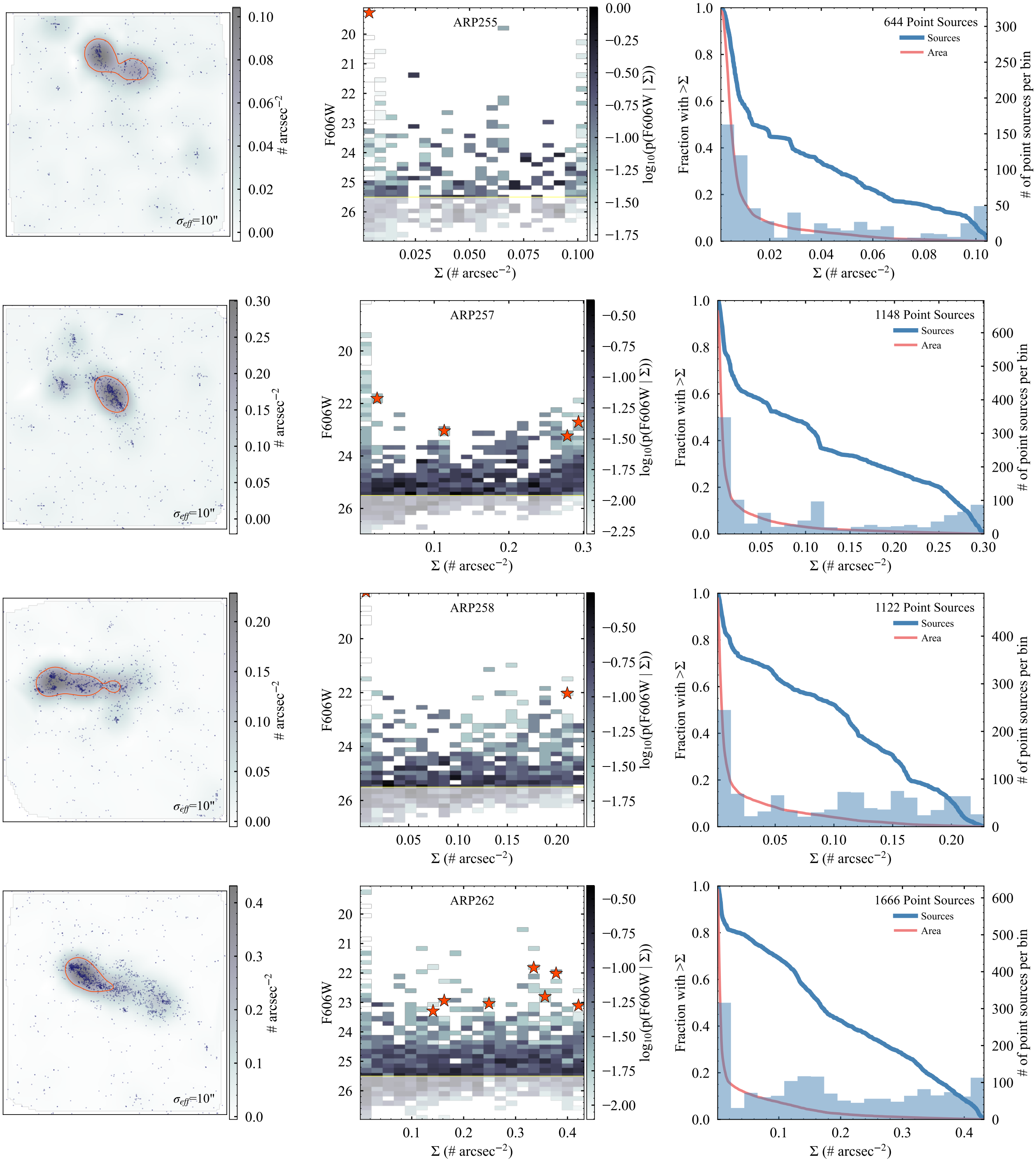

**Figure D3.** [Continued] ARP255, ARP257, ARP258, and ARP262 (top to bottom). [Left] The spatial distribution of point sources (points) and the smoothed background density field (greyscale); [Middle] the conditional luminosity function (greyscale) in bins of local surface density, with red stars marking the magnitude of the top 2% of the brightest sources in any density bin with more than 50 stars brighter than F606W=25.5; [Right] the histogram (shaded blue; right axis) and cumulative distributions (solid lines) of densities calculated at the location of point sources (blue thick line) or pixels (red thin line). Please see Section 2.5 and Figure 12 for a fuller description. [Continued]

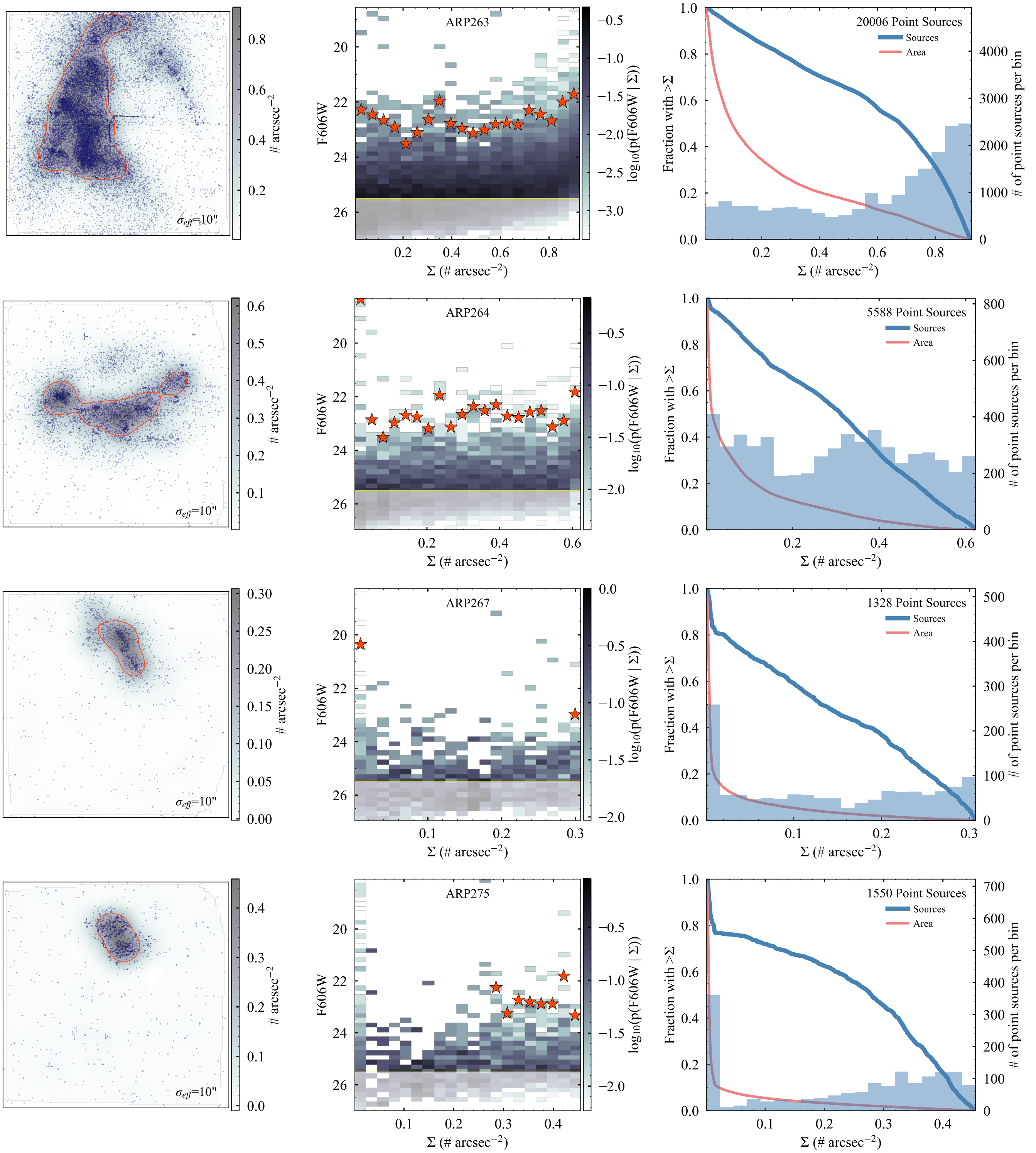

**Figure D3.** [Continued] ARP263, ARP264, ARP267, and ARP275 (top to bottom). [Left] The spatial distribution of point sources (points) and the smoothed background density field (greyscale); [Middle] the conditional luminosity function (greyscale) in bins of local surface density, with red stars marking the magnitude of the top 2% of the brightest sources in any density bin with more than 50 stars brighter than F606W=25.5; [Right] the histogram (shaded blue; right axis) and cumulative distributions (solid lines) of densities calculated at the location of point sources (blue thick line) or pixels (red thin line). Please see Section 2.5 and Figure 12 for a fuller description. [Continued]

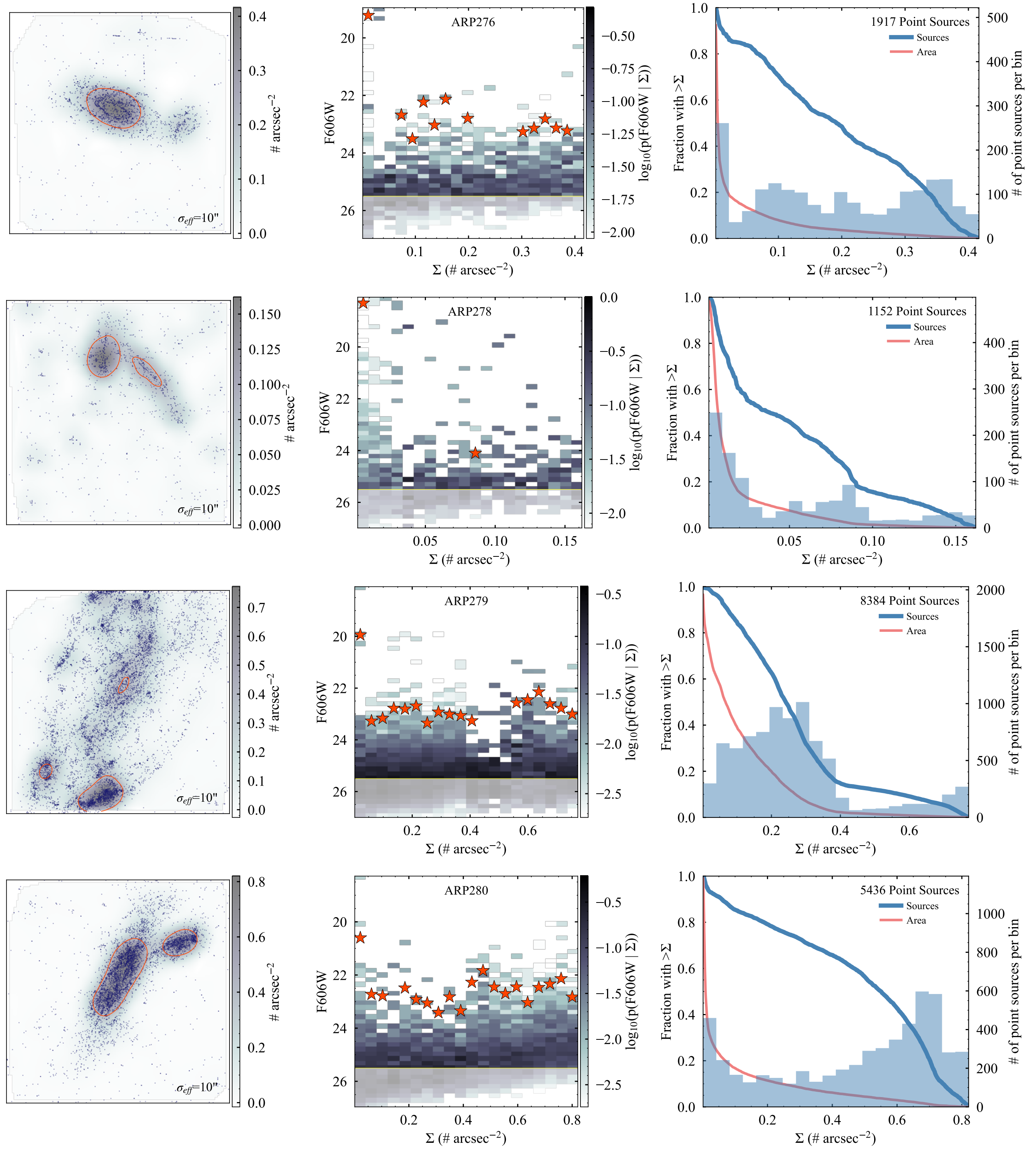

**Figure D3.** [Continued] ARP276, ARP278, ARP279, and ARP280 (top to bottom). [Left] The spatial distribution of point sources (points) and the smoothed background density field (greyscale); [Middle] the conditional luminosity function (greyscale) in bins of local surface density, with red stars marking the magnitude of the top 2% of the brightest sources in any density bin with more than 50 stars brighter than F606W=25.5; [Right] the histogram (shaded blue; right axis) and cumulative distributions (solid lines) of densities calculated at the location of point sources (blue thick line) or pixels (red thin line). Please see Section 2.5 and Figure 12 for a fuller description. [Continued]

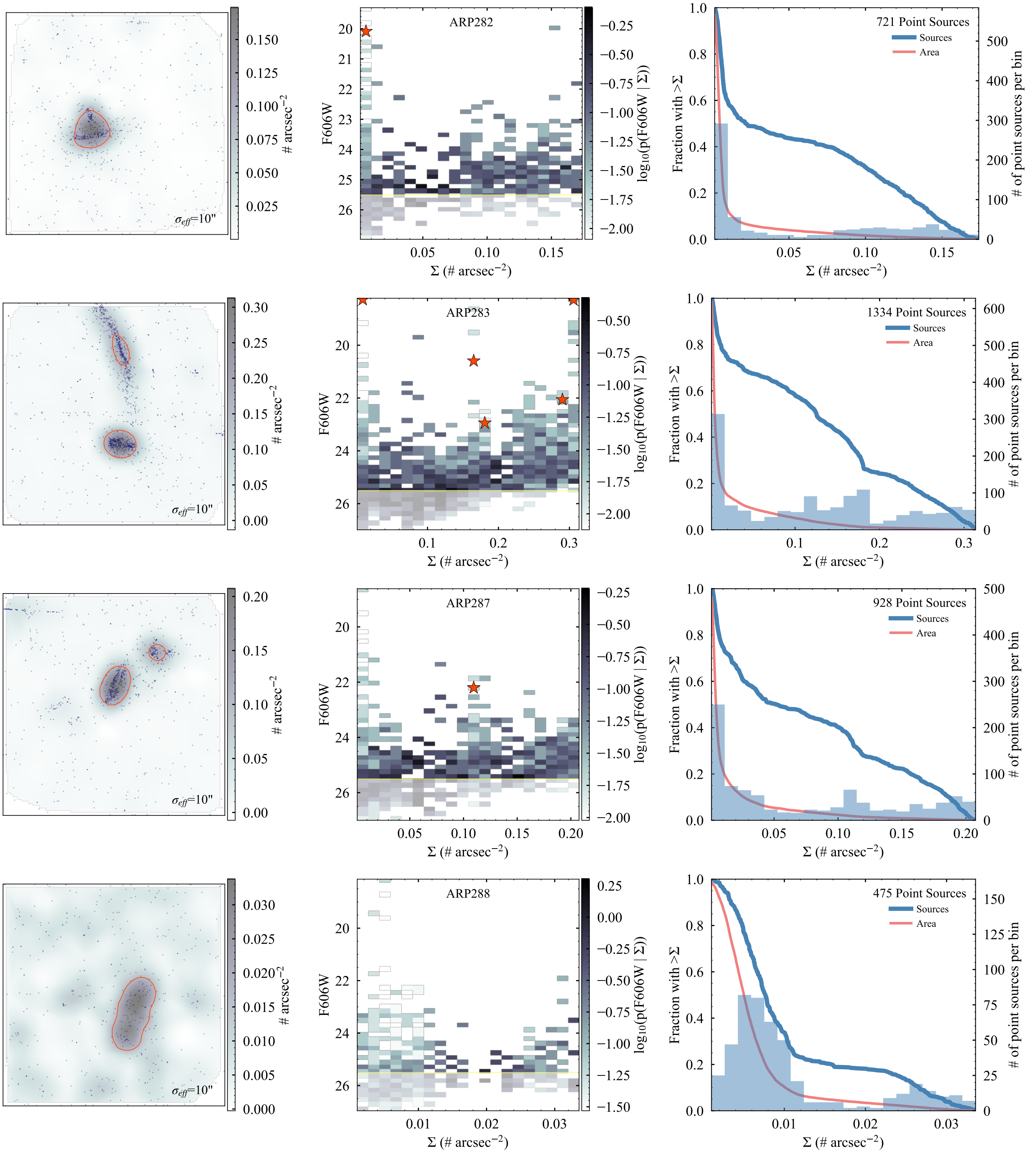

**Figure D3.** [Continued] ARP282, ARP283, ARP287, and ARP288 (top to bottom). [Left] The spatial distribution of point sources (points) and the smoothed background density field (greyscale); [Middle] the conditional luminosity function (greyscale) in bins of local surface density, with red stars marking the magnitude of the top 2% of the brightest sources in any density bin with more than 50 stars brighter than F606W=25.5; [Right] the histogram (shaded blue; right axis) and cumulative distributions (solid lines) of densities calculated at the location of point sources (blue thick line) or pixels (red thin line). Please see Section 2.5 and Figure 12 for a fuller description. [Continued]

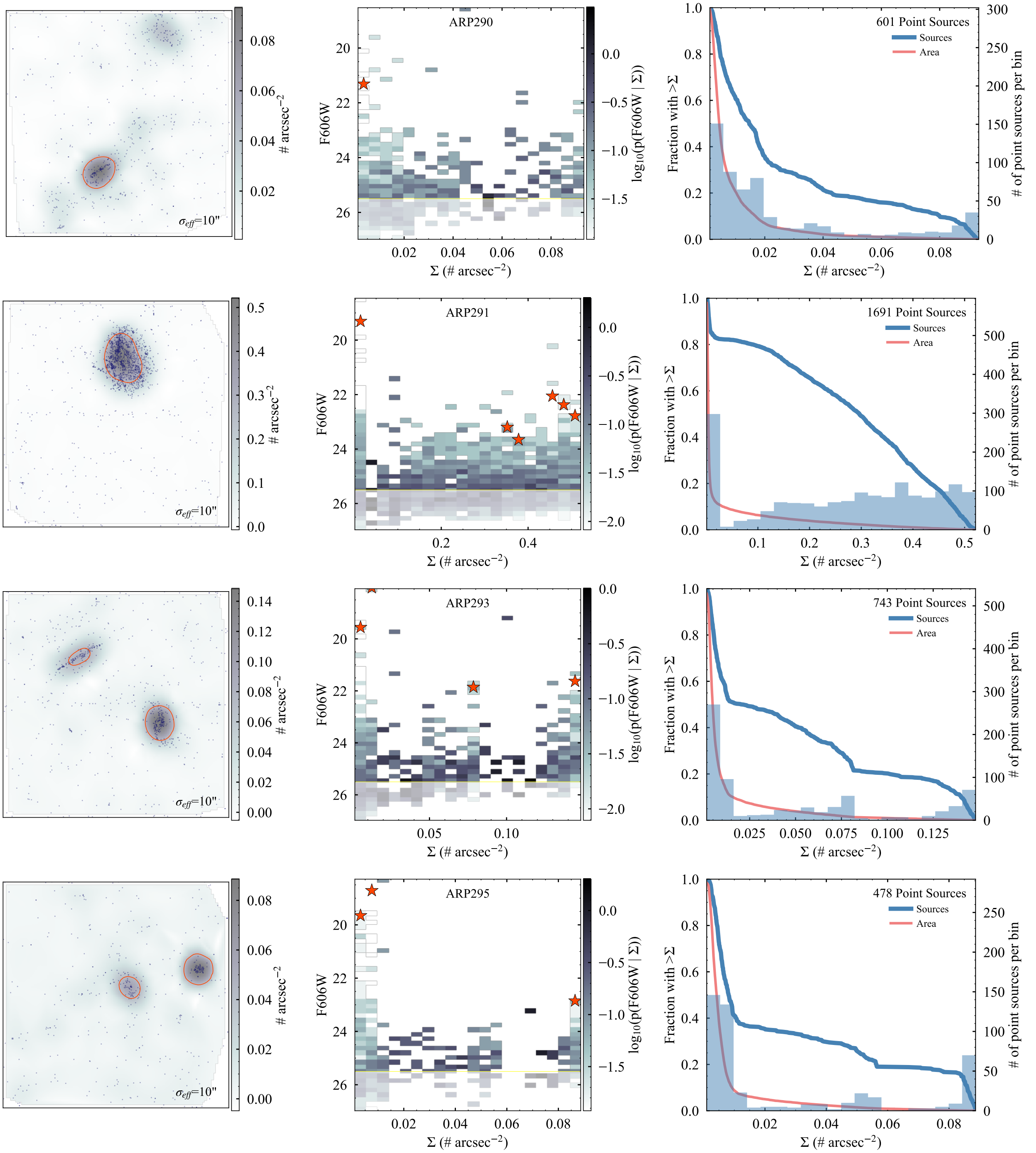

**Figure D3.** [Continued] ARP290, ARP291, ARP293, and ARP295 (top to bottom). [Left] The spatial distribution of point sources (points) and the smoothed background density field (greyscale); [Middle] the conditional luminosity function (greyscale) in bins of local surface density, with red stars marking the magnitude of the top 2% of the brightest sources in any density bin with more than 50 stars brighter than F606W=25.5; [Right] the histogram (shaded blue; right axis) and cumulative distributions (solid lines) of densities calculated at the location of point sources (blue thick line) or pixels (red thin line). Please see Section 2.5 and Figure 12 for a fuller description. [Continued]

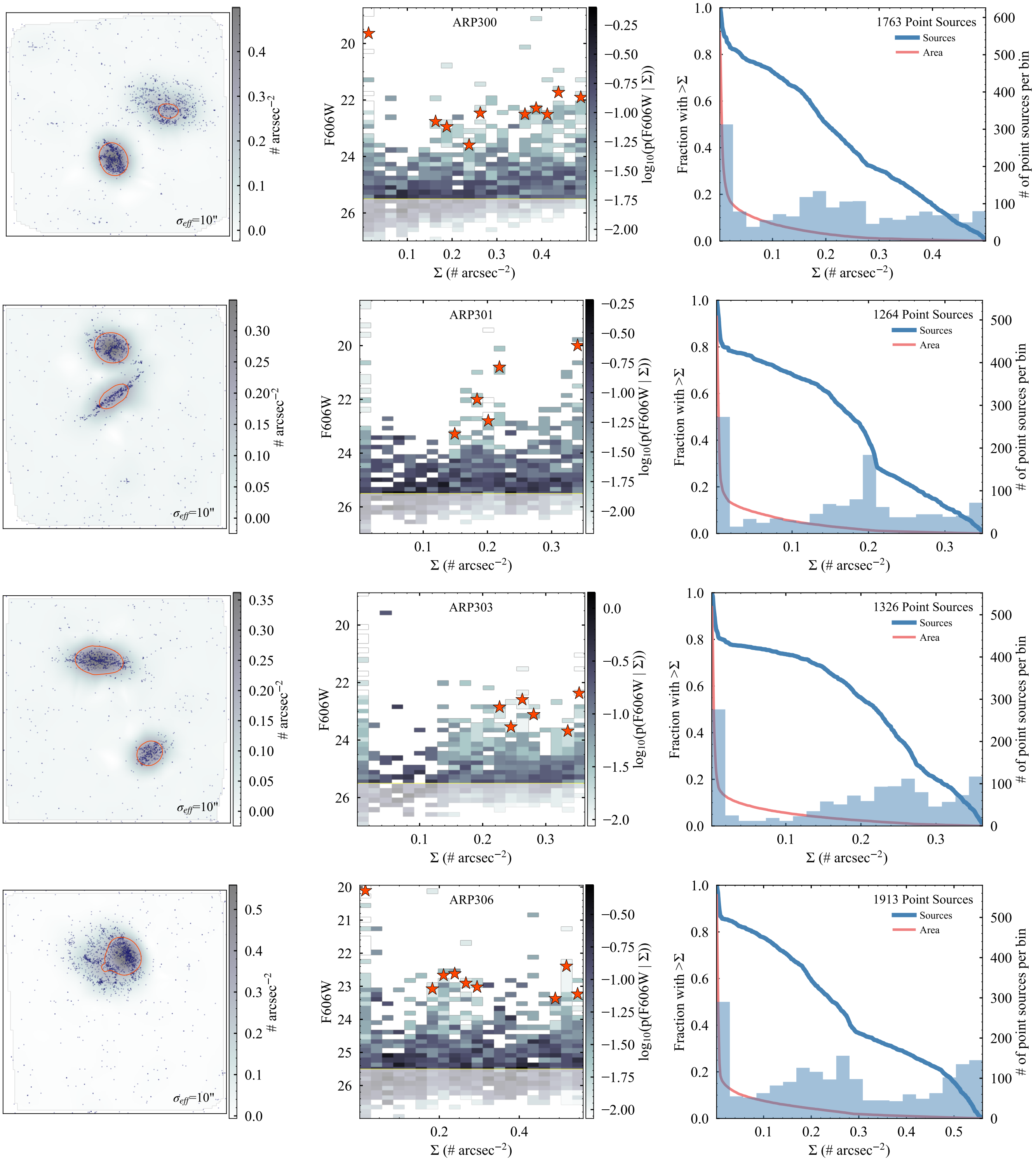

**Figure D3.** [Continued] ARP300, ARP301, ARP303, and ARP306 (top to bottom). [Left] The spatial distribution of point sources (points) and the smoothed background density field (greyscale); [Middle] the conditional luminosity function (greyscale) in bins of local surface density, with red stars marking the magnitude of the top 2% of the brightest sources in any density bin with more than 50 stars brighter than F606W=25.5; [Right] the histogram (shaded blue; right axis) and cumulative distributions (solid lines) of densities calculated at the location of point sources (blue thick line) or pixels (red thin line). Please see Section 2.5 and Figure 12 for a fuller description. [Continued]

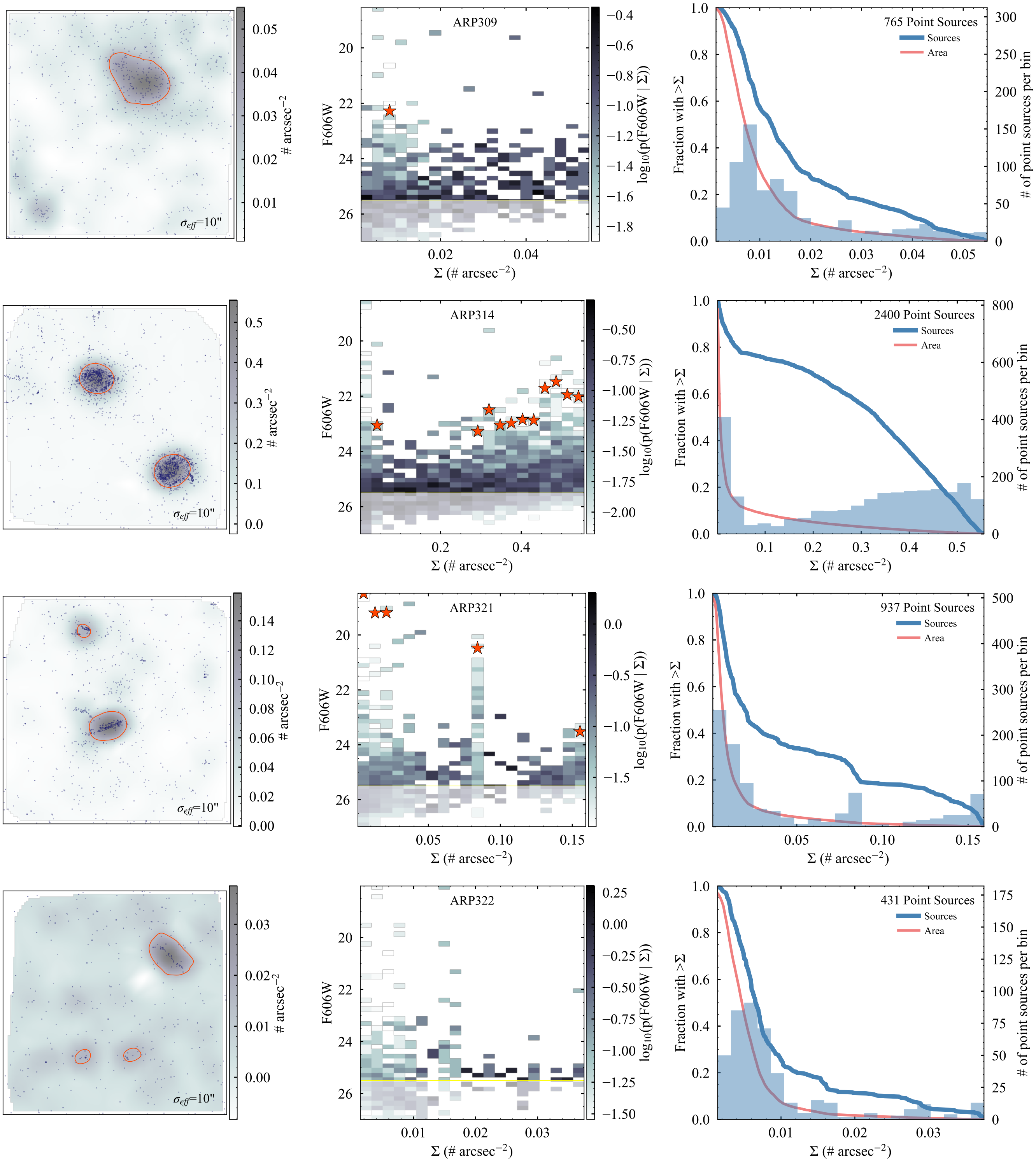

**Figure D3.** [Continued] ARP309, ARP314, ARP321, and ARP322 (top to bottom). [Left] The spatial distribution of point sources (points) and the smoothed background density field (greyscale); [Middle] the conditional luminosity function (greyscale) in bins of local surface density, with red stars marking the magnitude of the top 2% of the brightest sources in any density bin with more than 50 stars brighter than F606W=25.5; [Right] the histogram (shaded blue; right axis) and cumulative distributions (solid lines) of densities calculated at the location of point sources (blue thick line) or pixels (red thin line). Please see Section 2.5 and Figure 12 for a fuller description. [Continued]

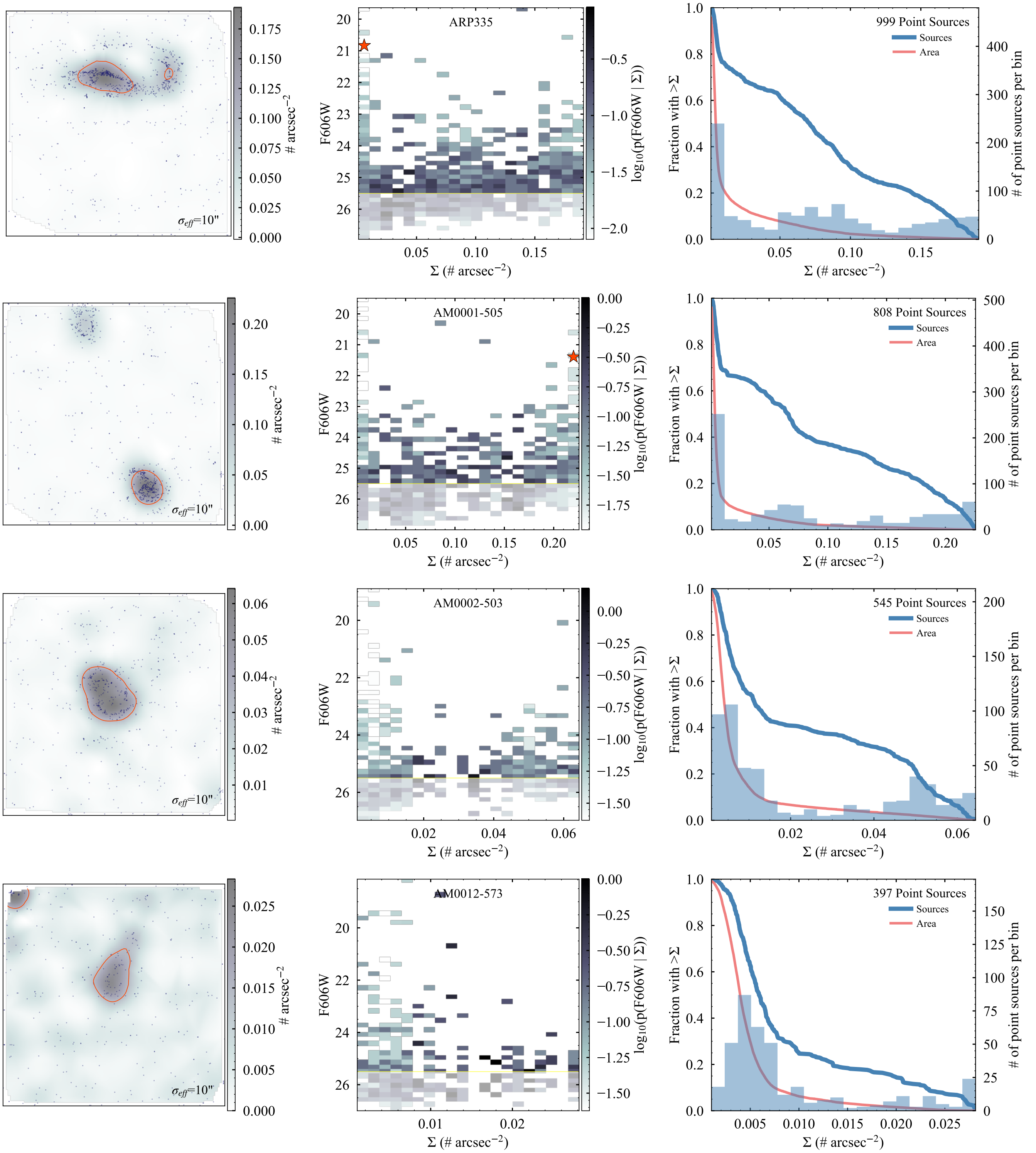

**Figure D3.** [Continued] ARP335, ARP-MADORE0001-505, ARP-MADORE0002-503, and ARP-MADORE0012-573 (top to bottom). [Left] The spatial distribution of point sources (points) and the smoothed background density field (greyscale); [Middle] the conditional luminosity function (greyscale) in bins of local surface density, with red stars marking the magnitude of the top 2% of the brightest sources in any density bin with more than 50 stars brighter than F606W=25.5; [Right] the histogram (shaded blue; right axis) and cumulative distributions (solid lines) of densities calculated at the location of point sources (blue thick line) or pixels (red thin line). Please see Section 2.5 and Figure 12 for a fuller description. [Continued]

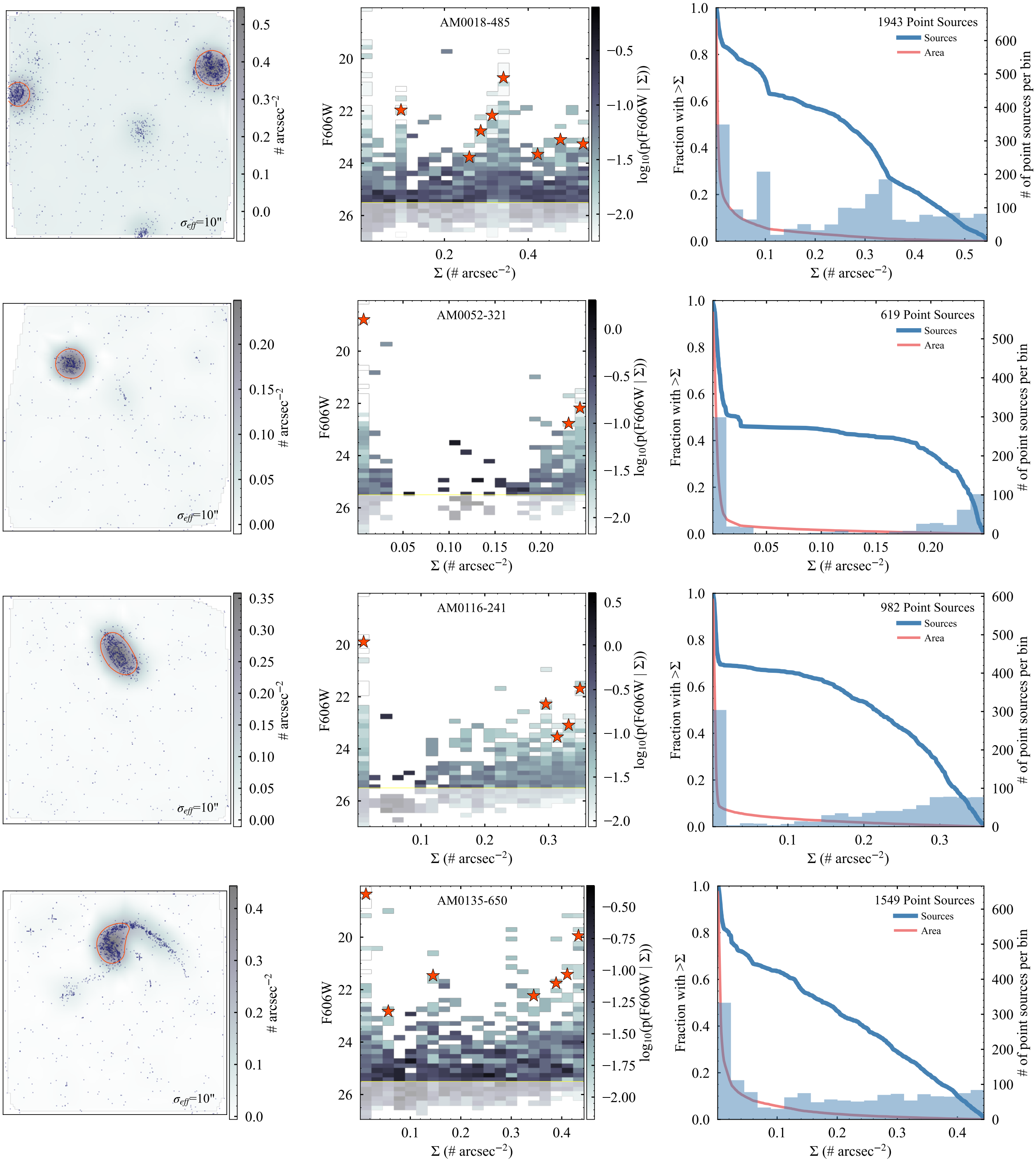

**Figure D3.** [Continued] ARP-MADORE0018-485, ARP-MADORE0052-321, ARP-MADORE0116-241, and ARP–MADORE0135-650 (top to bottom). [Left] The spatial distribution of point sources (points) and the smoothed background density field (greyscale); [Middle] the conditional luminosity function (greyscale) in bins of local surface density, with red stars marking the magnitude of the top 2% of the brightest sources in any density bin with more than 50 stars brighter than F606W=25.5; [Right] the histogram (shaded blue; right axis) and cumulative distributions (solid lines) of densities calculated at the location of point sources (blue thick line) or pixels (red thin line). Please see Section 2.5 and Figure 12 for a fuller description. [Continued]

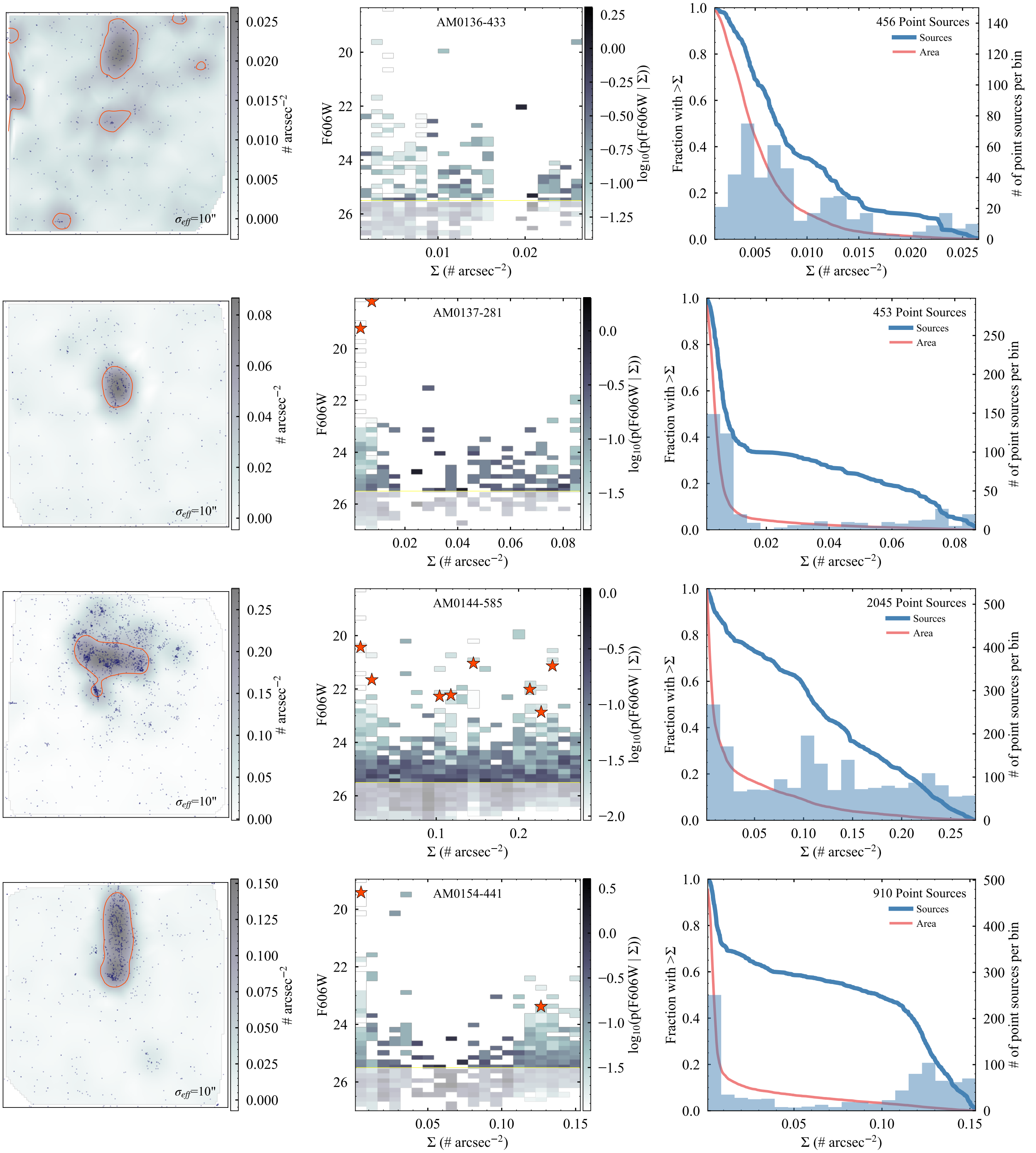

**Figure D3.** [Continued] ARP-MADORE0136-433, ARP-MADORE0137-281, ARP-MADORE0144-585, and ARP–MADORE0154-441 (top to bottom). [Left] The spatial distribution of point sources (points) and the smoothed background density field (greyscale); [Middle] the conditional luminosity function (greyscale) in bins of local surface density, with red stars marking the magnitude of the top 2% of the brightest sources in any density bin with more than 50 stars brighter than F606W=25.5; [Right] the histogram (shaded blue; right axis) and cumulative distributions (solid lines) of densities calculated at the location of point sources (blue thick line) or pixels (red thin line). Please see Section 2.5 and Figure 12 for a fuller description. [Continued]

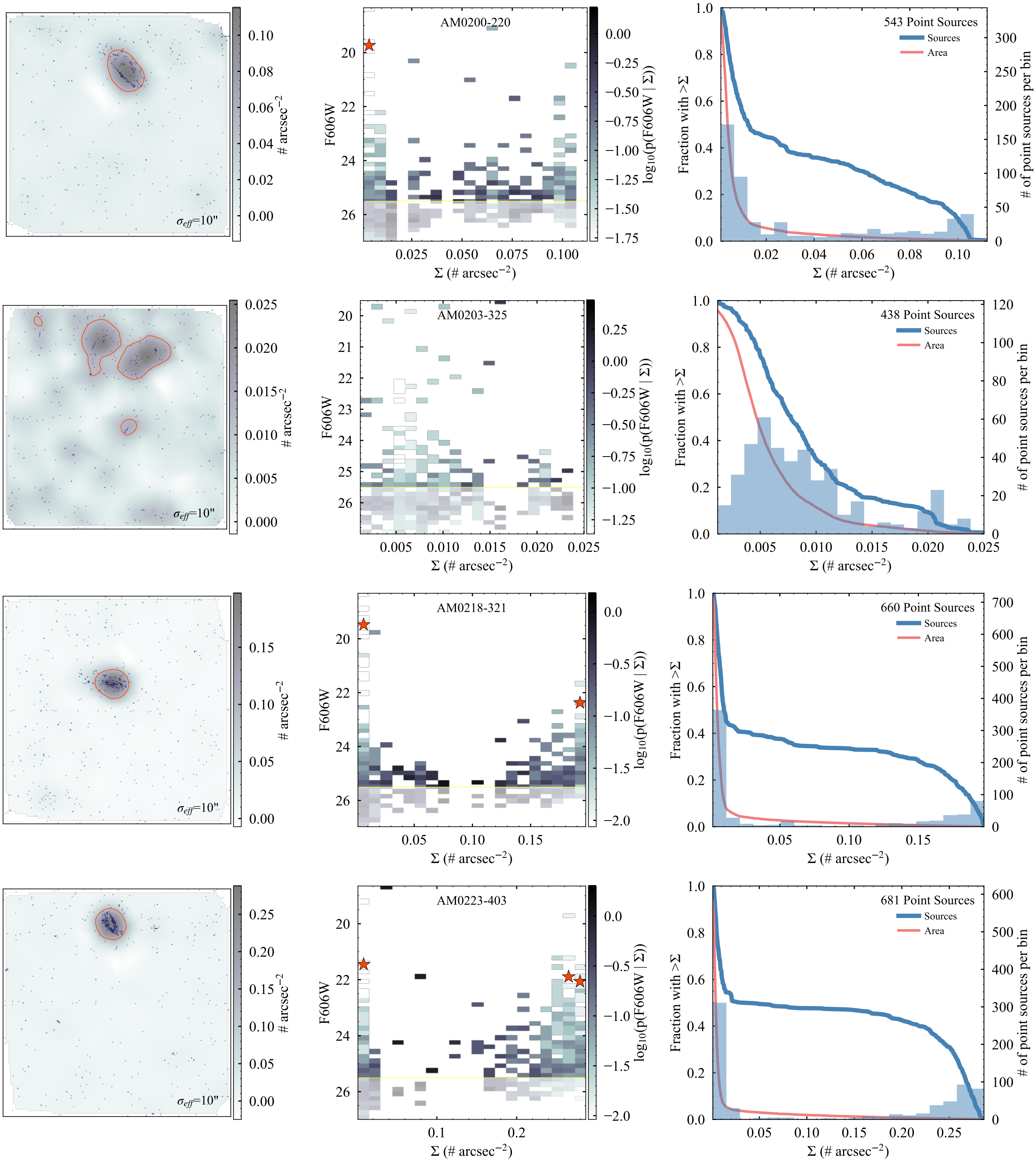

**Figure D3.** [Continued] ARP-MADORE0200-220, ARP-MADORE0203-325, ARP-MADORE0218-321, and ARP–MADORE0223-403 (top to bottom). [Left] The spatial distribution of point sources (points) and the smoothed background density field (greyscale); [Middle] the conditional luminosity function (greyscale) in bins of local surface density, with red stars marking the magnitude of the top 2% of the brightest sources in any density bin with more than 50 stars brighter than F606W=25.5; [Right] the histogram (shaded blue; right axis) and cumulative distributions (solid lines) of densities calculated at the location of point sources (blue thick line) or pixels (red thin line). Please see Section 2.5 and Figure 12 for a fuller description. [Continued]

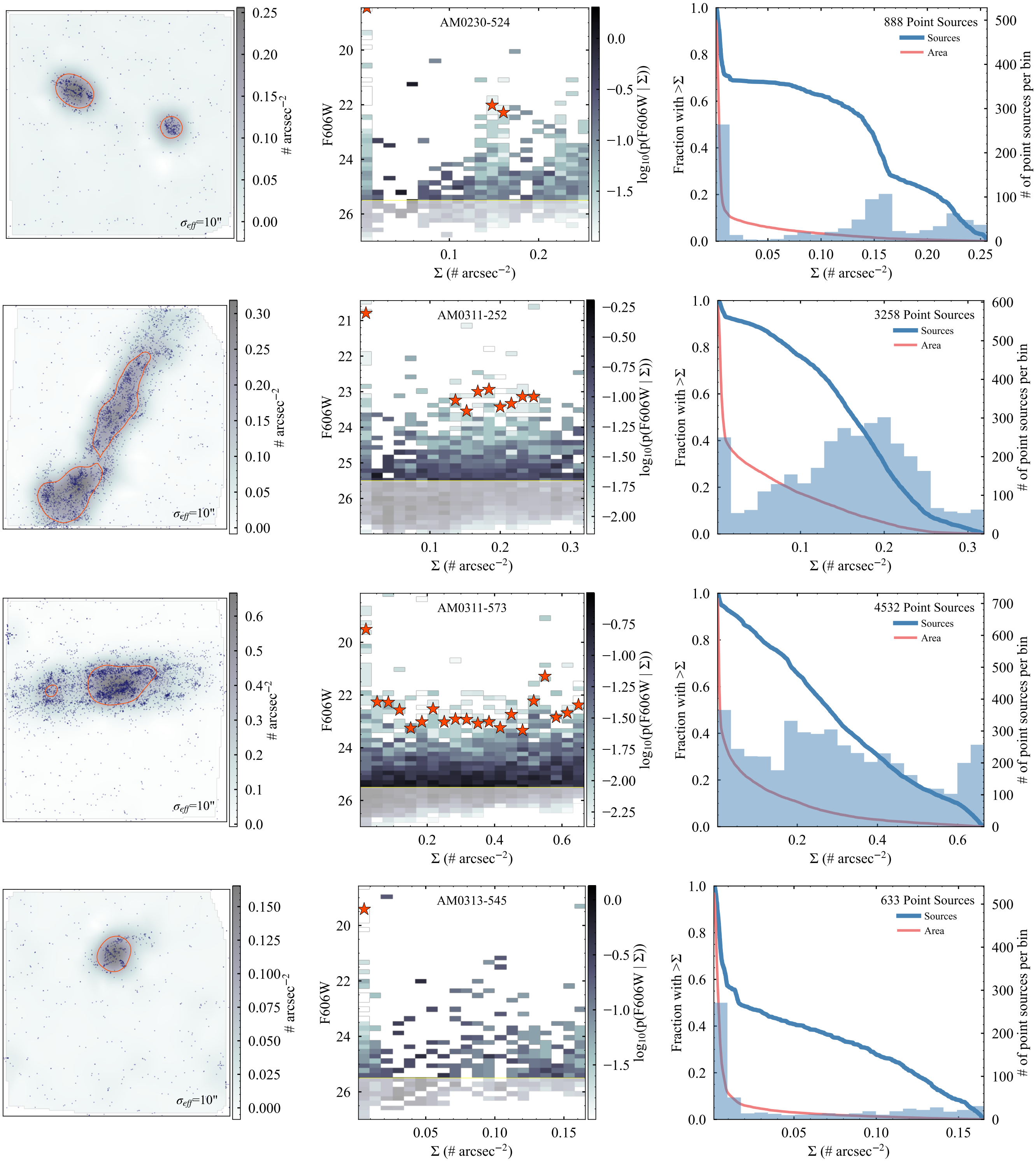

**Figure D3.** [Continued] ARP-MADORE0230-524, ARP-MADORE0311-252, ARP-MADORE0311-573, and ARP–MADORE0313-545 (top to bottom). [Left] The spatial distribution of point sources (points) and the smoothed background density field (greyscale); [Middle] the conditional luminosity function (greyscale) in bins of local surface density, with red stars marking the magnitude of the top 2% of the brightest sources in any density bin with more than 50 stars brighter than F606W=25.5; [Right] the histogram (shaded blue; right axis) and cumulative distributions (solid lines) of densities calculated at the location of point sources (blue thick line) or pixels (red thin line). Please see Section 2.5 and Figure 12 for a fuller description. [Continued]

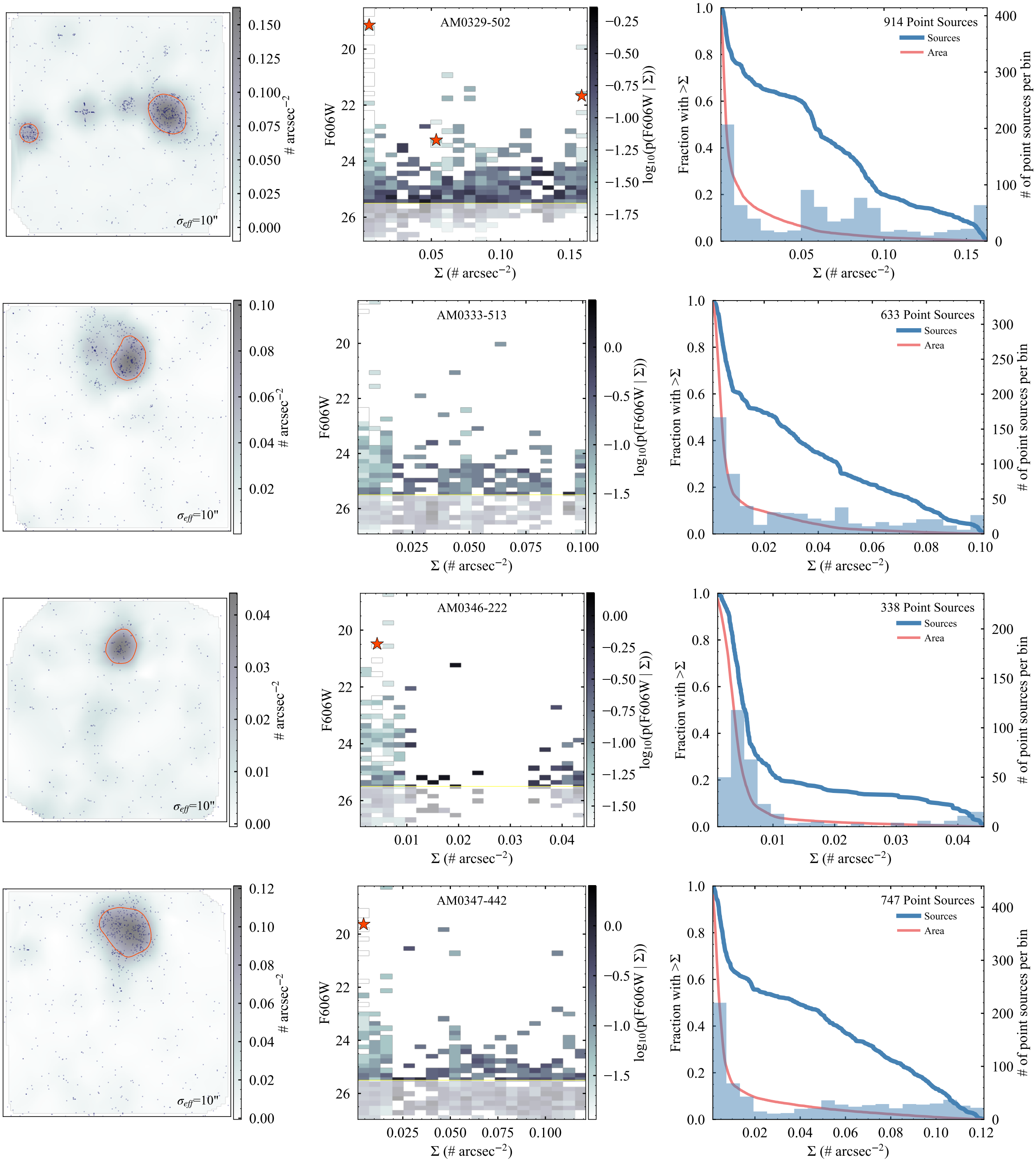

**Figure D3.** [Continued] ARP-MADORE0329-502, ARP-MADORE0333-513, ARP-MADORE0346-222, and ARP–MADORE0347-442 (top to bottom). [Left] The spatial distribution of point sources (points) and the smoothed background density field (greyscale); [Middle] the conditional luminosity function (greyscale) in bins of local surface density, with red stars marking the magnitude of the top 2% of the brightest sources in any density bin with more than 50 stars brighter than F606W=25.5; [Right] the histogram (shaded blue; right axis) and cumulative distributions (solid lines) of densities calculated at the location of point sources (blue thick line) or pixels (red thin line). Please see Section 2.5 and Figure 12 for a fuller description. [Continued]

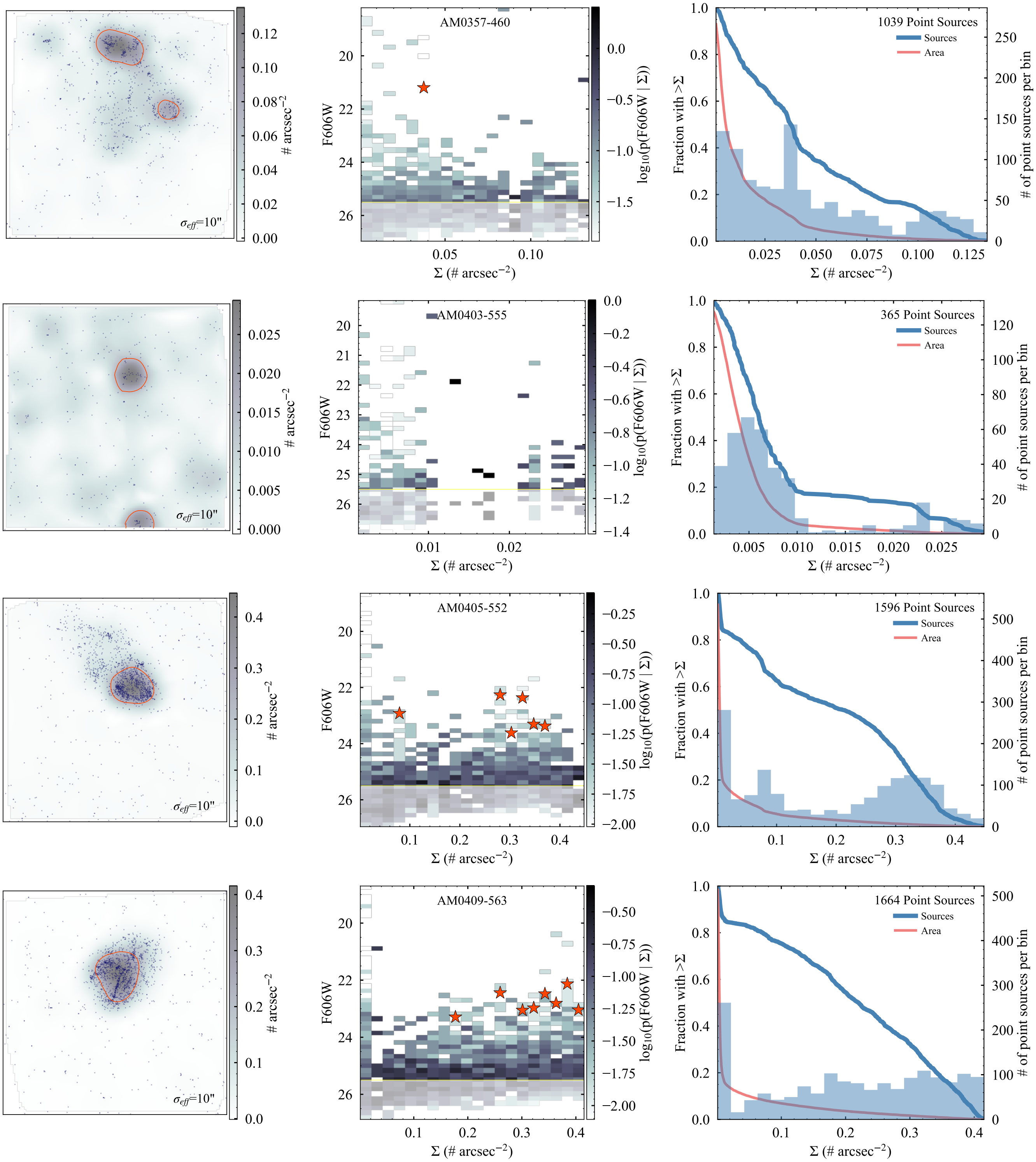

**Figure D3.** [Continued] ARP-MADORE0357-460, ARP-MADORE0403-555, ARP-MADORE0405-552, and ARP–MADORE0409-563 (top to bottom). [Left] The spatial distribution of point sources (points) and the smoothed background density field (greyscale); [Middle] the conditional luminosity function (greyscale) in bins of local surface density, with red stars marking the magnitude of the top 2% of the brightest sources in any density bin with more than 50 stars brighter than F606W=25.5; [Right] the histogram (shaded blue; right axis) and cumulative distributions (solid lines) of densities calculated at the location of point sources (blue thick line) or pixels (red thin line). Please see Section 2.5 and Figure 12 for a fuller description. [Continued]

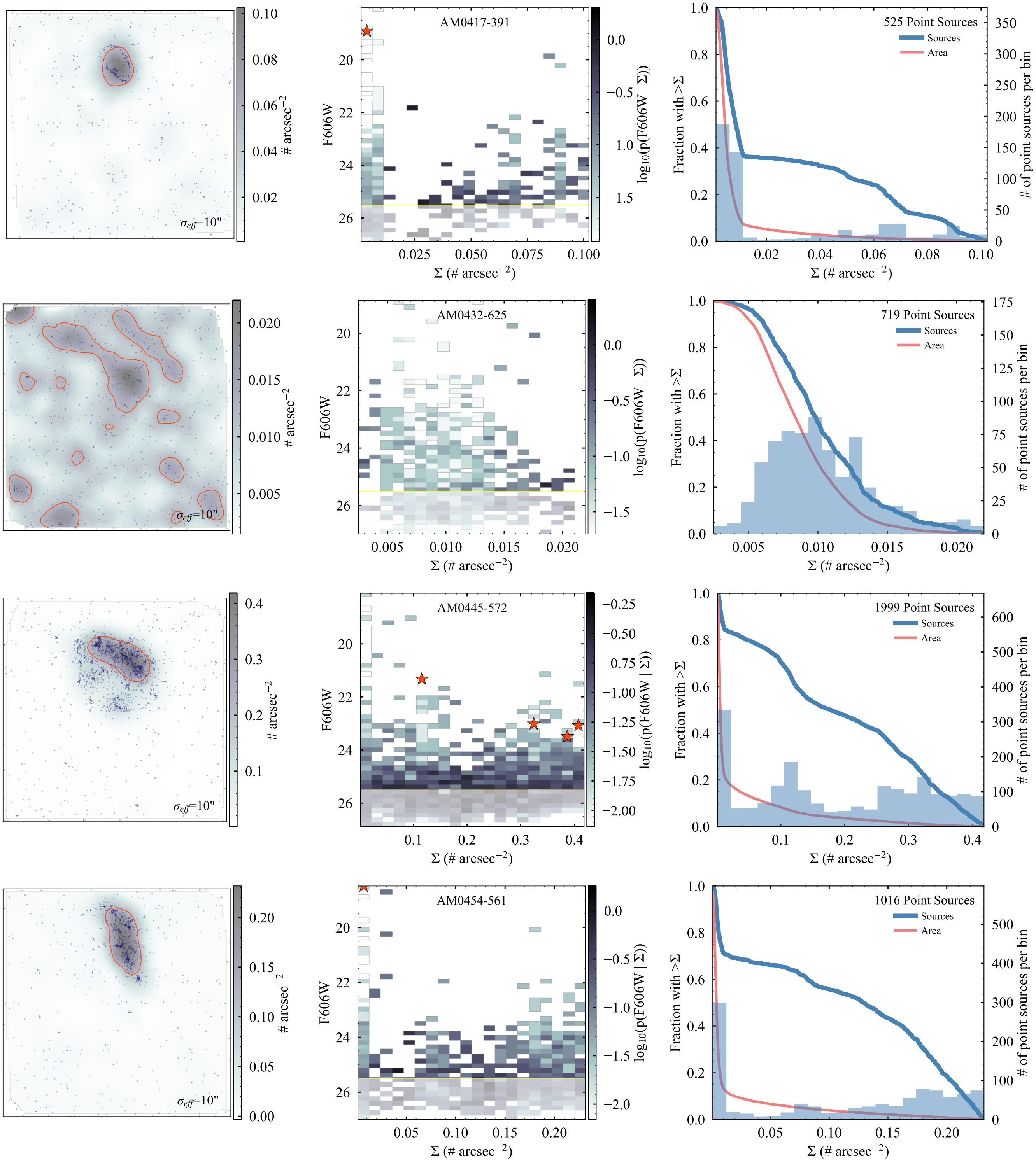

**Figure D3.** [Continued] ARP-MADORE0417-391, ARP-MADORE0432-625, ARP-MADORE0445-572, and ARP–MADORE0454-561 (top to bottom). [Left] The spatial distribution of point sources (points) and the smoothed background density field (greyscale); [Middle] the conditional luminosity function (greyscale) in bins of local surface density, with red stars marking the magnitude of the top 2% of the brightest sources in any density bin with more than 50 stars brighter than F606W=25.5; [Right] the histogram (shaded blue; right axis) and cumulative distributions (solid lines) of densities calculated at the location of point sources (blue thick line) or pixels (red thin line). Please see Section 2.5 and Figure 12 for a fuller description. [Continued]

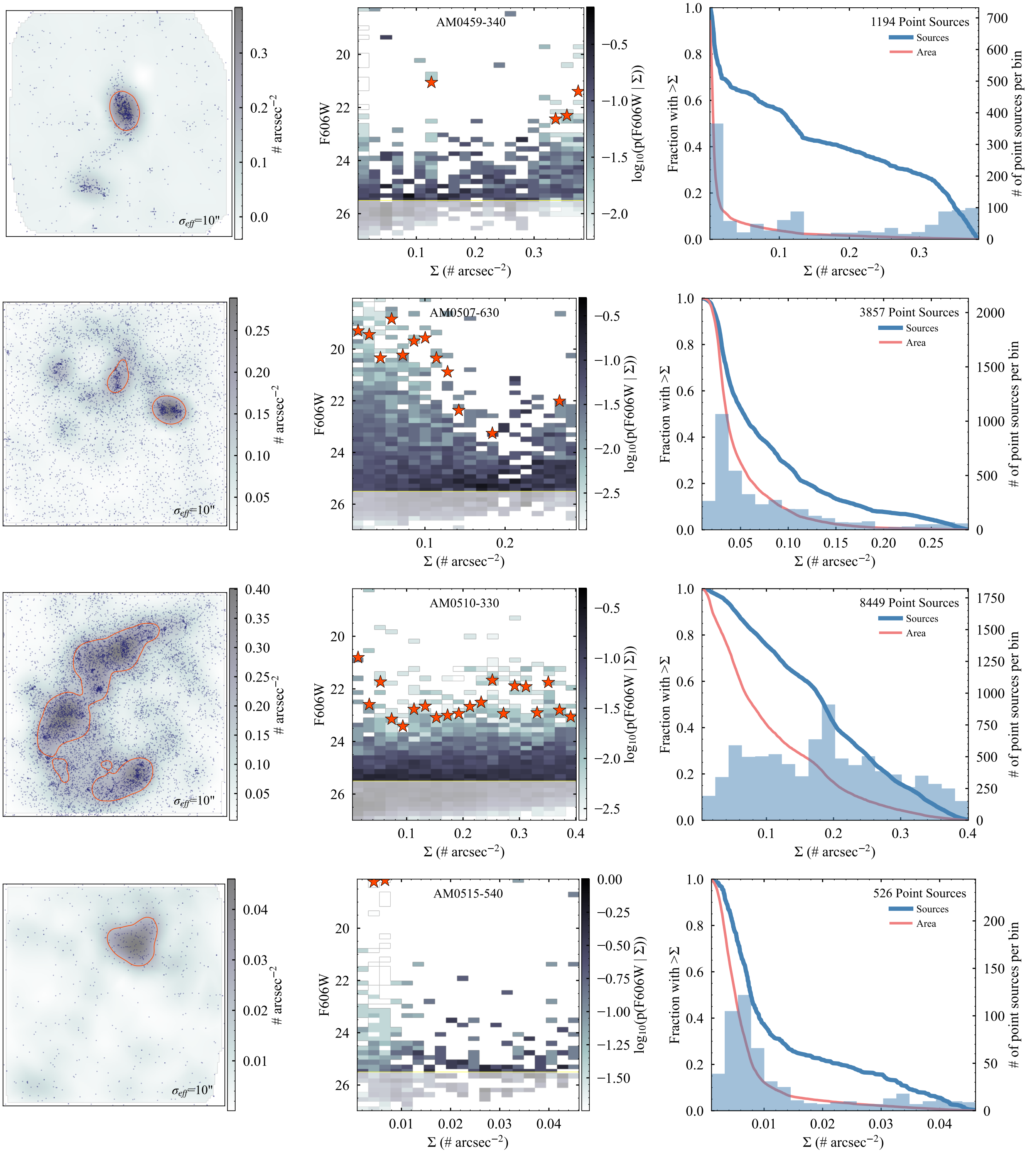

**Figure D3.** [Continued] ARP-MADORE0459-340, ARP-MADORE0507-630, ARP-MADORE0510-330, and ARP–MADORE0515-540 (top to bottom). [Left] The spatial distribution of point sources (points) and the smoothed background density field (greyscale); [Middle] the conditional luminosity function (greyscale) in bins of local surface density, with red stars marking the magnitude of the top 2% of the brightest sources in any density bin with more than 50 stars brighter than F606W=25.5; [Right] the histogram (shaded blue; right axis) and cumulative distributions (solid lines) of densities calculated at the location of point sources (blue thick line) or pixels (red thin line). Please see Section 2.5 and Figure 12 for a fuller description. [Continued]

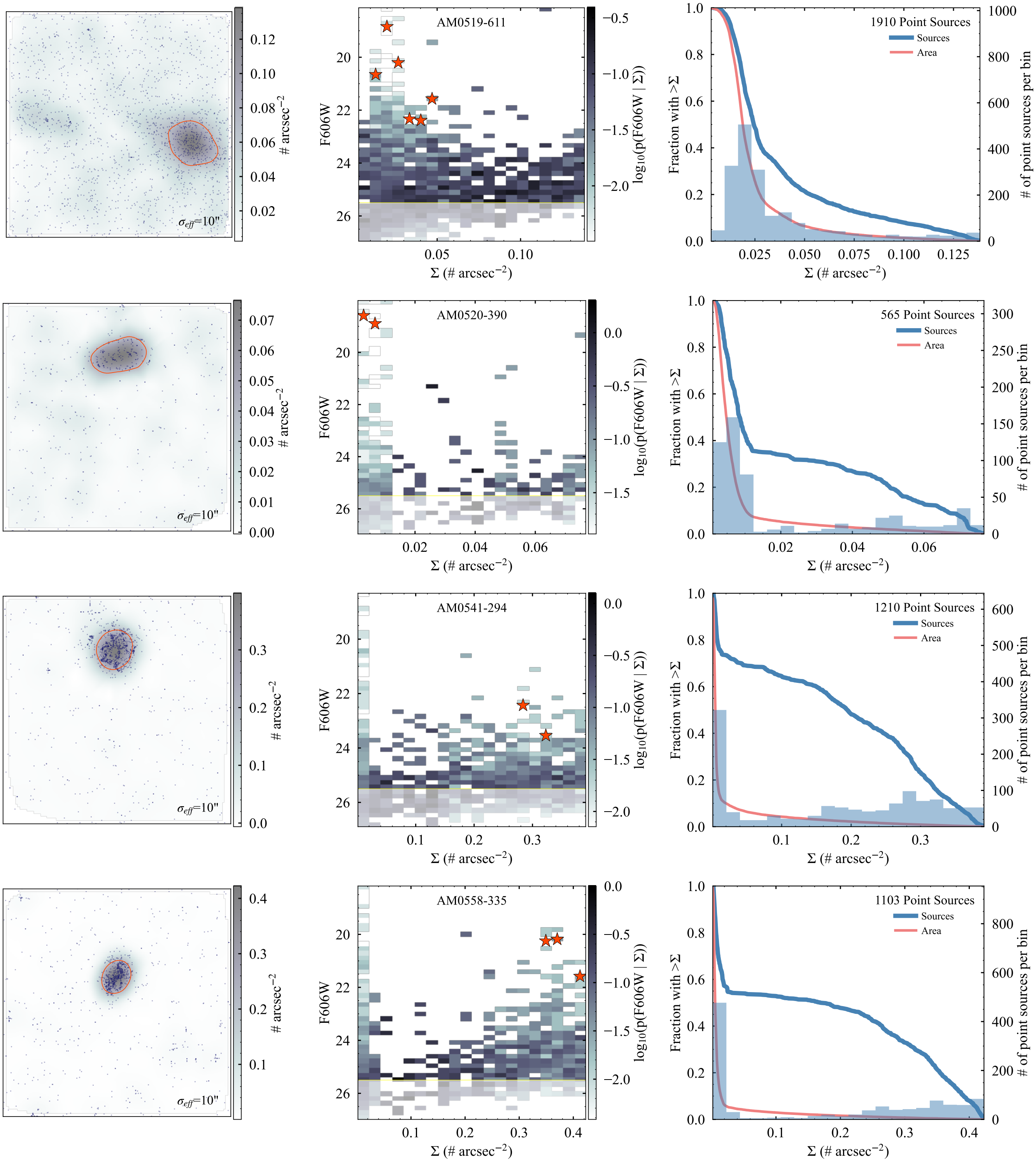

**Figure D3.** [Continued] ARP-MADORE0519-611, ARP-MADORE0520-390, ARP-MADORE0541-294, and ARP–MADORE0558-335 (top to bottom). [Left] The spatial distribution of point sources (points) and the smoothed background density field (greyscale); [Middle] the conditional luminosity function (greyscale) in bins of local surface density, with red stars marking the magnitude of the top 2% of the brightest sources in any density bin with more than 50 stars brighter than F606W=25.5; [Right] the histogram (shaded blue; right axis) and cumulative distributions (solid lines) of densities calculated at the location of point sources (blue thick line) or pixels (red thin line). Please see Section 2.5 and Figure 12 for a fuller description. [Continued]

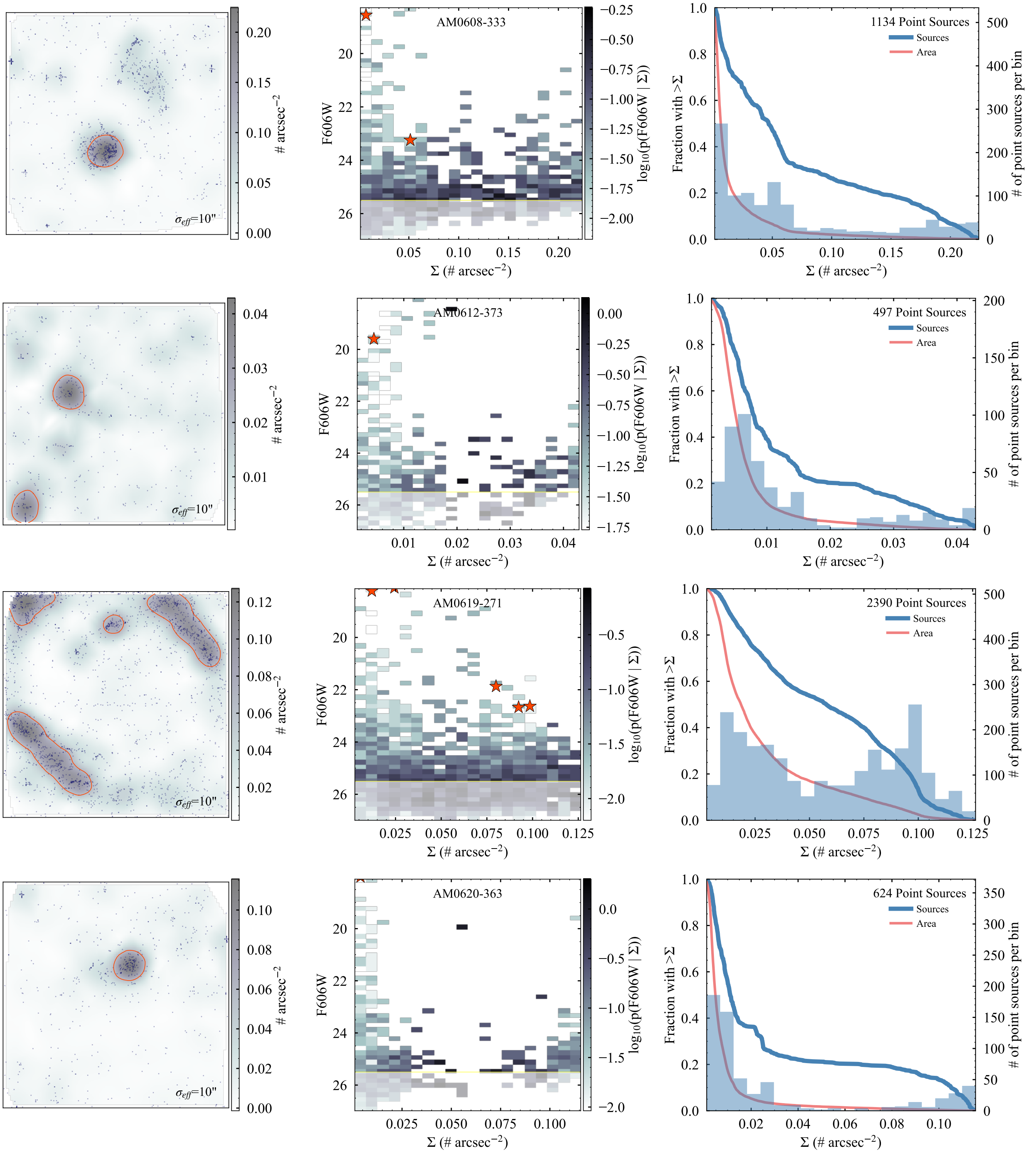

**Figure D3.** [Continued] ARP-MADORE0608-333, ARP-MADORE0612-373, ARP-MADORE0619-271, and ARP–MADORE0620-363 (top to bottom). [Left] The spatial distribution of point sources (points) and the smoothed background density field (greyscale); [Middle] the conditional luminosity function (greyscale) in bins of local surface density, with red stars marking the magnitude of the top 2% of the brightest sources in any density bin with more than 50 stars brighter than F606W=25.5; [Right] the histogram (shaded blue; right axis) and cumulative distributions (solid lines) of densities calculated at the location of point sources (blue thick line) or pixels (red thin line). Please see Section 2.5 and Figure 12 for a fuller description. [Continued]

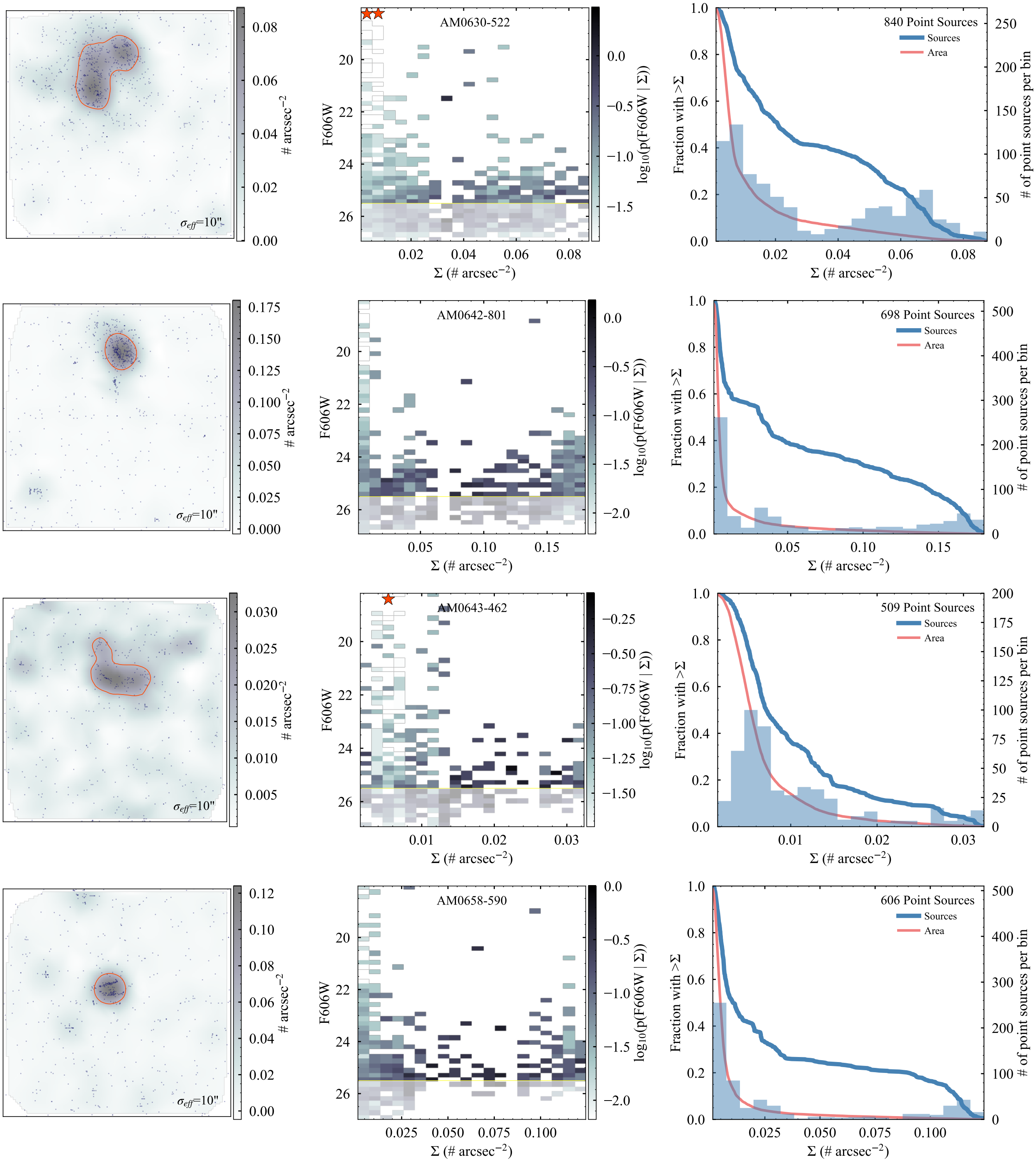

**Figure D3.** [Continued] ARP-MADORE0630-522, ARP-MADORE0642-801, ARP-MADORE0643-462, and ARP–MADORE0658-590 (top to bottom). [Left] The spatial distribution of point sources (points) and the smoothed background density field (greyscale); [Middle] the conditional luminosity function (greyscale) in bins of local surface density, with red stars marking the magnitude of the top 2% of the brightest sources in any density bin with more than 50 stars brighter than F606W=25.5; [Right] the histogram (shaded blue; right axis) and cumulative distributions (solid lines) of densities calculated at the location of point sources (blue thick line) or pixels (red thin line). Please see Section 2.5 and Figure 12 for a fuller description. [Continued]

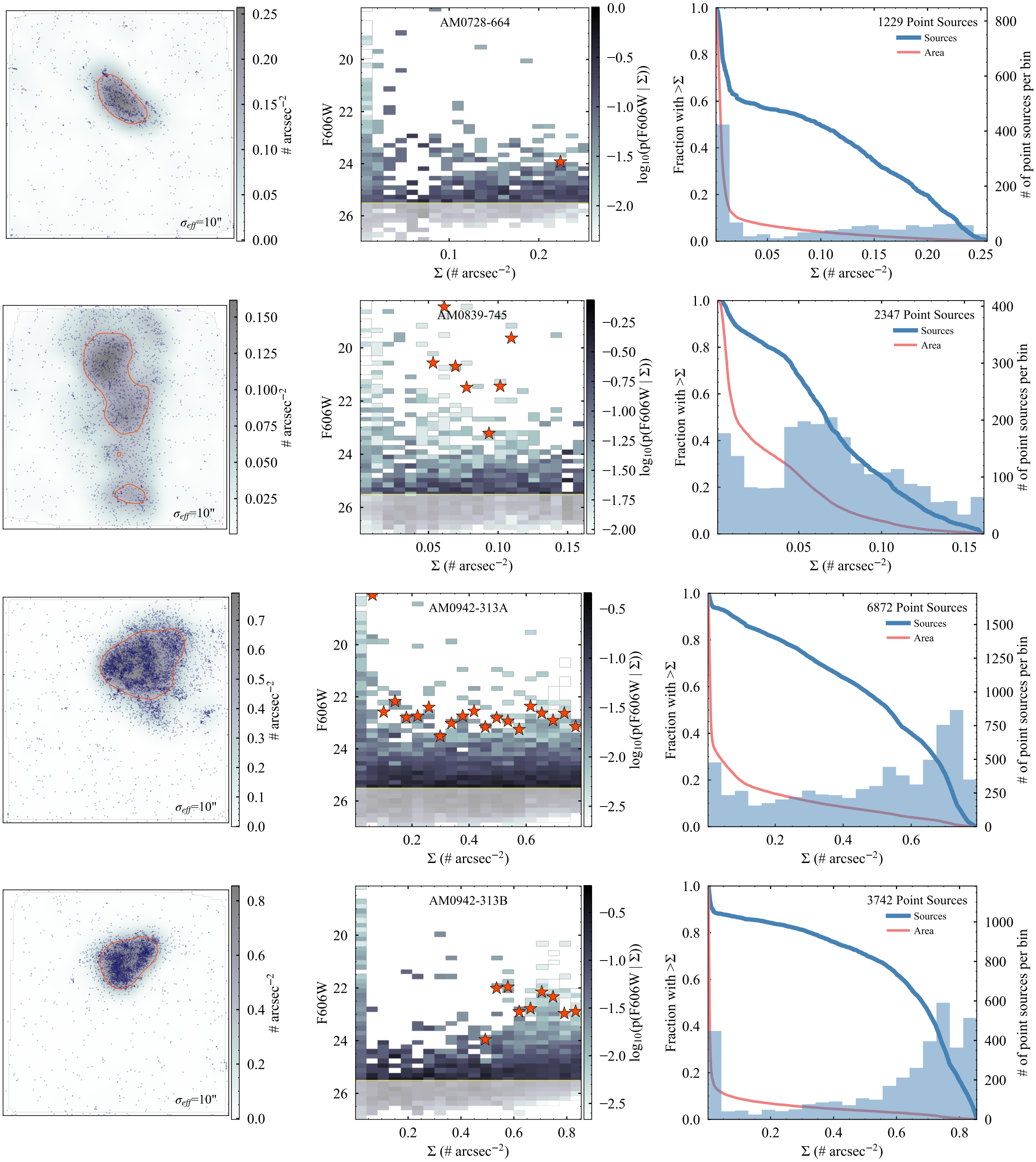

**Figure D3.** [Continued] ARP-MADORE0728-664, ARP-MADORE0839-745, ARP-MADORE0942-313A, and ARP–MADORE0942-313B (top to bottom). [Left] The spatial distribution of point sources (points) and the smoothed background density field (greyscale); [Middle] the conditional luminosity function (greyscale) in bins of local surface density, with red stars marking the magnitude of the top 2% of the brightest sources in any density bin with more than 50 stars brighter than F606W=25.5; [Right] the histogram (shaded blue; right axis) and cumulative distributions (solid lines) of densities calculated at the location of point sources (blue thick line) or pixels (red thin line). Please see Section 2.5 and Figure 12 for a fuller description. [Continued]

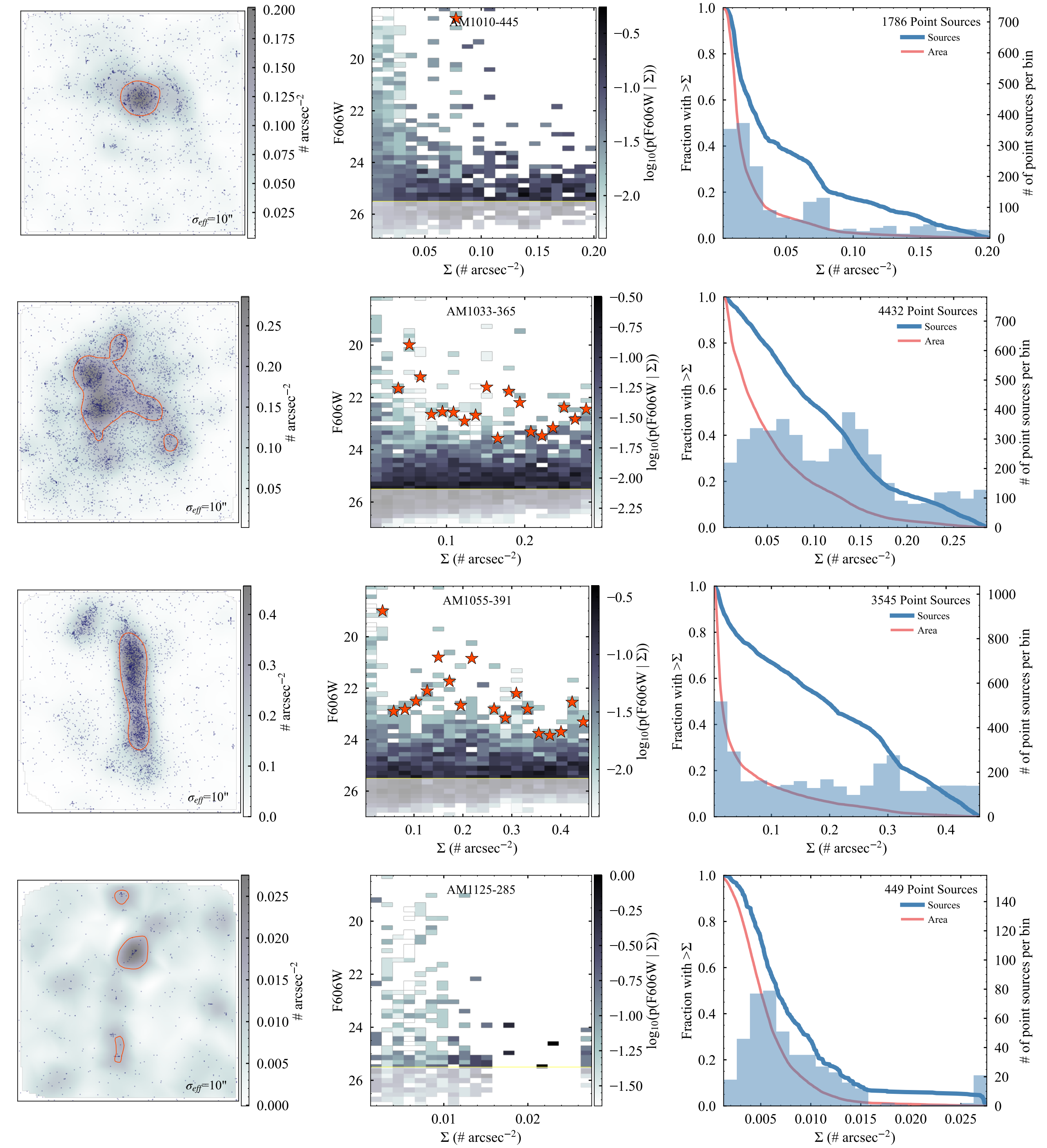

**Figure D3.** [Continued] ARP-MADORE1010-445, ARP-MADORE1033-365, ARP-MADORE1055-391, and ARP–MADORE1125-285 (top to bottom). [Left] The spatial distribution of point sources (points) and the smoothed background density field (greyscale); [Middle] the conditional luminosity function (greyscale) in bins of local surface density, with red stars marking the magnitude of the top 2% of the brightest sources in any density bin with more than 50 stars brighter than F606W=25.5; [Right] the histogram (shaded blue; right axis) and cumulative distributions (solid lines) of densities calculated at the location of point sources (blue thick line) or pixels (red thin line). Please see Section 2.5 and Figure 12 for a fuller description. [Continued]

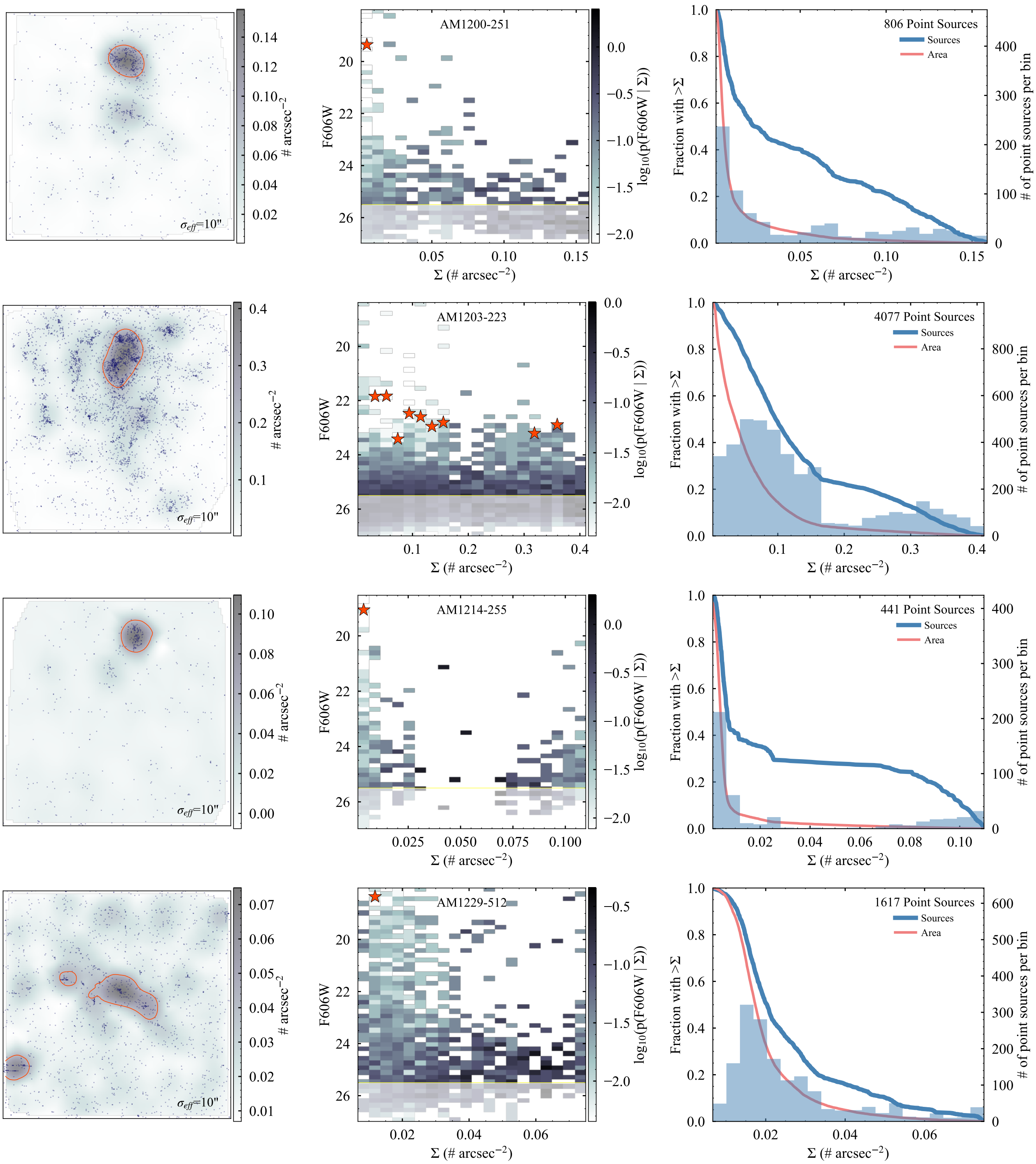

**Figure D3.** [Continued] ARP-MADORE1200-251, ARP-MADORE1203-223, ARP-MADORE1214-255, and ARP–MADORE1229-512 (top to bottom). [Left] The spatial distribution of point sources (points) and the smoothed background density field (greyscale); [Middle] the conditional luminosity function (greyscale) in bins of local surface density, with red stars marking the magnitude of the top 2% of the brightest sources in any density bin with more than 50 stars brighter than F606W=25.5; [Right] the histogram (shaded blue; right axis) and cumulative distributions (solid lines) of densities calculated at the location of point sources (blue thick line) or pixels (red thin line). Please see Section 2.5 and Figure 12 for a fuller description. [Continued]

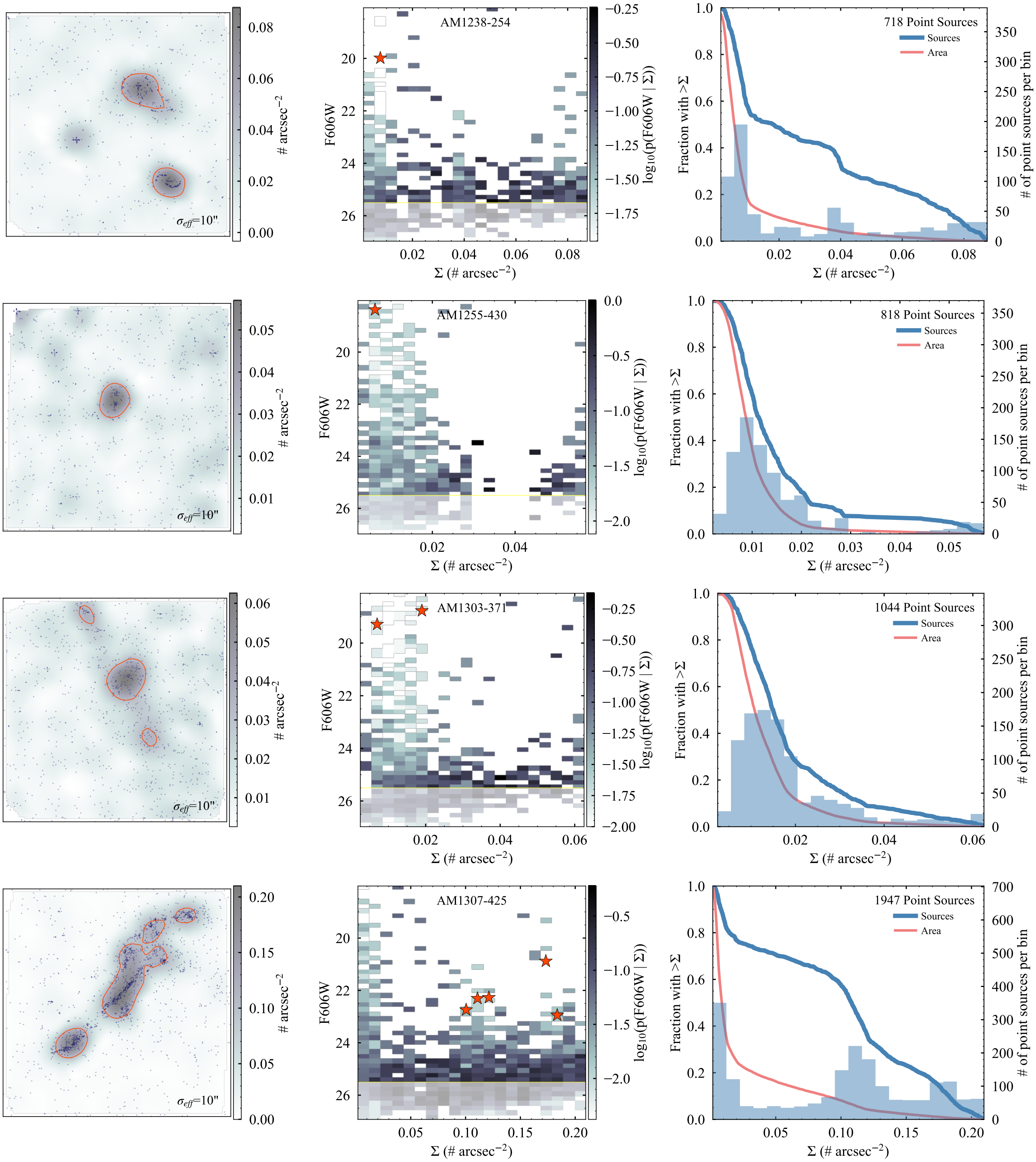

**Figure D3.** [Continued] ARP-MADORE1238-254, ARP-MADORE1255-430, ARP-MADORE1303-371, and ARP–MADORE1307-425 (top to bottom). [Left] The spatial distribution of point sources (points) and the smoothed background density field (greyscale); [Middle] the conditional luminosity function (greyscale) in bins of local surface density, with red stars marking the magnitude of the top 2% of the brightest sources in any density bin with more than 50 stars brighter than F606W=25.5; [Right] the histogram (shaded blue; right axis) and cumulative distributions (solid lines) of densities calculated at the location of point sources (blue thick line) or pixels (red thin line). Please see Section 2.5 and Figure 12 for a fuller description. [Continued]

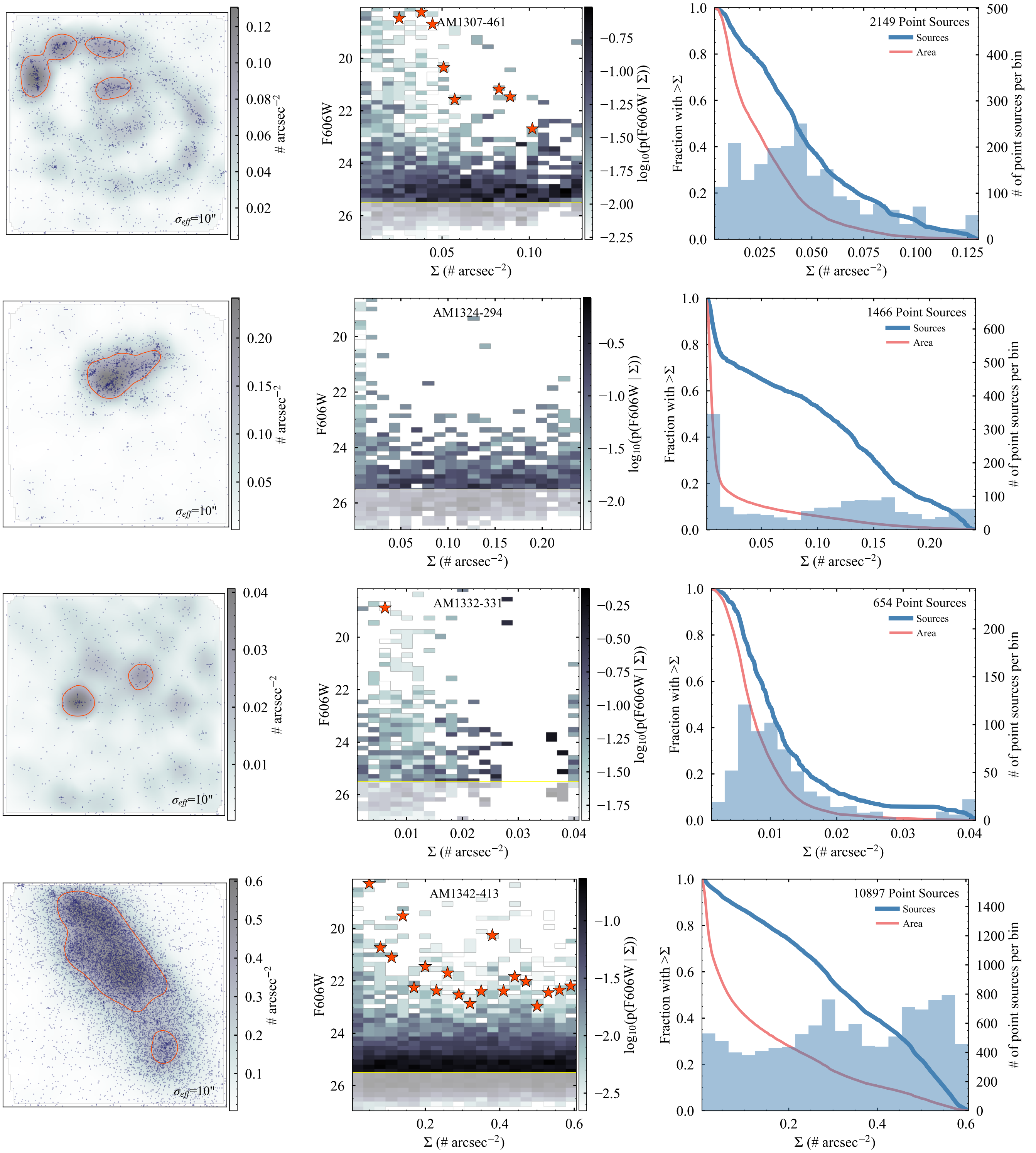

**Figure D3.** [Continued] ARP-MADORE1307-461, ARP-MADORE1324-294, ARP-MADORE1332-331, and ARP–MADORE1342-413 (top to bottom). [Left] The spatial distribution of point sources (points) and the smoothed background density field (greyscale); [Middle] the conditional luminosity function (greyscale) in bins of local surface density, with red stars marking the magnitude of the top 2% of the brightest sources in any density bin with more than 50 stars brighter than F606W=25.5; [Right] the histogram (shaded blue; right axis) and cumulative distributions (solid lines) of densities calculated at the location of point sources (blue thick line) or pixels (red thin line). Please see Section 2.5 and Figure 12 for a fuller description. [Continued]

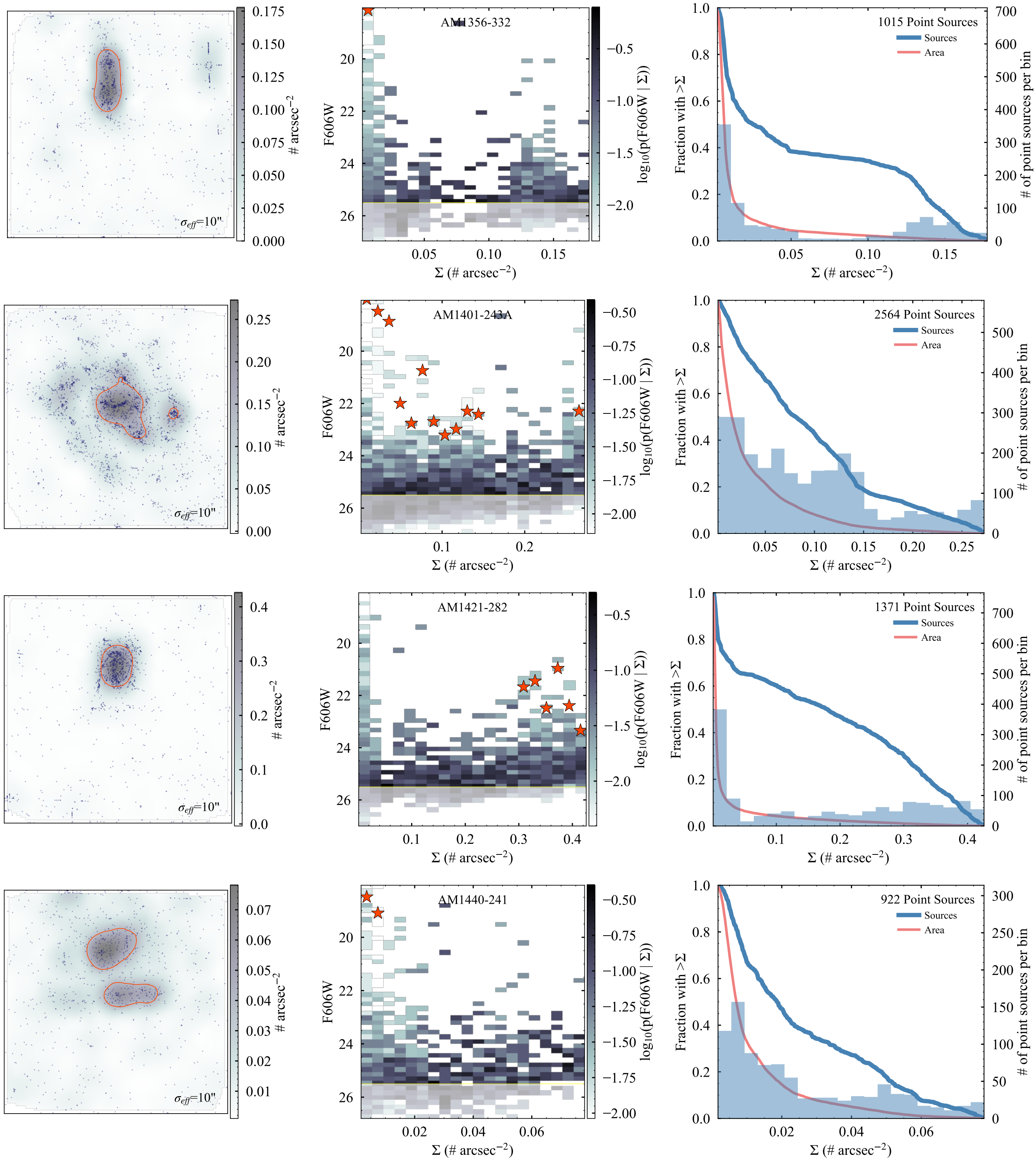

**Figure D3.** [Continued] ARP-MADORE1356-332, ARP-MADORE1401-243A, ARP-MADORE1421-282, and ARP–MADORE1440-241 (top to bottom). [Left] The spatial distribution of point sources (points) and the smoothed background density field (greyscale); [Middle] the conditional luminosity function (greyscale) in bins of local surface density, with red stars marking the magnitude of the top 2% of the brightest sources in any density bin with more than 50 stars brighter than F606W=25.5; [Right] the histogram (shaded blue; right axis) and cumulative distributions (solid lines) of densities calculated at the location of point sources (blue thick line) or pixels (red thin line). Please see Section 2.5 and Figure 12 for a fuller description. [Continued]

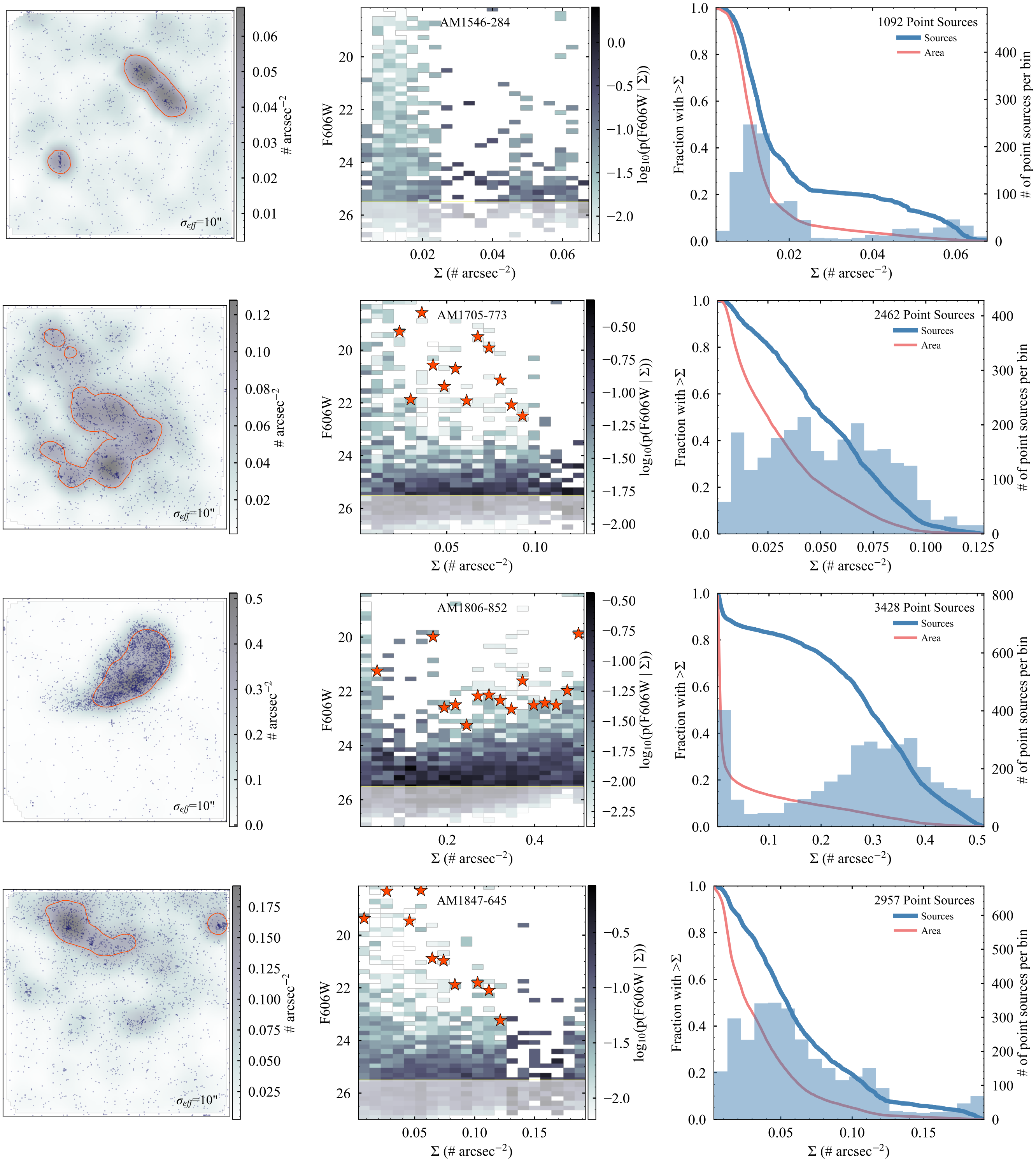

**Figure D3.** [Continued] ARP-MADORE1546-284, ARP-MADORE1705-773, ARP-MADORE1806-852, and ARP–MADORE1847-645 (top to bottom). [Left] The spatial distribution of point sources (points) and the smoothed background density field (greyscale); [Middle] the conditional luminosity function (greyscale) in bins of local surface density, with red stars marking the magnitude of the top 2% of the brightest sources in any density bin with more than 50 stars brighter than F606W=25.5; [Right] the histogram (shaded blue; right axis) and cumulative distributions (solid lines) of densities calculated at the location of point sources (blue thick line) or pixels (red thin line). Please see Section 2.5 and Figure 12 for a fuller description. [Continued]

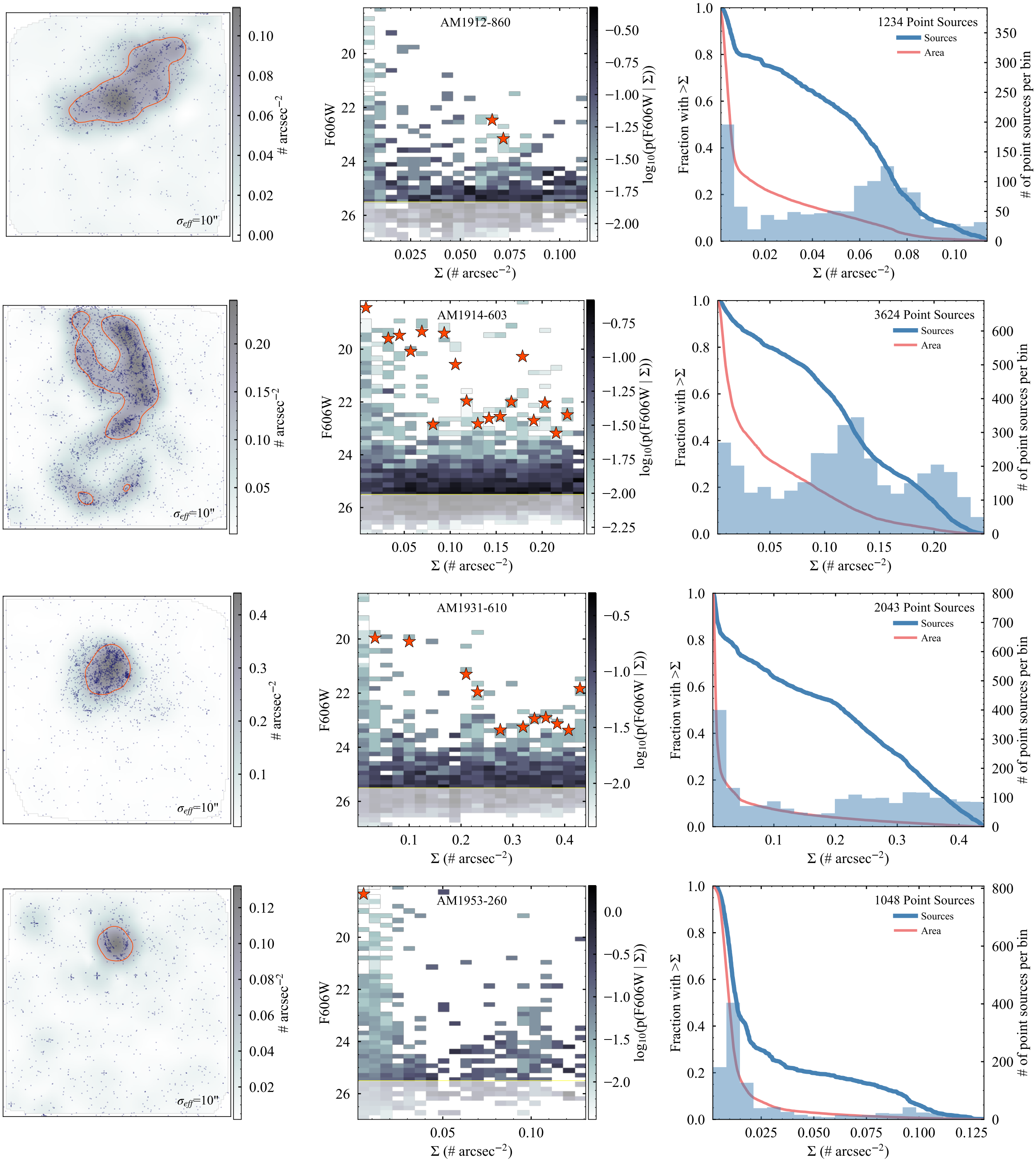

**Figure D3.** [Continued] ARP-MADORE1912-860, ARP-MADORE1914-603, ARP-MADORE1931-610, and ARP–MADORE1953-260 (top to bottom). [Left] The spatial distribution of point sources (points) and the smoothed background density field (greyscale); [Middle] the conditional luminosity function (greyscale) in bins of local surface density, with red stars marking the magnitude of the top 2% of the brightest sources in any density bin with more than 50 stars brighter than F606W=25.5; [Right] the histogram (shaded blue; right axis) and cumulative distributions (solid lines) of densities calculated at the location of point sources (blue thick line) or pixels (red thin line). Please see Section 2.5 and Figure 12 for a fuller description. [Continued]

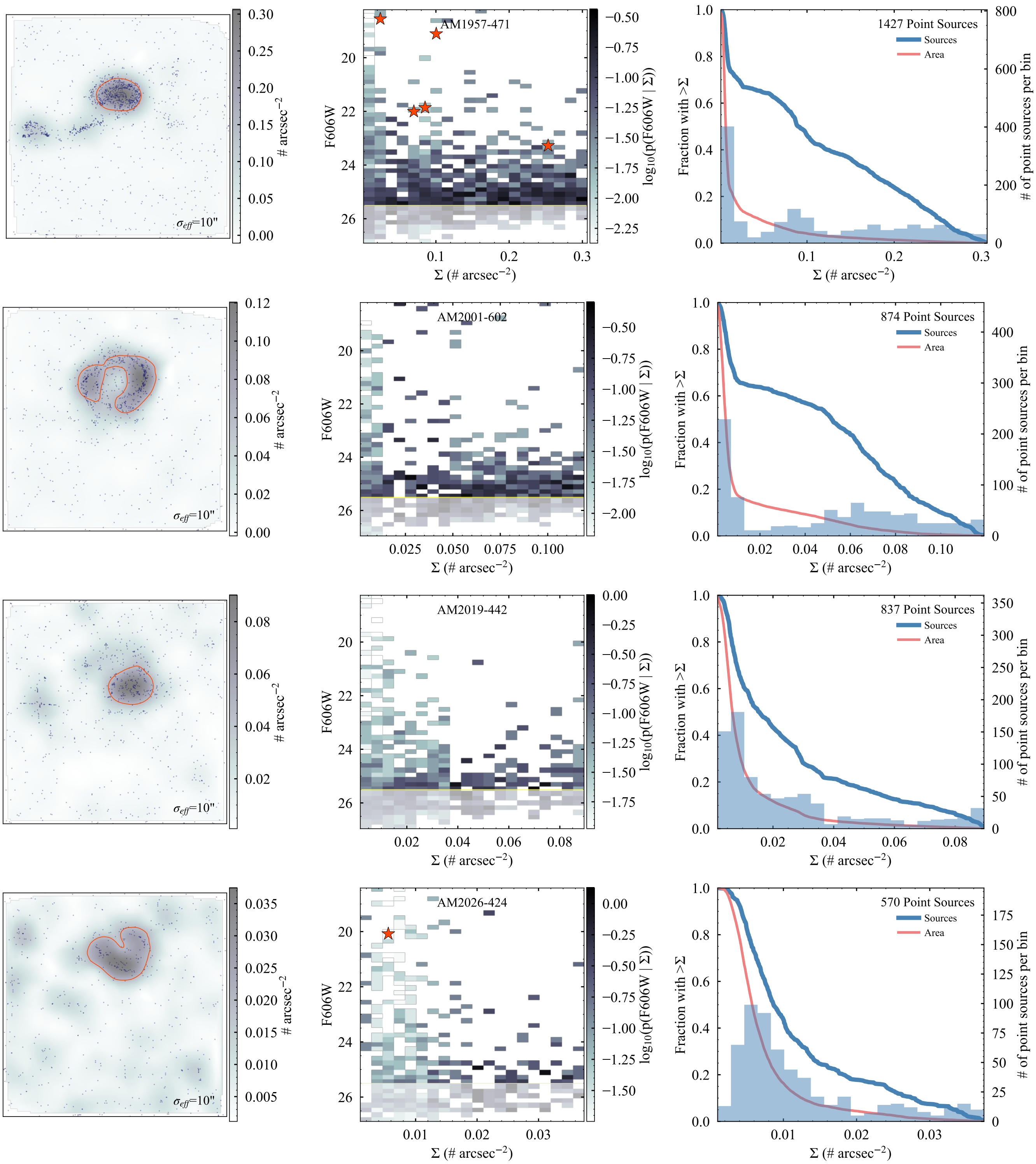

**Figure D3.** [Continued] ARP-MADORE1957-471, ARP-MADORE2001-602, ARP-MADORE2019-442, and ARP–MADORE2026-424 (top to bottom). [Left] The spatial distribution of point sources (points) and the smoothed background density field (greyscale); [Middle] the conditional luminosity function (greyscale) in bins of local surface density, with red stars marking the magnitude of the top 2% of the brightest sources in any density bin with more than 50 stars brighter than F606W=25.5; [Right] the histogram (shaded blue; right axis) and cumulative distributions (solid lines) of densities calculated at the location of point sources (blue thick line) or pixels (red thin line). Please see Section 2.5 and Figure 12 for a fuller description. [Continued]

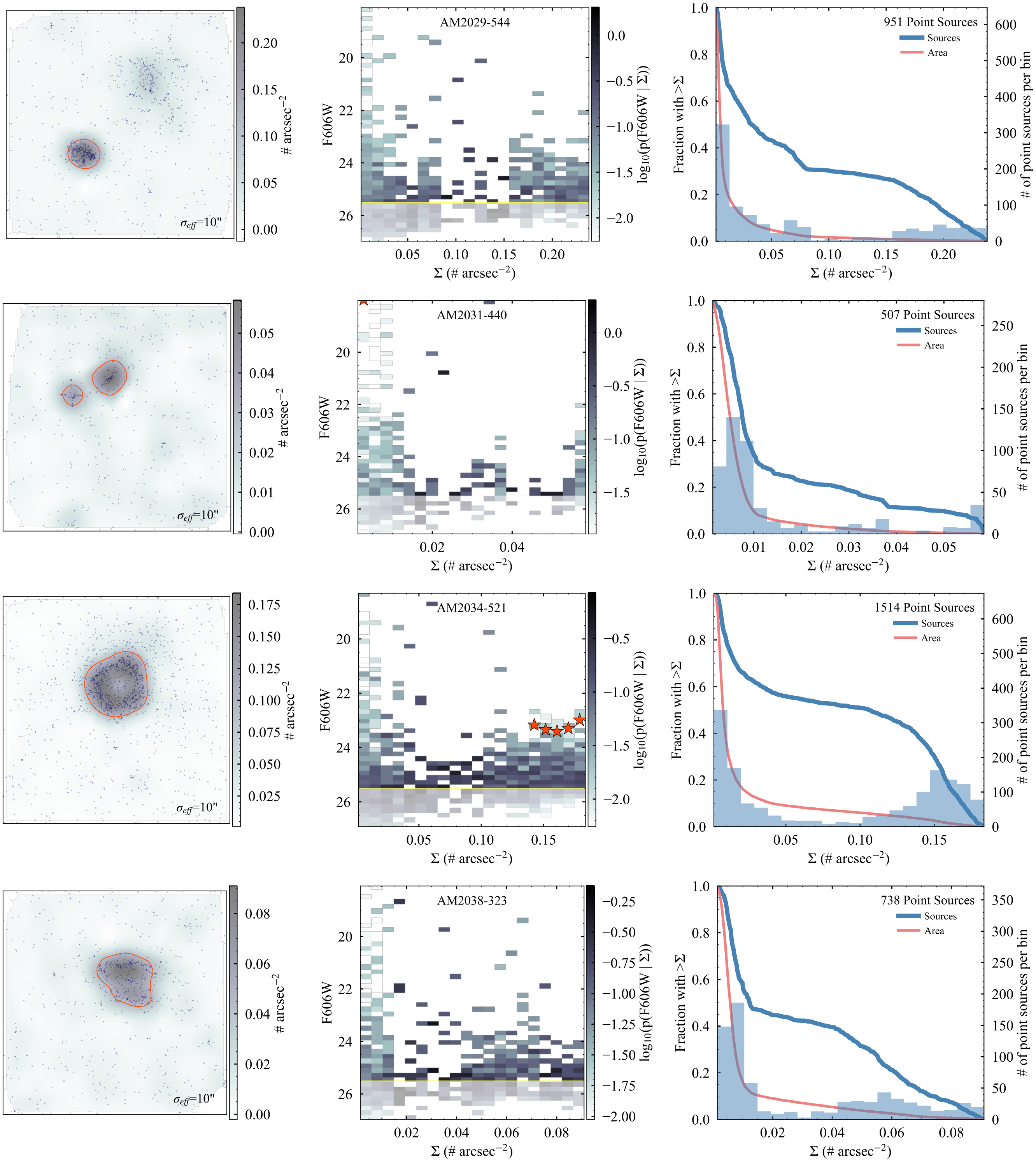

**Figure D3.** [Continued] ARP-MADORE2029-544, ARP-MADORE2031-440, ARP-MADORE2034-521, and ARP–MADORE2038-323 (top to bottom). [Left] The spatial distribution of point sources (points) and the smoothed background density field (greyscale); [Middle] the conditional luminosity function (greyscale) in bins of local surface density, with red stars marking the magnitude of the top 2% of the brightest sources in any density bin with more than 50 stars brighter than F606W=25.5; [Right] the histogram (shaded blue; right axis) and cumulative distributions (solid lines) of densities calculated at the location of point sources (blue thick line) or pixels (red thin line). Please see Section 2.5 and Figure 12 for a fuller description. [Continued]

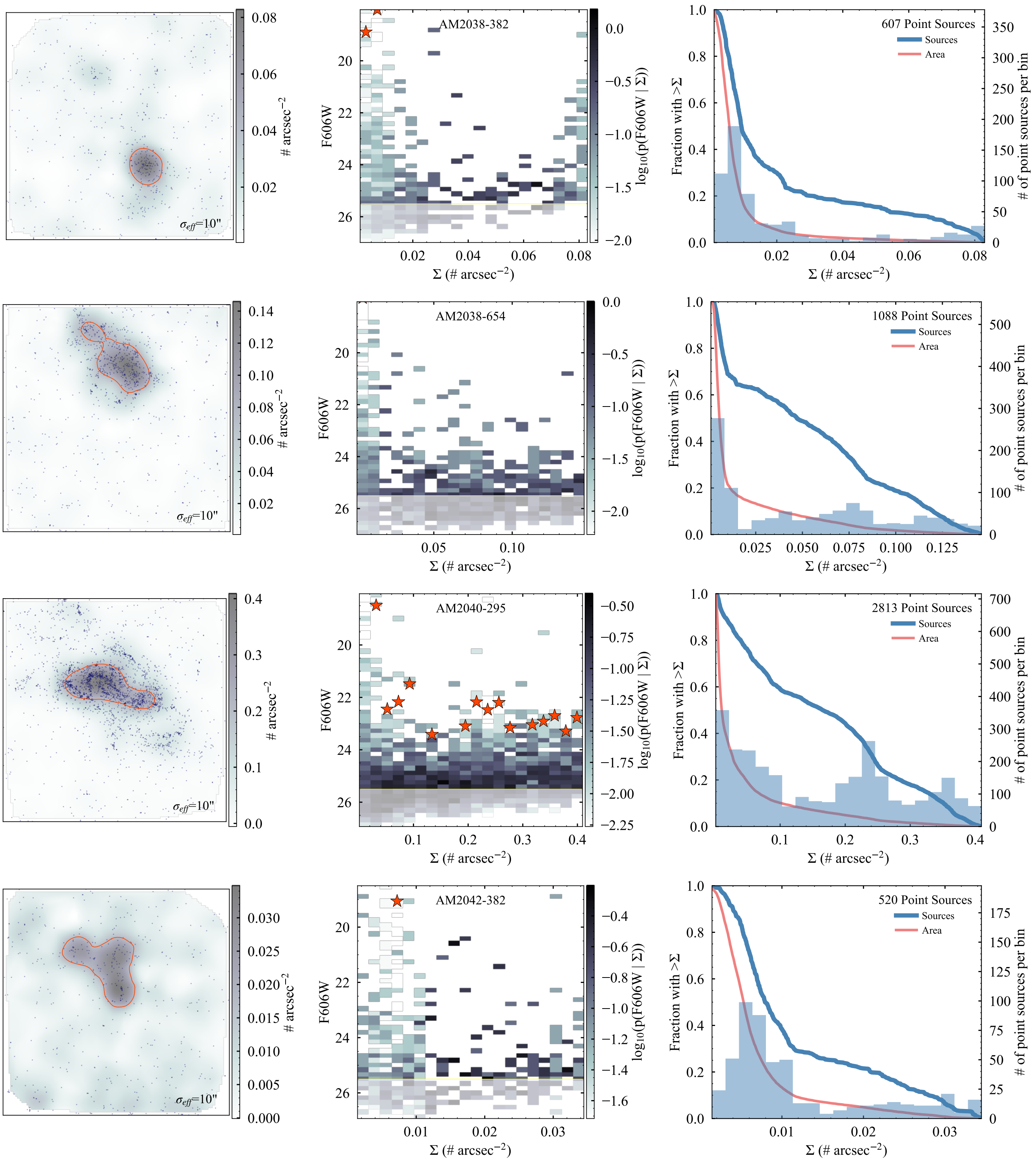

**Figure D3.** [Continued] ARP-MADORE2038-382, ARP-MADORE2038-654, ARP-MADORE2040-295, and ARP–MADORE2042-382 (top to bottom). [Left] The spatial distribution of point sources (points) and the smoothed background density field (greyscale); [Middle] the conditional luminosity function (greyscale) in bins of local surface density, with red stars marking the magnitude of the top 2% of the brightest sources in any density bin with more than 50 stars brighter than F606W=25.5; [Right] the histogram (shaded blue; right axis) and cumulative distributions (solid lines) of densities calculated at the location of point sources (blue thick line) or pixels (red thin line). Please see Section 2.5 and Figure 12 for a fuller description. [Continued]

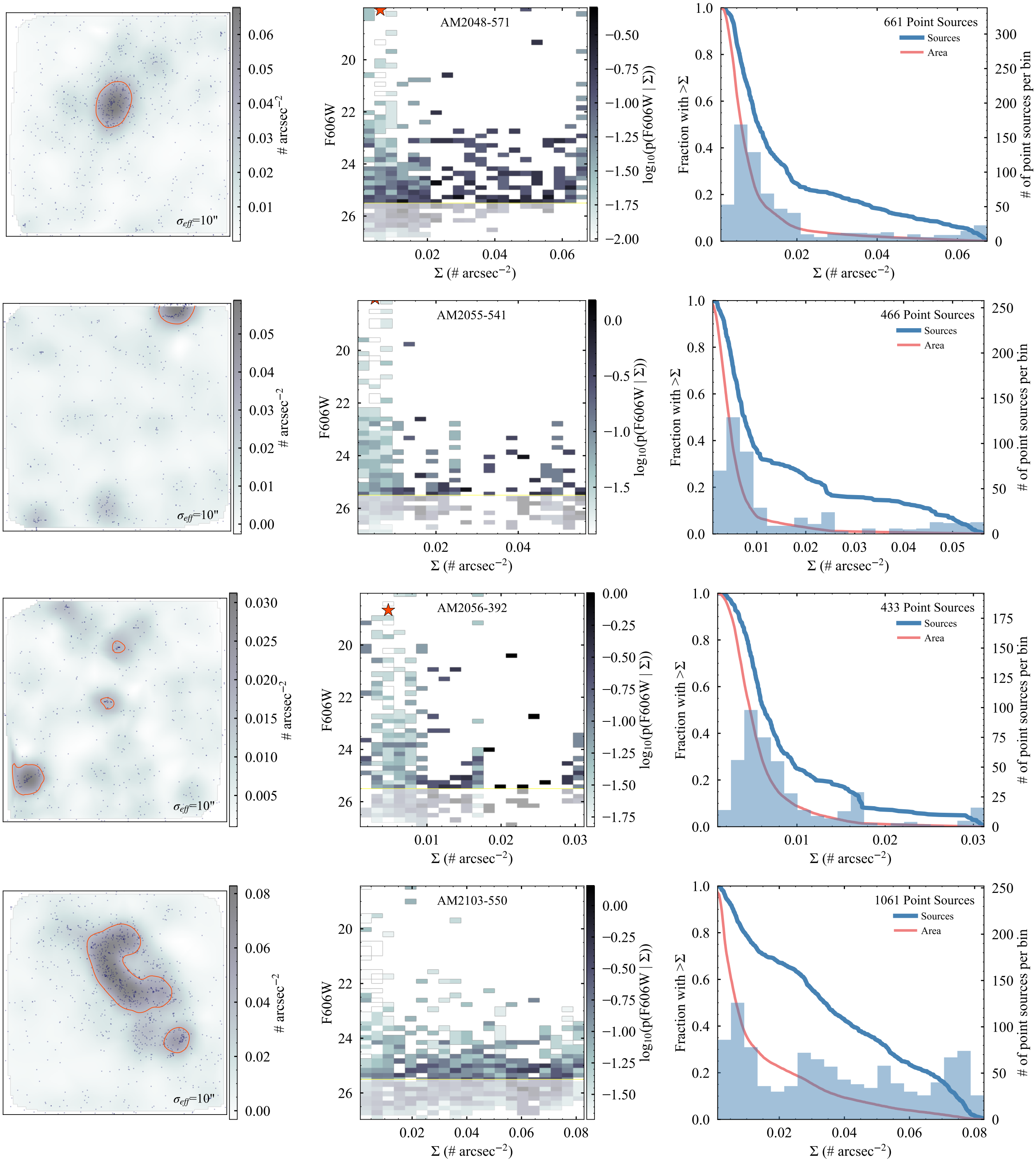

**Figure D3.** [Continued] ARP-MADORE2048-571, ARP-MADORE2055-541, ARP-MADORE2056-392, and ARP–MADORE2103-550 (top to bottom). [Left] The spatial distribution of point sources (points) and the smoothed background density field (greyscale); [Middle] the conditional luminosity function (greyscale) in bins of local surface density, with red stars marking the magnitude of the top 2% of the brightest sources in any density bin with more than 50 stars brighter than F606W=25.5; [Right] the histogram (shaded blue; right axis) and cumulative distributions (solid lines) of densities calculated at the location of point sources (blue thick line) or pixels (red thin line). Please see Section 2.5 and Figure 12 for a fuller description. [Continued]

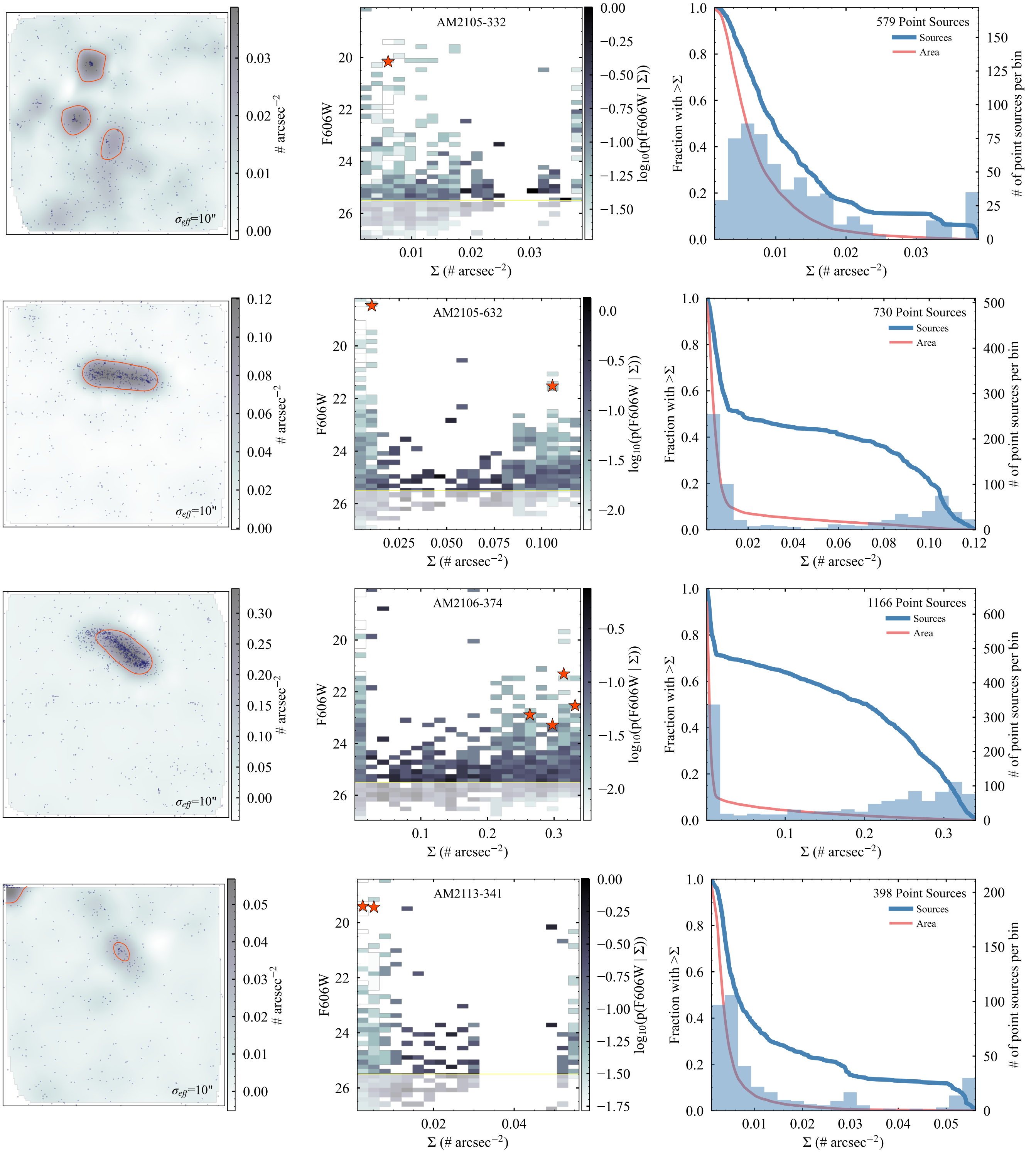

**Figure D3.** [Continued] ARP-MADORE2105-332, ARP-MADORE2105-632, ARP-MADORE2106-374, and ARP–MADORE2113-341 (top to bottom). [Left] The spatial distribution of point sources (points) and the smoothed background density field (greyscale); [Middle] the conditional luminosity function (greyscale) in bins of local surface density, with red stars marking the magnitude of the top 2% of the brightest sources in any density bin with more than 50 stars brighter than F606W=25.5; [Right] the histogram (shaded blue; right axis) and cumulative distributions (solid lines) of densities calculated at the location of point sources (blue thick line) or pixels (red thin line). Please see Section 2.5 and Figure 12 for a fuller description. [Continued]

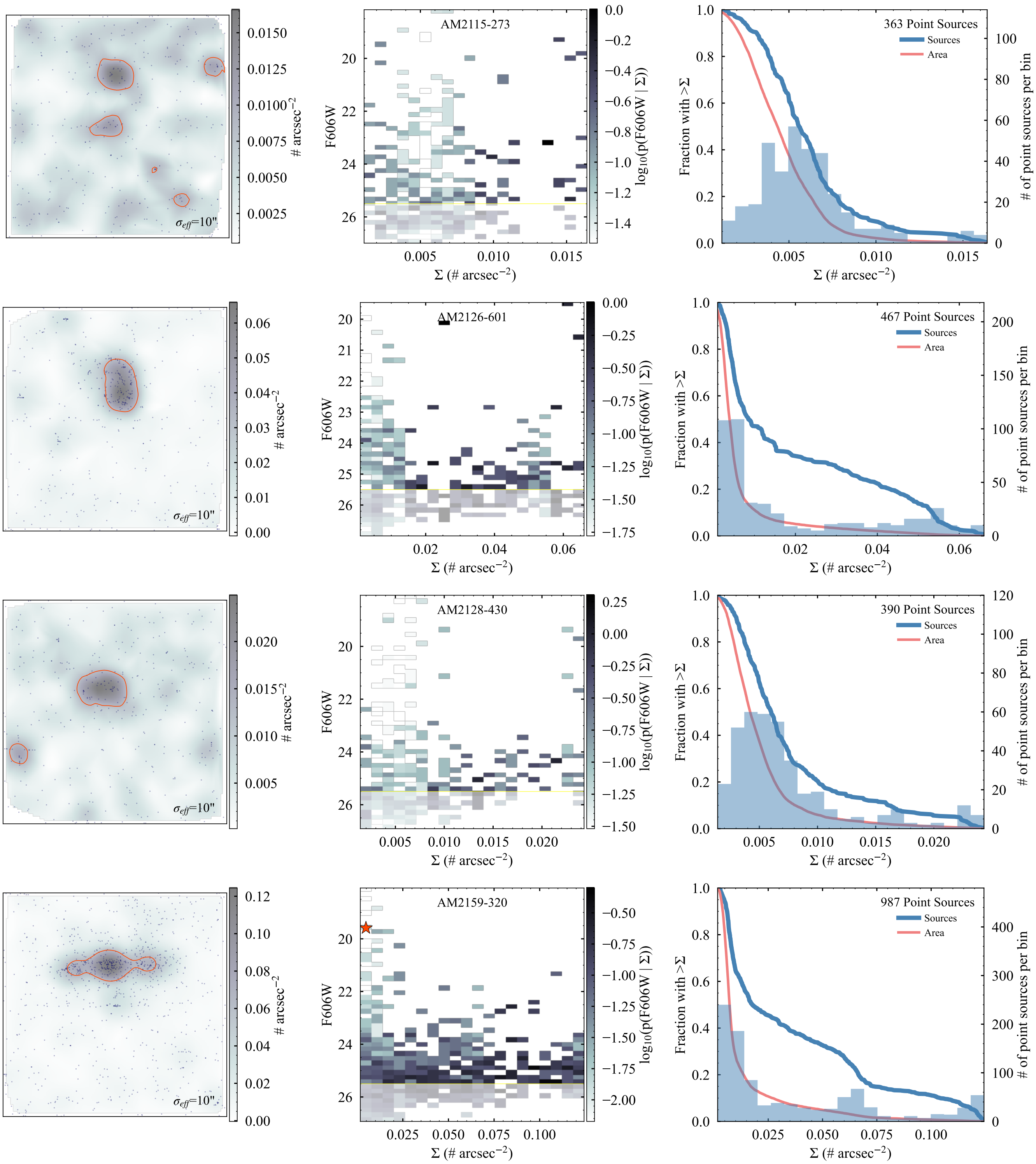

**Figure D3.** [Continued] ARP-MADORE2115-273, ARP-MADORE2126-601, ARP-MADORE2128-430, and ARP–MADORE2159-320 (top to bottom). [Left] The spatial distribution of point sources (points) and the smoothed background density field (greyscale); [Middle] the conditional luminosity function (greyscale) in bins of local surface density, with red stars marking the magnitude of the top 2% of the brightest sources in any density bin with more than 50 stars brighter than F606W=25.5; [Right] the histogram (shaded blue; right axis) and cumulative distributions (solid lines) of densities calculated at the location of point sources (blue thick line) or pixels (red thin line). Please see Section 2.5 and Figure 12 for a fuller description. [Continued]

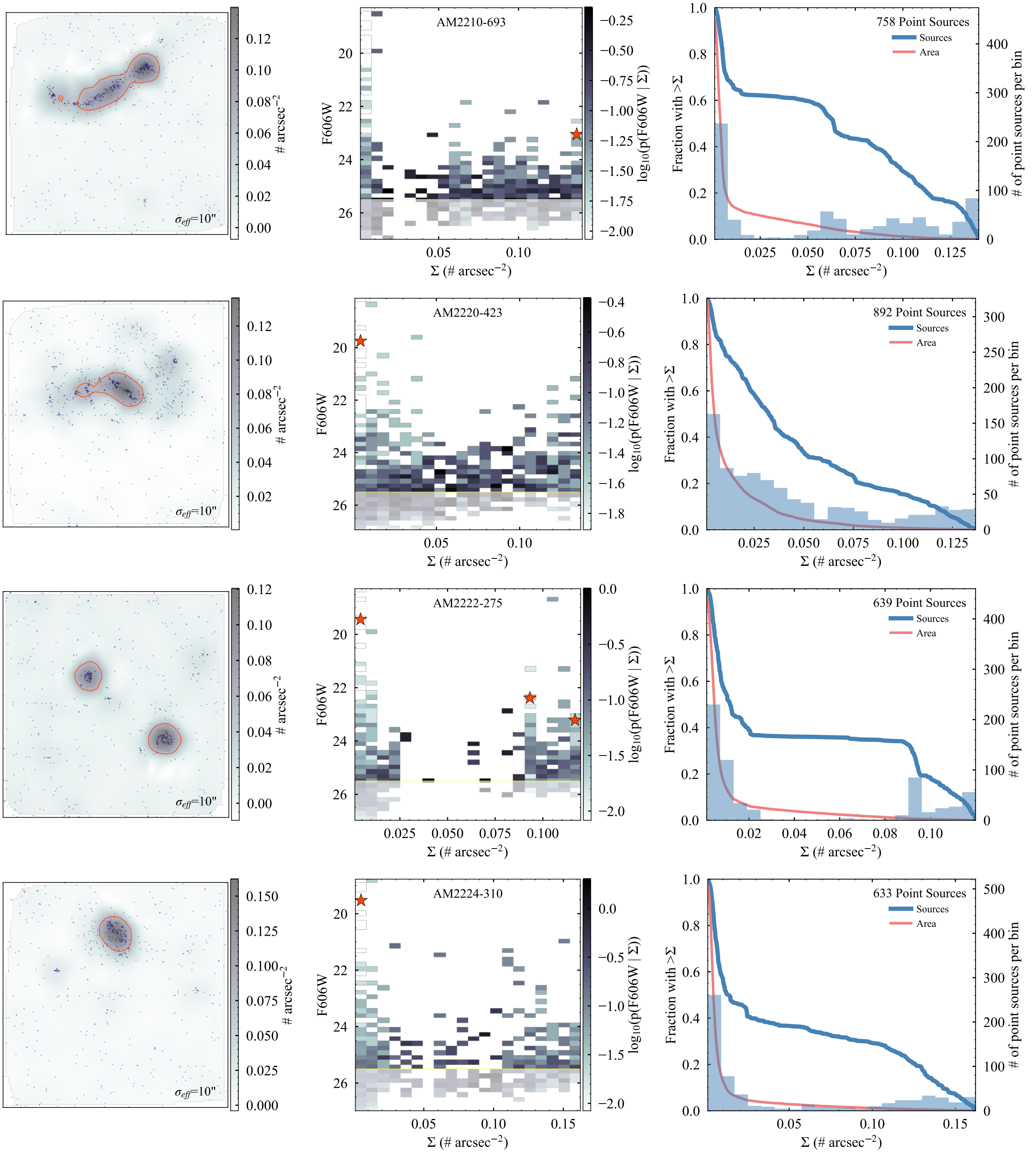

**Figure D3.** [Continued] ARP-MADORE2210-693, ARP-MADORE2220-423, ARP-MADORE2222-275, and ARP–MADORE2224-310 (top to bottom). [Left] The spatial distribution of point sources (points) and the smoothed background density field (greyscale); [Middle] the conditional luminosity function (greyscale) in bins of local surface density, with red stars marking the magnitude of the top 2% of the brightest sources in any density bin with more than 50 stars brighter than F606W=25.5; [Right] the histogram (shaded blue; right axis) and cumulative distributions (solid lines) of densities calculated at the location of point sources (blue thick line) or pixels (red thin line). Please see Section 2.5 and Figure 12 for a fuller description. [Continued]

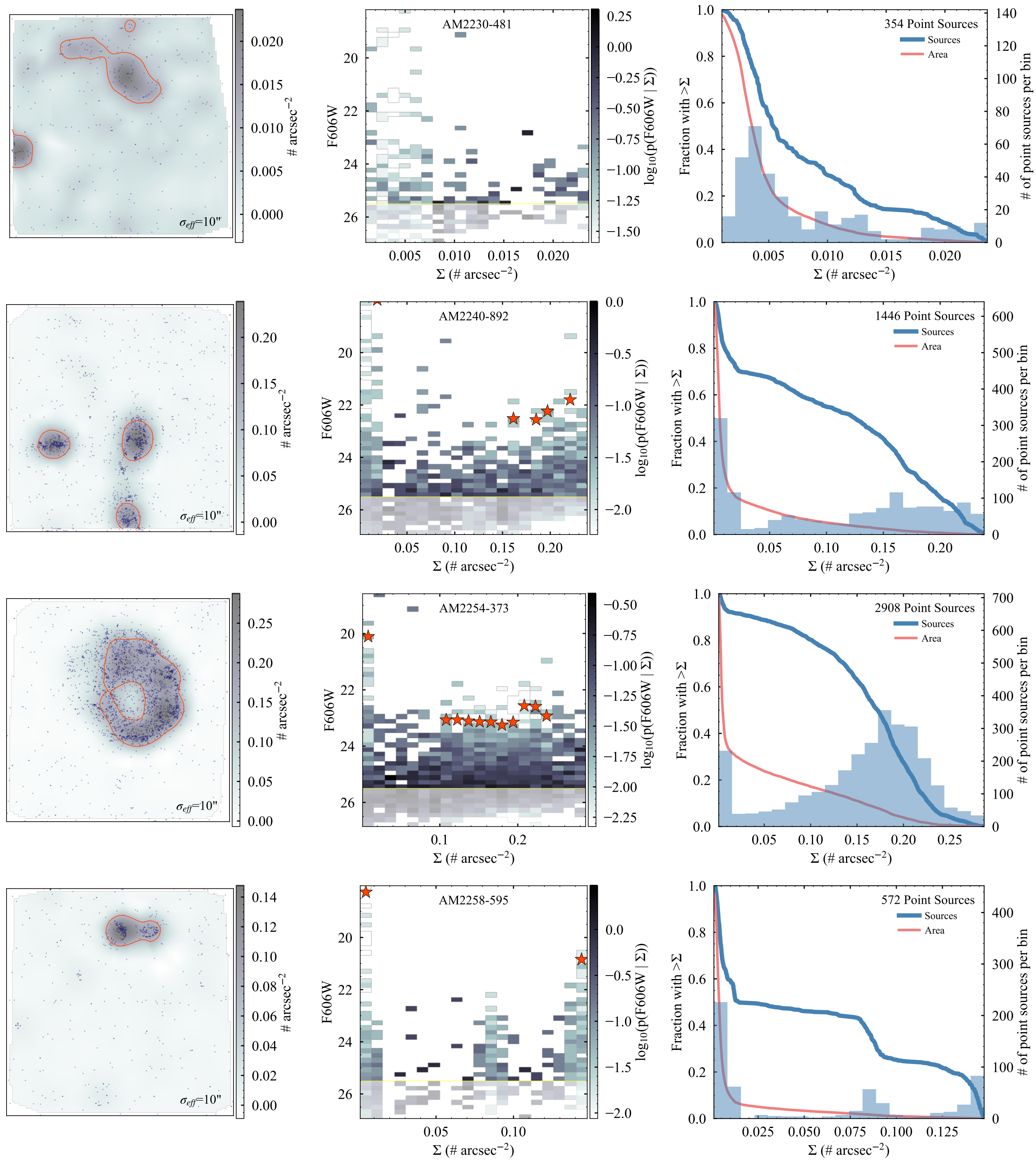

**Figure D3.** [Continued] ARP-MADORE2230-481, ARP-MADORE2240-892, ARP-MADORE2254-373, and ARP–MADORE2258-595 (top to bottom). [Left] The spatial distribution of point sources (points) and the smoothed background density field (greyscale); [Middle] the conditional luminosity function (greyscale) in bins of local surface density, with red stars marking the magnitude of the top 2% of the brightest sources in any density bin with more than 50 stars brighter than F606W=25.5; [Right] the histogram (shaded blue; right axis) and cumulative distributions (solid lines) of densities calculated at the location of point sources (blue thick line) or pixels (red thin line). Please see Section 2.5 and Figure 12 for a fuller description. [Continued]

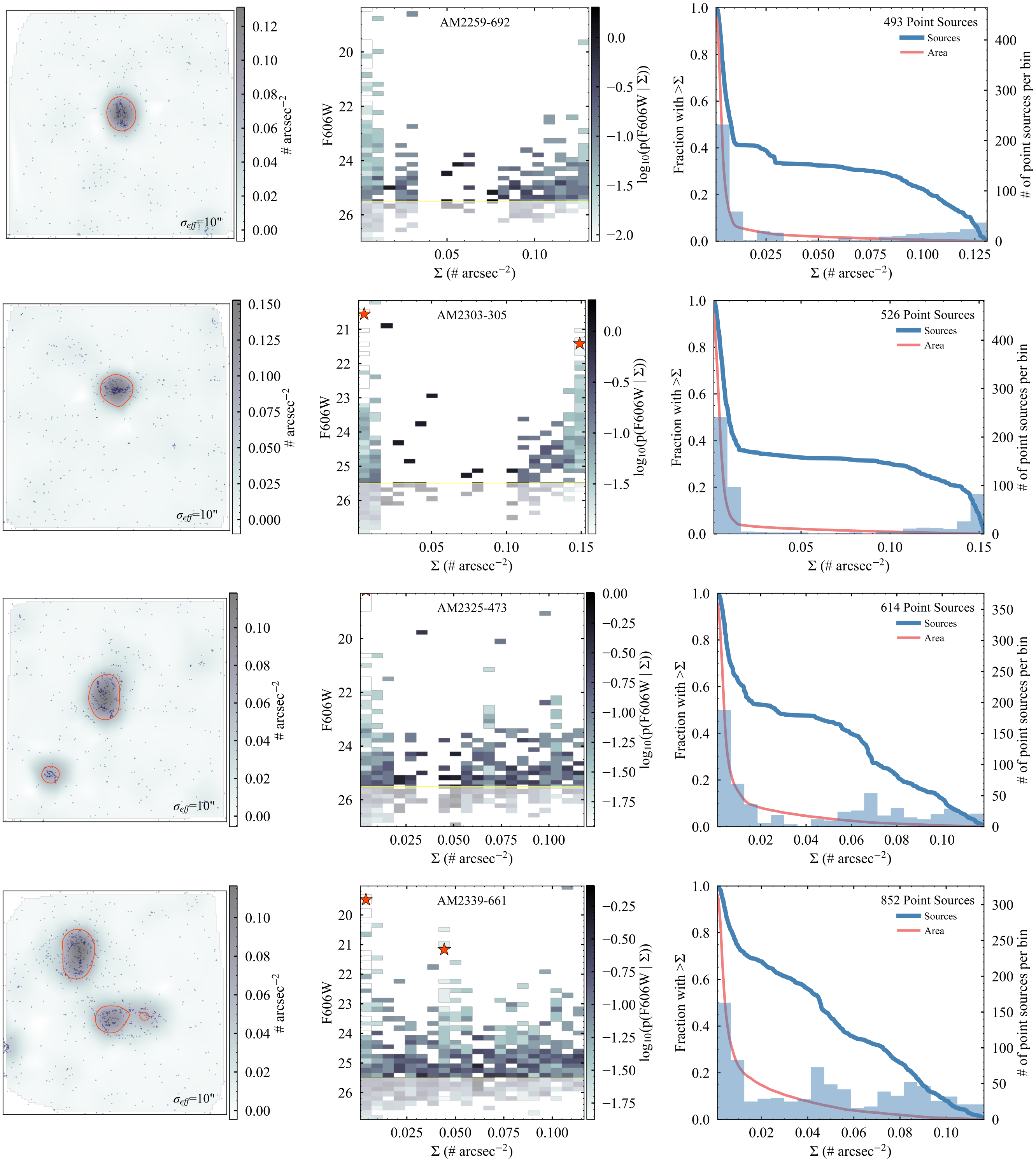

**Figure D3.** [Continued] ARP-MADORE2259-692, ARP-MADORE2303-305, ARP-MADORE2325-473, and ARP–MADORE2339-661 (top to bottom). [Left] The spatial distribution of point sources (points) and the smoothed background density field (greyscale); [Middle] the conditional luminosity function (greyscale) in bins of local surface density, with red stars marking the magnitude of the top 2% of the brightest sources in any density bin with more than 50 stars brighter than F606W=25.5; [Right] the histogram (shaded blue; right axis) and cumulative distributions (solid lines) of densities calculated at the location of point sources (blue thick line) or pixels (red thin line). Please see Section 2.5 and Figure 12 for a fuller description. [Continued]

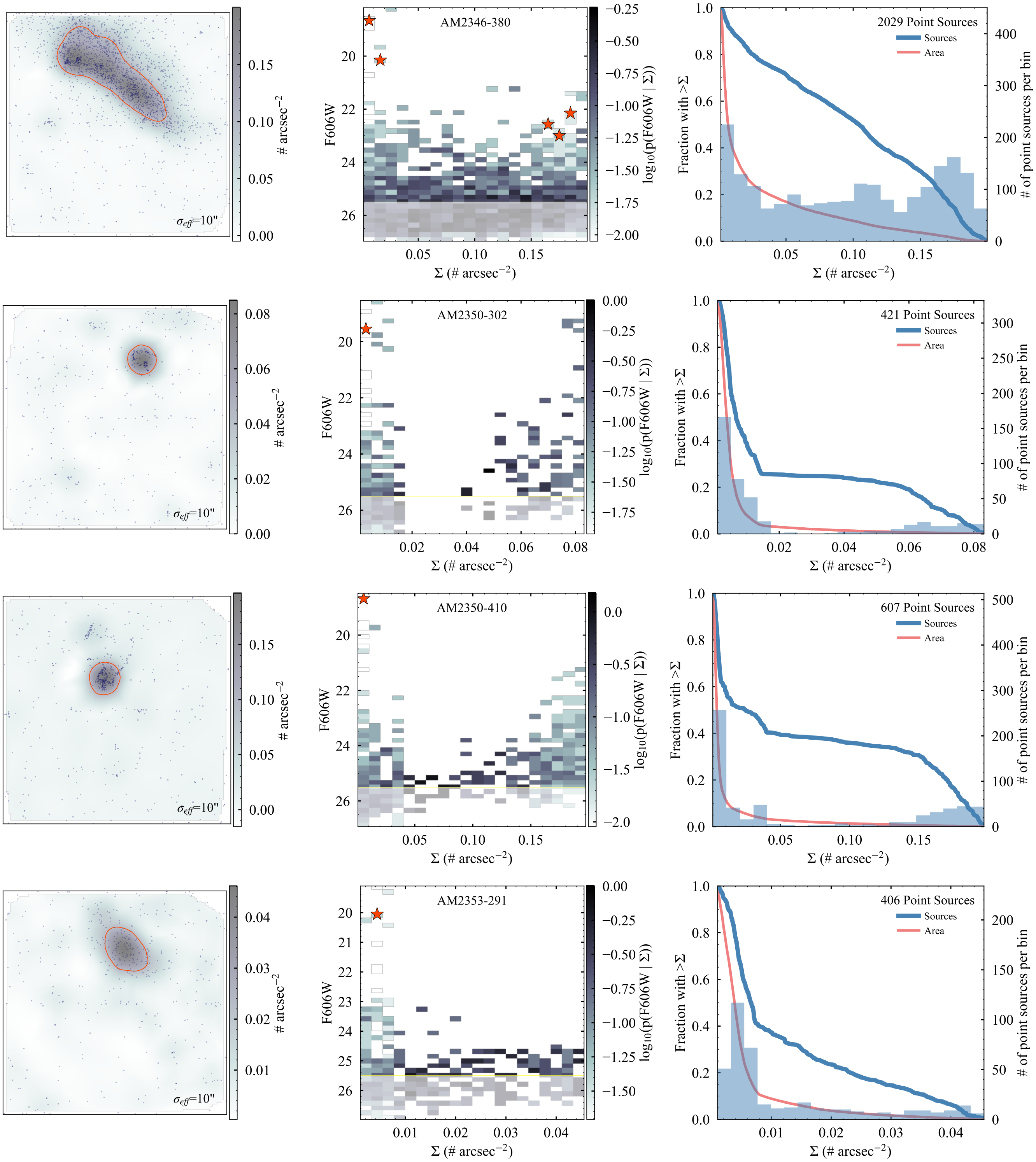

**Figure D3.** [Continued] ARP-MADORE2346-380, ARP-MADORE2350-302, ARP-MADORE2350-410, and ARP–MADORE2353-291 (top to bottom). [Left] The spatial distribution of point sources (points) and the smoothed background density field (greyscale); [Middle] the conditional luminosity function (greyscale) in bins of local surface density, with red stars marking the magnitude of the top 2% of the brightest sources in any density bin with more than 50 stars brighter than F606W=25.5; [Right] the histogram (shaded blue; right axis) and cumulative distributions (solid lines) of densities calculated at the location of point sources (blue thick line) or pixels (red thin line). Please see Section 2.5 and Figure 12 for a fuller description.

## E. AUXILIARY IMAGING FOR CENTRALLY ACTIVE & CENTRALLY QUIESCENT SYSTEMS

Based on Section 4.8, in Section 4.9 we collected systems in the Atlas that have strong, central, red WISE W3+W4 emission in at least one member galaxy. This type of emission is indicative of a central starburst or AGN activity. HST images of the systems were shown in Figure 23 and Figure 24, for strong and more moderate central emission, respectively. Here, we show the corresponding W3+W4 WISE images used to make the selection, presented in Figure E4 and Figure E5. We note that the WISE images do not have the same orientation as the HST thumbnails, which are plotted in image coordinates. Instead, the WISE images are oriented to align with sky coordinates, as are the full HST atlas images presented in Appendix B. The entire HST image footprint is contained within the WISE images.

These figures also include the corresponding W1+W2 images, whose higher resolution make them better for identifying the location and NIR color of the source of the W3+W4 emission. In general, very red colors in W1–W2 favors an AGN interpretation, especially when found in a highly luminous central point source, with W1–W2$\gtrsim$0.5–0.7 often taken as a dividing line for extragalactic samples with a wide range of redshift (see QSO selection criteria in T. H. Jarrett et al. 2011; D. Stern et al. 2012; R. J. Assef et al. 2018; S. Guo et al. 2018; K. Storey-Fisher et al. 2024, for WISE and comparable Spitzer colors). When considering W1–W2 colors of local sources only (i.e., those with spectral classifications available in surveys like SDSS), however, there is overlap in the color distributions of AGN and normal star forming galaxies with substantial hot dust contributions to W2, below the very reddest colors (W1–W2$\gtrsim$0.8–1) and particularly at lower luminosities (e.g., see distributions for SDSS-classified sources compiled in R. Nikutta et al. 2014; J. A. O'Connor et al. 2016; F. Z. Zeraatgari et al. 2024, for example). Thus more quantitative analysis and comparison with existing spectral classification is needed to better ascribe the central activity to a starburst or AGN (e.g., D. Asmus et al. 2020), especially for lower luminosity sources (K. N. Hainline et al. 2016). However, a source with a high luminosity point source in both WISE images, and noticeably red colors in W1+W2, is particularly likely to be a QSO.

In addition to sources with evidence for centrally concentrated activity, we also have compiled HST images that lack any such evidence, in either the dust-sensitive WISE W3+W4 images (available for all systems) or the 1.4Ghz radio bandpass of the VLASS survey (sensitive to synchrotron and/or free-free emission, but only available only for galaxies with declinations northward of $-40°$). HST thumbnails of these systems are shown in Figure E6. We note that some of these systems did show WISE point sources in the corresponding W3+W4 images, but not at a central location (c.f., the very well-resolved nearby galaxy Arp 263, where the emission was from a single star forming region, rather than a central burst), or, off the frame of the HST image.

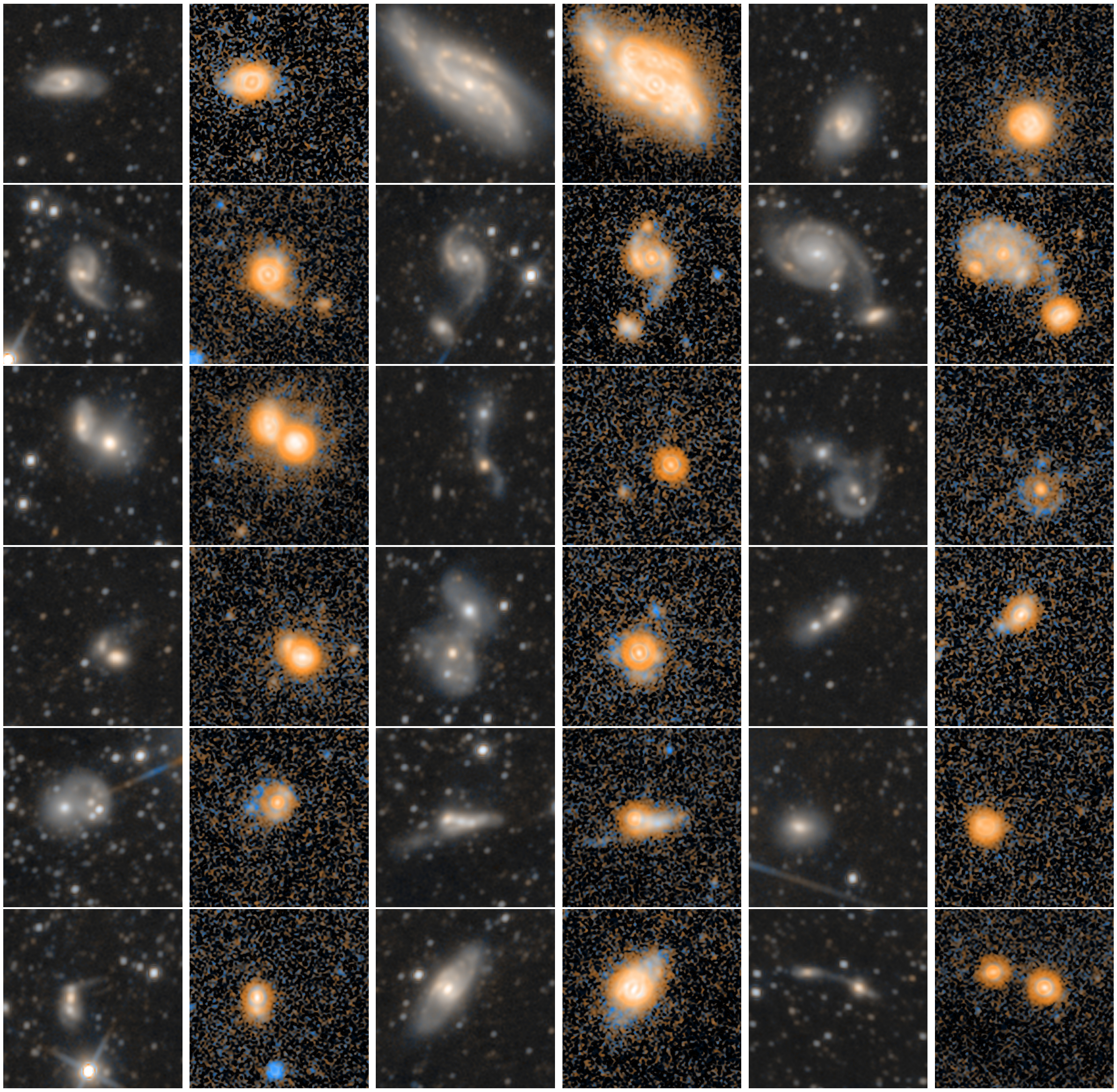

**Figure E4.** Pairs of WISE thumbnails (left:W1+W2; right:W3+W4) of systems containing at least one galaxy with strong, red, central WISE W3+W4 point source emission. All have detectable VLASS emission aligned with the WISE source if observed ($\delta > -40°$), with the exception of Arp 82, AM1421-282, and AM2350-302, which were observed but had no visible radio emission in VLASS. WISE images are aligned to the sky (north up), and should be compared to the full-page HST images in Appendix B, not the thumbnails in Figure 4 or Figure 23. From top left to bottom right, systems shown are: ARP15; ARP18; ARP49; ARP72; ARP82; ARP86; ARP91; ARP97; ARP107; ARP126; ARP143; ARP144; ARP145; ARP158; ARP163; ARP195; ARP200; ARP204 [Continued]

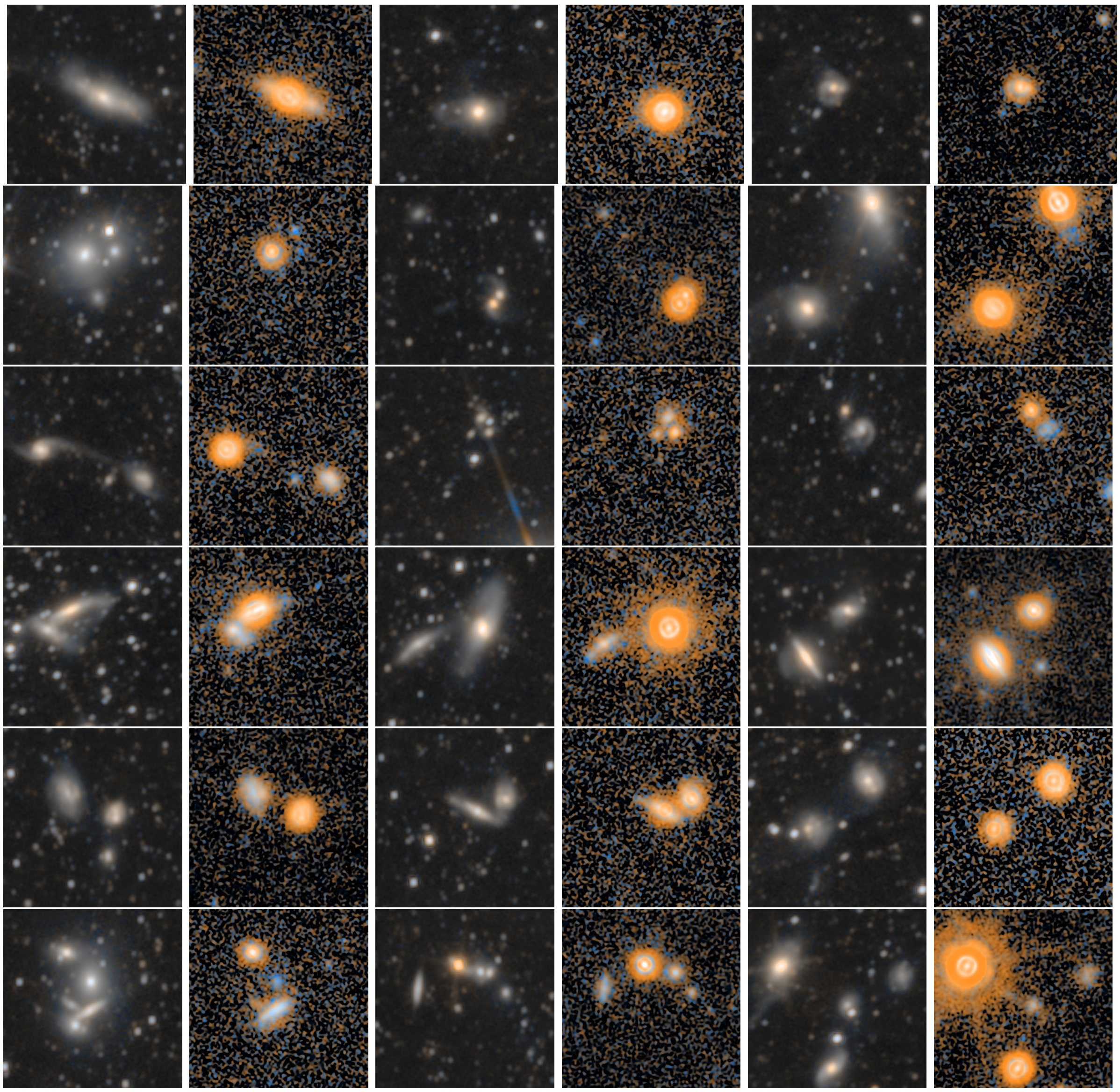

**Figure E4.** [Continued] Pairs of WISE thumbnails (left:W1+W2; right:W3+W4) of systems containing at least one galaxy with strong, red, central WISE W3+W4 point source emission. All have detectable VLASS emission aligned with the WISE source if observed ($\delta > -40°$), with the exception of Arp 82, AM1421-282, and AM2350-302, which were observed but had no visible radio emission in VLASS. WISE images are aligned to the sky (north up), and should be compared to the full-page HST images in Appendix B, not the thumbnails in Figure 4 or Figure 23. From top left to bottom right, systems shown are: ARP205; ARP216; ARP219; ARP221; ARP241; ARP245; ARP248; ARP251; ARP255; ARP278; ARP283; ARP293; ARP300; ARP301; ARP314; ARP321; ARP322; AM0018-485; [Continued]

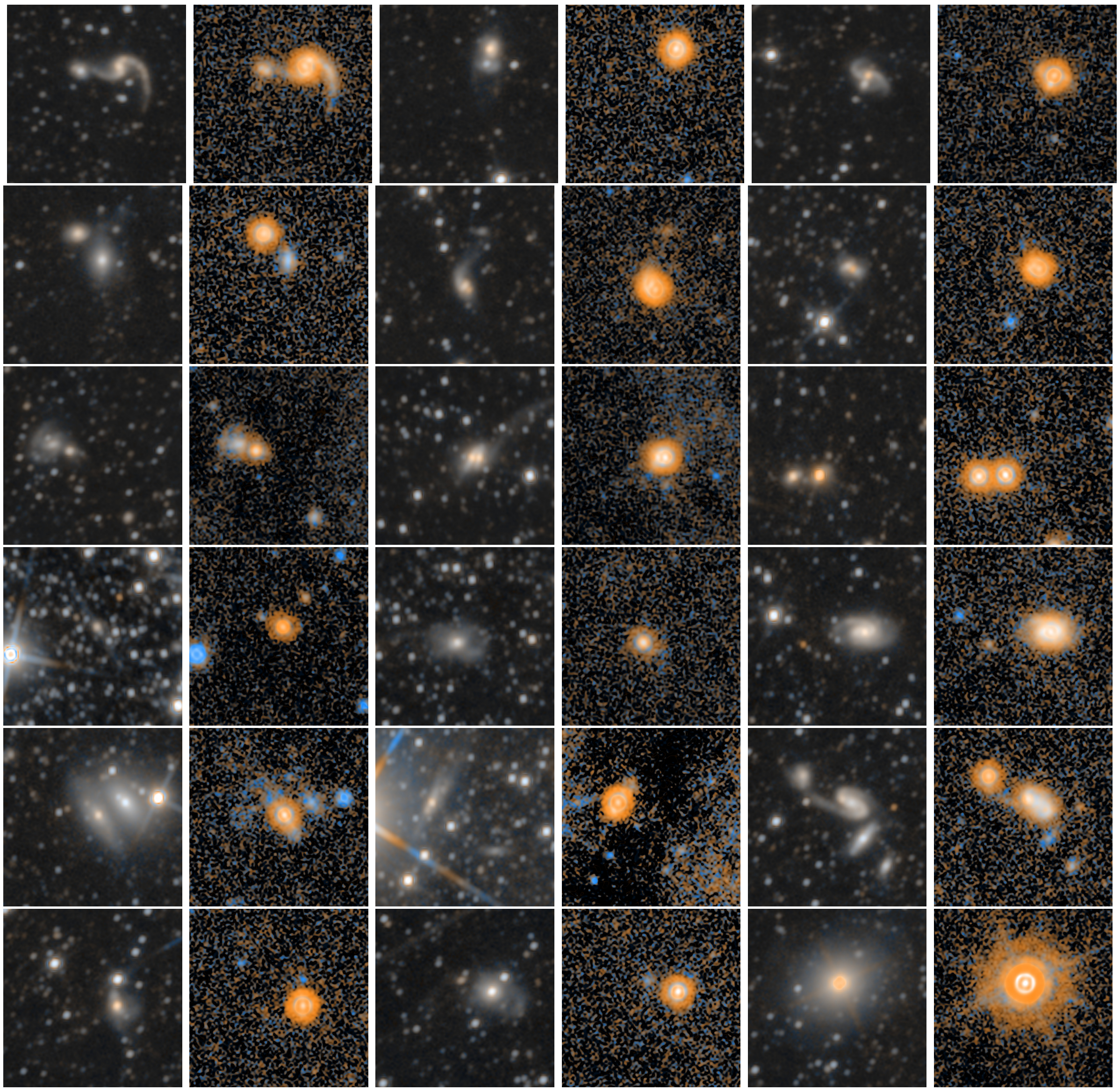

**Figure E4.** [Continued] Pairs of WISE thumbnails (left:W1+W2; right:W3+W4) of systems containing at least one galaxy with strong, red, central WISE W3+W4 point source emission. All have detectable VLASS emission aligned with the WISE source if observed ($\delta > -40^{\circ}$), with the exception of Arp 82, AM1421-282, and AM2350-302, which were observed but had no visible radio emission in VLASS. WISE images are aligned to the sky (north up), and should be compared to the full-page HST images in Appendix B, not the thumbnails in Figure 4 or Figure 23. From top left to bottom right, systems shown are: AM0135-650; AM0223-403; AM0313-545; AM0052-321; AM0459-340; AM0558-335; AM0642-801; AM0658-590; AM1214-255; AM1229-512; AM1255-430; AM1421-282; AM1440-241; AM1546-284; AM1957-471; AM2031-440; AM2038-382; AM2048-571;

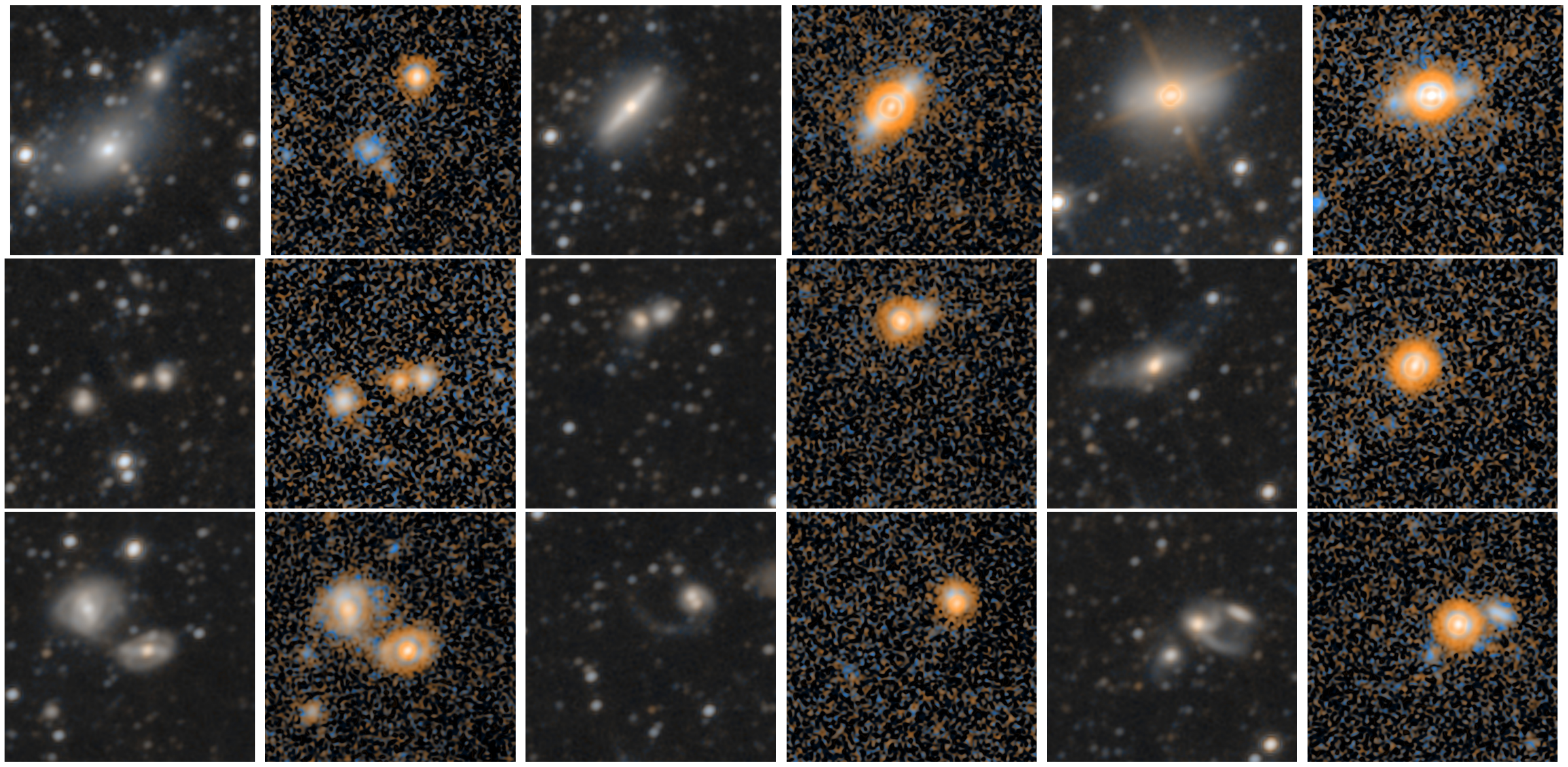

**Figure E4.** [Continued] Pairs of WISE thumbnails (left:W1+W2; right:W3+W4) of systems containing at least one galaxy with strong, red, central WISE W3+W4 point source emission. All have detectable VLASS emission aligned with the WISE source if observed ($\delta > -40°$), with the exception of Arp 82, AM1421-282, and AM2350-302, which were observed but had no visible radio emission in VLASS. WISE images are aligned to the sky (north up), and should be compared to the full-page HST images in Appendix B, not the thumbnails in Figure 4 or Figure 23. From top left to bottom right, systems shown are: AM2105-332; AM2105-632; AM2159-320; AM2222-275; AM2258-595; AM2303-305; AM2339-661; AM2350-302; AM2350-410

**Table E4.** Systems with Strong Central MIR Emission

| Target Name | Category | $V_r$ (km/s) | Target Name | Category | $V_r$ (km/s) | Target Name | Category | $V_r$ (km/s) |
|---|---|---|---|---|---|---|---|---|
| ARP15 | A:c | 3744 | ARP221 | D:k | 5495 | AM0642-801 | 6,8,15 | 4779 |
| ARP18 | A:c | 749 | ARP241 | D:m | 10431 | AM0658-590 | 2,15 | 8270 |
| ARP49 | B:b | 2218 | ARP245 | D:m | 2292 | AM1214-255 | 2,15 | 11465 |
| ARP72 | B:b | 3306 | ARP248 | D:m | 5275 | AM1229-512 | 7 | 2617 |
| ARP82 | B:c | 4067 | ARP251 | D:m | 21997 | AM1255-430 | 15 | 8964 |
| ARP86 | B:c | 4996 | ARP255 | D:m | 11997 | AM1421-282 | 24 | 4340 |
| ARP91 | B:c | 1980 | ARP278 | E:b | 4577 | AM1440-241 | 2 | 3528 |
| ARP97 | B:d | 6830 | ARP283 | E:c | 1695 | AM1546-284 | 16 | 4045 |
| ARP107 | C:a | 10167 | ARP293 | E:d | 5514 | AM1957-471 | 4 | 6359 |
| ARP126 | C:c | 5686 | ARP300 | E:f | 3736 | AM2031-440 | 15 | 8853 |
| ARP143 | C:e | 3955 | ARP301 | E:f | 5965 | AM2038-382 | 15,16 | 5996 |
| ARP144 | C:e | 5729 | ARP314 | F:a | 3704 | AM2048-571 | 8,14 | 3359 |
| ARP145 | C:e | 5326 | ARP321 | F:a | 6611 | AM2105-332 | 8,9 | 5305 |
| ARP158 | D:c | 4756 | ARP322 | F:b | 7985 | AM2105-632 | 14 | 3142 |
| ARP163 | D:d | 1072 | AM0018-485 | 4 | 3361 | AM2159-320 | 3,14 | 2541 |
| ARP195 | D:h | 16696 | AM0135-650 | 9 | 7927 | AM2222-275 | 1,2,9,24 | 14872 |
| ARP200 | D:h | 3656 | AM0223-403 | 2,15 | 6307 | AM2258-595 | 2 | 10108 |
| ARP204 | D:h | 4640 | AM0313-545 | 2,13 | 8201 | AM2303-305 | 8,15 | 8463 |
| ARP205 | D:h | 1378 | AM0052-321 | 2,14 | 9396 | AM2339-661 | 2 | 10013 |
| ARP216 | D:j | 5126 | AM0459-340 | 1,24 | 5269 | AM2350-302 | 8,10,24 | 14143 |
| ARP219 | D:j | 10436 | AM0558-335 | 2,15 | 2872 | AM2350-410 | 3 | 9162 |

Note—Category definitions are given in Table 2 and Table 3

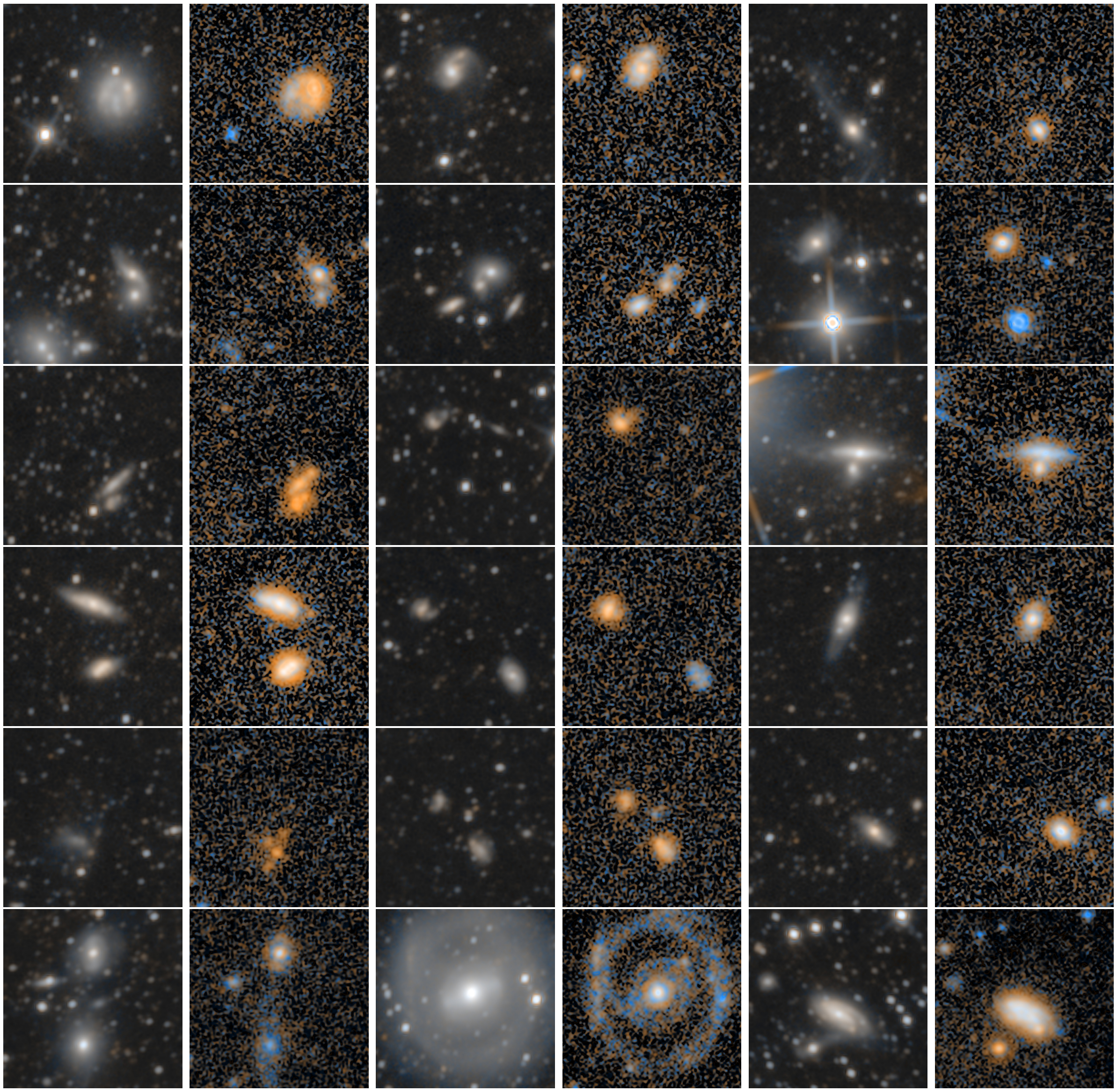

**Figure E5.** Pairs of WISE thumbnails (left:W1+W2; right:W3+W4) of systems containing at least one galaxy with moderate, red, central WISE W3+W4 emission, but weaker than Figure 23 and/or amiguous evidence for being a point source. The galaxies with WISE W3+W4 emission that were observed with VLASS (north of $\delta = -40°$) all have detectable VLASS emission aligned with the WISE source, with the exception of Arp 6, Arp 208, AM0942-313A, and AM1401-243A, which were observed but had no visible radio emission in VLASS. WISE images are aligned to the sky (north up), and should be compared to the full-page HST images in Appendix B, not the thumbnails in Figure 4 or Figure 24. From top left to bottom right, systems shown are: ARP6; ARP75; ARP101; ARP122; ARP150; ARP156; ARP202; ARP208; ARP282; ARP303; AM0001-505; AM0137-281; AM0144-585; AM0230-524; AM0346-222; AM0519-611; AM0619-271; AM0728-664; [Continued]

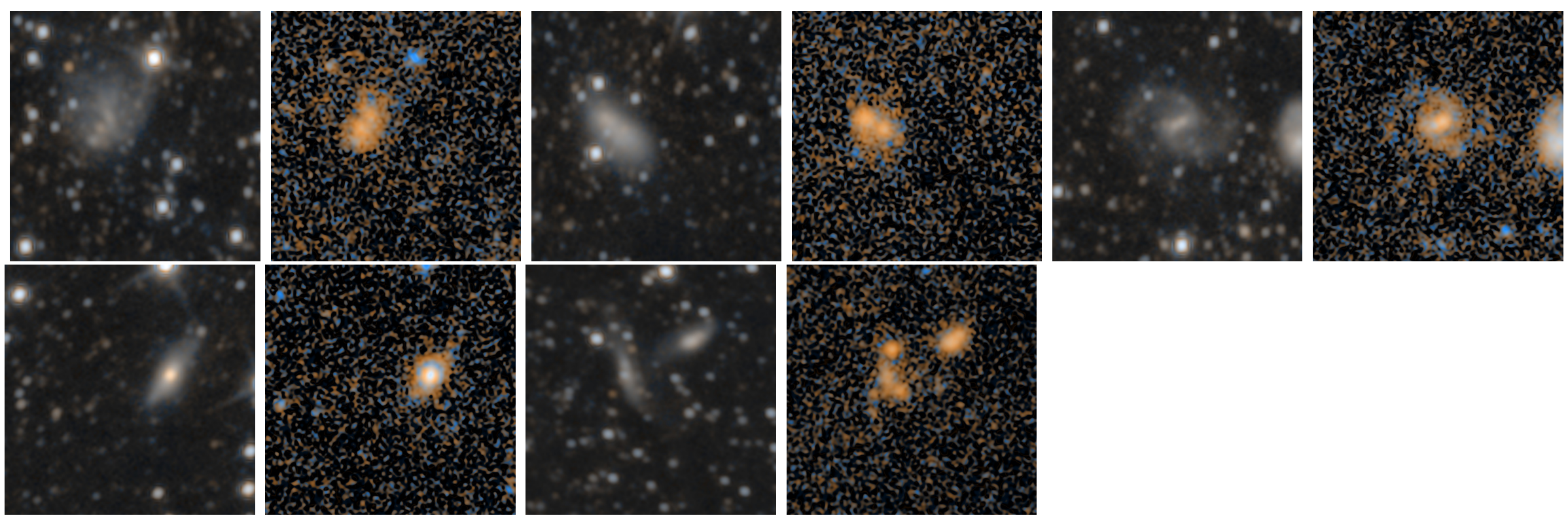

**Figure E5.** [Continued] Pairs of WISE thumbnails (left:W1+W2; right:W3+W4) of systems containing at least one galaxy with moderate, red, central WISE W3+W4 emission, but weaker than Figure 23 and/or amiguous evidence for being a point source. The galaxies with WISE W3+W4 emission that were observed with VLASS (north of $\delta = -40°$) all have detectable VLASS emission aligned with the WISE source, with the exception of Arp 6, Arp 208, AM0942-313A, and AM1401-243A, which were observed but had no visible radio emission in VLASS. WISE images are aligned to the sky (north up), and should be compared to the full-page HST images in Appendix B, not the thumbnails in Figure 4 or Figure 24. From top left to bottom right, systems shown are: AM0942-313A; AM0942-313B; AM1401-243A; AM2113-341; AM2240-892

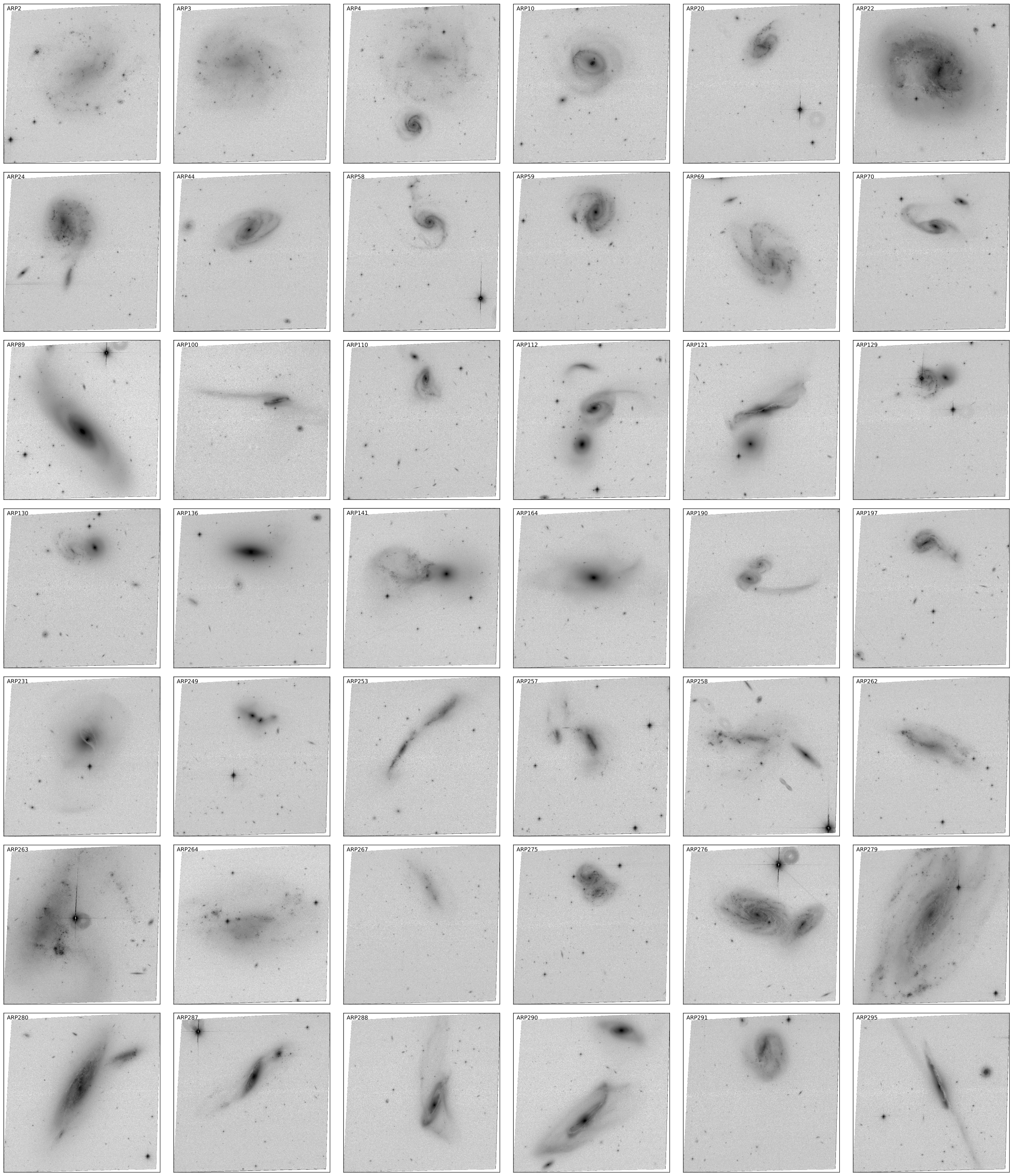

**Figure E6.** Thumbnails of systems containing no galaxies with significant compact, central ALLWISE or VLASS emission. All systems have VLASS radio imaging except for Arp-Madore systems further south than $-40^{\circ}$ (i.e., AMXXXX-YYY, where the first Y is 4 or greater). Systems shown are: ARP2; ARP3; ARP4; ARP10; ARP20; ARP22; ARP24; ARP44; ARP58; ARP59; ARP69; ARP70; ARP89; ARP100; ARP110; ARP112; ARP121; ARP129; ARP130; ARP136; ARP141; ARP164; ARP190; ARP197; ARP231; ARP249; ARP253; ARP257; ARP258; ARP262; ARP263; ARP264; ARP267; ARP275; ARP276; ARP279; ARP280; ARP287; ARP288; ARP290; ARP291; ARP295; [Continued]

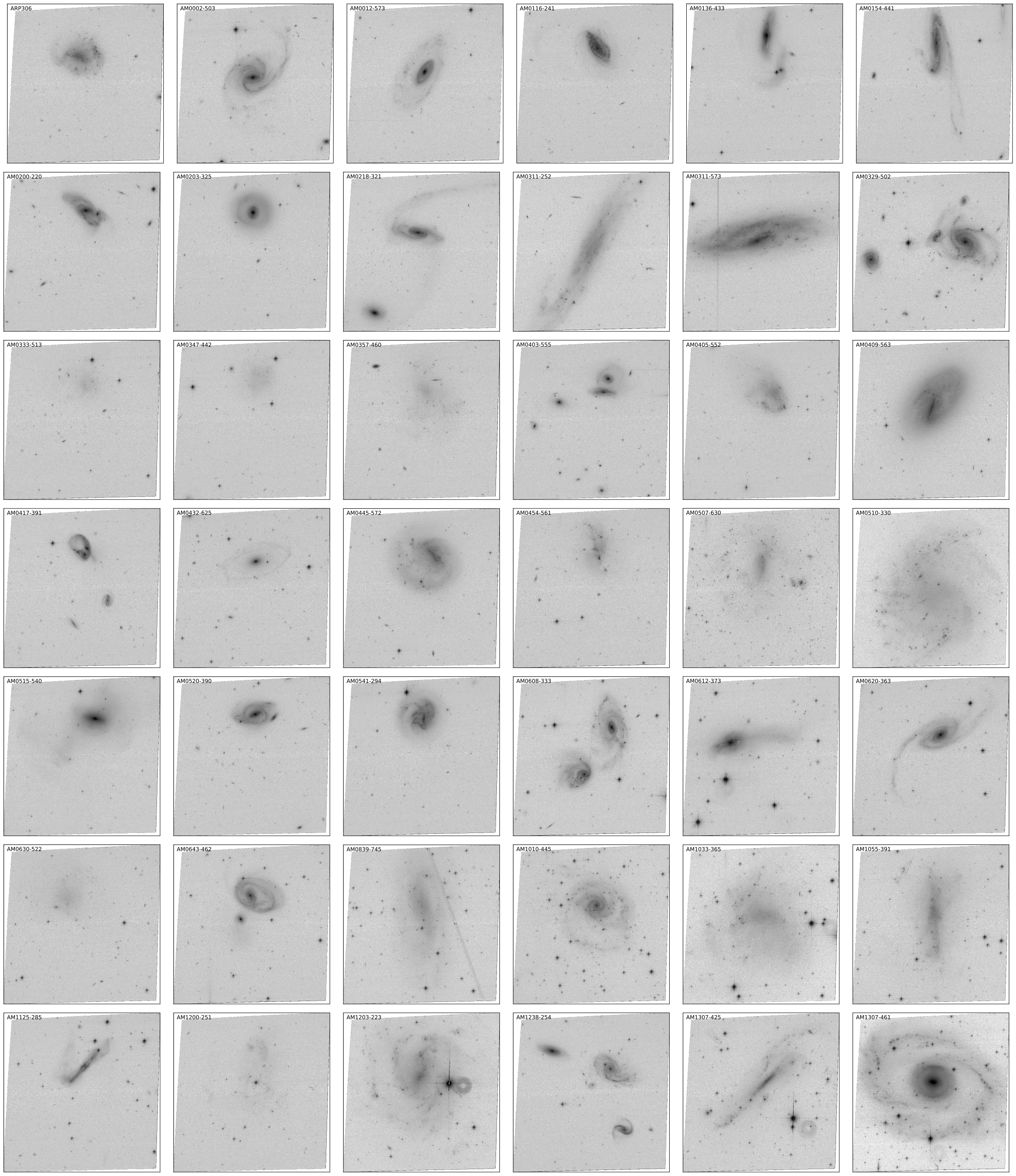

**Figure E6.** [Continued] Thumbnails of systems containing no galaxies with significant compact, central ALLWISE or VLASS emission. All systems have VLASS radio imaging except for Arp-Madore systems further south than $-40°$ (i.e., AMXXXX-YYY, where the first Y is 4 or greater). Systems shown are: ARP306; AM0002-503; AM0012-573; AM0116-241; AM0136-433; AM0154-441; AM0200-220; AM0203-325; AM0218-321; AM0311-252; AM0311-573; AM0329-502; AM0333-513; AM0347-442; AM0357-460; AM0403-555; AM0405-552; AM0409-563; AM0417-391; AM0432-625; AM0445-572; AM0454-561; AM0507-630; AM0510-330; AM0515-540; AM0520-390; AM0541-294; AM0608-333; AM0612-373; AM0620-363; AM0630-522; AM0643-462; AM0839-745; AM1010-445; AM1033-365; AM1055-391; AM1125-285; AM1200-251; AM1203-223; AM1238-254; AM1307-425; AM1307-461; [Continued]

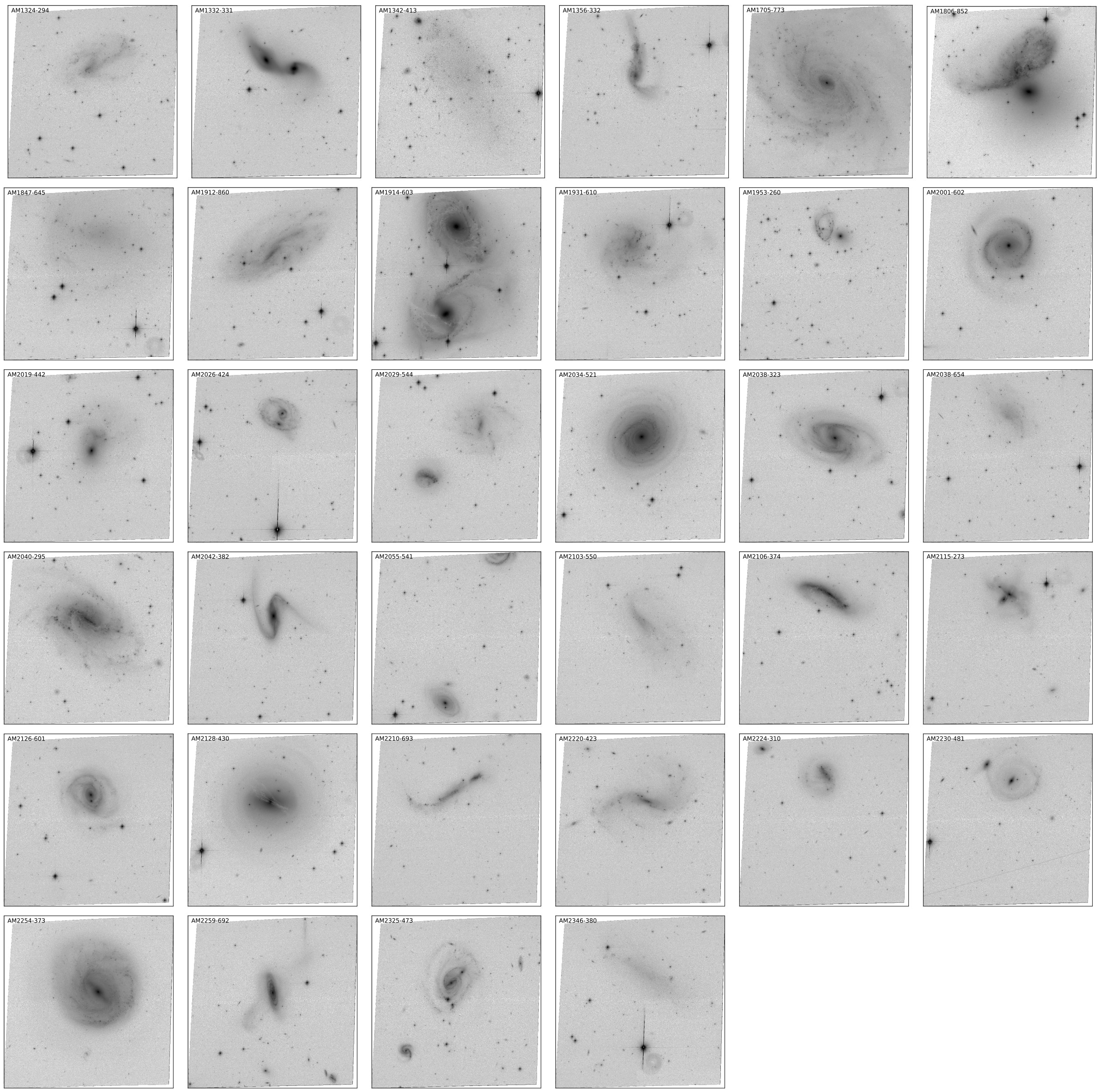

**Figure E6.** [Continued] Thumbnails of systems containing no galaxies with significant compact, central ALLWISE or VLASS emission. All systems have VLASS radio imaging except for Arp-Madore systems further south than −40° (i.e., AMXXXX-YYY, where the first Y is 4 or greater). Systems shown are: AM1324-294; AM1332-331; AM1342-413; AM1356-332; AM1705-773; AM1806-852; AM1847-645; AM1912-860; AM1914-603; AM1931-610; AM1953-260; AM2001-602; AM2019-442; AM2026-424; AM2029-544; AM2034-521; AM2038-323; AM2038-654; AM2040-295; AM2042-382; AM2055-541; AM2103-550; AM2106-374; AM2115-273; AM2126-601; AM2128-430; AM2210-693; AM2220-423; AM2224-310; AM2230-481; AM2254-373; AM2259-692; AM2325-473; AM2346-380

**Table E7**. Systems without Significant Central MIR or 1.4Gz Emission

| Target Name | Category | $V_r$ (km/s) | Target Name | Category | $V_r$ (km/s) | Target Name | Category | $V_r$ (km/s) |
|---|---|---|---|---|---|---|---|---|
| ARP2 | A:a | 712 | ARP291 | E:d | 1218 | AM1203-223 | 8 | 1722 |
| ARP3 | A:a | 1694 | ARP295 | E:e | 6846 | AM1238-254 | 8,12 | 16941 |
| ARP4 | A:a | 1614 | ARP306 | E:f | 1959 | AM1307-425 | 2,8,16 | 2135 |
| ARP10 | A:b | 9116 | AM0002-503 | 8 | 10195 | AM1307-461 | 6,10 | 3103 |
| ARP20 | A:d | 4209 | AM0012-573 | 6,10 | 15900 | AM1324-294 | 20 | 1902 |
| ARP22 | A:e | 1662 | AM0116-241 | 8,10 | 6963 | AM1332-331 | 2,17 | 3963 |
| ARP24 | A:e | 2053 | AM0136-433 | 2,15 | 6239 | AM1342-413 | 20 | 545 |
| ARP44 | B:a | 5505 | AM0154-441 | 8,10,13 | 5728 | AM1356-332 | 16 | 3706 |
| ARP58 | B:b | 10954 | AM0200-220 | 8,16 | 12872 | AM1705-773 | 8,20,23 | 2953 |
| ARP59 | B:b | 4549 | AM0203-325 | 6,8 | 5775 | AM1806-852 | 2 | 2541 |
| ARP69 | B:b | 3573 | AM0218-321 | 1,8,9,13,17 | 9322 | AM1847-645 | 20 | 1006 |
| ARP70 | B:b | 10494 | AM0311-252 | 8,12 | 1735 | AM1912-860 | 8 | 2444 |
| ARP89 | B:c | 2064 | AM0311-573 | 16 | 1140 | AM1914-603 | 3,8 | 3817 |
| ARP100 | B:d | 6071 | AM0329-502 | 1,2 | 11565 | AM1931-610 | 8,20 | 1796 |
| ARP110 | C:b | 9384 | AM0333-513 | 20 | 1030 | AM1953-260 | 6 | 14619 |
| ARP112 | C:b | 4789 | AM0347-442 | 20 | 1248 | AM2001-602 | 6,12 | 3587 |
| ARP121 | C:c | 5715 | AM0357-460 | 8,20 | 900 | AM2019-442 | 13,15 | 2963 |
| ARP129 | C:c | 6511 | AM0403-555 | 2,6 | 17093 | AM2026-424 | 1,6 | 15124 |
| ARP130 | C:c | 5832 | AM0405-552 | 16,20 | 1066 | AM2029-544 | 23 | 3403 |
| ARP136 | C:d | 3251 | AM0409-563 | 14,16 | 1310 | AM2034-521 | 23 | 4581 |
| ARP141 | C:e | 2684 | AM0417-391 | 6,8 | 15255 | AM2038-323 | 1,6,8 | 5551 |
| ARP164 | D:d | 5231 | AM0432-625 | 6 | 16063 | AM2038-654 | 20 | 1626 |
| ARP190 | D:g | 10216 | AM0445-572 | 10,13,16 | 1205 | AM2040-295 | 8,23 | 2709 |
| ARP197 | D:h | 6077 | AM0454-561 | 16 | 1779 | AM2042-382 | 1,8 | 6932 |
| ARP231 | D:l | 5646 | AM0507-630 | 20 | 1464 | AM2055-541 | 2,10 | 12864 |
| ARP249 | D:m | 11472 | AM0510-330 | 20 | 927 | AM2103-550 | 20 | 1402 |
| ARP253 | D:m | 1867 | AM0515-540 | 1,2,15,23 | 3877 | AM2106-374 | 12,17 | 2632 |
| ARP257 | D:n | 3323 | AM0520-390 | 1,6,23 | 14734 | AM2115-273 | 2,15 | 6458 |
| ARP258 | D:n | 4093 | AM0541-294 | 15,16,24 | 3818 | AM2126-601 | 10 | 8660 |
| ARP262 | D:n | 1772 | AM0608-333 | 2 | 8652 | AM2128-430 | 10,14,22 | 2362 |
| ARP263 | D:n | 755 | AM0612-373 | 2,14,15 | 9734 | AM2210-693 | 2 | 3155 |
| ARP264 | D:n | 602 | AM0620-363 | 8,10 | 9327 | AM2220-423 | 20 | 2420 |
| ARP267 | D:n | 582 | AM0630-522 | 20 | 1193 | AM2224-310 | 1 | 3974 |
| ARP275 | E:b | 5057 | AM0643-462 | 1,6 | 11774 | AM2230-481 | 1,6 | 8185 |
| ARP276 | E:b | 4148 | AM0839-745 | 8 | 1142 | AM2254-373 | 12,16 | 1801 |
| ARP279 | E:b | 1711 | AM1010-445 | 23 | 4113 | AM2259-692 | 15 | 3901 |
| ARP280 | E:b | 727 | AM1033-365 | 20 | 955 | AM2325-473 | 1,10 | 15238 |
| ARP287 | E:d | 2450 | AM1055-391 | 16,20 | 1002 | AM2346-380 | 20 | 647 |
| ARP288 | E:d | 7049 | AM1125-285 | 2,6,12 | 7127 | | | |
| ARP290 | E:d | 3614 | AM1200-251 | 20,22 | 1790 | | | |

NOTE—Category definitions are given in Table 2 and Table 3

**Table E5.** Systems with Moderate Central MIR Emission

| Target Name | Category | $V_r$ (km/s) | Target Name | Category | $V_r$ (km/s) |
|---|---|---|---|---|---|
| ARP6 | A:a | 447 | AM0144-585 | 20 | 2210 |
| ARP75 | B:b | 10668 | AM0230-524 | 2,15 | 6450 |
| ARP101 | B:d | 4599 | AM0346-222 | 7,16 | 12141 |
| ARP122 | C:c | 12082 | AM0519-611 | 2,8 | 4871 |
| ARP150 | D:b | 11888 | AM0619-271 | 8,10,13,14 | 1622 |
| ARP156 | D:c | 1875 | AM0728-664 | 1,8 | 5155 |
| ARP202 | D:h | 3144 | AM0942-313A | 16,23 | 964 |
| ARP208 | D:h | 9025 | AM0942-313B | 16,23 | 1253 |
| ARP282 | E:c | 4569 | AM1401-243A | 12,23 | 2332 |
| ARP303 | E:f | 5964 | AM2113-341 | 14,18 | 8798 |
| AM0001-505 | 23 | 11649 | AM2240-892 | 2,10 | 2525 |
| AM0137-281 | 23 | 5815 | | | |

Note—Category definitions are given in Table 2 and Table 3

**Table E6.** Systems with Significant Central 1.4Gz Emission, but no MIR emission

| Target Name | Category | $V_r$ (km/s) |
|---|---|---|
| ARP113 | C:c | 7201 |
| ARP123 | C:c | 2300 |
| ARP165 | D:d | 5043 |
| ARP172 | D:e | 10043 |
| ARP176 | D:f | 3107 |
| ARP180 | D:g | 4289 |
| ARP184 | D:g | 3888 |
| ARP309 | E:f | 4658 |
| ARP335 | F:c | 7574 |
| AM1303-371 | 8,16,17 | 4897 |
| AM2056-392 | 6,13,24 | 13500 |
| AM2353-291 | 2,6 | 8828 |

Note—Category definitions are given in Table 2 and Table 3